\documentclass[11pt]{article}
\usepackage{makeidx}
\usepackage{amsfonts}
\usepackage{amssymb}
\usepackage{graphicx}
\usepackage{amsmath}
\usepackage{geometry}
\usepackage{appendix}

\input{tcilatex}
\DeclareMathSizes{10.95}{10}{7}{6}

\begin{document}

\title{Statistical Field Theory and Neural Structures Dynamics III:
Effective Action for Connectivities, Interactions and Emerging Collective
States}
\author{Pierre Gosselin\thanks{%
Pierre Gosselin: Institut Fourier, UMR 5582 CNRS-UGA, Universit\'{e}
Grenoble Alpes, BP 74, 38402 Saint Martin d'H\`{e}res, France\ E-Mail:
Pierre.Gosselin@univ-grenoble-alpes.fr} \and A\"{\i}leen Lotz\thanks{%
A\"{\i}leen Lotz: Cerca Trova, BP 114, 38001 Grenoble Cedex 1, France.\
E-mail: a.lotz@cercatrova.eu}}
\maketitle

\begin{abstract}
This paper elaborates on the effective field theory for the connectivity
field previously introduced in (\cite{GLs}). We demonstrate that dynamic
interactions among connectivities induce modifications in the background
state. These modifications can be understood as the emergence of interacting
collective states above the background state. The emergence of such states
is contingent on both interactions and the shape of the static or
quasi-static background, which acts as a conditioning factor for potential
emerging states.
\end{abstract}

\section{Introduction}

In this series of papers, we present a field-theoretic approach to model the
dynamics of connectivity within a system of interacting spiking neurons. To
accomplish this, we have developed a two-field model in Part 1 that
describes the dynamics of both neural activity and the connectivities
between points in the thread. The first field, akin to the one introduced in
(\cite{GL}), characterizes the neural assembly, while the second field
delineates the dynamics of connectivity between cells. Both fields interact
with themselves, describing interactions within the network, and with each
other, encapsulating the intricate interplay between neural activities and
network connectivities. This field-based description encompasses the
collective and individual aspects of the system. The two-field system is
delineated by a field action functional that portrays the system's
interactions at the microscopic level. This action functional
comprehensively accounts for the system's dynamics as a whole.

This field-theoretic framework has enabled us to determine the effective
action of the system, as well as the associated background field, which
represents the minimum of the effective action. This background field
characterizes the collective state of the system. The field framework allows
for the computation of shifting rates, i.e., neural activity, at each point
within the system for a given background state. It also serves as a suitable
framework for deriving the propagation of perturbations in neural activity
from one point to another. As previously demonstrated in (\cite{GL}), we
established the existence of persistent nonlinear traveling waves along the
thread by considering the field action for neurons alone.

In (\cite{GLr}), when considering the field for connectivity functions, our
framework enabled us to derive both the background fields for neural
interactions and connectivities that minimize the action functional. These
background fields represent the collective configurations of the system and
determine the potential static equilibria for neural activities and
connectivities. These equilibria serve as the structural foundation of the
system, governing fluctuations and the propagation of signals within it.
They are contingent upon internal parameters of the system and external
stimuli. We demonstrated that several background states and their associated
connectivities are feasible, with the thread primarily organized into groups
of interconnected points. Based on these findings, we qualitatively explored
the mechanisms of emergence of collective states resulting from
perturbations or signals along the thread, as well as transitions between
such states.

However, this study was limited to an examination at a static level. In (%
\cite{GLs}), we provide dynamic foundations for these transitions by
considering effective actions within a specified background field.
Initially, recognizing that the timescale of connectivities is slower than
that of individual cells, we focused on the action related to the
connectivity field. We elucidated how recurrent activations at specific
points can propagate across the thread, gradually altering the connectivity
functions. For oscillatory perturbations, oscillatory responses may exhibit
interference phenomena. At points of constructive interference, both the
background state for connectivities and average connectivities undergo
modifications. These long-term alterations manifest as emerging states with
enhanced connectivities between certain points. These states are reflective
of external activations and can be regarded as records of such activations.
They persist over time and can be reactivated by external perturbations.
Furthermore, the association of these emerging states is possible when their
activation occurs at proximate times. The resultant state is thus a
composite of two states, describable as a modification of an initial
background state at several points. Activating one of the two states
reactivates their combination. Consequently, regardless of the cause of
their activation, these states with enhanced connectivity exhibit the
characteristics of interacting partial neuronal assemblies.

This effective formalism allows us to comprehend the system's connectivity
dynamics as modifications of the connectivity field induced by external
perturbations. We interpret the system's fluctuations due to specific
external perturbations as transitions between initial and final states of
this field. The results from (\cite{GLr}), such as the emergence of combined
structures and the reactivation of one structure by another, occur within a
coherent field description of the system of connectivities. In this context,
the system's dynamics arise as a consequence of a fluctuating background
state influenced by external perturbations.

However, these results were obtained by exclusively working with the
connectivities field and by considering the dynamics of perturbations
resulting from exogenous signals. We did not base our findings on the
interactions between the neuronal field and the connectivity field, thereby
excluding the internal dynamics of the system.

The present paper extends this approach by examining the system of
connectivities for itself. By replacing the individual cell field as an
effective quantity dependent on connectivities, we described the effective
dynamics for the connectivity fields, with the integration of the cell field
giving rise to self-interactions for the connectivity field.

As a result, certain internal dynamics come into play, potentially altering
the static background state at specific points. These self-interactions,
induced by perturbations, may initiate internal patterns of connections
between certain cells. Depending on internal parameters, we observe that
permanent shifts in connectivity background states may occur in certain
regions of the thread while leaving others unaffected. This effective theory
can also be applied to describe the mechanisms of connectivity reinforcement
between several cells and the emergence of groups with altered
connectivities. These collective shifts can be understood as the emergence
of additional structures whose formation is contingent upon the background
state.

This paper is organized as follows: In Section 2, we provide a
field-theoretic description of the spiking neuron system. Section 3 is
dedicated to deriving the effective action for the system within the
background state outlined in (\cite{GLr}). In Section 4, we calculate the
modification of this background field resulting from connectivity
interactions. We establish the conditions for shifted states based on the
characteristics of the background. An alternative approximated approach is
introduced in Section 5, which will be well-suited for future developments
involving large sets of interacting collective states. In Section 6, we
extend the approach to $n$ types of different interacting fields. Finally,
in Section 7, we use the previous formalism to compute, as an example, the
interactions and dynamics between two connectivity states.

\section{Field theoretic description of the system}

Based on \cite{GL1}\cite{GL2}\cite{GL3}\cite{GL4}, we gave in (\cite{GLr},
resume in \cite{GLs}) a statistical field formalism to describe both cells
and connectivities dynamics. This decription relies on two fields, $\Psi $
for cells, and $\Gamma $ for connectivities. The field action for the system
is:

\begin{eqnarray}
S_{full} &=&-\frac{1}{2}\Psi ^{\dagger }\left( \theta ,Z,\omega \right)
\nabla \left( \frac{\sigma _{\theta }^{2}}{2}\nabla -\omega ^{-1}\left(
J,\theta ,Z,\left\vert \Psi \right\vert ^{2}\right) \right) \Psi \left(
\theta ,Z\right) +V\left( \Psi \right)  \label{flt} \\
&&+\frac{1}{2\eta ^{2}}\left( S_{\Gamma }^{\left( 0\right) }+S_{\Gamma
}^{\left( 1\right) }+S_{\Gamma }^{\left( 2\right) }+S_{\Gamma }^{\left(
3\right) }+S_{\Gamma }^{\left( 4\right) }\right) +U\left( \left\{ \left\vert
\Gamma \left( \theta ,Z,Z^{\prime },C,D\right) \right\vert ^{2}\right\}
\right)  \notag
\end{eqnarray}%
where activity satisfies:%
\begin{equation}
\omega ^{-1}\left( J,\theta ,Z,\left\vert \Psi \right\vert ^{2}\right)
=G\left( J\left( \theta \right) +\frac{\kappa }{N}\int T\left(
Z,Z_{1},\theta \right) \frac{\omega \left( \theta -\frac{\left\vert
Z-Z_{1}\right\vert }{c},Z_{1}\right) }{\omega \left( \theta ,Z\right) }%
\left\vert \Psi \left( \theta -\frac{\left\vert Z-Z_{1}\right\vert }{c}%
,Z_{1}\right) \right\vert ^{2}dZ_{1}\right)  \label{dng}
\end{equation}%
with $S_{\Gamma }^{\left( 1\right) }$, $S_{\Gamma }^{\left( 2\right) }$, $%
S_{\Gamma }^{\left( 3\right) }$, $S_{\Gamma }^{\left( 4\right) }$ now given
by: 
\begin{equation}
S_{\Gamma }^{\left( 1\right) }=\int \Gamma ^{\dag }\left( T,\hat{T},\theta
,Z,Z^{\prime },C,D\right) \nabla _{T}\left( \frac{\sigma _{T}^{2}}{2}\nabla
_{T}-\left( -\frac{1}{\tau \omega }T+\frac{\lambda }{\omega }\hat{T}\right)
\right) \Gamma \left( T,\hat{T},\theta ,Z,Z^{\prime },C,D\right)  \label{wGD}
\end{equation}%
\begin{eqnarray}
S_{\Gamma }^{\left( 2\right) } &=&\int \Gamma ^{\dag }\left( T,\hat{T}%
,\theta ,Z,Z^{\prime },C,D\right)  \label{wGT} \\
&&\times \nabla _{\hat{T}}\left( \frac{\sigma _{\hat{T}}^{2}}{2}\nabla _{%
\hat{T}}-\frac{\rho }{\omega \left( J,\theta ,Z,\left\vert \Psi \right\vert
^{2}\right) }\left( \left( h\left( Z,Z^{\prime }\right) -\hat{T}\right)
C\left\vert \Psi \left( \theta ,Z\right) \right\vert ^{2}h_{C}\left( \omega
\left( J,\theta ,Z,\left\vert \Psi \right\vert ^{2}\right) \right) \right.
\right.  \notag \\
&&\left. \left. -D\hat{T}\left\vert \Psi \left( \theta -\frac{\left\vert
Z-Z^{\prime }\right\vert }{c},Z^{\prime }\right) \right\vert ^{2}h_{D}\left(
\omega \left( J,\theta -\frac{\left\vert Z-Z^{\prime }\right\vert }{c}%
,Z^{\prime },\left\vert \Psi \right\vert ^{2}\right) \right) \right) \right)
\Gamma \left( T,\hat{T},\theta ,Z,Z^{\prime },C,D\right)  \notag
\end{eqnarray}%
\begin{eqnarray}
S_{\Gamma }^{\left( 3\right) } &=&\Gamma ^{\dag }\left( T,\hat{T},\theta
,Z,Z^{\prime },C,D\right) \nabla _{C}\left( \frac{\sigma _{C}^{2}}{2}\nabla
_{C}+\left( \frac{C}{\tau _{C}\omega \left( J,\theta ,Z,\left\vert \Psi
\right\vert ^{2}\right) }\right. \right.  \label{wGQ} \\
&&\left. \left. -\alpha _{C}\left( 1-C\right) \frac{\omega \left( J,\theta -%
\frac{\left\vert Z-Z^{\prime }\right\vert }{c},Z^{\prime },\left\vert \Psi
\right\vert ^{2}\right) \left\vert \Psi \left( \theta -\frac{\left\vert
Z-Z^{\prime }\right\vert }{c},Z^{\prime }\right) \right\vert ^{2}}{\omega
\left( J,\theta ,Z,\left\vert \Psi \right\vert ^{2}\right) }\right) \right)
\Gamma \left( T,\hat{T},\theta ,Z,Z^{\prime },C,D\right)  \notag
\end{eqnarray}%
\begin{eqnarray}
S_{\Gamma }^{\left( 4\right) } &=&\Gamma ^{\dag }\left( T,\hat{T},\theta
,Z,Z^{\prime },C,D\right) \nabla _{D}\left( \frac{\sigma _{D}^{2}}{2}\nabla
_{D}+\left( \frac{D}{\tau _{D}\omega \left( J,\theta ,Z,\left\vert \Psi
\right\vert ^{2}\right) }-\alpha _{D}\left( 1-D\right) \left\vert \Psi
\left( \theta ,Z\right) \right\vert ^{2}\right) \right)  \label{wGC} \\
&&\times \Gamma \left( T,\hat{T},\theta ,Z,Z^{\prime },C,D\right)  \notag
\end{eqnarray}%
In (\ref{flt}), we added a potential:%
\begin{equation*}
U\left( \left\{ \left\vert \Gamma \left( \theta ,Z,Z^{\prime },C,D\right)
\right\vert ^{2}\right\} \right) =U\left( \int T\left\vert \Gamma \left( T,%
\hat{T},\theta ,Z,Z^{\prime },C,D\right) \right\vert ^{2}dTd\hat{T}\right)
\end{equation*}%
that models the constraint about the number of active connections in the
system.

\section{Effective action for $\Gamma $ above the background and
interactions between connectivities}

In this section, we leverage the findings from (\cite{GLr}) and (\cite{GLs})
to construct an effective action for the connectivity field, denoted as $%
\Gamma $ above its background. This is accomplished by expressing the
activity $\omega \left( J,\theta ,Z,\left\vert \Psi \right\vert ^{2}\right) $
of the background state in (\ref{flt}) as a function of the connectivity
field. The computation of $\omega \left( J,\theta ,Z,\left\vert \Psi
\right\vert ^{2}\right) $ was detailed in (\cite{GLs}) and leads to the
development of an interacting effective action for the connectivity field $%
\Gamma $,.which includes self-interacting dynamics among the connectivities.
The resulting effective action will present some minima that modifies the
connectivities relative to the initial static background state derived in (%
\cite{GLr}). These states represent emerging states above the background. In
this section, we examine these modifications in a quasi-static context.
Subsequently, a dynamic version of these modifications will be developed to
establish an effective description of assemblies interactions.

\subsection{Replacing auxiliary variables by averages}

To write the effective action for $\Gamma \left( T,\hat{T},C,D\right) $ we
start with a simplification by replacing, as in (\cite{GLs}) $C$ and $D$ by
averages: 
\begin{eqnarray}
C &\rightarrow &\left\langle C\left( \theta \right) \right\rangle =\frac{%
\alpha _{C}\omega ^{\prime }\left\langle \left\vert \Psi \left( \theta -%
\frac{\left\vert Z-Z^{\prime }\right\vert }{c},Z^{\prime }\right)
\right\vert ^{2}\right\rangle }{\frac{1}{\tau _{C}}+\alpha _{C}\omega
^{\prime }\left\langle \left\vert \Psi \left( \theta -\frac{\left\vert
Z-Z^{\prime }\right\vert }{c},Z^{\prime }\right) \right\vert
^{2}\right\rangle }\equiv C\left( \theta \right) \\
D &\rightarrow &\left\langle D\left( \theta \right) \right\rangle =\frac{%
\alpha _{D}\omega \left\langle \left\vert \Psi \left( \theta ,Z\right)
\right\vert ^{2}\right\rangle }{\frac{1}{\tau _{D}}+\alpha _{D}\omega
\left\langle \left\vert \Psi \left( \theta ,Z\right) \right\vert
^{2}\right\rangle }\equiv D\left( \theta \right)
\end{eqnarray}

We willalso disgard the threshold term $\eta H\left( \delta -T\right) $ in
the sequel. We also choose:%
\begin{eqnarray*}
h_{C}\left( \omega \left( \theta ,Z,\left\vert \Psi \right\vert ^{2}\right)
\right) &=&\omega \left( \theta ,Z,\left\vert \Psi \right\vert ^{2}\right) \\
h_{D}\left( \omega \left( \theta -\frac{\left\vert Z-Z^{\prime }\right\vert 
}{c},Z^{\prime },\left\vert \Psi \right\vert ^{2}\right) \right) &=&\omega
\left( \theta -\frac{\left\vert Z-Z^{\prime }\right\vert }{c},Z^{\prime
},\left\vert \Psi \right\vert ^{2}\right)
\end{eqnarray*}%
and the action considered is thus:%
\begin{eqnarray}
&&\Gamma ^{\dag }\left( T,\hat{T},\theta ,Z,Z^{\prime }\right) \left( \nabla
_{T}\left( \nabla _{T}-\left( -\frac{1}{\tau \omega }T+\frac{\lambda }{%
\omega }\hat{T}\right) \left\vert \Psi \left( \theta ,Z\right) \right\vert
^{2}\right) \right) \Gamma \left( T,\hat{T},\theta ,Z,Z^{\prime }\right)
\label{tCD} \\
&&+\Gamma ^{\dag }\left( T,\hat{T},\theta ,Z,Z^{\prime }\right) \left(
\nabla _{\hat{T}}\left( \nabla _{\hat{T}}-\frac{\rho }{\omega \left(
J,\theta ,Z,\left\vert \Psi \right\vert ^{2}\right) }\left( \left( h\left(
Z,Z^{\prime }\right) -\hat{T}\right) C\left( \theta \right) \left\vert \Psi
\left( \theta ,Z\right) \right\vert ^{2}h_{C}\left( \omega \left( \theta
,Z,\left\vert \Psi \right\vert ^{2}\right) \right) \right. \right. \right. 
\notag \\
&&\times \left. \left. \left. -D\left( \theta \right) \hat{T}\left\vert \Psi
\left( \theta -\frac{\left\vert Z-Z^{\prime }\right\vert }{c},Z^{\prime
}\right) \right\vert ^{2}h_{D}\left( \omega \left( \theta -\frac{\left\vert
Z-Z^{\prime }\right\vert }{c},Z^{\prime },\left\vert \Psi \right\vert
^{2}\right) \right) \right) \right) \right) \Gamma \left( T,\hat{T},\theta
,Z,Z^{\prime }\right)  \notag
\end{eqnarray}

\subsection{Activities as functions of connectivities}

At this stage, we take into account the dependency of $\omega \left( \theta
,Z,\left\vert \Psi \right\vert ^{2}\right) $ in the connectivity field.
Actually, in (\ref{tCD}) activity $\omega \left( J,\theta ,Z,\left\vert \Psi
\right\vert ^{2}\right) $ is itself a functional of $\Gamma \left( T,\hat{T}%
,\theta ,Z,Z^{\prime }\right) $ whose derivation was given in (\cite{GLs}).
We decompose $\omega \left( J,\theta ,Z,\left\vert \Psi \right\vert
^{2}\right) $ as:%
\begin{equation}
\omega \left( J,\theta ,Z,\left\vert \Psi \right\vert ^{2}\right) =\omega
_{0}\left( Z\right) +\delta \omega \left( \theta ,Z,\left\vert \Psi
\right\vert ^{2}\right)  \label{sdn}
\end{equation}%
where $\omega _{0}\left( Z\right) $ is the static background activity. The
source-dependent part of activity $\delta \omega \left( \theta ,Z,\left\vert
\Psi \right\vert ^{2}\right) $ is given by: 
\begin{equation}
\delta \omega \left( \theta ,Z,\left\vert \Psi \right\vert ^{2}\right)
\equiv \sum_{i}\int K\left( Z,\theta ,Z_{i},\theta _{i}\right) \left\{
\sum_{i}a\left( Z_{i},\theta \right) \frac{\omega _{0}\left( J,\theta
,Z_{i}\right) }{\Lambda ^{2}}\right\} d\theta _{i}  \label{MC}
\end{equation}%
To consider an internal dynamics of the system rather than an
external-source driven perturbation we replace the particular sources in (%
\ref{MC}):%
\begin{equation*}
\sum_{i}a\left( Z_{i},\theta \right) \frac{\omega _{0}\left( J,\theta
,Z_{i}\right) }{\Lambda ^{2}}
\end{equation*}%
by the non contracted field:%
\begin{equation}
\left\vert \Psi \left( Z,\theta _{i}\right) \right\vert ^{2}\frac{\omega
_{0}\left( \theta _{i},Z\right) }{\Lambda ^{2}}  \label{RM}
\end{equation}%
and $\frac{1}{\Lambda }$ by $\left\vert \Psi \right\vert ^{2}$\ to remain in
a general case.

In (\cite{GLs}), we showed that in first approximation:%
\begin{equation}
K\left( Z,\theta ,Z_{i},\theta _{i}\right) =\int^{\theta _{i}}\check{T}%
\left( 1-\left( 1+\left\vert \Psi \left( Z,\theta \right) \right\vert ^{2}-%
\frac{\frac{\check{T}}{\left( 1-\left( 1+\left\vert \Psi \right\vert
^{2}\right) \check{T}\right) }\left[ \left\vert \Psi \left( Z,\theta \right)
\right\vert ^{2}\frac{\omega _{0}\left( \theta ,Z\right) }{\Lambda ^{2}}%
\right] }{\omega _{0}\left( Z\right) +\frac{\check{T}}{\left( 1-\left(
1+\left\vert \Psi \right\vert ^{2}\right) \check{T}\right) }\left[
\left\vert \Psi \left( Z,\theta \right) \right\vert ^{2}\frac{\omega
_{0}\left( \theta ,Z\right) }{\Lambda ^{2}}\right] }\right) \check{T}\right)
^{-1}\left( Z,\theta ,Z_{i},\theta _{i}\right)  \label{rnll}
\end{equation}

\subsection{Decomposition of $\Gamma \left( T,\hat{T},\protect\theta %
,Z,Z^{\prime }\right) $ as background state plus fluctuation}

Starting with (\ref{tCD}), our aim is to obtain an effective action for
fluctuations of $\Gamma $ around some quasi-static background fld. For this
purpose, we decompose the field into the background field and the
fluctuations:%
\begin{eqnarray}
\Gamma \left( T,\hat{T},\theta ,Z,Z^{\prime }\right) &=&\Gamma _{0}\left( T,%
\hat{T},\theta ,Z,Z^{\prime }\right) +\Delta \Gamma \left( T,\hat{T},\theta
,Z,Z^{\prime }\right)  \label{dcp} \\
\Gamma ^{\dag }\left( T,\hat{T},\theta ,Z,Z^{\prime }\right) &=&\Gamma
_{0}^{\dag }\left( T,\hat{T},\theta ,Z,Z^{\prime }\right) +\Delta \Gamma
^{\dag }\left( T,\hat{T},\theta ,Z,Z^{\prime }\right)  \notag
\end{eqnarray}

\subsection{Activity induced by fluctuations $\Delta \Gamma \left( T,\hat{T},%
\protect\theta ,Z,Z^{\prime }\right) $}

Since our focus in this section is on the connectivity-induced change of
activities, we introduce the expression $\delta \omega \left( \theta
,Z,\left\vert \Psi \right\vert ^{2}\right) $ for the part of $\delta \omega
\left( \theta ,Z,\left\vert \Psi \right\vert ^{2}\right) $ that depends on $%
\Delta \Gamma \left( T,\hat{T},\theta ,Z,Z^{\prime }\right) $ and $\Delta
\Gamma ^{\dag }\left( T,\hat{T},\theta ,Z,Z^{\prime }\right) $. In other
words:%
\begin{equation*}
\delta \omega \left( \theta ,Z,\left\vert \Psi \right\vert ^{2}\right)
-\delta \omega \left( \theta ,Z,\left\vert \Psi \right\vert ^{2},\Delta
\Gamma =\Delta \Gamma ^{\dag }=0\right) \rightarrow \delta \omega \left(
\theta ,Z,\left\vert \Psi \right\vert ^{2}\right)
\end{equation*}%
The connectivity independent part of $\delta \omega \left( \theta
,Z,\left\vert \Psi \right\vert ^{2}\right) $ is written: 
\begin{equation*}
\delta \omega _{0}=\delta \omega \left( \theta ,Z,\left\vert \Psi
\right\vert ^{2},\Delta \Gamma =\Delta \Gamma ^{\dag }=0\right)
\end{equation*}%
We studied in the previous section the impact of this term $\delta \omega
_{0}$. Consequently, we incorporate this contribution into the definition of 
$\omega _{0}$ and redefine: 
\begin{equation*}
\omega _{0}+\delta \omega _{0}\rightarrow \omega _{0}
\end{equation*}%
This is consistent with our objective in this section: we aim to isolate the
self-interaction of the connectivity functions so that we can incorporate
external fluctuations $\delta \omega _{0}$ in the definition of $\omega _{0}$%
.

In conclusion, it is noteworthy that the variation $\delta \omega \left(
\theta ,Z,\left\vert \Psi \right\vert ^{2}\right) $ can be viewed as a
series expansion:%
\begin{eqnarray*}
\delta \omega \left( \theta ,Z,\left\vert \Psi \right\vert ^{2}\right)
&=&\int \left( \frac{\delta \left( \delta \omega \left( \theta ,Z,\left\vert
\Psi \right\vert ^{2}\right) \right) }{\delta \left\vert \Delta \Gamma
\left( T,\hat{T},\theta ,Z_{1},Z_{1}^{\prime }\right) \right\vert ^{2}}%
\right) _{\Delta \Gamma \left( T,\hat{T},\theta ,Z,Z^{\prime }\right)
=0}\left\vert \Delta \Gamma \left( T,\hat{T},\theta ,Z_{1},Z_{1}^{\prime
}\right) \right\vert ^{2} \\
&&+\int \left( \frac{\delta \left( \delta \omega \left( \theta ,Z,\left\vert
\Psi \right\vert ^{2}\right) \right) }{\delta \left\vert \Delta \Gamma
\left( T,\hat{T},\theta ,Z_{1},Z_{1}^{\prime }\right) \right\vert ^{2}\delta
\left\vert \Delta \Gamma \left( T,\hat{T},\theta ,Z_{1},Z_{1}^{\prime
}\right) \right\vert ^{2}}\right) _{\Delta \Gamma =0} \\
&&\times \left\vert \Delta \Gamma \left( T,\hat{T},\theta
,Z_{1},Z_{1}^{\prime }\right) \right\vert ^{2}\left\vert \Delta \Gamma
\left( T,\hat{T},\theta ,Z_{2},Z_{2}^{\prime }\right) \right\vert ^{2} \\
&&+...
\end{eqnarray*}%
as in the second part of this work (\cite{GLs}), but now, the series depend
explicitely on the perturbations in the connectivities.

\subsection{Effective action for $\Delta \Gamma \left( T,\hat{T},\protect%
\theta ,Z,Z^{\prime }\right) $}

In this paragraph, we expand the action (\ref{tCD}) using the averages
values of $T$ and $\hat{T}$ in the background state, we show in appendix 1
that the fluctuation part of action (\ref{tCD}) around the background state:%
\begin{equation*}
\Gamma _{0}\left( T,\hat{T},\theta ,Z,Z^{\prime }\right) ,\Gamma _{0}^{\dag
}\left( T,\hat{T},\theta ,Z,Z^{\prime }\right)
\end{equation*}%
is then decomposed in three contributions:%
\begin{equation*}
\Gamma _{0}^{\dag }\left( T,\hat{T},\theta ,Z,Z^{\prime }\right) \Xi \Gamma
_{0}\left( T,\hat{T},\theta ,Z,Z^{\prime }\right) +S_{f}+V\left( \Delta
\Gamma ,\Delta \Gamma ^{\dag }\right)
\end{equation*}%
with:%
\begin{eqnarray}
&&S_{f}=\Delta \Gamma ^{\dag }\left( T,\hat{T},\theta ,Z,Z^{\prime }\right)
\left( \nabla _{T}\left( \nabla _{T}-\left( -\frac{1}{\tau \omega _{0}\left(
Z\right) }T+\frac{\lambda }{\omega _{0}}\hat{T}\right) \right) \right)
\Delta \Gamma \left( T,\hat{T},\theta ,Z,Z^{\prime }\right)  \label{trm} \\
&&+\Delta \Gamma ^{\dag }\left( T,\hat{T},\theta ,Z,Z^{\prime }\right)
\left( \nabla _{\hat{T}}\left( \nabla _{\hat{T}}-\Delta \right) \right)
\Delta \Gamma \left( T,\hat{T},\theta ,Z,Z^{\prime }\right)  \notag
\end{eqnarray}%
\begin{equation}
V\left( \Delta \Gamma ,\Delta \Gamma ^{\dag }\right) =\Delta \Gamma ^{\dag
}\left( T,\hat{T},\theta ,Z,Z^{\prime }\right) \left( \nabla _{T}\left( \Phi
\right) -\nabla _{\hat{T}}\left( \Xi -\Upsilon \right) \right) \Delta \Gamma
\left( T,\hat{T},\theta ,Z,Z^{\prime }\right)  \label{trl}
\end{equation}%
and where we defined:%
\begin{eqnarray*}
\Xi &=&\frac{\rho }{\omega _{0}^{2}\left( Z\right) }\left( D\left( \theta
\right) \left\langle \hat{T}\right\rangle \left\vert \Psi _{0}\left(
Z^{\prime }\right) \right\vert ^{2}\left( \omega _{0}\left( Z\right) \delta
\omega \left( \theta -\frac{\left\vert Z-Z^{\prime }\right\vert }{c}%
,Z^{\prime },\left\vert \Psi \right\vert ^{2}\right) -\omega _{0}\left(
Z^{\prime }\right) \delta \omega \left( \theta ,Z,\left\vert \Psi
\right\vert ^{2}\right) \right) \right) \\
\Delta &=&\frac{\rho }{\omega _{0}\left( Z\right) }\left( \left( h\left(
Z,Z^{\prime }\right) -\hat{T}\right) C\left( \theta \right) \left\vert \Psi
_{0}\left( Z\right) \right\vert ^{2}\omega _{0}\left( Z\right) -D\left(
\theta \right) \hat{T}\left\vert \Psi _{0}\left( Z^{\prime }\right)
\right\vert ^{2}\omega _{0}\left( Z^{\prime }\right) \right) \\
\Upsilon &=&\frac{\rho }{\omega _{0}\left( Z\right) }\left( \left( C\left(
\theta \right) \delta \omega \left( \theta ,Z,\left\vert \Psi \right\vert
^{2}\right) +D\left( \theta \right) \delta \omega \left( \theta -\frac{%
\left\vert Z-Z^{\prime }\right\vert }{c},Z^{\prime },\left\vert \Psi
\right\vert ^{2}\right) \right) \left\vert \Psi _{0}\left( Z\right)
\right\vert ^{2}\left( \hat{T}-\left\langle \hat{T}\right\rangle \right)
\left\vert \Psi _{0}\left( Z^{\prime }\right) \right\vert ^{2}\right) \\
\Phi &=&\left( -\frac{\delta \omega \left( \theta ,Z,\left\vert \Psi
\right\vert ^{2}\right) }{\tau \omega _{0}^{2}\left( Z\right) }\left(
T-\left\langle T\right\rangle \right) +\frac{\lambda \delta \omega \left(
\theta ,Z,\left\vert \Psi \right\vert ^{2}\right) }{\omega _{0}^{2}}\left( 
\hat{T}-\left\langle \hat{T}\right\rangle \right) \right)
\end{eqnarray*}%
The first contribution:%
\begin{equation*}
\Gamma _{0}^{\dag }\left( T,\hat{T},\theta ,Z,Z^{\prime }\right) \Xi \Gamma
_{0}\left( T,\hat{T},\theta ,Z,Z^{\prime }\right)
\end{equation*}%
translates the modifications in the background state dynamics due to
fluctuations in the connectivities. This can be neglected in first
approximation. The second contribution $S_{f}$ computes the free transition
functions in the backgroundstate, i.e. the transitions due to internal
fluctuations in absence of interactions. The third contribution $V\left(
\Delta \Gamma ,\Delta \Gamma ^{\dag }\right) $\ is an interaction term.

The term (\ref{trm}) is the free part of the effective action, while (\ref%
{trl}) includes the interaction terms describing the self interaction of the
connectivities system, through the fluctuations in activities.

We demonstrate in appendix 1 that the first term in (\ref{trl}) may be
neglected and the second term can be approximated using estimates of $\delta
\omega \left( \theta -\frac{\left\vert Z-Z^{\prime }\right\vert }{c}%
,Z^{\prime },\left\vert \Psi \right\vert ^{2}\right) $ provided in appendix
1 and 2 of (\cite{GLs}). Therefore, we can alternatively use for
interactions:%
\begin{equation}
V\left( \Delta \Gamma ,\Delta \Gamma ^{\dag }\right) =-\Delta \Gamma ^{\dag
}\left( T,\hat{T},\theta ,Z,Z^{\prime }\right) \left( \nabla _{\hat{T}}\Xi
\right) \Delta \Gamma \left( T,\hat{T},\theta ,Z,Z^{\prime }\right)
\label{Tlr}
\end{equation}%
or its continuous approximation:%
\begin{equation}
V\left( \Delta \Gamma ,\Delta \Gamma ^{\dag }\right) =\Delta \Gamma ^{\dag
}\left( T,\hat{T},\theta ,Z,Z^{\prime }\right) \left( \nabla _{\hat{T}}\hat{%
\Xi}\right) \Delta \Gamma \left( T,\hat{T},\theta ,Z,Z^{\prime }\right)
\label{Trm}
\end{equation}%
where:%
\begin{equation*}
\hat{\Xi}=\frac{\rho }{\omega _{0}\left( Z\right) }\left( D\left( \theta
\right) \left\langle \hat{T}\right\rangle \left\vert \Psi _{0}\left(
Z^{\prime }\right) \right\vert ^{2}\left( \left( \left( Z-Z^{\prime }\right)
\left( \nabla _{Z}+\nabla _{Z}\omega _{0}\left( Z\right) \right) +\frac{%
\left\vert Z-Z^{\prime }\right\vert }{c}\right) \delta \omega \left( \theta
,Z,\left\vert \Psi \right\vert ^{2}\right) \right) \right)
\end{equation*}

Ultimately, appendix 1 shows that the free action can be rewritten using the
background state equations, so that the effective actin for $\Gamma \left( T,%
\hat{T},\theta ,Z,Z^{\prime }\right) $ becomes:%
\begin{eqnarray}
&&S\left( \Delta \Gamma \left( T,\hat{T},\theta ,Z,Z^{\prime }\right) \right)
\label{fcgg} \\
&=&\Delta \Gamma ^{\dag }\left( T,\hat{T},\theta ,Z,Z^{\prime }\right)
\left( \nabla _{T}\left( \nabla _{T}+\frac{\left( T-\left\langle
T\right\rangle \right) }{\tau \omega _{0}\left( Z\right) }\left\vert \Psi
\left( \theta ,Z\right) \right\vert ^{2}\right) \right) \Delta \Gamma \left(
T,\hat{T},\theta ,Z,Z^{\prime }\right)  \notag \\
&&+\Delta \Gamma ^{\dag }\left( T,\hat{T},\theta ,Z,Z^{\prime }\right)
\nabla _{\hat{T}}\left( \nabla _{\hat{T}}+\rho \left\vert \bar{\Psi}%
_{0}\left( Z,Z^{\prime }\right) \right\vert ^{2}\left( \hat{T}-\left\langle 
\hat{T}\right\rangle \right) \right) \Delta \Gamma \left( T,\hat{T},\theta
,Z,Z^{\prime }\right) +V\left( \Delta \Gamma ,\Delta \Gamma ^{\dag }\right) 
\notag
\end{eqnarray}%
where $\left\vert \bar{\Psi}_{0}\left( Z,Z^{\prime }\right) \right\vert ^{2}$
is a weighted sum of the field over the two connected points: 
\begin{equation}
\left\vert \bar{\Psi}_{0}\left( Z,Z^{\prime }\right) \right\vert ^{2}=\frac{%
C\left( \theta \right) \left\vert \Psi _{0}\left( Z\right) \right\vert
^{2}\omega _{0}\left( Z\right) +D\left( \theta \right) \left\vert \Psi
_{0}\left( Z^{\prime }\right) \right\vert ^{2}\omega _{0}\left( Z^{\prime
}\right) }{\omega _{0}\left( Z\right) }  \label{PS}
\end{equation}

\subsection{Replacing $\protect\delta \protect\omega \left( \protect\theta %
,Z,\left\vert \Psi \right\vert ^{2}\right) $ in the effective action as a
function of $\Delta \Gamma \left( T,\hat{T},\protect\theta ,Z,Z^{\prime
}\right) $}

As seen from equation (\ref{rnll}) $\delta \omega \left( \theta
,Z,\left\vert \Psi \right\vert ^{2}\right) $ depends on $T\left\vert \Gamma
\left( T,\hat{T},\theta ,Z,Z^{\prime }\right) \right\vert ^{2}$. In a
fluctuating state:%
\begin{equation*}
\Gamma _{0}\left( T,\hat{T},\theta ,Z,Z^{\prime }\right) +\Delta \Gamma
\left( T,\hat{T},\theta ,Z,Z^{\prime }\right)
\end{equation*}%
the activities $\delta \omega \left( \theta ,Z,\left\vert \Psi \right\vert
^{2}\right) $ are modified by $\Delta \Gamma \left( T,\hat{T},\theta
,Z,Z^{\prime }\right) $ and in each operator $\check{T}$, the averages:%
\begin{equation*}
T\left\vert \Gamma \left( T,\hat{T},\theta ,Z,Z^{\prime }\right) \right\vert
^{2}
\end{equation*}%
may be replaced by:%
\begin{equation*}
\left\langle T\right\rangle \left\vert \Gamma _{0}\left( T,\hat{T},\theta
,Z,Z^{\prime }\right) \right\vert ^{2}\left( 1+\frac{\left( T-\left\langle
T\right\rangle \right) \left\vert \Delta \Gamma \left( T,\hat{T},\theta
,Z,Z^{\prime }\right) \right\vert ^{2}}{\left\langle T\right\rangle
\left\vert \Gamma _{0}\left( T,\hat{T},\theta ,Z,Z^{\prime }\right)
\right\vert ^{2}}\right)
\end{equation*}

To isolate the dependency of $\delta \omega \left( \theta ,Z,\left\vert \Psi
\right\vert ^{2}\right) $ in $\left\vert \Delta \Gamma \left( T,\hat{T}%
,\theta ,Z,Z^{\prime }\right) \right\vert ^{2}$ we start with the expansion (%
\ref{rnll}) including higher order corrections evaluated at $\left\vert
\Gamma _{0}\right\vert ^{2}$. Then we sum over all the possible
modifications obtained by inserting at least at one point a factor:%
\begin{equation*}
\left( 1+\frac{\left( T-\left\langle T\right\rangle \right) \left\vert
\Delta \Gamma \left( T,\hat{T},\theta ,Z,Z^{\prime }\right) \right\vert ^{2}%
}{\left\langle T\right\rangle \left\vert \Gamma _{0}\left( T,\hat{T},\theta
,Z,Z^{\prime }\right) \right\vert ^{2}}\right)
\end{equation*}%
Using again estimations of appendix 1 and 2 of (\cite{GLs}) for the operator 
$\frac{\check{T}}{\left( 1-\left( 1+\right) \check{T}\right) }$, we show
that the sum of these modifications is obtained by first considering the
following contribution for any line $Z,Z_{1},...,Z_{n}$ of any number $n$ of
points:%
\begin{eqnarray}
&&G\left( \theta -\theta _{1},Z-Z_{1}\right) \left[ \prod \frac{\Delta
T\left\vert \Delta \Gamma \left( \theta _{j},Z_{j},Z_{j+1}\right)
\right\vert ^{2}}{T}\left\vert \Psi \left( \theta _{j},Z_{j}\right)
\right\vert ^{2}G\left( \theta _{j}-\theta _{j+1},Z_{j}-Z_{j+1}\right) %
\right]  \label{lnc} \\
&&\times G\left( \theta _{n-1}-\theta _{n},Z_{n-1}-Z_{n}\right) \left\vert
\Psi _{0}\left( Z_{n},\theta _{n}\right) \right\vert ^{2}\frac{\omega
_{0}\left( \theta _{n},Z_{n}\right) }{\Lambda ^{2}}
\end{eqnarray}

where:%
\begin{equation}
G\left( \theta _{j}-\theta _{j+1},Z_{j}-Z_{j+1}\right) =\frac{\exp \left(
-c\left( \theta _{j}-\theta _{j+1}\right) -\alpha \left( \left( c\left(
\theta _{j}-\theta _{j+1}\right) \right) ^{2}-\left\vert
Z_{j}-Z_{j+1}\right\vert ^{2}\right) \right) }{D}  \label{prg}
\end{equation}%
Then, we branch such series at some points $\left( \theta _{k},Z_{k}\right)
_{k}$ to produce a tree with an arbitrary number of nodes. Each node can
have an arbitrary number of branches originating from this point. The
contribution associated with such a tree is the product of the terms
associated with each branch. Finally, we sum over all possible branching
points.

The contribution associated to this sum is thus:%
\begin{eqnarray}
&&\sum \prod\limits_{p}\underset{\left( Z_{j}^{\left( p\right) },\theta
_{j}^{\left( p\right) }\right) }{\ast }\left[ G\left( \theta -\theta
_{1}^{\left( p\right) },Z-Z_{1}^{\left( p\right) }\right) \right.
\label{vnt} \\
&&\times \left[ \prod \frac{\Delta T\left\vert \Delta \Gamma \left( \theta
_{j}^{\left( p\right) },Z_{j}^{\left( p\right) },Z_{j+1}^{\left( p\right)
}\right) \right\vert ^{2}}{T}\left\vert \Psi \left( \theta _{j}^{\left(
p\right) },Z_{j}^{\left( p\right) }\right) \right\vert ^{2}G\left( \theta
_{j}^{\left( p\right) }-\theta _{j+1}^{\left( p\right) },Z_{j}^{\left(
p\right) }-Z_{j+1}^{\left( p\right) }\right) \right]  \notag \\
&&\left. \times G\left( \theta _{n}^{\left( p\right) }-\theta _{i}^{\left(
p\right) },Z_{n}^{\left( p\right) }-Z_{i}^{\left( p\right) }\right)
\left\vert \Psi _{\Gamma }\left( Z_{i}^{\left( p\right) },\theta
_{i}^{\left( p\right) }\right) \right\vert ^{2}\frac{\omega _{0}\left(
\theta _{i}^{\left( p\right) },Z_{i}^{\left( p\right) }\right) }{\Lambda ^{2}%
}\right]  \notag \\
&=&\sum \prod\limits_{p}\underset{\left( Z_{j}^{\left( p\right) },\theta
_{j}^{\left( p\right) }\right) }{\ast }V\left( \left( Z_{j}^{\left( p\right)
},\theta _{j}^{\left( p\right) }\right) \right)  \notag
\end{eqnarray}%
and the symbol $\underset{\left( Z_{j}^{\left( p\right) },\theta
_{j}^{\left( p\right) }\right) }{\ast }$ denotes the branching of lines at
any points. The sum takes into acount all the possibility of branching lines.

Ultimately, we replace $\delta \omega \left( \theta ,Z,\left\vert \Psi
\right\vert ^{2}\right) $ in (\ref{Tlr}) and (\ref{Trm})\ by:%
\begin{equation}
\sum \prod\limits_{p}\underset{\left( Z_{j}^{\left( p\right) },\theta
_{j}^{\left( p\right) }\right) }{\ast }V\left( \left( Z_{j}^{\left( p\right)
},\theta _{j}^{\left( p\right) }\right) \right)  \label{nrss}
\end{equation}%
given in (\ref{vnt}).

\subsection{Graphs expansion for the effective action}

The interaction terms (\ref{nrss}) terms allow to compute the transitions
from one state $\left( \Delta T_{j}^{\left( i\right) }\left( Z_{j}^{\left(
i\right) },Z_{j}^{\prime \left( i\right) }\right) \right) _{j\leqslant n}$
of $n$ connections to an other $\left( \Delta T_{j}^{\left( f\right) }\left(
Z_{j}^{\left( f\right) },Z_{j}^{\prime \left( f\right) }\right) \right)
_{j\leqslant n}$.

The amplitudes are given by the sum over $k$ of products of $k$ vertices:%
\begin{eqnarray}
&&\left\langle \left( \Delta T_{j}^{\left( i\right) }\left( Z_{j}^{\left(
i\right) },Z_{j}^{\prime \left( i\right) }\right) \right) _{j\leqslant
n}\right\vert \left\{ \int \Delta \Gamma ^{\dag }\left( T,\hat{T},\theta
,Z,Z^{\prime }\right) \right. \\
&&\times \left( \nabla _{\hat{T}}\left( \frac{\rho }{\omega _{0}\left(
Z\right) }\left( D\left( \theta \right) \left\langle \hat{T}\right\rangle
\left\vert \Psi _{0}\left( Z^{\prime }\right) \right\vert ^{2}\left( \left(
\alpha \left\vert Z-Z_{1}\right\vert ^{2}+\frac{\left\vert Z-Z^{\prime
}\right\vert }{c}\right) +\left( Z^{\prime }-Z\right) \nabla _{Z}\omega
_{0}\left( Z\right) \right) \right) \right) \right.  \notag \\
&&\left. \times \left( \sum \prod\limits_{p}\underset{\left( Z_{j}^{\left(
p\right) },\theta _{j}^{\left( p\right) }\right) }{\ast }V\left( \left(
Z_{j}^{\left( p\right) },\theta _{j}^{\left( p\right) }\right) \right)
\right) \right) \left. \Delta \Gamma \left( T,\hat{T},\theta ,Z,Z^{\prime
}\right) \right\} ^{k}\left\vert \left( \Delta T_{j}^{\left( f\right)
}\left( Z_{j}^{\left( f\right) },Z_{j}^{\prime \left( f\right) }\right)
\right) _{j\leqslant n}\right\rangle  \notag
\end{eqnarray}

The computation can be expressed in terms of graphs,as explained in appendix
1. Due to the form of the propagators, some simplifications arise. We show
that the sum of vertices can be simplified and that the interaction term $%
\delta \omega \left( \theta ,Z,\left\vert \Psi \right\vert ^{2}\right) $ in (%
\ref{trm}) writes recursively.

At the first order

\begin{eqnarray}
&&\delta \omega \left( \theta ,Z,\left\vert \Psi \right\vert ^{2}\right) =%
\check{T}\left( 1-\left( 1+\left\langle \left\vert \Psi _{\Gamma
}\right\vert ^{2}\right\rangle \right) \check{T}\right) ^{-1}\left( Z,\theta
,Z_{1},\theta _{1}\right) \left[ \frac{\Delta T\left\vert \Delta \Gamma
\left( \theta _{1},Z_{1},Z_{1}\right) \right\vert ^{2}}{T}d\theta _{1}\right]
\label{fsdd} \\
&\equiv &\sum_{i}\int K\left( Z,\theta ,Z_{1},\theta _{1}\right) \left\{ 
\frac{\Delta T\left\vert \Delta \Gamma \left( \theta _{1},Z_{1},Z_{1}\right)
\right\vert ^{2}}{T}\right\} d\theta _{1}  \notag
\end{eqnarray}%
and recursively, the corrections are obtained order by order by replacing:%
\begin{equation*}
\left( 1+\left\langle \left\vert \Psi _{\Gamma }\right\vert
^{2}\right\rangle \right) \check{T}
\end{equation*}%
by:%
\begin{equation*}
\left( 1+\left\langle \left\vert \Psi _{\Gamma }\right\vert
^{2}\right\rangle -\frac{\frac{\check{T}}{\left( 1-\left( 1+\left\langle
\left\vert \Psi _{\Gamma }\right\vert ^{2}\right\rangle \right) \check{T}%
\right) }\left[ \frac{\Delta T\left\vert \Delta \Gamma \left( \theta
_{1},Z_{1},Z_{1}\right) \right\vert ^{2}}{T}\right] }{\omega _{0}\left(
Z\right) +\frac{\check{T}}{\left( 1-\left( 1+\left\langle \left\vert \Psi
_{\Gamma }\right\vert ^{2}\right\rangle \right) \check{T}\right) }\left[ 
\frac{\Delta T\left\vert \Delta \Gamma \left( \theta _{1},Z_{1},Z_{1}\right)
\right\vert ^{2}}{T}\right] }\right) \check{T}
\end{equation*}

\subsection{Effective action at the first order in perturbation}

The effective action for $\Delta \Gamma \left( T,\hat{T},\theta ,Z,Z^{\prime
}\right) $ can be rewritten at the lower order in perturbation using
approximation (\ref{Trm}): 
\begin{eqnarray}
&&S\left( \Delta \Gamma \left( T,\hat{T},\theta ,Z,Z^{\prime }\right) \right)
\label{fcp} \\
&=&-\Delta \Gamma ^{\dag }\left( T,\hat{T},\theta ,Z,Z^{\prime }\right)
\left( \nabla _{T}\left( \nabla _{T}+\frac{\left( T-\left\langle
T\right\rangle \right) -\lambda \left( \hat{T}-\left\langle \hat{T}%
\right\rangle \right) }{\tau \omega _{0}\left( Z\right) }\left\vert \Psi
\left( \theta ,Z\right) \right\vert ^{2}\right) \right) \Delta \Gamma \left(
T,\hat{T},\theta ,Z,Z^{\prime }\right)  \notag \\
&&-\Delta \Gamma ^{\dag }\left( T,\hat{T},\theta ,Z,Z^{\prime }\right)
\nabla _{\hat{T}}\left( \nabla _{\hat{T}}+\rho \left\vert \bar{\Psi}%
_{0}\left( Z,Z^{\prime }\right) \right\vert ^{2}\left( \hat{T}-\left\langle 
\hat{T}\right\rangle \right) \right) \Delta \Gamma \left( T,\hat{T},\theta
,Z,Z^{\prime }\right)  \notag \\
&&-V\left( \Delta \Gamma ,\Delta \Gamma ^{\dag }\right)  \notag
\end{eqnarray}%
with $\left\vert \bar{\Psi}_{0}\left( Z,Z^{\prime }\right) \right\vert ^{2}$
defined in (\ref{PS}) and:%
\begin{eqnarray}
&&V\left( \Delta \Gamma ,\Delta \Gamma ^{\dag }\right) =\Delta \Gamma ^{\dag
}\left( T,\hat{T},\theta ,Z,Z^{\prime }\right)  \label{pnl} \\
&&\times \left( \nabla _{\hat{T}}\left( \frac{\rho D\left( \theta \right)
\left\langle \hat{T}\right\rangle \left\vert \Psi _{0}\left( Z^{\prime
}\right) \right\vert ^{2}}{\omega _{0}\left( Z\right) }\check{T}\left(
1-\left( 1+\left\langle \left\vert \Psi _{\Gamma }\right\vert
^{2}\right\rangle \right) \check{T}\right) ^{-1}\left[ O\frac{\Delta
T\left\vert \Delta \Gamma \left( \theta _{1},Z_{1},Z_{1}^{\prime }\right)
\right\vert ^{2}}{T\Lambda ^{2}}\right] \right) \right) \Delta \Gamma \left(
T,\hat{T},\theta ,Z,Z^{\prime }\right)  \notag
\end{eqnarray}%
and:%
\begin{equation}
O\left( Z,Z^{\prime },Z_{1}\right) =-\frac{\left\vert Z-Z^{\prime
}\right\vert }{c}\nabla _{\theta _{1}}+\frac{\left( Z^{\prime }-Z\right) ^{2}%
}{2}\left( \frac{\nabla _{Z_{1}}^{2}}{2}+\frac{\nabla _{\theta _{1}}^{2}}{%
2c^{2}}-\frac{\nabla _{Z}^{2}\omega _{0}\left( Z\right) }{2}\right)
\label{fms}
\end{equation}%
We close this section by considering the potential for connectivities:%
\begin{equation*}
U\left( \left\{ \left\vert \Gamma \left( T,\hat{T},Z,Z^{\prime },C,D\right)
\right\vert ^{2}\right\} \right)
\end{equation*}%
This potential adds an additional term to (\ref{fcp}). Assuming the
variation $\Delta \Gamma \left( T,\hat{T},\theta ,Z,Z^{\prime }\right) $ to
be orthogonal to $\Gamma _{0}\left( T,\hat{T},\theta ,Z,Z^{\prime
},C,D\right) $, we write the effective potential: 
\begin{eqnarray*}
&&U_{\Delta \Gamma }\left( \left\{ \Delta \left\vert \Gamma \left( T,\hat{T}%
,Z,Z^{\prime }\right) \right\vert ^{2}\right\} \right) \\
&=&\sum_{k}\int \left( \frac{\delta ^{k}U\left( \left\{ \left\vert \Gamma
\left( T,\hat{T},Z,Z^{\prime },C,D\right) \right\vert ^{2}\right\} \right) }{%
\delta ^{k}\left\vert \Gamma \left( T_{i},\hat{T}_{i},Z_{i},Z_{i}^{\prime
}\right) \right\vert ^{2}}\right) _{\Gamma _{0}\left( T_{i},\hat{T}%
_{i},Z_{i},Z_{i}^{\prime }\right) }\prod\limits_{i\leqslant k}\left\vert
\Delta \Gamma \left( T_{i},\hat{T}_{i},Z_{i},Z_{i}^{\prime }\right)
\right\vert ^{2}
\end{eqnarray*}%
We assume a slowly increasing function of $\left\Vert \Delta \Gamma \left(
Z,Z^{\prime }\right) \right\Vert ^{2}$, i.e. a function of the global
activity at each point $\left( Z,Z^{\prime }\right) $:%
\begin{equation}
U_{\Delta \Gamma }\left( \left\Vert \Delta \Gamma \left( Z,Z^{\prime
}\right) \right\Vert ^{2}\right) =\sum_{k}\int \left( \frac{\delta
^{k}U\left( \left\Vert \Gamma \left( Z,Z^{\prime }\right) \right\Vert
^{2}\right) }{\delta ^{k}\left\Vert \Gamma \left( Z,Z^{\prime }\right)
\right\Vert ^{2}}\right) _{\Gamma _{0}\left( Z,Z^{\prime }\right)
}\left\Vert \Delta \Gamma \left( Z,Z^{\prime }\right) \right\Vert ^{2k}
\label{Ptm}
\end{equation}%
with:%
\begin{equation*}
\left\Vert \Delta \Gamma \left( Z,Z^{\prime }\right) \right\Vert ^{2}=\int
\left\vert \Delta \Gamma \left( T,\hat{T},\theta ,Z,Z^{\prime }\right)
\right\vert ^{2}d\left( T,\hat{T}\right)
\end{equation*}%
where the coefficients satisfy: 
\begin{equation}
\left\vert \left( \frac{\delta ^{k}U\left( \left\Vert \Gamma \left(
Z,Z^{\prime }\right) \right\Vert ^{2}\right) }{\delta ^{k}\left\Vert \Gamma
\left( Z,Z^{\prime }\right) \right\Vert ^{2}}\right) _{\Gamma _{0}\left(
Z,Z^{\prime }\right) }\right\vert <<1  \label{Ptnnn}
\end{equation}%
It allows to discard the potential term in the sequel, but it ensures that
the norm at point $\left( Z,Z^{\prime }\right) $ of $\Delta \Gamma \left( T,%
\hat{T},\theta ,Z,Z^{\prime }\right) $:%
\begin{equation*}
\left\Vert \Delta \Gamma \left( Z,Z^{\prime }\right) \right\Vert ^{2}
\end{equation*}%
written $\left\Vert \Delta \Gamma \right\Vert $, is bounded by a maximum,
representing a maximum in the shift in activity. We will also assume that:%
\begin{eqnarray}
\left( \frac{\delta U\left( \left\Vert \Gamma \left( Z,Z^{\prime }\right)
\right\Vert ^{2}\right) }{\delta \left\Vert \Gamma \left( Z,Z^{\prime
}\right) \right\Vert ^{2}}\right) _{\Gamma _{0}\left( Z,Z^{\prime }\right) }
&<&0  \label{Ptt} \\
\left( \frac{\delta ^{k}U\left( \left\Vert \Gamma \left( Z,Z^{\prime
}\right) \right\Vert ^{2}\right) }{\delta ^{k}\left\Vert \Gamma \left(
Z,Z^{\prime }\right) \right\Vert ^{2}}\right) _{\Gamma _{0}\left(
Z,Z^{\prime }\right) } &>&0\text{ for }k>2  \notag
\end{eqnarray}%
so that $U\left( \left\Vert \Gamma \left( Z,Z^{\prime }\right) \right\Vert
^{2}\right) $ has a minimum.

We will also assume that the global connectivity:%
\begin{equation*}
\left\Vert \Delta \Gamma \right\Vert ^{2}=\int \left\vert \Delta \Gamma
\left( T,\hat{T},\theta ,Z,Z^{\prime }\right) \right\vert ^{2}d\left( T,\hat{%
T},Z,Z^{\prime }\right)
\end{equation*}%
is constrained to be closed to some value $\overline{\left\Vert \Delta
\Gamma \right\Vert }^{2}$. This corresponds to some overall modifications
corresponding to some external modification.

\section{Application 1: Interactions and modifications of the background}

The introduction of effective interactions among connectivities in (\ref{fcp}%
) should modify the background state of the system. This section is
dedicated to the computation and description of these modifications. The
modified states closely resemble fundamental modifications occurring above
the background state, which results from self-interactions.

\subsection{Equations for background field in interaction}

The saddle point equation derived from (\ref{fcp}) writes:

\begin{eqnarray}
0 &=&\left( \nabla _{T}\left( \sigma _{T}^{2}\nabla _{T}+\frac{\left(
T-\left\langle T\right\rangle \right) -\lambda \left( \hat{T}-\left\langle 
\hat{T}\right\rangle \right) }{\tau \omega _{0}\left( Z\right) }\left\vert
\Psi \left( \theta ,Z\right) \right\vert ^{2}\right) \right) \Delta \Gamma
\left( T,\hat{T},\theta ,Z,Z^{\prime }\right)  \label{Sdt} \\
&&+\nabla _{\hat{T}}\left( \sigma _{\hat{T}}^{2}\nabla _{\hat{T}}+\rho
\left\vert \bar{\Psi}_{0}\left( Z,Z^{\prime }\right) \right\vert ^{2}\left( 
\hat{T}-\left\langle \hat{T}\right\rangle \right) \right) \Delta \Gamma
\left( T,\hat{T},\theta ,Z,Z^{\prime }\right)  \notag \\
&&+\left( V_{0}\left( \theta ,Z,Z^{\prime },\Delta \Gamma \right) +\left(
V_{1}\left( \theta ,Z,Z^{\prime },\Delta \Gamma \right) \left( 1+V_{2}\left(
\theta ,Z,Z^{\prime },\Delta \Gamma \right) \right) \right) \Delta T+\alpha
\left( \left\Vert \Delta \Gamma \left( Z,Z^{\prime }\right) \right\Vert
^{2}\right) \right) \Delta \Gamma \left( T,\hat{T},\theta ,Z,Z^{\prime
}\right)  \notag
\end{eqnarray}

with:%
\begin{equation}
\left\vert \bar{\Psi}_{0}\left( Z,Z^{\prime }\right) \right\vert ^{2}=\frac{%
C\left( \theta \right) \left\vert \Psi _{0}\left( Z\right) \right\vert
^{2}\omega _{0}\left( Z\right) +D\left( \theta \right) \left\vert \Psi
_{0}\left( Z^{\prime }\right) \right\vert ^{2}\omega _{0}\left( Z^{\prime
}\right) }{\omega _{0}\left( Z\right) }
\end{equation}%
\begin{equation}
V_{0}\left( \theta ,Z,Z^{\prime },\Delta \Gamma \right) =\nabla _{\hat{T}%
}\left( \frac{\rho D\left( \theta \right) \left\langle \hat{T}\right\rangle
\left\vert \Psi _{0}\left( Z^{\prime }\right) \right\vert ^{2}}{\omega
_{0}\left( Z\right) }\check{T}\left( 1-\left( 1+\left\langle \left\vert \Psi
_{\Gamma }\right\vert ^{2}\right\rangle \right) \check{T}\right) ^{-1}\left[
O\frac{\Delta T\left\vert \Delta \Gamma \left( \theta
_{1},Z_{1},Z_{1}^{\prime }\right) \right\vert ^{2}}{T}\right] \right)
\label{Vfr}
\end{equation}%
\begin{eqnarray}
&&V_{1}\left( \theta ,Z,Z^{\prime },\Delta \Gamma \right)  \label{Vfn} \\
&=&\int \Delta \Gamma ^{\dag }\left( T_{2},\hat{T}_{2},\theta
_{2},Z_{2},Z_{2}^{\prime }\right) \nabla _{\hat{T}_{2}}\left( \frac{\rho
D\left( \theta \right) \left\langle \hat{T}\right\rangle \left\vert \Psi
_{0}\left( Z_{2}^{\prime }\right) \right\vert ^{2}}{\omega _{0}\left(
Z_{2}\right) }\left[ \check{T}\left( 1-\left( 1+\left\langle \left\vert \Psi
_{\Gamma }\right\vert ^{2}\right\rangle \right) \check{T}\right) ^{-1}O%
\right] _{\left( T,\hat{T},\theta ,Z,Z^{\prime }\right) }^{\left( T_{2},\hat{%
T}_{2},\theta _{2},Z_{2},Z_{2}^{\prime }\right) }\right)  \notag \\
&&\times \Delta \Gamma \left( T_{2},\hat{T}_{2},\theta
_{2},Z_{2},Z_{2}^{\prime }\right) d\left( T_{2},\hat{T}_{2},\theta
_{2},Z_{2},Z_{2}^{\prime }\right)  \notag
\end{eqnarray}

\begin{equation}
V_{2}\left( \theta ,Z,Z^{\prime },\Delta \Gamma \right) =\int \left[ \check{T%
}\left( 1-\left( 1+\left\langle \left\vert \Psi _{\Gamma }\right\vert
^{2}\right\rangle \right) \check{T}\right) ^{-1}\right] _{\left( T_{1},\hat{T%
}_{1},\theta _{1},Z_{1},Z_{1}^{\prime }\right) }^{\left( T,\hat{T},\theta
,Z,Z^{\prime }\right) }\left[ \frac{\Delta T\left\vert \Delta \Gamma \left(
\theta _{1},Z_{1},Z_{1}^{\prime }\right) \right\vert ^{2}}{T}\right] d\left(
T_{1},\hat{T}_{1},\theta _{1},Z_{1},Z_{1}^{\prime }\right)  \label{Vft}
\end{equation}%
and:%
\begin{equation*}
\alpha \left( \left\Vert \Delta \Gamma \left( Z,Z^{\prime }\right)
\right\Vert ^{2}\right) =-\frac{\delta U_{\Delta \Gamma }\left( \left\Vert
\Delta \Gamma \left( Z,Z^{\prime }\right) \right\Vert ^{2}\right) }{\delta
\left\Vert \Delta \Gamma \left( Z,Z^{\prime }\right) \right\Vert ^{2}}%
+\alpha _{0}
\end{equation*}%
$\alpha $ implements the potential described in the previous paragraph and $%
\alpha _{0}$ implements the constraint . Given our assumptions (see (\ref%
{Ptt})), $\alpha \left( \left\Vert \Delta \Gamma \left( Z,Z^{\prime }\right)
\right\Vert ^{2}\right) $ increases from zero to a maximum, then decreases.
We will write $\alpha $\ for $\alpha \left( \left\Vert \Delta \Gamma \left(
Z,Z^{\prime }\right) \right\Vert ^{2}\right) +\alpha _{0}$ and thus consider 
$\alpha >0$.

Operator $O$ has a kernel\ given by (\ref{fms}):%
\begin{equation}
O\left( Z,Z^{\prime },Z_{1}\right) =-\frac{\left\vert Z-Z^{\prime
}\right\vert }{c}\nabla _{\theta _{1}}+\frac{\left( Z^{\prime }-Z\right) ^{2}%
}{2}\left( \frac{\nabla _{Z_{1}}^{2}}{2}+\frac{\nabla _{\theta _{1}}^{2}}{%
2c^{2}}-\frac{\nabla _{Z}^{2}\omega _{0}\left( Z\right) }{2}\right)
\end{equation}

Note that the kernel $\left[ \check{T}\left( 1-\left( 1+\left\langle
\left\vert \Psi _{\Gamma }\right\vert ^{2}\right\rangle \right) \check{T}%
\right) O\right] _{\left( T,\hat{T},\theta ,Z,Z^{\prime }\right) }^{\left(
T_{i},\hat{T}_{i},\theta _{i},Z_{i},Z_{i}^{\prime }\right) }$ and $\left[ 
\check{T}\left( 1-\left( 1+\left\langle \left\vert \Psi _{\Gamma
}\right\vert ^{2}\right\rangle \right) \check{T}\right) \right] _{\left(
T_{i},\hat{T}_{i},\theta _{i},Z_{i},Z_{i}^{\prime }\right) }^{\left( T,\hat{T%
},\theta ,Z,Z^{\prime }\right) }$ are computed with the average values of $%
\hat{T}$. As a consequence they do not depend on $T$ and $\hat{T}$. That's
why $V_{1}\left( \theta ,Z,Z^{\prime },\Delta \Gamma \right) $ and $%
V_{2,i}\left( \theta ,Z,Z^{\prime },\Delta \Gamma \right) $ do not depend on 
$T$ and $\hat{T}$. We will define:%
\begin{equation}
V\left( \theta ,Z,Z^{\prime },\Delta \Gamma \right) =V_{1}\left( \theta
,Z,Z^{\prime },\Delta \Gamma \right) +V_{1}\left( \theta ,Z,Z^{\prime
},\Delta \Gamma \right) V_{2}\left( \theta ,Z,Z^{\prime },\Delta \Gamma
\right)  \label{DV}
\end{equation}

\subsection{Formal solutions}

We solve (\ref{Sdt}) by the same method as in appendix 2 in (\cite{GLr}).
The detailled computations are given in appendix 7. Starting by writing (\ref%
{Sdt}):%
\begin{equation}
\left( \mathbf{\nabla }^{2}+\left( \mathbf{\nabla }\right) ^{t}\left( \gamma 
\mathbf{\Delta T}+V_{0}\mathbf{a}_{0}\right) +V\left( \mathbf{a}\right) ^{t}%
\mathbf{\Delta T}+\alpha \right) \Gamma \left( T,\hat{T},\theta ,Z,Z^{\prime
}\right) =0  \label{mdt}
\end{equation}%
with:%
\begin{equation*}
\left( \mathbf{\Delta T}\right) ^{t}=\left( \Delta T,\Delta \hat{T}\right) 
\text{, }\left( \mathbf{a}_{0}\right) ^{t}=\left( 0,1\right) \text{, }\left( 
\mathbf{a}\right) ^{t}=\left( 1,0\right)
\end{equation*}%
\begin{equation*}
\gamma =\left( 
\begin{array}{cc}
u & s \\ 
0 & v%
\end{array}%
\right)
\end{equation*}%
and:%
\begin{eqnarray*}
u &=&\frac{\left\vert \Psi _{0}\left( Z\right) \right\vert ^{2}}{\tau \omega
_{0}\left( Z\right) } \\
v &=&\rho C\frac{\left\vert \Psi _{0}\left( Z\right) \right\vert
^{2}h_{C}\left( \omega _{0}\left( Z\right) \right) }{\omega _{0}\left(
Z\right) }+\rho D\frac{\left\vert \Psi _{0}\left( Z^{\prime }\right)
\right\vert ^{2}h_{D}\left( \omega _{0}\left( Z^{\prime }\right) \right) }{%
\omega _{0}\left( Z\right) } \\
s &=&-\frac{\lambda \left\vert \Psi _{0}\left( Z\right) \right\vert ^{2}}{%
\omega _{0}\left( Z\right) }
\end{eqnarray*}%
The potential $V_{0}\left( Z,Z^{\prime }\right) $ isgiven by:%
\begin{eqnarray}
V_{0}\left( Z,Z^{\prime }\right) &=&\left( \frac{\rho D\left( \theta \right)
\left\langle \hat{T}\right\rangle \left\vert \Psi _{0}\left( Z^{\prime
}\right) \right\vert ^{2}}{\omega _{0}\left( Z\right) }\check{T}\left(
1-\left( 1+\left\langle \left\vert \Psi _{\Gamma }\right\vert
^{2}\right\rangle \right) \check{T}\right) ^{-1}\left[ O\frac{\Delta
T\left\vert \Delta \Gamma \left( \theta _{1},Z_{1},Z_{1}^{\prime }\right)
\right\vert ^{2}}{T}\right] \right)  \label{Fmv} \\
&\simeq &A_{0}\left( Z,Z^{\prime }\right) \frac{\left\langle \Delta
T\right\rangle }{\left\langle T\right\rangle }\left\Vert \Delta \Gamma
\right\Vert ^{2}  \notag
\end{eqnarray}%
where:%
\begin{equation*}
A_{0}\left( Z,Z^{\prime }\right) =\left\langle \frac{\rho D\left( \theta
\right) \left\langle \hat{T}\right\rangle \left\vert \Psi _{0}\left(
Z^{\prime }\right) \right\vert ^{2}}{\omega _{0}\left( Z\right) }\check{T}%
\left( 1-\left( 1+\left\langle \left\vert \Psi _{\Gamma }\right\vert
^{2}\right\rangle \right) \check{T}\right) ^{-1}O\right\rangle ^{Z,Z^{\prime
}}
\end{equation*}%
We solve equation (\ref{mdt}) by shifting the variables:%
\begin{eqnarray}
\mathbf{\Delta T}+\gamma ^{-1}V_{0}\mathbf{a}_{0} &\rightarrow &\mathbf{%
\Delta T}  \label{svb} \\
-V\left( \mathbf{a}\right) ^{t}\gamma ^{-1}V_{0}\mathbf{a}_{0}+\alpha
&\rightarrow &\alpha  \notag
\end{eqnarray}%
so that (\ref{mdt}) writes:%
\begin{equation}
\left( \mathbf{\nabla }^{2}+\left( \mathbf{\nabla }\right) ^{t}\gamma 
\mathbf{\Delta T}+V\left( \mathbf{a}\right) ^{t}\mathbf{\Delta T}+\alpha
\right) \Gamma \left( T,\hat{T},\theta ,Z,Z^{\prime }\right) =0
\end{equation}%
Thisequation is solved by considering the Fourier transform of this equation:%
\begin{equation}
\left( -\mathbf{k}^{2}-\left( \mathbf{k}\right) ^{t}\gamma \mathbf{\nabla }_{%
\mathbf{k}}-iV\left( \mathbf{a}\right) ^{t}\mathbf{\nabla }_{\mathbf{k}%
}\right) \Gamma \left( \mathbf{k},\theta ,Z,Z^{\prime }\right) =0
\label{Fsp}
\end{equation}%
with solution:%
\begin{eqnarray*}
\Gamma _{\delta }\left( \mathbf{k},\theta ,Z,Z^{\prime }\right) &=&\exp
\left( -\frac{1}{2}\mathbf{k}^{t}N\mathbf{k}\right) \left( k_{1}^{2}+\left( 
\frac{V}{u}\right) ^{2}\right) ^{\frac{\delta }{2\alpha u}}\left( \left(
k_{1}+\frac{v-u}{s}k_{2}\right) ^{2}+\left( \frac{V}{v}\right) ^{2}\right) ^{%
\frac{\left( 1-\delta \right) \alpha }{2v}} \\
&&\times \exp \left( -i\left( \frac{\alpha \delta }{u}\arctan \left( \frac{%
k_{1}u}{V}\right) +\frac{\left( 1-\delta \right) \alpha }{v}\arctan \left( 
\frac{\left( k_{1}+\frac{v-u}{s}k_{2}\right) v}{V}\right) \right) \right)
\end{eqnarray*}%
where:%
\begin{equation*}
N=\left( 
\begin{array}{cc}
\frac{1}{u}\left( 1+\frac{s^{2}}{v\left( u+v\right) }\right) & -\frac{s}{%
v\left( u+v\right) } \\ 
-\frac{s}{v\left( u+v\right) } & \frac{1}{v}%
\end{array}%
\right)
\end{equation*}%
We impose that $\alpha >0$ to ensure the solutions are well defined.

In the limit of large interactions, the solution to (\ref{Sdt}) is, up to
some constant:

\begin{eqnarray*}
\Gamma _{\delta }\left( T,\hat{T},\theta ,Z,Z^{\prime }\right) &=&i^{\frac{%
\alpha \delta }{u}+\frac{\left( 1-\delta \right) \alpha }{v}}\int \exp
\left( -\frac{1}{2}\mathbf{k}^{t}N\mathbf{k}-i\mathbf{k}\left( \mathbf{%
\Delta T-}\overline{\mathbf{\Delta T}}\right) \right) \\
&&\times \left( k_{1}^{2}+\left( \frac{V}{u}\right) ^{2}\right) ^{\frac{%
\alpha \delta }{2u}}\left( \left( k_{1}+\frac{v-u}{s}k_{2}\right)
^{2}+\left( \frac{V}{v}\right) ^{2}\right) ^{\frac{\left( 1-\delta \right)
\alpha }{2v}}\frac{d\mathbf{k}}{2\pi }
\end{eqnarray*}%
with:%
\begin{eqnarray}
\mathbf{\Delta T} &\mathbf{=}&\left( 
\begin{array}{c}
T-\left\langle T\right\rangle \\ 
\hat{T}-\left\langle \hat{T}\right\rangle%
\end{array}%
\right)  \label{svl} \\
\overline{\mathbf{\Delta T}} &=&\left( 
\begin{array}{c}
-\frac{\alpha }{V} \\ 
-\frac{\left( 1-\delta \right) \left( v-u\right) \alpha }{Vs}%
\end{array}%
\right)  \notag
\end{eqnarray}%
The estimation of the integral is presented in appendix 1. It uses the
diagonalization of $N=PDP^{-1}$. It thus implies that $\Gamma _{\delta
}\left( T,\hat{T},\theta ,Z,Z^{\prime }\right) $ is given by:

\begin{eqnarray*}
\Gamma _{\delta }\left( T,\hat{T},\theta ,Z,Z^{\prime }\right) &=&i^{\frac{%
\alpha \delta }{u}+\frac{\left( 1-\delta \right) \alpha }{v}}\int \exp
\left( -\frac{1}{2}\mathbf{k}^{t}D\mathbf{k}-i\mathbf{k}\left( \mathbf{%
\Delta T}^{\prime }-\overline{\mathbf{\Delta T}}^{\prime }\right) \right)
\times \\
&&\times \left( k_{1}\cos x-k_{2}\sin x\right) ^{\frac{\alpha \delta }{u}%
}\left( k_{1}\left( \cos x+\frac{v-u}{s}\sin x\right) +\left( \frac{v-u}{s}%
\cos x-\sin x\right) k_{2}\right) ^{\frac{\left( 1-\delta \right) \alpha }{v}%
}\frac{d\mathbf{k}}{2\pi }
\end{eqnarray*}%
with:%
\begin{eqnarray*}
\mathbf{\Delta T}^{\prime }-\overline{\mathbf{\Delta T}}^{\prime }
&=&P^{t}\left( \mathbf{\Delta T-}\overline{\mathbf{\Delta T}}\right) \\
D &=&\left( 
\begin{array}{cc}
\lambda _{+} & 0 \\ 
0 & \lambda _{-}%
\end{array}%
\right) ,P=\left( 
\begin{array}{cc}
\cos x & -\sin x \\ 
\sin x & \cos x%
\end{array}%
\right)
\end{eqnarray*}

\begin{eqnarray*}
\lambda _{\pm } &=&\frac{\frac{1}{u}\left( 1+\frac{s^{2}}{v\left( u+v\right) 
}\right) +\frac{1}{v}}{2}\pm \sqrt{\left( \frac{\frac{1}{u}\left( 1+\frac{%
s^{2}}{v\left( u+v\right) }\right) -\frac{1}{v}}{2}\right) ^{2}+\left( \frac{%
s}{v\left( u+v\right) }\right) ^{2}} \\
x &=&-\frac{1}{2}\arctan \left( \frac{4su}{v^{2}-u^{2}+s^{2}}\right)
\end{eqnarray*}

In the approximation given in the text, we have $s<<1$ so that $x<<1$ and
the computations of appendix 2 in (\cite{GLr}) apply. We find for relatively
large interaction $V>1$:

\begin{eqnarray}
&&\Gamma _{\delta }\left( T,\hat{T},\theta ,Z,Z^{\prime }\right)
\label{Slgg} \\
&\simeq &\left( \frac{v-u}{s}\right) ^{\frac{\left( 1-\delta \right) \alpha 
}{v}}2^{\frac{\alpha }{u}+1}\prod\limits_{i=1}^{2}\exp \left( -\left(
\left( \frac{D^{-\frac{1}{2}}P^{t}\left( \mathbf{\Delta T}-\overline{\mathbf{%
\Delta T}}\right) }{4}\right) _{i}\right) ^{2}\right)  \notag \\
&&\times \left\{ \prod\limits_{i=1}^{2}\hat{D}_{p_{i}}^{m_{i}}\left( \left( 
\frac{D^{-\frac{1}{2}}P^{t}\left( \mathbf{\Delta T}-\overline{\mathbf{\Delta
T}}\right) }{4}\right) _{i}\right) \right.  \notag \\
&&\left. +\nabla _{\left( \Delta T^{\prime }\right) _{1}}\nabla _{\left(
\Delta T^{\prime }\right) _{2}}\left\{ x\alpha \frac{\delta
\prod\limits_{i=1}^{2}\hat{D}_{p_{i}^{\left( 1\right) }}^{m_{i}}\left(
\left( \frac{D^{-\frac{1}{2}}P^{t}\left( \mathbf{\Delta T}-\overline{\mathbf{%
\Delta T}}\right) }{4}\right) _{i}\right) }{u}-\frac{s\alpha \left( 1-\delta
\right) \prod\limits_{i=1}^{2}\hat{D}_{p_{i}^{\left( 1\right)
}}^{m_{i}}\left( \left( \frac{D^{-\frac{1}{2}}P^{t}\left( \mathbf{\Delta T}-%
\overline{\mathbf{\Delta T}}\right) }{4}\right) _{i}\right) }{v\left(
u-v\right) }\right\} \right\}  \notag
\end{eqnarray}%
where:%
\begin{eqnarray*}
p_{1} &=&\frac{\alpha \delta }{u},p_{2}=\frac{\left( 1-\delta \right) \alpha 
}{v} \\
p_{1}^{\left( 1\right) } &=&\frac{\alpha \delta }{u}-1,p_{2}^{\left(
1\right) }=\frac{\left( 1-\delta \right) \alpha }{v}+1 \\
p_{1}^{\left( 1\right) } &=&\frac{\alpha \delta }{u}+1,p_{2}^{\left(
1\right) }=\frac{\left( 1-\delta \right) \alpha }{v}-1
\end{eqnarray*}%
Ultimately, equation (\ref{mdt}) has an equivalent for $\Gamma ^{\dagger
}\left( T,\hat{T},\theta ,Z,Z^{\prime }\right) $, obtained by transposition:%
\begin{equation}
\left( \mathbf{\nabla }^{2}-\left( \gamma \mathbf{\Delta T}+V_{0}\mathbf{a}%
_{0}\right) ^{t}\left( \mathbf{\nabla }\right) +V\left( \mathbf{a}\right)
^{t}\mathbf{\Delta T}+\alpha \right) \Gamma ^{\dagger }\left( T,\hat{T}%
,\theta ,Z,Z^{\prime }\right) =0  \label{mtd}
\end{equation}%
that can be solved similarly:%
\begin{eqnarray}
&&\Gamma _{\delta }^{\dagger }\left( T,\hat{T},\theta ,Z,Z^{\prime }\right)
\label{Sld} \\
&\simeq &\left( \frac{v-u}{s}\right) ^{\frac{\left( 1-\delta \right) \alpha 
}{v}}2^{\frac{\alpha }{u}+1}\prod\limits_{i=1}^{2}\exp \left( \left( \left( 
\frac{D^{-\frac{1}{2}}P^{t}\left( \mathbf{\Delta T}-\overline{\mathbf{\Delta
T}}\right) }{4}\right) _{i}\right) ^{2}\right)  \notag \\
&&\times \left\{ \prod\limits_{i=1}^{2}\hat{D}_{p_{i}}^{m_{i}}\left( \left( 
\frac{D^{-\frac{1}{2}}P^{t}\left( \mathbf{\Delta T}-\overline{\mathbf{\Delta
T}}\right) }{4}\right) _{i}\right) \right.  \notag \\
&&\left. +\nabla _{\left( \Delta T^{\prime }\right) _{1}}\nabla _{\left(
\Delta T^{\prime }\right) _{2}}\left\{ x\alpha \frac{\delta
\prod\limits_{i=1}^{2}\hat{D}_{p_{i}^{\left( 1\right) }}^{m_{i}}\left(
\left( \frac{D^{-\frac{1}{2}}P^{t}\left( \mathbf{\Delta T}-\overline{\mathbf{%
\Delta T}}\right) }{4}\right) _{i}\right) }{u}-\frac{s\alpha \left( 1-\delta
\right) \prod\limits_{i=1}^{2}\hat{D}_{p_{i}^{\left( 1\right)
}}^{m_{i}}\left( \left( \frac{D^{-\frac{1}{2}}P^{t}\left( \mathbf{\Delta T}-%
\overline{\mathbf{\Delta T}}\right) }{4}\right) _{i}\right) }{v\left(
u-v\right) }\right\} \right\}  \notag
\end{eqnarray}

Since $\Delta T$ and $\Delta \hat{T}$ can be either positive or negative, to
ensure the solutions to be integrable over $%
%TCIMACRO{\U{211d} }%
%BeginExpansion
\mathbb{R}
%EndExpansion
$, the parameters $p_{1}=\frac{\alpha \delta }{u},p_{2}=\frac{\left(
1-\delta \right) \alpha }{v}$ have have the form $\frac{1}{2}+k$ and $\frac{1%
}{2}+l$ respectively. This implies some constraint on the background state
as well as on the repartition of shift in connectivity functions in the
thread.

Note that equations (\ref{Slgg}) and (\ref{Sld}) are defined up to a
normalization factor at each point $\left( Z,Z^{\prime }\right) $, written $%
\left\Vert \Delta \Gamma _{\delta }\left( Z,Z^{\prime }\right) \right\Vert
^{2}$. If this normalization factor is nul, the solutions are trivial at
this point:

\begin{equation*}
\Gamma _{\delta }=\Gamma _{\delta }^{\dagger }=0
\end{equation*}%
and no shift occurs. To find the normalization $\left\Vert \Delta \Gamma
_{\delta }\left( Z,Z^{\prime }\right) \right\Vert ^{2}$, we need to find at
which condition the state $\Gamma _{\delta }\left( T,\hat{T},\theta
,Z,Z^{\prime }\right) $ is a minimum of the action. Doing so, we need to
compute the shift in average connectivity induced by states $\Gamma _{\delta
}\left( T,\hat{T},\theta ,Z,Z^{\prime }\right) $.

Note that, for $V>>1$, solutions (\ref{Slgg}) and (\ref{Sld}) further
simplify:%
\begin{eqnarray*}
&&\Gamma _{\delta }\left( T,\hat{T},\theta ,Z,Z^{\prime }\right) \\
&\simeq &\left( \frac{V}{u}\right) ^{\frac{\alpha \delta }{u}}\left( \frac{V%
}{v}\right) ^{\frac{\left( 1-\delta \right) \alpha }{v}}\exp \left( -\frac{1%
}{4}\left( \mathbf{\Delta T-}\overline{\mathbf{\Delta T}}\right)
N^{-1}\left( \mathbf{\Delta T-}\overline{\mathbf{\Delta T}}\right) \right)
\int \exp \left( -\frac{1}{4}\mathbf{k}^{t}N\mathbf{k}-i\mathbf{k}\left( 
\mathbf{\Delta T-}\overline{\mathbf{\Delta T}}\right) \right) \frac{d\mathbf{%
k}}{2\pi } \\
&=&\left( \frac{V}{u}\right) ^{\frac{\alpha \delta }{u}}\left( \frac{V}{v}%
\right) ^{\frac{\left( 1-\delta \right) \alpha }{v}}\exp \left( -\frac{1}{2}%
\left( \mathbf{\Delta T-}\overline{\mathbf{\Delta T}}\right) N^{-1}\left( 
\mathbf{\Delta T-}\overline{\mathbf{\Delta T}}\right) \right)
\end{eqnarray*}%
and:%
\begin{eqnarray*}
&&\Gamma _{\delta }^{\dagger }\left( T,\hat{T},\theta ,Z,Z^{\prime }\right)
\\
&\simeq &\left( \frac{V}{u}\right) ^{\frac{\alpha \delta }{u}}\left( \frac{V%
}{v}\right) ^{\frac{\left( 1-\delta \right) \alpha }{v}}\exp \left( \frac{1}{%
4}\left( \mathbf{\Delta T-}\overline{\mathbf{\Delta T}}\right) N^{-1}\left( 
\mathbf{\Delta T-}\overline{\mathbf{\Delta T}}\right) \right) \int \exp
\left( -\frac{1}{2}\mathbf{k}^{t}N\mathbf{k}-i\mathbf{k}\left( \mathbf{%
\Delta T-}\overline{\mathbf{\Delta T}}\right) \right) \frac{d\mathbf{k}}{%
2\pi } \\
&=&\left( \frac{V}{u}\right) ^{\frac{\alpha \delta }{u}}\left( \frac{V}{v}%
\right) ^{\frac{\left( 1-\delta \right) \alpha }{v}}
\end{eqnarray*}

\subsection{Equation for shift in connectivity functions}

We derive in appendix 1 the equations for the shifts in average connectivity
at each point $\left( Z,Z^{\prime }\right) $. This shift is given by $%
\overline{\mathbf{\Delta T}}$ as defined in (\ref{svl}), plus additional
contributions arising from the successive change of variables. We find:%
\begin{eqnarray}
\Delta T\left( Z,Z^{\prime }\right) &=&-\frac{\alpha }{V\left( Z,Z^{\prime
}\right) }  \label{hfn} \\
\Delta \hat{T}\left( Z,Z^{\prime }\right) &=&-\left( \frac{1}{v}+\frac{%
\left( 1-\delta \right) \left( v-u\right) }{uv}\right) V_{0}-\frac{\left(
1-\delta \right) \left( v-u\right) \alpha }{V\left( Z,Z^{\prime }\right) s} 
\notag
\end{eqnarray}

Given our assumptions, the terms $V_{0}$ and $V$ are relatively large.
Moreover, $V_{0}$ measures the modification due to sources terms, and $V$
the backreaction of the system on the sources.

\subsection{Average shift}

\subsubsection{Equations for the average shift}

To solve (\ref{hfn}) we first compute the averages over space:%
\begin{eqnarray*}
\left\langle \Delta T\right\rangle &=&\left\langle \Delta T\left(
Z_{i},Z_{i}^{\prime }\right) \right\rangle _{\left( Z_{i},Z_{i}^{\prime
}\right) } \\
\left\langle \Delta \hat{T}\right\rangle &=&\left\langle \Delta \hat{T}%
\left( Z_{i},Z_{i}^{\prime }\right) \right\rangle _{\left(
Z_{i},Z_{i}^{\prime }\right) }
\end{eqnarray*}%
by averaging all quantities in (\ref{hfn}) over space. Appendix 1 shows that
these functions satify a system of equations:%
\begin{eqnarray}
\left\langle \Delta T\right\rangle &=&\frac{d}{\left\langle \Delta \hat{T}%
\right\rangle \left( 1+f\left\langle \Delta T\right\rangle \right) }
\label{vrq} \\
\left\langle \Delta \hat{T}\right\rangle &=&g\left\langle \Delta
T\right\rangle +\frac{h}{\left\langle \Delta \hat{T}\right\rangle \left(
1+f\left\langle \Delta T\right\rangle \right) }  \notag
\end{eqnarray}%
where the parameters are:%
\begin{eqnarray*}
d &=&-\frac{\alpha }{A_{1}\left\Vert \Delta \Gamma \right\Vert ^{2}} \\
f &=&A_{2}\frac{\left\Vert \Delta \Gamma \right\Vert ^{2}}{\left\langle
T\right\rangle } \\
g &=&-\left( \frac{1}{v}+\frac{\left( 1-\delta \right) \left( v-u\right) }{uv%
}\right) A\frac{\left\Vert \Delta \Gamma \right\Vert ^{2}}{\left\langle
T\right\rangle } \\
h &=&-\left\langle \frac{\left( 1-\delta \right) \left( v-u\right) }{s}%
\right\rangle \frac{\alpha }{A_{1}\left\Vert \Delta \Gamma \right\Vert ^{2}}%
=\left\langle \frac{\left( 1-\delta \right) \left( v-u\right) }{s}%
\right\rangle d
\end{eqnarray*}%
and the constants $A$, $A_{1}$, $A_{2}$ are defined by:%
\begin{eqnarray}
\left\langle V_{0}\left( Z,Z^{\prime }\right) \right\rangle &=&A\frac{%
\left\langle \Delta T\right\rangle }{\left\langle T\right\rangle }\left\Vert
\Delta \Gamma \right\Vert ^{2}  \label{dtv} \\
\left\langle V_{1}\left( Z,Z^{\prime }\right) \right\rangle
&=&A_{1}\left\langle \Delta \hat{T}\right\rangle \left\Vert \Delta \Gamma
\right\Vert ^{2}\text{, }\left\langle V_{2}\left( Z,Z^{\prime }\right)
\right\rangle =A_{2}\frac{\left\langle \Delta T\right\rangle }{\left\langle
T\right\rangle }\left\Vert \Delta \Gamma \right\Vert ^{2}  \notag
\end{eqnarray}%
The appendix provides estimations for $A$ and $A_{i}$ $i=1,2$.

Equations (\ref{vrq}) can be solved for $\left\langle \Delta \hat{T}%
\right\rangle $ as a function of $\left\langle \Delta T\right\rangle $: 
\begin{equation}
\left\langle \Delta \hat{T}\right\rangle =\left\langle \Delta T\right\rangle 
\frac{h+dg}{d}  \label{Rsll}
\end{equation}%
and $\left\langle \Delta T\right\rangle $ satifies:%
\begin{equation}
\left\langle \Delta \tilde{T}\right\rangle ^{3}+\left\langle \Delta \tilde{T}%
\right\rangle ^{2}-\frac{d^{2}f^{2}}{\left( h+dg\right) }=0  \label{qnc}
\end{equation}%
with:%
\begin{equation*}
\Delta \tilde{T}=f\Delta T
\end{equation*}

\subsubsection{Several type of solutions}

From equation (\ref{qnc}) we find the conditions for the solutions. A
particular case arises when $\left\Vert \Delta \Gamma \right\Vert ^{2}>>1$.
In such case:%
\begin{equation*}
-\frac{d^{2}f^{2}}{\left( h+dg\right) }<0
\end{equation*}%
and equation (\ref{qnc}) has a single negative root. Too many fluctuations
in connectivities leads ultimately to a lower shift in this variable. For $%
\left\Vert \Delta \Gamma \right\Vert ^{2}$ of order $1$ there are three
cases depending on $\frac{d^{2}}{\left( h+dg\right) }f^{2}$

\paragraph{1. Lowered connectivity}

If:%
\begin{equation*}
\frac{d^{2}}{\left( h+dg\right) }f^{2}<0
\end{equation*}%
equation (\ref{qnc}) has a single negative root. We can check that given our
assumption $s<<1$, this case corresponds to:%
\begin{eqnarray*}
&&\left\langle \check{T}\left( 1-\left( 1+\left\langle \left\vert \Psi
_{0}\right\vert ^{2}\right\rangle \frac{\left\langle \Delta T\right\rangle }{%
\left\langle T\right\rangle }\left\Vert \Delta \Gamma \right\Vert
^{2}\right) \check{T}\right) ^{-1}O\right\rangle \\
&=&\left\langle \left\langle \check{T}\left( 1-\left( 1+\left\langle
\left\vert \Psi _{0}\right\vert ^{2}\right\rangle \frac{\left\langle \Delta
T\right\rangle }{\left\langle T\right\rangle }\left\Vert \Delta \Gamma
\right\Vert ^{2}\right) \check{T}\right) ^{-1}O\right\rangle ^{\left( T,\hat{%
T},\theta ,Z,Z^{\prime }\right) }\right\rangle >0
\end{eqnarray*}

We show in appendix 1, that this quantity computes, in the continuous
approximation, the opposite effect between output and input spikes in the
connection process:%
\begin{equation*}
-\left( \omega _{0}\left( Z\right) \delta \omega \left( \theta -\frac{%
\left\vert Z-Z^{\prime }\right\vert }{c},Z^{\prime },\left\vert \Psi
\right\vert ^{2}\right) -\omega _{0}\left( Z^{\prime }\right) \delta \omega
\left( \theta ,Z,\left\vert \Psi \right\vert ^{2}\right) \right)
\end{equation*}%
When this term is positive, the following inequality is satisfied in
average: 
\begin{equation*}
\omega _{0}\left( Z^{\prime }\right) \delta \omega \left( \theta
,Z,\left\vert \Psi \right\vert ^{2}\right) >\omega _{0}\left( Z\right)
\delta \omega \left( \theta -\frac{\left\vert Z-Z^{\prime }\right\vert }{c}%
,Z^{\prime },\left\vert \Psi \right\vert ^{2}\right)
\end{equation*}%
It implies that incoming spikes are not matched by an equivalent amount of
output spikes: points $Z$ and $Z^{\prime }$ decorrelate and the connectivity
function is shifted to an lower value. Dynamically, the variations in
connectivity did not exhibit positive associations, leading to a decrease in
connections.

\paragraph{2. System backreaction and multiple solutions}

If:

\begin{equation*}
\frac{d^{2}}{\left( h+dg\right) }f^{2}>0\text{, }\frac{4}{27}-\frac{d^{2}}{%
\left( h+dg\right) }f^{2}>0
\end{equation*}%
There are three different real roots. Two of them are negative, and one is
postive. This case\ corresponds to a mild modification. The variation in
connectivities my induce a positive or negative shift, due to the
backreaction of the system, which tends to counteract any variation. Several
equilibrium shifts may result from the interactions.

\paragraph{3. Increased connectivity}

If:

\begin{equation*}
\frac{4}{27}-\frac{d^{2}}{\left( h+dg\right) }f^{2}<0
\end{equation*}%
there is a single positive root. This corresponds to the case:%
\begin{equation*}
\left\langle \check{T}\left( 1-\left( 1+\left\langle \left\vert \Psi
_{0}\right\vert ^{2}\right\rangle \frac{\left\langle \Delta T\right\rangle }{%
\left\langle T\right\rangle }\left\Vert \Delta \Gamma \right\Vert
^{2}\right) \check{T}\right) ^{-1}O\right\rangle <0
\end{equation*}%
which implies that the variations in connectivity induce a higher
correlations between input and output spikes. Points $Z$ and $Z^{\prime }$
bind and the connectivity function is shifted to a higher value. However,
this situation is not symetric with the first case.

The values of the shifts in case 1 and 3 are.

\begin{eqnarray}
\left\langle \Delta T\right\rangle &=&\frac{1}{f}\left( \frac{1}{3}+\sqrt[3]{%
\frac{d^{2}f^{2}}{\left( h+dg\right) }+\frac{1}{27}+\sqrt{\frac{\left( \frac{%
d^{2}f^{2}}{\left( h+dg\right) }+\frac{1}{27}\right) ^{2}}{4}+\left( \frac{1%
}{27}\right) ^{2}}}\right)  \label{Vr} \\
&&+\frac{1}{f}\left( \sqrt[3]{\frac{d^{2}f^{2}}{\left( h+dg\right) }+\frac{1%
}{27}-\sqrt{\frac{\left( \frac{d^{2}f^{2}}{\left( h+dg\right) }+\frac{1}{27}%
\right) ^{2}}{4}+\left( \frac{1}{27}\right) ^{2}}}\right)  \notag \\
\left\langle \Delta \hat{T}\right\rangle &=&\frac{h+dg}{d}  \notag
\end{eqnarray}

\subsection{Solving equation (\protect\ref{hfn}) for $\Delta T\left(
Z,Z^{\prime }\right) $ and $\Delta \hat{T}\left( Z,Z^{\prime }\right) $}

We use (\ref{DV}) and (\ref{Vfr}), (\ref{Vfn}), (\ref{Vft}), to write (\ref%
{hfn}):%
\begin{eqnarray}
\Delta T\left( Z,Z^{\prime }\right) &=&-\frac{\alpha }{\left( 1+\frac{%
A_{2}\left( Z,Z^{\prime }\right) \left\langle \Delta T\right\rangle }{%
\left\langle T\right\rangle }\left\Vert \Delta \Gamma \right\Vert
^{2}\right) A_{1}\left( Z,Z^{\prime }\right) \left\langle \Delta \hat{T}%
\right\rangle \left\Vert \Delta \Gamma \right\Vert ^{2}}  \label{pdtt} \\
\Delta \hat{T}\left( Z,Z^{\prime }\right) &=&-\left( \frac{1}{v}+\frac{%
\left( 1-\delta \right) \left( v-u\right) }{uv}\right) \frac{F\left(
Z,Z^{\prime }\right) A_{0}\left( Z,Z^{\prime }\right) \left\langle \Delta
T\right\rangle }{\left\langle T\right\rangle }\left\Vert \Delta \Gamma
\right\Vert ^{2}  \notag \\
&&-\frac{\left( 1-\delta \right) \left( v-u\right) \alpha }{s\left( 1+\frac{%
A_{2}\left( Z,Z^{\prime }\right) \left\langle \Delta T\right\rangle }{%
\left\langle T\right\rangle }\left\Vert \Delta \Gamma \right\Vert
^{2}\right) A_{1}\left( Z,Z^{\prime }\right) \left\langle \Delta \hat{T}%
\right\rangle \left\Vert \Delta \Gamma \right\Vert ^{2}}  \notag
\end{eqnarray}%
with $\left\langle \Delta T\right\rangle $ and $\left\langle \Delta \hat{T}%
\right\rangle $ given by (\ref{Vr}). The coefficients involvd in (\ref{pdtt}%
) are:%
\begin{eqnarray}
F\left( Z,Z^{\prime }\right) &=&\frac{\rho D\left( \theta \right)
\left\langle \hat{T}\right\rangle \left\vert \Psi _{0}\left( Z^{\prime
}\right) \right\vert ^{2}}{\omega _{0}\left( Z\right) }  \label{Vzp} \\
A_{0}\left( Z,Z^{\prime }\right) &=&\left\langle \check{T}\left( 1-\left(
1+\left\langle \left\vert \Psi _{0}\right\vert ^{2}\right\rangle \left( 1+%
\frac{\Delta T}{\left\langle T\right\rangle }\right) \left\Vert \Delta
\Gamma \right\Vert ^{2}\right) \check{T}\right) ^{-1}O\right\rangle ^{\left(
T,\hat{T},\theta ,Z,Z^{\prime }\right) }  \notag
\end{eqnarray}%
\begin{eqnarray}
&&V_{1}\left( Z,Z^{\prime },\Delta \Gamma \right)  \label{Vwp} \\
&\simeq &-k\left\langle \Delta \hat{T}\right\rangle \left\langle \left[
F\left( Z_{2},Z_{2}^{\prime }\right) \left[ \check{T}\left( 1-\left\langle
\left\vert \Psi _{0}\right\vert ^{2}\right\rangle \frac{\left\langle \Delta
T\right\rangle }{T}\left\Vert \Delta \Gamma \right\Vert ^{2}\right) ^{-1}O%
\right] \right] _{\left( T,\hat{T},\theta ,Z,Z^{\prime }\right)
}\right\rangle \left\Vert \Delta \Gamma \right\Vert ^{2}  \notag \\
&=&A_{1}\left( Z,Z^{\prime }\right) \left\langle \Delta \hat{T}\right\rangle
\left\Vert \Delta \Gamma \right\Vert ^{2}  \notag
\end{eqnarray}%
\begin{eqnarray}
V_{2}\left( Z,Z^{\prime },\Delta \Gamma \right) &=&\left\langle \left[ 
\check{T}\left( 1-\left( 1+\left\langle \left\vert \Psi _{0}\right\vert
^{2}\right\rangle \frac{\left\langle \Delta T\right\rangle }{T}\left\Vert
\Delta \Gamma \right\Vert ^{2}\right) \check{T}\right) ^{-1}\right] ^{\left(
T,\hat{T},\theta ,Z,Z^{\prime }\right) }\right\rangle \frac{\left\langle
\Delta T\right\rangle }{\left\langle T\right\rangle }\left\Vert \Delta
\Gamma \right\Vert ^{2}  \label{Vxp} \\
&=&A_{2}\left( Z,Z^{\prime }\right) \frac{\left\langle \Delta T\right\rangle 
}{\left\langle T\right\rangle }\left\Vert \Delta \Gamma \right\Vert ^{2} 
\notag
\end{eqnarray}%
and where the notation:%
\begin{equation*}
\left\langle \left[ O\right] _{\left( X\right) }\right\rangle \text{, }%
\left\langle \left[ O\right] ^{\left( X\right) }\right\rangle
\end{equation*}%
for an operator with kernel $O\left( X,Y\right) $ denotes $\int O\left(
X,Y\right) dY$ and $\int O\left( Y,X\right) dY$ respectively.

\subsection{Condition for existence of shifted state and associated shift}

\subsubsection{Condition for shifted state}

The stability of a shifted state depends on the sign of the associated
action in this state. For states (\ref{Slgg}) and (\ref{Sld}) the action is
given by:%
\begin{equation}
S\left( \Delta \Gamma \left( T,\hat{T},\theta ,Z,Z^{\prime }\right) \right)
+U_{\Delta \Gamma }\left( \left\Vert \Delta \Gamma \left( Z,Z^{\prime
}\right) \right\Vert ^{2}\right)  \label{CTN}
\end{equation}%
where the first term is given by (\ref{fcp}) and the potential by (\ref{Ptm}%
). A stable state is possible if (\ref{CTN}) is negative.

Given (\ref{Sdt}), the action functional (\ref{CTN}) reduces to:%
\begin{eqnarray}
&&S\left( \Delta \Gamma \left( T,\hat{T},\theta ,Z,Z^{\prime }\right)
\right) +U_{\Delta \Gamma }\left( \left\Vert \Delta \Gamma \left(
Z,Z^{\prime }\right) \right\Vert ^{2}\right)  \label{cnff} \\
&=&\int \Delta \Gamma ^{\dagger }\left( T,\hat{T},\theta ,Z,Z^{\prime
}\right) \left( \left( V_{1}\left( \theta ,Z,Z^{\prime },\Delta \Gamma
\right) \left( 1+V_{2}\left( \theta ,Z,Z^{\prime },\Delta \Gamma \right)
\right) \right) \Delta T\right) \Delta \Gamma \left( T,\hat{T},\theta
,Z,Z^{\prime }\right)  \notag \\
&&+U_{\Delta \Gamma }\left( \left\Vert \Delta \Gamma \left( Z,Z^{\prime
}\right) \right\Vert ^{2}\right) +\int \left( \alpha _{0}-\frac{\delta
U_{\Delta \Gamma }\left( \left\Vert \Delta \Gamma \left( Z,Z^{\prime
}\right) \right\Vert ^{2}\right) }{\delta \left\Vert \Delta \Gamma \left(
Z,Z^{\prime }\right) \right\Vert ^{2}}\right) \left\Vert \Delta \Gamma
\left( Z,Z^{\prime }\right) \right\Vert ^{2}  \notag
\end{eqnarray}%
Using computations similar to the previous paragraphs, we show in appendix 1
that this simplifies ultimately:%
\begin{equation*}
S\left( \Delta \Gamma \left( T,\hat{T},\theta ,Z,Z^{\prime }\right) \right)
=\int U_{\Delta \Gamma }\left( \left\Vert \Delta \Gamma \left( Z,Z^{\prime
}\right) \right\Vert ^{2}\right)
\end{equation*}%
and the minimization of:%
\begin{equation}
\int U_{\Delta \Gamma }\left( \left\Vert \Delta \Gamma \left( Z,Z^{\prime
}\right) \right\Vert ^{2}\right)  \label{ptg}
\end{equation}%
yields: 
\begin{equation*}
\left\Vert \Delta \Gamma \left( Z,Z^{\prime }\right) \right\Vert
^{2}=\left\Vert \Delta \Gamma \left( Z,Z^{\prime }\right) \right\Vert _{\min
}^{2}
\end{equation*}%
This value determines the norm $\Delta \Gamma \left( Z,Z^{\prime }\right) $
and, consequently, the values of the shifted connectivities. Since the
potential is negative at its minimum, this corresponds to a stable state.

However, a constraint has to be included to obtain admissible solutions for
the states (\ref{Slgg}) and (\ref{Sld}). Since $\Delta T$ and $\Delta \hat{T}
$ can be both positive or negative, and the parabolic cylinder functions are
not bounded for a negative argument and non integer parameters, we impose: 
\begin{equation*}
p_{1}=\frac{\alpha \delta }{u},p_{2}=\frac{\left( 1-\delta \right) \alpha }{v%
}
\end{equation*}%
to belong to $\frac{1}{2}+%
%TCIMACRO{\U{2115} }%
%BeginExpansion
\mathbb{N}
%EndExpansion
$. This allow to obtain integrable solutions $\Delta \Gamma \left( T,\hat{T}%
,Z,Z^{\prime }\right) $ over $%
%TCIMACRO{\U{211d} }%
%BeginExpansion
\mathbb{R}
%EndExpansion
^{2}$, we have the condition:%
\begin{equation}
\frac{\left( \alpha +\left( V\frac{s}{uv}V_{0}\right) \right) \delta }{u},%
\frac{\left( 1-\delta \right) \left( \alpha +\left( V\frac{s}{uv}%
V_{0}\right) \right) }{v}\in \frac{1}{2}+%
%TCIMACRO{\U{2115} }%
%BeginExpansion
\mathbb{N}
%EndExpansion
\label{cng}
\end{equation}%
The minimization of (\ref{ptg}) under constraint (\ref{cng}) is computed in
appendix 1. It yields the multiplier $\alpha $, the normalization factor $%
\left\Vert \Delta \Gamma \left( Z,Z^{\prime }\right) \right\Vert ^{2}$ and
the condition for a shifted state at $\left( Z,Z^{\prime }\right) $: 
\begin{equation}
\left\Vert \Delta \Gamma \left( Z,Z^{\prime }\right) \right\Vert ^{2}\simeq 
\frac{\left\Vert \Delta \Gamma \right\Vert ^{2}}{V}+\Delta \left\Vert \Delta
\Gamma \left( Z,Z^{\prime }\right) \right\Vert _{\min }^{2}-\frac{\Delta
\left( \frac{ku+lv}{1-\frac{sA\left\Vert \Delta \Gamma \right\Vert ^{2}}{%
\left\langle T\right\rangle }}\right) }{U_{\Delta \Gamma }^{\prime \prime
}\left( \left\Vert \Delta \Gamma \left( Z,Z^{\prime }\right) \right\Vert
_{\min }^{2}\right) }  \label{DL}
\end{equation}%
where $\left\Vert \Delta \Gamma \left( Z,Z^{\prime }\right) \right\Vert
_{\min }^{2}$ is the minimum of the potential $U_{\Delta \Gamma }$ at $%
\left( Z,Z^{\prime }\right) $, and:%
\begin{eqnarray*}
\Delta \left\Vert \Delta \Gamma \left( Z,Z^{\prime }\right) \right\Vert
_{\min }^{2} &=&\left\Vert \Delta \Gamma \left( Z,Z^{\prime }\right)
\right\Vert _{\min }^{2}-\left\langle \left\Vert \Delta \Gamma \left(
Z,Z^{\prime }\right) \right\Vert _{\min }^{2}\right\rangle \\
\Delta \left( \frac{ku+lv}{1-\frac{sA\left\Vert \Delta \Gamma \right\Vert
^{2}}{\left\langle T\right\rangle }}\right) &=&\frac{ku+lv}{1-\frac{%
sA\left\Vert \Delta \Gamma \right\Vert ^{2}}{\left\langle T\right\rangle }}%
-\left\langle \frac{ku+lv}{1-\frac{sA\left\Vert \Delta \Gamma \right\Vert
^{2}}{\left\langle T\right\rangle }}\right\rangle
\end{eqnarray*}%
are the deviations of these quantities from their averages over the entire
space. Formula(\ref{DL}) shows that due to the constraint, the existence of
a shifted state depends on the full system, through the overall norm $\frac{%
\left\Vert \Delta \Gamma \right\Vert ^{2}}{V}$.

The potential (\ref{ptg})\ for this value is equal to:%
\begin{eqnarray}
&&\int U_{\Delta \Gamma }\left( \left\Vert \Delta \Gamma \left( Z,Z^{\prime
}\right) \right\Vert ^{2}\right)  \label{fsp} \\
&=&\int U_{\Delta \Gamma }\left( \left\Vert \Delta \Gamma \left( Z,Z^{\prime
}\right) \right\Vert _{\min }^{2}\right)  \notag \\
&&+\frac{1}{2}\int U_{\Delta \Gamma }^{\prime \prime }\left( \left\Vert
\Delta \Gamma \left( Z,Z^{\prime }\right) \right\Vert _{\min }^{2}\right)
\left( \frac{\left\Vert \Delta \Gamma \right\Vert ^{2}}{V}-\left\langle
\left\Vert \Delta \Gamma \left( Z,Z^{\prime }\right) \right\Vert _{\min
}^{2}\right\rangle -\frac{\Delta \left( \frac{ku+lv}{1-\frac{sA\left\Vert
\Delta \Gamma \right\Vert ^{2}}{\left\langle T\right\rangle }}\right) }{%
U_{\Delta \Gamma }^{\prime \prime }\left( \left\Vert \Delta \Gamma \left(
Z,Z^{\prime }\right) \right\Vert _{\min }^{2}\right) }\right) ^{2}  \notag
\end{eqnarray}

Assuming a $U$ shape form for the potential so that $U_{\Delta \Gamma
}^{\prime \prime }\left( \left\Vert \Delta \Gamma \left( Z,Z^{\prime
}\right) \right\Vert _{\min }^{2}\right) >0$, implies that states with $%
\left\Vert \Delta \Gamma \left( Z,Z^{\prime }\right) \right\Vert ^{2}>0$
exists if the quantity (\ref{fsp}) is negative i.e.:%
\begin{equation}
\left\vert \frac{\left\Vert \Delta \Gamma \right\Vert ^{2}}{V}-\left\langle
\left\Vert \Delta \Gamma \left( Z,Z^{\prime }\right) \right\Vert _{\min
}^{2}\right\rangle -\frac{\Delta \left( \frac{ku+lv}{1-\frac{sA\left\Vert
\Delta \Gamma \right\Vert ^{2}}{\left\langle T\right\rangle }}\right) }{%
U_{\Delta \Gamma }^{\prime \prime }\left( \left\Vert \Delta \Gamma \left(
Z,Z^{\prime }\right) \right\Vert _{\min }^{2}\right) }\right\vert <\sqrt{-%
\frac{2U_{\Delta \Gamma }\left( \left\Vert \Delta \Gamma \left( Z,Z^{\prime
}\right) \right\Vert _{\min }^{2}\right) }{U_{\Delta \Gamma }^{\prime \prime
}\left( \left\Vert \Delta \Gamma \left( Z,Z^{\prime }\right) \right\Vert
_{\min }^{2}\right) }}  \label{CD}
\end{equation}%
Based on (\ref{CD}) and assuming by consistency that $\frac{\left\Vert
\Delta \Gamma \right\Vert ^{2}}{V}-\left\langle \left\Vert \Delta \Gamma
\left( Z,Z^{\prime }\right) \right\Vert _{\min }^{2}\right\rangle =0$,
appendix 1 shows that points for which:%
\begin{equation}
\left\vert \left( u+v\right) -\left\langle u+v\right\rangle \right\vert <%
\sqrt{-8U_{\Delta \Gamma }\left( \left\Vert \Delta \Gamma \left( Z,Z^{\prime
}\right) \right\Vert _{\min }^{2}\right) U_{\Delta \Gamma }^{\prime \prime
}\left( \left\Vert \Delta \Gamma \left( Z,Z^{\prime }\right) \right\Vert
_{\min }^{2}\right) }  \label{CP}
\end{equation}%
have a shifted states, others present $\Delta \Gamma \left( Z,Z^{\prime
}\right) =0$. Appendix 1 also studies the case for which the consistency is
not satisfied, i.e.:%
\begin{equation*}
\frac{\left\Vert \Delta \Gamma \right\Vert ^{2}}{V}-\left\langle \left\Vert
\Delta \Gamma \left( Z,Z^{\prime }\right) \right\Vert _{\min
}^{2}\right\rangle \neq 0
\end{equation*}

\subsubsection{Value of the shift}

Ultimately, we obtain, the value for $\alpha $:%
\begin{eqnarray*}
\alpha &=&\alpha _{0}-U_{\Delta \Gamma }^{\prime }\left( \Delta \Gamma
\left( Z,Z^{\prime }\right) \right) \\
&=&\frac{ku+lv}{1-\frac{sA\left\Vert \Delta \Gamma \right\Vert ^{2}}{%
\left\langle T\right\rangle }}
\end{eqnarray*}

and as a consequence the shifts are:%
\begin{eqnarray}
\left\langle \Delta T\right\rangle &\simeq &-\frac{\left( ku+lv\right) }{%
\left( 1-\frac{sA\left\Vert \Delta \Gamma \right\Vert ^{2}}{\left\langle
T\right\rangle }\right) A_{1}\left\langle \Delta \hat{T}\right\rangle
\left\Vert \Delta \Gamma \right\Vert ^{2}\left( 1+A_{2}\frac{\left\langle
\Delta T\right\rangle }{\left\langle T\right\rangle }\left\Vert \Delta
\Gamma \right\Vert ^{2}\right) } \\
\left\langle \Delta \hat{T}\right\rangle &\simeq &-\left( \frac{1}{%
\left\langle v\right\rangle }+\frac{\left( 1-\delta \right) \left(
v-u\right) }{\left\langle u\right\rangle \left\langle v\right\rangle }%
\right) A\frac{\left\langle \Delta T\right\rangle }{\left\langle
T\right\rangle }\left\Vert \Delta \Gamma \right\Vert ^{2}-\frac{ku+lv}{1-%
\frac{sA\left\Vert \Delta \Gamma \right\Vert ^{2}}{\left\langle
T\right\rangle }}\frac{\left\langle \frac{\left( 1-\delta \right) \left(
v-u\right) }{s}\right\rangle }{A_{1}\left\langle \Delta \hat{T}\right\rangle
\left\Vert \Delta \Gamma \right\Vert ^{2}\left( 1+A_{2}\frac{\left\langle
\Delta T\right\rangle }{\left\langle T\right\rangle }\left\Vert \Delta
\Gamma \right\Vert ^{2}\right) }  \notag
\end{eqnarray}%
\bigskip

\subsection{Dynamics for shifts}

To conclude this section, we restore the time dependency in the point
averages of the connectivity shift $\Delta T\left( Z,Z^{\prime }\right) $
and derive a propagation type equation for the perturbations in these
variables.

Start with (\ref{hfn}):%
\begin{eqnarray}
\Delta T\left( Z,Z^{\prime }\right) &=&-\frac{\alpha }{V\left( Z,Z^{\prime
}\right) } \\
\Delta \hat{T}\left( Z,Z^{\prime }\right) &=&-\left( \frac{1}{v}+\frac{%
\left( 1-\delta \right) \left( v-u\right) }{uv}\right) V_{0}-\frac{\left(
1-\delta \right) \left( v-u\right) \alpha }{V\left( Z,Z^{\prime }\right) s} 
\notag
\end{eqnarray}%
and expand this system around $\left\langle \Delta T\right\rangle $ and $%
\left\langle \Delta \hat{T}\right\rangle $ by redefining:%
\begin{eqnarray*}
\left( \Delta T\left( Z,Z^{\prime }\right) -\left\langle \Delta
T\right\rangle \right) \left\vert \Delta \Gamma \left( Z,Z^{\prime }\right)
\right\vert ^{2} &\rightarrow &\Delta T\left( Z,Z^{\prime }\right) \\
\left( \Delta \hat{T}\left( Z,Z^{\prime }\right) -\left\langle \Delta \hat{T}%
\right\rangle \right) \left\vert \Delta \Gamma \left( Z,Z^{\prime }\right)
\right\vert ^{2} &\rightarrow &\Delta \hat{T}\left( Z,Z^{\prime }\right)
\end{eqnarray*}%
and this expansion leads to\footnote{%
The integrals are implicitely over $Z_{1}$and $Z_{1}^{\prime }$.}:%
\begin{eqnarray}
\Delta T\left( Z,Z^{\prime }\right) &=&-\int K_{2}\left( Z,Z^{\prime
},Z_{1},Z_{1}^{\prime }\right) \Delta T\left( Z_{1},Z_{1}^{\prime }\right)
+\int K_{1}\left( Z,Z^{\prime },Z_{1},Z_{1}^{\prime }\right) \Delta \hat{T}%
\left( Z_{1},Z_{1}^{\prime }\right)  \label{DNs} \\
\Delta \hat{T}\left( Z,Z^{\prime }\right) &=&-\int \left( cK_{0}\left(
Z,Z^{\prime },Z_{1},Z_{1}^{\prime }\right) +\int dK_{2}\left( Z,Z^{\prime
},Z_{1},Z_{1}^{\prime }\right) \right) \Delta T\left( Z_{1},Z_{1}^{\prime
}\right)  \notag \\
&&+c\int K_{1}\left( Z,Z^{\prime },Z_{1},Z_{1}^{\prime }\right) \Delta \hat{T%
}\left( Z_{1},Z_{1}^{\prime }\right)  \notag
\end{eqnarray}%
whith kernels defined by:%
\begin{eqnarray}
K_{0}\left( Z,Z^{\prime },Z_{1},Z_{1}^{\prime }\right) &=&\frac{\rho D\left(
\theta \right) \left\langle \hat{T}\right\rangle \left\vert \Psi _{0}\left(
Z^{\prime }\right) \right\vert ^{2}}{\omega _{0}\left( Z\right) \left\langle
T\right\rangle \left\vert \Delta \Gamma \left( Z,Z^{\prime }\right)
\right\vert ^{2}}\left[ \check{T}\left( 1-\left( 1+\left\langle \left\vert
\Psi _{\Gamma }\right\vert ^{2}\right\rangle \right) \check{T}\right) ^{-1}O%
\right] _{\left( T_{1},\hat{T}_{1},\theta _{1},Z_{1},Z_{1}^{\prime }\right)
}^{\left( T,\hat{T},\theta ,Z,Z^{\prime }\right) }  \label{KNl} \\
K_{1}\left( Z,Z^{\prime },Z_{1},Z_{1}^{\prime }\right) &=&-k\frac{\alpha
\rho D\left( \theta \right) \left\langle \hat{T}\right\rangle \left\vert
\Psi _{0}\left( Z_{1}^{\prime }\right) \right\vert ^{2}\left[ \check{T}%
\left( 1-\left( 1+\left\langle \left\vert \Psi _{\Gamma }\right\vert
^{2}\right\rangle \right) \check{T}\right) ^{-1}O\right] _{\left( T,\hat{T}%
,\theta ,Z,Z^{\prime }\right) }^{\left( T_{1},\hat{T}_{1},\theta
_{1},Z_{1},Z_{1}^{\prime }\right) }}{\omega _{0}\left( Z_{1}\right) \left( 1+%
\frac{A_{2}\left\langle \Delta T\right\rangle }{\left\langle T\right\rangle }%
\left\Vert \Delta \Gamma \right\Vert ^{2}\right) \left( A_{1}\left\langle
\Delta \hat{T}\right\rangle \left\Vert \Delta \Gamma \right\Vert ^{2}\right)
^{2}\left\vert \Delta \Gamma \left( Z,Z^{\prime }\right) \right\vert ^{2}} 
\notag \\
K_{2}\left( Z,Z^{\prime },Z_{1},Z_{1}^{\prime }\right) &=&\frac{\alpha \left[
\check{T}\left( 1-\left( 1+\left\langle \left\vert \Psi _{\Gamma
}\right\vert ^{2}\right\rangle \right) \check{T}\right) ^{-1}\right]
_{\left( T_{1},\hat{T}_{1},\theta _{1},Z_{1},Z_{1}^{\prime }\right)
}^{\left( T,\hat{T},\theta ,Z,Z^{\prime }\right) }}{\left( 1+\frac{%
A_{2}\left\langle \Delta T\right\rangle }{\left\langle T\right\rangle }%
\left\Vert \Delta \Gamma \right\Vert ^{2}\right) ^{2}A_{1}\left\langle
\Delta \hat{T}\right\rangle \left\Vert \Delta \Gamma \right\Vert
^{2}\left\vert \Delta \Gamma \left( Z,Z^{\prime }\right) \right\vert ^{2}} 
\notag
\end{eqnarray}%
and constants:%
\begin{eqnarray*}
c &=&\left( \frac{1}{v}+\frac{\left( 1-\delta \right) \left( v-u\right) }{uv}%
\right) \\
d &=&\frac{\alpha \left( 1-\delta \right) \left( v-u\right) }{s}
\end{eqnarray*}%
Operator $O$ is local and is defined by:%
\begin{equation}
O=-\frac{\left\vert Z-Z^{\prime }\right\vert }{c}\nabla _{\theta _{1}}+\frac{%
\left( Z^{\prime }-Z\right) ^{2}}{2}\left( \frac{\nabla _{Z_{1}}^{2}}{2}+%
\frac{\nabla _{\theta _{1}}^{2}}{2c^{2}}-\frac{\nabla _{Z}^{2}\omega
_{0}\left( Z\right) }{2}\right)  \label{RT}
\end{equation}%
Kernels $K_{0}\left( Z,Z^{\prime },Z_{1},Z_{1}^{\prime }\right) $ and $%
K_{2}\left( Z,Z^{\prime },Z_{1},Z_{1}^{\prime }\right) $ are both backward
looking, so that we combine them to define:%
\begin{equation*}
K_{0,2}\left( Z,Z^{\prime },Z_{1},Z_{1}^{\prime }\right) =cK_{0}\left(
Z,Z^{\prime },Z_{1},Z_{1}^{\prime }\right) +dK_{2}\left( Z,Z^{\prime
},Z_{1},Z_{1}^{\prime }\right)
\end{equation*}%
and (\ref{DNs}) writes: 
\begin{eqnarray}
\Delta T\left( Z,Z^{\prime }\right) &=&-\int K_{2}\left( Z,Z^{\prime
},Z_{1},Z_{1}^{\prime }\right) \Delta T\left( Z_{1},Z_{1}^{\prime }\right)
+\int K_{1}\left( Z,Z^{\prime },Z_{1},Z_{1}^{\prime }\right) \Delta \hat{T}%
\left( Z_{1},Z_{1}^{\prime }\right)  \label{DNt} \\
\Delta \hat{T}\left( Z,Z^{\prime }\right) &=&-\int K_{0,2}\left( Z,Z^{\prime
},Z_{1},Z_{1}^{\prime }\right) \Delta T\left( Z_{1},Z_{1}^{\prime }\right)
+c\int K_{1}\left( Z,Z^{\prime },Z_{1},Z_{1}^{\prime }\right) \Delta \hat{T}%
\left( Z_{1},Z_{1}^{\prime }\right)  \notag
\end{eqnarray}%
The various kernels define operators $K_{1}$, $K_{2}$ and $K_{0,2}$. Both $%
K_{2}$ and $K_{0,2}$ backward looking, and $K_{1}$ forward looking. We write
(\ref{DNt}) as a system: 
\begin{eqnarray*}
\left( 1+K_{2}\right) \Delta T-K_{1}\Delta \hat{T} &=&0 \\
-K_{0,2}\Delta T+\left( 1+cK_{1}\right) \Delta \hat{T} &=&0
\end{eqnarray*}%
and replace:%
\begin{equation*}
\Delta \hat{T}=\left( 1+cK_{1}\right) ^{-1}K_{0,2}\Delta T
\end{equation*}%
to find the dynamics for connectivities: 
\begin{equation}
\left( \left( 1+K_{2}\right) -K_{1}\left( 1+cK_{1}\right)
^{-1}K_{0,2}\right) \Delta T=0  \label{DNc}
\end{equation}%
the kernel $K_{1}\left( 1+cK_{1}\right) ^{-1}$ is forward-looking and:%
\begin{equation*}
K_{1}\left( 1+cK_{1}\right) ^{-1}K_{0,2}\Delta T
\end{equation*}%
has the form:%
\begin{equation*}
\int d\left( Z_{2},Z_{2}^{\prime }\right) \left[ K_{1}\left( 1+cK_{1}\right)
^{-1}\right] \left( Z_{2},Z_{2}^{\prime },Z,Z^{\prime }\right) \int
K_{0,2}\left( Z,Z^{\prime },Z_{1},Z_{1}^{\prime }\right) \Delta T\left(
Z_{1},Z_{1}^{\prime }\right) d\left( Z_{1},Z_{1}^{\prime }\right)
\end{equation*}%
Given (\ref{KNl}), the first integral has the form:%
\begin{eqnarray*}
\int d\left( Z_{2},Z_{2}^{\prime }\right) \left[ K_{1}\left( 1+cK_{1}\right)
^{-1}\right] \left( Z_{2},Z_{2}^{\prime },Z,Z^{\prime }\right) &=&\int
d\left( Z_{2},Z_{2}^{\prime }\right) d\left( \hat{Z}_{2},\hat{Z}_{2}^{\prime
}\right) \left[ K_{1}\left( 1+cK_{1}\right) ^{-1}\right] \left( \hat{Z}_{2},%
\hat{Z}_{2}^{\prime },Z_{2},Z_{2}^{\prime }\right) \\
&&\times G\left( Z_{2},Z_{2}^{\prime },Z,Z^{\prime }\right) O\left(
Z,Z^{\prime }\right)
\end{eqnarray*}%
where $G\left( Z_{2},Z_{2}^{\prime },Z,Z^{\prime }\right) $ is the kernel of
the inverse operator of $O$. Writing:%
\begin{equation*}
C\left( Z,Z^{\prime }\right) =\int d\left( Z_{2},Z_{2}^{\prime }\right)
d\left( \hat{Z}_{2},\hat{Z}_{2}^{\prime }\right) \left[ K_{1}\left(
1+cK_{1}\right) ^{-1}\right] \left( \hat{Z}_{2},\hat{Z}_{2}^{\prime
},Z_{2},Z_{2}^{\prime }\right) G\left( Z_{2},Z_{2}^{\prime },Z,Z^{\prime
}\right)
\end{equation*}%
Equation (\ref{DNc}) reduces to:%
\begin{equation}
\left( 1+K_{2}-COK_{0,2}\right) \Delta T=0  \label{DNf}
\end{equation}%
As shown in appendix 1, the operator $O$ can be moved on the left of $\check{%
T}\left( 1-\left( 1+\left\langle \left\vert \Psi _{\Gamma }\right\vert
^{2}\right\rangle \right) \check{T}\right) ^{-1}$.Then factoring:%
\begin{equation*}
K_{\eta }\left( Z,Z^{\prime },Z_{1},Z_{1}^{\prime }\right) =\bar{K}_{\eta
}\left( Z,Z^{\prime }\right) \left[ \left( 1-\left( 1+\left\langle
\left\vert \Psi _{\Gamma }\right\vert ^{2}\right\rangle \right) \check{T}%
\right) ^{-1}\right] _{\left( T_{1},\hat{T}_{1},\theta
_{1},Z_{1},Z_{1}^{\prime }\right) }^{\left( T,\hat{T},\theta ,Z,Z^{\prime
}\right) }
\end{equation*}%
and multiplying (\ref{DNf}) by $\left( 1-\left( 1+\left\langle \left\vert
\Psi _{\Gamma }\right\vert ^{2}\right\rangle \right) \check{T}\right) $, we
obtain the following equation for:%
\begin{equation*}
\left( 1-\left( 1+\left\langle \left\vert \Psi _{\Gamma }\right\vert
^{2}\right\rangle \right) \check{T}\right) ^{-1}\Delta T\rightarrow \Delta T
\end{equation*}%
\begin{equation*}
\left( 1+\left( -\left( 1+\left\langle \left\vert \Psi _{\Gamma }\right\vert
^{2}\right\rangle \right) \hat{T}+\bar{K}_{2}\hat{T}-CO\bar{K}_{0,2}\hat{T}%
\right) \right) \Delta T=0
\end{equation*}%
\begin{eqnarray}
\bar{K}_{0,2}\left( Z,Z^{\prime }\right) &=&\frac{c\rho D\left( \theta
\right) \left\langle \hat{T}\right\rangle \left\vert \Psi _{0}\left(
Z^{\prime }\right) \right\vert ^{2}}{\omega _{0}\left( Z\right) \left\langle
T\right\rangle \left\vert \Delta \Gamma \left( Z,Z^{\prime }\right)
\right\vert ^{2}}O\left( Z,Z^{\prime }\right) +\frac{d\alpha }{\left( 1+%
\frac{A_{2}\left\langle \Delta T\right\rangle }{\left\langle T\right\rangle }%
\left\Vert \Delta \Gamma \right\Vert ^{2}\right) ^{2}A_{1}\left\langle
\Delta \hat{T}\right\rangle \left\Vert \Delta \Gamma \right\Vert
^{2}\left\vert \Delta \Gamma \left( Z,Z^{\prime }\right) \right\vert ^{2}} \\
\bar{K}_{2}\left( Z,Z^{\prime }\right) &=&\frac{\alpha }{\left( 1+\frac{%
A_{2}\left\langle \Delta T\right\rangle }{\left\langle T\right\rangle }%
\left\Vert \Delta \Gamma \right\Vert ^{2}\right) ^{2}A_{1}\left\langle
\Delta \hat{T}\right\rangle \left\Vert \Delta \Gamma \right\Vert
^{2}\left\vert \Delta \Gamma \left( Z,Z^{\prime }\right) \right\vert ^{2}} 
\notag
\end{eqnarray}

Ultimately, we use that in the local approximation, the kernel $\hat{T}$ can
be replaced by a differential operator of the form: 
\begin{equation*}
M=\tau -\tau _{1}\frac{\left\vert Z-Z^{\prime }\right\vert }{c}\nabla
_{\theta _{1}}+\frac{\left( Z^{\prime }-Z\right) ^{2}}{2}\left( \tau _{1}^{2}%
\frac{\nabla _{Z_{1}}^{2}}{2}+\tau _{1}^{2}\frac{\nabla _{\theta _{1}}^{2}}{%
2c^{2}}\right)
\end{equation*}%
and we obtain the differential equation:%
\begin{equation*}
\left( 1+\left( \bar{K}_{2}-\left( 1+\left\langle \left\vert \Psi _{\Gamma
}\right\vert ^{2}\right\rangle \right) -C_{2}\left( Z,Z^{\prime }\right)
O-C_{1}\left( Z,Z^{\prime }\right) O^{2}\right) M\right) \Delta T=0
\end{equation*}%
The factors are defined as:%
\begin{eqnarray*}
C_{1}\left( Z,Z^{\prime }\right) &=&\frac{d\alpha }{\left( 1+\frac{%
A_{2}\left\langle \Delta T\right\rangle }{\left\langle T\right\rangle }%
\left\Vert \Delta \Gamma \right\Vert ^{2}\right) ^{2}A_{1}\left\langle
\Delta \hat{T}\right\rangle \left\Vert \Delta \Gamma \right\Vert
^{2}\left\vert \Delta \Gamma \left( Z,Z^{\prime }\right) \right\vert ^{2}}%
C\left( Z,Z^{\prime }\right) \\
C_{2}\left( Z,Z^{\prime }\right) &=&\frac{c\rho D\left( \theta \right)
\left\langle \hat{T}\right\rangle \left\vert \Psi _{0}\left( Z^{\prime
}\right) \right\vert ^{2}}{\omega _{0}\left( Z\right) \left\langle
T\right\rangle \left\vert \Delta \Gamma \left( Z,Z^{\prime }\right)
\right\vert ^{2}}C\left( Z,Z^{\prime }\right)
\end{eqnarray*}%
This is a propagation equation, including fourth order corrections.

\section{Application 1 continued: First approximation approach}

In the perspective of developping an effective theory for large number of
collective states, i.e. groupes of connected cells, it is usefull te
reconsider the modification in the background states in the limit of
relatively small interactions. This will simplify the computations, and
provide some hints about the emergence of collective states. The saddle
points equations for the modifications of the background become second order
differential equations plus some potential. Condition of existence, shift in
connectivities as well as the form of the background states are easier to
derive compared to the previous section. This more convenient
representations will allow in the fourth paper of the series to develop the
field theory for interactions of large number of such backgrounds.

\subsection{Equation for background state}

We can consider the case $s<<1$ directly. Performing a change of variables
along with a shift in variables, we show in appendix 1 that (\ref{Sdt})
writes in first approximation:%
\begin{eqnarray}
&&0=\left( -\sigma _{\hat{T}}^{2}\nabla _{\hat{T}}^{2}+\frac{1}{4\sigma _{%
\hat{T}}^{2}}\left( \rho \left\vert \bar{\Psi}_{0}\left( Z,Z^{\prime
}\right) \right\vert ^{2}\Delta \hat{T}+\frac{\rho \left( V_{0}-\frac{\sigma
_{\hat{T}}^{2}}{\sigma _{T}^{2}}\lambda \Delta T\right) }{\omega _{0}\left(
Z\right) }\right) ^{2}\right) \Delta \Gamma \left( T,\hat{T},\theta
,Z,Z^{\prime }\right)  \label{fcn} \\
&&+\left( -\sigma _{T}^{2}\nabla _{T}^{2}+\frac{1}{4\sigma _{T}^{2}}\left( 
\frac{\Delta T-\lambda \tau \Delta \hat{T}}{\tau \omega _{0}\left( Z\right) }%
\right) ^{2}\right) \Delta \Gamma \left( T,\hat{T},\theta ,Z,Z^{\prime
}\right)  \notag \\
&&-\left( \frac{\rho \left\vert \bar{\Psi}_{0}\left( Z,Z^{\prime }\right)
\right\vert ^{2}}{2}+\frac{\left\vert \Psi \left( Z\right) \right\vert ^{2}}{%
2\tau \omega _{0}\left( Z\right) }+V\left( \theta ,Z,Z^{\prime },\Delta
\Gamma \right) \Delta T-\alpha \right) \Delta \Gamma \left( T,\hat{T},\theta
,Z,Z^{\prime }\right)  \notag
\end{eqnarray}%
with:%
\begin{equation}
V_{0}\left( Z,Z^{\prime }\right) =\left( \frac{\rho D\left( \theta \right)
\left\langle \hat{T}\right\rangle \left\vert \Psi _{0}\left( Z^{\prime
}\right) \right\vert ^{2}}{\omega _{0}\left( Z\right) }\hat{T}\left(
1-\left( 1+\left\langle \left\vert \Psi _{\Gamma }\right\vert
^{2}\right\rangle \right) \hat{T}\right) ^{-1}\left[ O\frac{\Delta
T\left\vert \Delta \Gamma \left( \theta _{1},Z_{1},Z_{1}^{\prime }\right)
\right\vert ^{2}}{T}\right] \right)  \label{cfh}
\end{equation}%
\begin{equation*}
V\left( \theta ,Z,Z^{\prime },\Delta \Gamma \right) =V_{1}\left( \theta
,Z,Z^{\prime },\Delta \Gamma \right) \left( 1+V_{2}\left( \theta
,Z,Z^{\prime },\Delta \Gamma \right) \right)
\end{equation*}%
with $V_{1}$\ and $V_{2}$\ given by (\ref{Vfn}) and (\ref{Vft})
respectively. Operator $O$ is defined by (\ref{RT}). Recall that $\alpha $
implements the constraint $\left\Vert \Delta \Gamma \right\Vert =\overline{%
\left\Vert \Delta \Gamma \right\Vert }$. As in the previous paragraph $%
\alpha $ stands for:%
\begin{equation}
\alpha _{0}+U^{\prime }\left( \left\vert \Delta \Gamma \left( Z,Z^{\prime
}\right) \right\vert ^{2}\right)  \label{lpg}
\end{equation}%
where $\alpha _{0}$ is the Lagrange multiplier for the overall constraint,
and $U\left( \Delta \Gamma \left( Z,Z^{\prime }\right) \right) $ is the
potential. However, it should be noted that here, we are considering the
case of weak interactions. As a consequence, we could omit any global
contraint on $\left\Vert \Delta \Gamma \right\Vert $ and set $\alpha _{0}=0$%
. This will be discussed at the end of the paragraph.

After diagonalization of the potential by a matrix $P=\left( 
\begin{array}{cc}
w_{1} & w_{2} \\ 
w_{1}^{\prime } & w_{2}^{\prime }%
\end{array}%
\right) $, whose components are given in the appendix, we show that this
background state equation (\ref{fcn}) becomes: 
\begin{eqnarray}
0 &=&\left( -\sigma _{\hat{T}}^{2}\nabla _{\hat{T}^{\prime }}^{2}+\frac{%
\lambda _{+}^{2}}{4\sigma _{\hat{T}}^{2}}\left( \Delta \hat{T}^{\prime
}-\Delta \hat{T}_{0}^{\prime }-\frac{w_{2}}{\lambda _{+}}V\right)
^{2}\right) \Delta \Gamma \left( T,\hat{T},\theta ,Z,Z^{\prime }\right)
\label{sdl1} \\
&&+\left( -\sigma _{T}^{2}\nabla _{T^{\prime }}^{2}+\frac{\lambda _{-}^{2}}{%
\sigma _{T}^{2}}\left( \Delta T^{\prime }-\Delta T_{0}^{\prime }-\frac{w_{1}%
}{\lambda _{-}}V\right) ^{2}\right) \Delta \Gamma \left( T,\hat{T},\theta
,Z,Z^{\prime }\right)  \notag \\
&&-\left( u+v+\left( \frac{w_{1}^{2}}{\lambda _{+}}V^{2}+\frac{w_{2}^{2}}{%
\lambda _{-}}V^{2}\right) -\alpha \right) \Delta \Gamma \left( T,\hat{T}%
,\theta ,Z,Z^{\prime }\right)  \notag
\end{eqnarray}%
where the coefficients are:%
\begin{equation*}
\lambda _{\pm }=\sqrt{\frac{1}{2}\left( u^{2}+v^{2}\right) +s^{2}\pm \frac{%
\left( u+v\right) }{2}\sqrt{\left( u-v\right) ^{2}+4s^{2}}}
\end{equation*}%
\begin{eqnarray*}
u &=&\frac{\left\vert \Psi _{0}\left( Z\right) \right\vert ^{2}}{\tau \omega
_{0}\left( Z\right) } \\
v &=&\rho \left\vert \bar{\Psi}_{0}\left( Z,Z^{\prime }\right) \right\vert
^{2} \\
s &=&-\frac{\lambda \left\vert \Psi _{0}\left( Z\right) \right\vert ^{2}}{%
\omega _{0}\left( Z\right) }\frac{\sigma _{\hat{T}}}{\sigma _{T}}
\end{eqnarray*}%
and:%
\begin{equation}
\left( \Delta T_{0},\Delta \hat{T}_{0}\right) \simeq \left( -\frac{\lambda
\tau V_{0}}{\sigma _{T}\omega _{0}\left( Z\right) \left\vert \bar{\Psi}%
_{0}\left( Z,Z^{\prime }\right) \right\vert ^{2}},\frac{\Delta T_{0}}{%
\lambda \tau }\frac{\sigma _{T}}{\sigma _{\hat{T}}}\right)  \label{Sfh}
\end{equation}%
The notation $\left( X^{\prime },\hat{X}^{\prime }\right) $ stands for the
coordinates of any vector in the diagonal basis of the potential:%
\begin{equation*}
\left( X^{\prime },\hat{X}^{\prime }\right) ^{t}=P^{-1}\left( X,\hat{X}%
\right)
\end{equation*}%
Note that:%
\begin{equation*}
\lambda _{+}+\lambda _{-}=u+v
\end{equation*}

\subsection{Solutions of (\protect\ref{sdl1})}

\subsubsection{Condition for non-trivial solutions}

The operators:%
\begin{equation*}
\left( -\sigma _{\hat{T}}^{2}\nabla _{\hat{T}^{\prime }}^{2}+\frac{\lambda
_{+}}{4\sigma _{\hat{T}}^{2}}\left( \Delta \hat{T}^{\prime }-\Delta \hat{T}%
_{0}^{\prime }-\frac{w_{2}}{\lambda _{+}}V\right) ^{2}\right)
\end{equation*}%
and:%
\begin{equation*}
\left( -\sigma _{T}^{2}\nabla _{T^{\prime }}^{2}+\frac{\lambda _{-}}{\sigma
_{T}^{2}}\left( \Delta T^{\prime }-\Delta T_{0}^{\prime }-\frac{w_{1}}{%
\lambda _{-}}V\right) ^{2}\right)
\end{equation*}%
are positive and the saddle point equation (\ref{sdl1}) has a non trivial
solution if:%
\begin{equation*}
u+v+\left( \frac{w_{1}^{2}}{\lambda _{+}}+\frac{w_{2}^{2}}{\lambda _{-}}%
\right) V^{2}\geqslant \frac{\lambda _{+}+\lambda _{-}}{2}+\alpha
\end{equation*}%
that is, if:%
\begin{equation*}
\frac{u+v}{2}+\left( \frac{w_{1}^{2}}{\lambda _{+}}+\frac{w_{2}^{2}}{\lambda
_{-}}\right) V^{2}\geqslant \alpha
\end{equation*}%
This establishes the condition for a shift in connectivity functions at each
point $\left( Z,Z^{\prime }\right) $. Considering that $u$ and $v$ depend on
the point, but $V^{2}$ depends on the field $\Delta \Gamma $ over the entire
space, this condition may not be achievable.

Specifically, for points such that:%
\begin{equation*}
\frac{u+v}{2}+\left( \frac{w_{1}^{2}}{\lambda _{+}}+\frac{w_{2}^{2}}{\lambda
_{-}}\right) V^{2}<\alpha
\end{equation*}%
then:%
\begin{equation*}
\Delta \Gamma \left( T,\hat{T},\theta ,Z,Z^{\prime }\right) =0
\end{equation*}%
whereas, for points such that:%
\begin{equation}
\frac{u+v}{2}+\left( \frac{w_{1}^{2}}{\lambda _{+}}+\frac{w_{2}^{2}}{\lambda
_{-}}\right) V^{2}-\alpha >0  \label{cnd}
\end{equation}%
there are several solutions to (\ref{sdl}).

\subsubsection{Form of the solutions}

Since the variables in (\ref{sdl}) are separated, a solution is a product of
solutions of:%
\begin{equation}
0=\left( -\sigma _{\hat{T}}^{2}\nabla _{\hat{T}^{\prime }}^{2}+\frac{\lambda
_{+}}{4\sigma _{\hat{T}}^{2}}\left( \Delta \hat{T}^{\prime }-\Delta \hat{T}%
_{0}^{\prime }-\frac{w_{2}}{\lambda _{+}}V\right) ^{2}-\left( \frac{1}{2}%
+p_{1}\right) \lambda _{+}\right) \Delta \Gamma \left( T,\hat{T},\theta
,Z,Z^{\prime }\right)  \notag
\end{equation}%
and:%
\begin{equation*}
0=\left( -\sigma _{T}^{2}\nabla _{T^{\prime }}^{2}+\frac{\lambda _{-}}{%
\sigma _{T}^{2}}\left( \Delta T^{\prime }-\Delta T_{0}^{\prime }-\frac{w_{1}%
}{\lambda _{-}}V\right) ^{2}-\left( \frac{1}{2}+p_{2}\right) \lambda
_{-}\right) \Delta \Gamma \left( T,\hat{T},\theta ,Z,Z^{\prime }\right)
\end{equation*}%
with $p_{1}>0$, $p_{2}>0$ and :%
\begin{equation*}
p_{1}\lambda _{+}+p_{2}\lambda _{-}=\frac{u+v}{2}+\left( \frac{w_{1}^{2}}{%
\lambda _{+}}V^{2}+\frac{w_{2}^{2}}{\lambda _{-}}V^{2}\right) -\alpha
\end{equation*}%
As a consequence, taking into account (\ref{Cgv}) and (\ref{Cgr}), Let $%
0<\delta <p_{1}\lambda _{+}+p_{2}\lambda _{-}$, the solutions have the form:%
\begin{eqnarray}
&&\Delta \Gamma _{\delta }\left( T,\hat{T},\theta ,Z,Z^{\prime }\right)
\label{Stn} \\
&=&\exp \left( -\frac{1}{4}\left( \mathbf{\Delta T-\Delta \bar{T}}\right)
^{t}\hat{U}\left( \mathbf{\Delta T-\Delta \bar{T}}\right) \right)  \notag \\
&&\times D_{\delta }\left( \left( \mathbf{\Delta T}^{\prime }\mathbf{-\Delta 
\bar{T}}^{\prime }\right) _{2}\frac{\sigma _{T}\lambda _{+}}{2}\left( 
\mathbf{\Delta T}^{\prime }\mathbf{-\Delta \bar{T}}^{\prime }\right)
_{2}\right) D_{p-\delta }\left( \left( \mathbf{\Delta T}^{\prime }\mathbf{%
-\Delta \bar{T}}^{\prime }\right) _{2}\frac{\sigma _{\hat{T}}\lambda _{-}}{2}%
\left( \mathbf{\Delta T}^{\prime }\mathbf{-\Delta \bar{T}}^{\prime }\right)
_{2}\right)  \notag
\end{eqnarray}%
\begin{eqnarray}
&&\Delta \Gamma _{\delta }^{\dagger }\left( T,\hat{T},\theta ,Z,Z^{\prime
}\right)  \label{Stt} \\
&=&\exp \left( \frac{1}{4}\left( \mathbf{\Delta T-\Delta \bar{T}}\right) ^{t}%
\hat{U}\left( \mathbf{\Delta T-\Delta \bar{T}}\right) \right)  \notag \\
&&\times D_{\delta }\left( \left( \mathbf{\Delta T}^{\prime }\mathbf{-\Delta 
\bar{T}}^{\prime }\right) _{2}\frac{\sigma _{T}\lambda _{+}}{2}\left( 
\mathbf{\Delta T}^{\prime }\mathbf{-\Delta \bar{T}}^{\prime }\right)
_{2}\right) D_{p-\delta }\left( \left( \mathbf{\Delta T}^{\prime }\mathbf{%
-\Delta \bar{T}}^{\prime }\right) _{2}\frac{\sigma _{\hat{T}}\lambda _{-}}{2}%
\left( \mathbf{\Delta T}^{\prime }\mathbf{-\Delta \bar{T}}^{\prime }\right)
_{2}\right)  \notag
\end{eqnarray}%
where the variables are:%
\begin{eqnarray}
\mathbf{\Delta T-\Delta \bar{T}} &=&\left( 
\begin{array}{c}
\Delta T-\Delta T_{0}-\Delta T_{1} \\ 
\Delta \hat{T}-\Delta \hat{T}_{0}-\Delta \hat{T}_{1}%
\end{array}%
\right)  \label{shn} \\
\mathbf{\Delta T}^{\prime }\mathbf{-\Delta \bar{T}}^{\prime } &=&\left( 
\begin{array}{c}
\Delta T^{\prime }-\Delta T_{0}^{\prime }-\frac{w_{1}}{\lambda _{+}}V \\ 
\Delta \hat{T}^{\prime }-\Delta \hat{T}_{0}^{\prime }-\frac{w_{2}}{\lambda
_{-}}V%
\end{array}%
\right) \mathbf{=}P^{-1}\left( \mathbf{\Delta T-\Delta \bar{T}}\right) 
\notag
\end{eqnarray}%
with parameters:%
\begin{eqnarray*}
\Delta T_{0} &\simeq &-\frac{\lambda \tau V_{0}}{\omega _{0}\left( Z\right)
\left\vert \bar{\Psi}_{0}\left( Z,Z^{\prime }\right) \right\vert ^{2}} \\
\Delta \hat{T}_{0} &\simeq &\frac{\Delta T_{0}}{\lambda \tau }
\end{eqnarray*}%
\begin{equation*}
\left( 
\begin{array}{c}
\Delta T_{1} \\ 
\Delta \hat{T}_{1}%
\end{array}%
\right) =PD^{-1}P^{-1}\left( 
\begin{array}{c}
V \\ 
0%
\end{array}%
\right) =U^{-1}\left( 
\begin{array}{c}
V \\ 
0%
\end{array}%
\right)
\end{equation*}%
and the matrix $\hat{U}$ given by:%
\begin{equation*}
\hat{U}=\left( 
\begin{array}{cc}
\frac{1}{\sigma _{T}} & 0 \\ 
0 & \frac{1}{\sigma _{\hat{T}}}%
\end{array}%
\right) U\left( 
\begin{array}{cc}
\frac{1}{\sigma _{T}} & 0 \\ 
0 & \frac{1}{\sigma _{\hat{T}}}%
\end{array}%
\right) =\left( 
\begin{array}{cc}
\frac{s^{2}+u^{2}}{\sigma _{T}^{2}} & -\frac{s\left( u+v\right) }{\sigma
_{T}\sigma _{\hat{T}}} \\ 
-\frac{s\left( u+v\right) }{\sigma _{T}\sigma _{\hat{T}}} & \frac{s^{2}+v^{2}%
}{\sigma _{\hat{T}}^{2}}%
\end{array}%
\right)
\end{equation*}

The solutions $\Delta \Gamma _{\delta }$and $\Delta \Gamma _{\delta
}^{\dagger }$ are defined for a pair of points $\left( Z,Z^{\prime }\right) $%
. We can now describe the potential modified background state globally.
Given (\ref{cnd}), these modifications of the system are thus defined by
considering the set:%
\begin{equation}
W=\left\{ \left( Z,Z^{\prime }\right) ,p_{1}\left( Z,Z^{\prime }\right)
\lambda _{+}+p_{2}\left( Z,Z^{\prime }\right) \lambda _{-}=\frac{u+v}{2}%
+\left( \frac{w_{1}^{2}}{\lambda _{+}}+\frac{w_{2}^{2}}{\lambda _{-}}\right)
V^{2}-\alpha >0\right\}  \label{WSt}
\end{equation}%
and by associating to each function $\delta \left( Z,Z^{\prime }\right)
:W\rightarrow $ $\left[ 0,p\left( Z,Z^{\prime }\right) \right] $, the
potential background state:%
\begin{equation*}
\prod\limits_{W}\Delta \Gamma _{\delta \left( Z,Z^{\prime }\right) }\left(
T,\hat{T},\theta ,Z,Z^{\prime }\right)
\end{equation*}%
and:%
\begin{equation*}
\prod\limits_{W}\Delta \Gamma _{\delta \left( Z,Z^{\prime }\right)
}^{\dagger }\left( T,\hat{T},\theta ,Z,Z^{\prime }\right)
\end{equation*}%
At each point of $W$, the shift in connectivity$\mathbf{\Delta \bar{T}}$ is
defined by (\ref{shn}).

\subsection{Restrictions to integer values of $p$}

However, since the variables $\Delta T$ and $\Delta \hat{T}^{\prime }$ can
be both positive or negative, solutions (\ref{Stn}) are not suitable for $%
\Delta T<<0$ or $\Delta \hat{T}<<0$. It implies that we have to reduce the
solutions to feasible ones. This corresponds in first approximation, to
impose $p_{1}$ and $p_{2}\in 
%TCIMACRO{\U{2115} }%
%BeginExpansion
\mathbb{N}
%EndExpansion
$. It is equivalent to impose $p\in 
%TCIMACRO{\U{2115} }%
%BeginExpansion
\mathbb{N}
%EndExpansion
$ and $\delta \in 
%TCIMACRO{\U{2115} }%
%BeginExpansion
\mathbb{N}
%EndExpansion
$ with $p-\delta \geqslant 0$. We will see below that this approximation can
be partially relaxed.

The solutions (\ref{Stn}) become:%
\begin{eqnarray}
&&\Delta \Gamma _{\delta }\left( T,\hat{T},\theta ,Z,Z^{\prime }\right)
\label{Sdn} \\
&=&\exp \left( -\frac{1}{2}\left( \mathbf{\Delta T-\Delta \bar{T}}\right)
^{t}\hat{U}\left( \mathbf{\Delta T-\Delta \bar{T}}\right) \right)  \notag \\
&&\times H_{p}\left( \left( \mathbf{\Delta T}^{\prime }\mathbf{-\Delta \bar{T%
}}^{\prime }\right) _{2}\frac{\sigma _{T}\lambda _{+}}{2\sqrt{2}}\left( 
\mathbf{\Delta T}^{\prime }\mathbf{-\Delta \bar{T}}^{\prime }\right)
_{2}\right) H_{p-\delta }\left( \left( \mathbf{\Delta T}^{\prime }\mathbf{%
-\Delta \bar{T}}^{\prime }\right) _{2}\frac{\sigma _{\hat{T}}\lambda _{-}}{2%
\sqrt{2}}\left( \mathbf{\Delta T}^{\prime }\mathbf{-\Delta \bar{T}}^{\prime
}\right) _{2}\right)  \notag
\end{eqnarray}%
and (\ref{Stt}):%
\begin{eqnarray}
&&\Delta \Gamma _{\delta }^{\dagger }\left( T,\hat{T},\theta ,Z,Z^{\prime
}\right)  \label{Sdd} \\
&=&H_{p}\left( \left( \mathbf{\Delta T}^{\prime }\mathbf{-\Delta \bar{T}}%
^{\prime }\right) _{2}\frac{\sigma _{T}\lambda _{+}}{2\sqrt{2}}\left( 
\mathbf{\Delta T}^{\prime }\mathbf{-\Delta \bar{T}}^{\prime }\right)
_{2}\right) H_{p-\delta }\left( \left( \mathbf{\Delta T}^{\prime }\mathbf{%
-\Delta \bar{T}}^{\prime }\right) _{2}\frac{\sigma _{\hat{T}}\lambda _{-}}{2%
\sqrt{2}}\left( \mathbf{\Delta T}^{\prime }\mathbf{-\Delta \bar{T}}^{\prime
}\right) _{2}\right)  \notag
\end{eqnarray}%
where $H_{p}$ and $H_{p-\delta }$ are Hermite polynomials.

We conclude by considering an example of solutions of (\ref{sdl}), and
consider $\left( Z,Z^{\prime }\right) $ such that:%
\begin{equation*}
\frac{u+v}{2}+\left( \frac{w_{1}^{2}}{\lambda _{+}}+\frac{w_{2}^{2}}{\lambda
_{-}}\right) V^{2}-\alpha =0
\end{equation*}%
Taking into account (\ref{Cgv}) and (\ref{Cgr}), the background state for (%
\ref{sdl}) is:%
\begin{equation*}
\Delta \Gamma \left( T,\hat{T},Z,Z^{\prime }\right) =\exp \left( -\frac{1}{2}%
\left( 
\begin{array}{c}
\Delta T^{\prime }-\Delta T_{0}^{\prime }-\frac{w_{2}}{\lambda _{+}}V \\ 
\Delta \hat{T}^{\prime }-\Delta \hat{T}_{0}^{\prime }-\frac{w_{1}}{\lambda
_{-}}V%
\end{array}%
\right) ^{t}D\left( 
\begin{array}{c}
\Delta T^{\prime }-\Delta T_{0}^{\prime }-\frac{w_{2}}{\lambda _{+}}V \\ 
\Delta \hat{T}^{\prime }-\Delta \hat{T}_{0}^{\prime }-\frac{w_{1}}{\lambda
_{-}}V%
\end{array}%
\right) \right)
\end{equation*}%
Coming back to the initial variables and reintroducing $\sigma _{T}$ and $%
\sigma _{\hat{T}}$, it yields:%
\begin{equation}
\Delta \Gamma \left( T,\hat{T},Z,Z^{\prime }\right) =\exp \left( -\frac{1}{2}%
\left( 
\begin{array}{c}
\Delta T-\Delta T_{0}-\Delta T_{1} \\ 
\Delta \hat{T}-\Delta \hat{T}_{0}-\Delta \hat{T}_{1}%
\end{array}%
\right) ^{t}\hat{U}\left( 
\begin{array}{c}
\Delta T-\Delta T_{0}-\Delta T_{1} \\ 
\Delta \hat{T}-\Delta \hat{T}_{0}-\Delta \hat{T}_{1}%
\end{array}%
\right) \right)  \label{bcs}
\end{equation}%
and for $\Delta \Gamma ^{\dagger }\left( T,\hat{T},Z,Z^{\prime }\right) $: 
\begin{subequations}
\begin{equation}
\Delta \Gamma ^{\dagger }\left( T,\hat{T},Z,Z^{\prime }\right) =1
\label{bcj}
\end{equation}

\subsection{Stability and condition for shifting state}

The possibility for a shifted state, i.e. a state for which $\left\vert
\Delta \Gamma \left( Z,Z^{\prime }\right) \right\vert ^{2}>0$ depends on the
value of the action for this state. If the corresponding action is negative,
the state $\Delta \Gamma \left( T,\hat{T},Z,Z^{\prime }\right) $ is the
minimum of the system. Otherwise, the state $\Delta \Gamma \left( T,\hat{T}%
,Z,Z^{\prime }\right) =0$ is the background state at point $\left(
Z,Z^{\prime }\right) $.

In states (\ref{Stn}) and (\ref{Stt}) the value of (\ref{fcp}) at $\left(
Z,Z^{\prime }\right) $ is obtained by a change of variables given in
appendix 2. It yields: 
\end{subequations}
\begin{eqnarray}
&&\Delta \Gamma ^{\dagger }\left( T,\hat{T},\theta ,Z,Z^{\prime }\right)
\left( -\sigma _{\hat{T}}^{2}\nabla _{\hat{T}}^{2}\right. \\
&&\left. +\frac{1}{4\sigma _{\hat{T}}^{2}}\left( \rho \left\vert \bar{\Psi}%
_{0}\left( Z,Z^{\prime }\right) \right\vert ^{2}\Delta \hat{T}+\frac{\rho
\left( V_{0}-\frac{\sigma _{\hat{T}}^{2}}{\sigma _{T}^{2}}\lambda \Delta
T\right) }{\omega _{0}\left( Z\right) }\right) ^{2}\right) \Delta \Gamma
\left( T,\hat{T},\theta ,Z,Z^{\prime }\right)  \notag \\
&&+\left( -\sigma _{T}^{2}\nabla _{T}^{2}+\frac{1}{4\sigma _{T}^{2}}\left( 
\frac{\Delta T-\lambda \tau \Delta \hat{T}}{\tau \omega _{0}\left( Z\right) }%
\right) ^{2}\right) \Delta \Gamma \left( T,\hat{T},\theta ,Z,Z^{\prime
}\right)  \notag \\
&&-\left( \frac{\rho \left\vert \bar{\Psi}_{0}\left( Z,Z^{\prime }\right)
\right\vert ^{2}}{2}+\frac{\left\vert \Psi \left( Z\right) \right\vert ^{2}}{%
2\tau \omega _{0}\left( Z\right) }-U\left( \left\vert \Gamma \left( \theta
,Z,Z^{\prime }\right) \right\vert ^{2}\right) \right) \Delta \Gamma \left( T,%
\hat{T},\theta ,Z,Z^{\prime }\right)  \notag
\end{eqnarray}%
and given the saddle point equation (see (\ref{fcn})), this reduces to:%
\begin{equation}
\Delta \Gamma ^{\dagger }\left( T,\hat{T},\theta ,Z,Z^{\prime }\right)
\left( V\left( \theta ,Z,Z^{\prime },\Delta \Gamma \right) \Delta T-\alpha
\right) \Delta \Gamma \left( T,\hat{T},\theta ,Z,Z^{\prime }\right)
\label{cnw}
\end{equation}%
Using (\ref{shn}):%
\begin{equation}
\left( 
\begin{array}{c}
\Delta T_{0} \\ 
\Delta \hat{T}_{0}%
\end{array}%
\right) +P\left( 
\begin{array}{c}
\frac{w_{1}}{\lambda _{+}}V \\ 
\frac{w_{2}}{\lambda _{-}}V%
\end{array}%
\right) =\left( 
\begin{array}{c}
\Delta T_{0} \\ 
\Delta \hat{T}_{0}%
\end{array}%
\right) +\left( \frac{w_{1}^{2}}{\lambda _{-}}+\frac{w_{2}^{2}}{\lambda _{+}}%
\right) V  \label{shf}
\end{equation}%
action (\ref{cnw}) is equal to:%
\begin{equation*}
\Delta \Gamma ^{\dagger }\left( T,\hat{T},\theta ,Z,Z^{\prime }\right)
\left( U\left( \left\vert \Gamma \left( \theta ,Z,Z^{\prime }\right)
\right\vert ^{2}\right) +\Delta T_{0}+\left( \frac{w_{1}^{2}}{\lambda _{-}}+%
\frac{w_{2}^{2}}{\lambda _{+}}\right) V^{2}-\alpha \right) \Delta \Gamma
\left( T,\hat{T},\theta ,Z,Z^{\prime }\right)
\end{equation*}%
Using (\ref{sdl}) and (\ref{WSt}), the action at point $\left( Z,Z^{\prime
}\right) $ reduces to:%
\begin{eqnarray}
&&\int \left( U\left( \left\vert \Gamma \left( \theta ,Z,Z^{\prime }\right)
\right\vert ^{2}\right) +\left( \frac{w_{1}^{2}}{\lambda _{+}}V^{2}+\frac{%
w_{2}^{2}}{\lambda _{-}}V^{2}\right) +\left( \Delta T_{0}-\alpha \right)
\right) \left\vert \Delta \Gamma \left( T,\hat{T},Z,Z^{\prime }\right)
\right\vert ^{2}d\left( T,\hat{T}\right)  \label{fvm} \\
&=&\left( U\left( \left\vert \Gamma \left( \theta ,Z,Z^{\prime }\right)
\right\vert ^{2}\right) +\left( \frac{1}{2}+p_{1}\right) \lambda _{+}+\left( 
\frac{1}{2}+p_{2}\right) \lambda _{-}+\Delta T_{0}-\left( u+v\right) \right)
\left\vert \Delta \Gamma \left( Z,Z^{\prime }\right) \right\vert ^{2}  \notag
\end{eqnarray}%
and over the whole space:%
\begin{equation}
S=\int \left( U\left( \left\vert \Gamma \left( \theta ,Z,Z^{\prime }\right)
\right\vert ^{2}\right) +\Delta T_{0}-\frac{u+v}{2}\right) \left\vert \Delta
\Gamma \left( Z,Z^{\prime }\right) \right\vert ^{2}  \label{fvn}
\end{equation}%
As a consequence of (\ref{fvm}), a state with $\left\vert \Delta \Gamma
\left( Z,Z^{\prime }\right) \right\vert ^{2}\neq 0$ exists and is stable, if
at point $\left( Z,Z^{\prime }\right) $ where (\ref{fvm}) is minimum,
condition:%
\begin{equation}
\left( u\left( Z,Z^{\prime }\right) +v\left( Z,Z^{\prime }\right) \right)
-\Delta T_{0}-U\left( \left\vert \Delta \Gamma \left( Z,Z^{\prime }\right)
\right\vert ^{2}\right) >\frac{u+v}{2}  \label{CDn}
\end{equation}%
is realized. In such cases, the minimum is reached for $p=0$, and the
background state is given by expressions (\ref{bcs}) and (\ref{bcj}).

On the contrary, if 
\begin{equation*}
\left( u\left( Z,Z^{\prime }\right) +v\left( Z,Z^{\prime }\right) \right)
-\Delta T_{0}-U\left( \left\vert \Delta \Gamma \left( Z,Z^{\prime }\right)
\right\vert ^{2}\right) <\frac{u+v}{2}
\end{equation*}%
then, $\left\vert \Delta \Gamma \left( Z,Z^{\prime }\right) \right\vert
^{2}=0$.

The minimization of (\ref{fvm}) is implemented under the constraint (\ref%
{cnd}) with $p=0$:%
\begin{eqnarray}
0 &=&\frac{u+v}{2}+\left( \frac{w_{1}^{2}}{\lambda _{+}}+\frac{w_{2}^{2}}{%
\lambda _{-}}\right) V^{2}-\alpha  \label{cnT} \\
&=&\frac{u+v}{2}+\left( \frac{w_{1}^{2}}{\lambda _{+}}+\frac{w_{2}^{2}}{%
\lambda _{-}}\right) V^{2}-\alpha _{0}+U^{\prime }\left( \left\vert \Delta
\Gamma \left( Z,Z^{\prime }\right) \right\vert ^{2}\right)  \notag
\end{eqnarray}

Given that $\Delta T_{0}$ in (\ref{fvm}) depends on $\Delta T$ and thus on $%
\left\vert \Delta \Gamma \left( Z,Z^{\prime }\right) \right\vert ^{2}$, we
have to compute this quantity to derive the shifted states minimizing (\ref%
{fvm}).

\subsection{Estimation of $\left\langle \Delta T\right\rangle $, $%
\left\langle \Delta \hat{T}\right\rangle $}

To find $\Delta T$, we first compute its average $\left\langle \Delta
T\right\rangle $ over all space. We use (\ref{dtv}) to estimate $%
\left\langle \Delta T\right\rangle $: 
\begin{eqnarray}
\left\langle V_{0}\left( Z,Z^{\prime }\right) \right\rangle &=&A\frac{%
\left\langle \Delta T\right\rangle }{\left\langle T\right\rangle }\left\Vert
\Delta \Gamma \right\Vert ^{2} \\
\left\langle V_{1}\left( Z,Z^{\prime }\right) \right\rangle
&=&A_{1}\left\langle \Delta \hat{T}\right\rangle \left\Vert \Delta \Gamma
\right\Vert ^{2}\text{, }\left\langle V_{2}\left( Z,Z^{\prime }\right)
\right\rangle =A_{2}\frac{\left\langle \Delta T\right\rangle }{\left\langle
T\right\rangle }\left\Vert \Delta \Gamma \right\Vert ^{2}  \notag
\end{eqnarray}%
Using (\ref{cfh}) and (\ref{Sfh}) leads to:%
\begin{eqnarray}
\Delta T_{0}\left( Z,Z^{\prime }\right) &=&\frac{s}{uv}V_{0} \\
\Delta \hat{T}_{0}\left( Z,Z^{\prime }\right) &=&-\frac{1}{v}V_{0}  \notag
\end{eqnarray}%
Taking the average over space of (\ref{shf}) yields the equation for the
average shift:%
\begin{eqnarray*}
\left\langle \Delta T\right\rangle &\simeq &\left\langle \Delta
T_{0}\right\rangle +\left\langle \frac{w_{1}^{2}}{\lambda _{-}}+\frac{%
w_{2}^{2}}{\lambda _{+}}\right\rangle \left\langle V\right\rangle \\
&\simeq &\left\langle \frac{s}{uv}\right\rangle A\frac{\left\langle \Delta
T\right\rangle }{\left\langle T\right\rangle }\left\Vert \Delta \Gamma
\right\Vert ^{2}+\left\langle \frac{w_{1}^{2}}{\lambda _{-}}+\frac{w_{2}^{2}%
}{\lambda _{+}}\right\rangle A_{1}\left\langle \Delta \hat{T}\right\rangle
\left\Vert \Delta \Gamma \right\Vert ^{2}\left( 1+A_{2}\frac{\left\langle
\Delta T\right\rangle }{\left\langle T\right\rangle }\left\Vert \Delta
\Gamma \right\Vert ^{2}\right)
\end{eqnarray*}%
\begin{eqnarray*}
\left\langle \Delta \hat{T}\right\rangle &=&\left\langle \Delta \hat{T}%
_{0}\right\rangle +\left( \frac{w_{1}w_{1}^{\prime }}{\lambda _{-}}+\frac{%
w_{2}w_{2}^{\prime }}{\lambda _{+}}\right) \left\langle V\right\rangle \\
&\simeq &-\frac{1}{v}A\frac{\left\langle \Delta T\right\rangle }{%
\left\langle T\right\rangle }\left\Vert \Delta \Gamma \right\Vert
^{2}+\left\langle \frac{w_{1}w_{1}^{\prime }}{\lambda _{-}}+\frac{%
w_{2}w_{2}^{\prime }}{\lambda _{+}}\right\rangle A_{1}\left\langle \Delta 
\hat{T}\right\rangle \left\Vert \Delta \Gamma \right\Vert ^{2}\left( 1+A_{2}%
\frac{\left\langle \Delta T\right\rangle }{\left\langle T\right\rangle }%
\left\Vert \Delta \Gamma \right\Vert ^{2}\right)
\end{eqnarray*}%
Combination of these equations yields $\left\langle \Delta \hat{T}%
\right\rangle $\ as a function of $\left\langle \Delta T\right\rangle $: 
\begin{equation}
\left\langle \Delta \hat{T}\right\rangle =\hat{A}\left\langle \Delta
T\right\rangle  \label{dlt}
\end{equation}%
with:%
\begin{equation}
\hat{A}=\left( \frac{\left\langle \frac{w_{1}w_{1}^{\prime }}{\lambda _{-}}+%
\frac{w_{2}w_{2}^{\prime }}{\lambda _{+}}\right\rangle }{\left\langle \frac{%
w_{1}^{2}}{\lambda _{-}}+\frac{w_{2}^{2}}{\lambda _{+}}\right\rangle }\left(
1-\left\langle \frac{s}{uv}\right\rangle A\frac{\left\Vert \Delta \Gamma
\right\Vert ^{2}}{\left\langle T\right\rangle }\right) -\frac{1}{v}A\frac{%
\left\Vert \Delta \Gamma \right\Vert ^{2}}{\left\langle T\right\rangle }%
\right)  \label{HT}
\end{equation}%
At our oder of approximation, since $s<<\left( u,v\right) $, we have:%
\begin{eqnarray*}
\frac{w_{1}w_{1}^{\prime }}{\lambda _{-}}+\frac{w_{2}w_{2}^{\prime }}{%
\lambda _{+}} &=&-\sqrt{\frac{1}{2}\left( 1+\sqrt{\frac{\left( \frac{\left(
v-u\right) }{2s}\right) ^{2}}{1+\left( \frac{\left( v-u\right) }{2s}\right)
^{2}}}\right) }\sqrt{\frac{1}{2}\left( 1-\sqrt{\frac{\left( \frac{\left(
v-u\right) }{2s}\right) ^{2}}{1+\left( \frac{\left( v-u\right) }{2s}\right)
^{2}}}\right) }\left( \lambda _{+}-\lambda _{-}\right) \\
&=&-\frac{1}{2}\sqrt{\left( 1-\frac{\left( \frac{\left( v-u\right) }{2s}%
\right) ^{2}}{1+\left( \frac{\left( v-u\right) }{2s}\right) ^{2}}\right) }%
\left\vert v-u\right\vert \simeq s
\end{eqnarray*}%
and:%
\begin{equation}
\hat{A}\simeq -\frac{1}{v}A\frac{\left\Vert \Delta \Gamma \right\Vert ^{2}}{%
\left\langle T\right\rangle }  \label{fht}
\end{equation}%
Equation (\ref{dlt}) this leads to the following first approximation for $%
\left\langle \Delta T\right\rangle $:%
\begin{equation}
\left\langle \Delta T\right\rangle \simeq \left( \frac{1-\frac{\left\langle 
\frac{s}{uv}\right\rangle A\left\Vert \Delta \Gamma \right\Vert ^{2}}{%
\left\langle T\right\rangle }}{\left\vert A_{1}\hat{A}\right\vert
\left\langle \frac{w_{1}^{2}}{\lambda _{-}}+\frac{w_{2}^{2}}{\lambda _{+}}%
\right\rangle \left\Vert \Delta \Gamma \right\Vert ^{2}}-1\right) \frac{%
\left\langle T\right\rangle }{A_{2}\left\Vert \Delta \Gamma \right\Vert ^{2}}
\label{ntd}
\end{equation}%
Given that:%
\begin{equation*}
\left\langle \frac{w_{1}^{2}}{\lambda _{-}}+\frac{w_{2}^{2}}{\lambda _{+}}%
\right\rangle \simeq \frac{1}{\lambda _{-}}\simeq \frac{1}{\max \left(
u,v\right) }
\end{equation*}%
and using (\ref{fht}), formula (\ref{ntd}) reduces to:%
\begin{equation*}
\left\langle \Delta T\right\rangle \simeq \left( \frac{\left( 1-\left\langle 
\frac{s}{uv}\right\rangle A\frac{\left\Vert \Delta \Gamma \right\Vert ^{2}}{%
\left\langle T\right\rangle }\right) v\max \left( u,v\right) }{\left\vert
A_{1}\right\vert A\left\Vert \Delta \Gamma \right\Vert ^{4}}\left\langle
T\right\rangle -1\right) \frac{\left\langle T\right\rangle }{A_{2}\left\Vert
\Delta \Gamma \right\Vert ^{2}}
\end{equation*}%
Given our order of approximation this is positive, except if: 
\begin{equation*}
\frac{\left( 1-\left\langle \frac{s}{uv}\right\rangle A\frac{\left\Vert
\Delta \Gamma \right\Vert ^{2}}{\left\langle T\right\rangle }\right) v\max
\left( u,v\right) }{\left\vert A_{1}\right\vert A\left\Vert \Delta \Gamma
\right\Vert ^{4}}\left\langle T\right\rangle <1
\end{equation*}%
In most case $u<v$, so that the dependency in the background parameters are
of order:%
\begin{equation}
\left\langle \Delta T\right\rangle \simeq \frac{\omega _{0}\left( Z\right)
\left\langle T\right\rangle }{\rho D\left( \theta \right) \left\langle \hat{T%
}\right\rangle \left\vert \Psi _{0}\left( Z^{\prime }\right) \right\vert
^{2}k\underline{A_{1}}\left\Vert \Delta \Gamma \right\Vert ^{6}}\left\langle
\rho \frac{\left\vert \bar{\Psi}_{0}\left( Z,Z^{\prime }\right) \right\vert
^{2}}{A}\right\rangle ^{2}\left\langle T\right\rangle  \label{VRP}
\end{equation}%
with:%
\begin{equation*}
A_{1}\left( Z,Z^{\prime }\right) =\frac{\rho D\left( \theta \right)
\left\langle \hat{T}\right\rangle \left\vert \Psi _{0}\left( Z^{\prime
}\right) \right\vert ^{2}}{\omega _{0}\left( Z\right) }\underline{A_{1}}%
\left( Z,Z^{\prime }\right)
\end{equation*}%
and:%
\begin{equation*}
\underline{A_{1}}\left( Z,Z^{\prime }\right) =\left\langle \left[ F\left(
Z_{2},Z_{2}^{\prime }\right) \left[ \check{T}\left( 1-\left\langle
\left\vert \Psi _{0}\right\vert ^{2}\right\rangle \frac{\left\langle \Delta
T\right\rangle }{T}\left\Vert \Delta \Gamma \right\Vert ^{2}\right) ^{-1}O%
\right] \right] _{\left( T,\hat{T},\theta ,Z,Z^{\prime }\right)
}\right\rangle
\end{equation*}

\subsection{Estimation of $\Delta T\left( Z,Z^{\prime }\right) $ and $\Delta 
\hat{T}\left( Z,Z^{\prime }\right) $}

Using (\ref{Vzp}), (\ref{Vwp}), (\ref{Vxp}), we obtain $\Delta T\left(
Z,Z^{\prime }\right) $ and $\Delta \hat{T}\left( Z,Z^{\prime }\right) $:%
\begin{eqnarray*}
\Delta T\left( Z,Z^{\prime }\right) &=&\Delta T_{0}\left( Z,Z^{\prime
}\right) +\left( \frac{w_{1}^{2}}{\lambda _{-}}+\frac{w_{2}^{2}}{\lambda _{+}%
}\right) V \\
&=&\frac{s}{uv}V_{0}+\left( \frac{w_{1}^{2}}{\lambda _{-}}+\frac{w_{2}^{2}}{%
\lambda _{+}}\right) V_{1}\left( 1+V_{2}\right) \\
\Delta \hat{T}\left( Z,Z^{\prime }\right) &=&-\frac{1}{v}V_{0}+\left( \frac{%
w_{1}w_{1}^{\prime }}{\lambda _{-}}+\frac{w_{2}w_{2}^{\prime }}{\lambda _{+}}%
\right) V_{1}\left( 1+V_{2}\right)
\end{eqnarray*}%
Given (\ref{HT}) and (\ref{ntd}) writes:%
\begin{eqnarray}
\Delta T\left( Z,Z^{\prime }\right) &=&\left( \frac{sA_{0}\left( Z,Z^{\prime
}\right) }{uv\left\langle T\right\rangle }+\left( \frac{w_{1}^{2}}{\lambda
_{-}}+\frac{w_{2}^{2}}{\lambda _{+}}\right) \frac{A_{1}\left( Z,Z^{\prime
}\right) \hat{A}}{\left\langle A_{1}\hat{A}\right\rangle }\left( \frac{1-%
\frac{\left\langle \frac{s}{uv}\right\rangle A\left\Vert \Delta \Gamma
\right\Vert ^{2}}{\left\langle T\right\rangle }}{\left\langle \frac{w_{1}^{2}%
}{\lambda _{-}}+\frac{w_{2}^{2}}{\lambda _{+}}\right\rangle \left\Vert
\Delta \Gamma \right\Vert ^{2}}\right) \right) \left\Vert \Delta \Gamma
\right\Vert ^{2}\left\langle \Delta T\right\rangle  \label{Vrs} \\
\Delta \hat{T}\left( Z,Z^{\prime }\right) &=&\left( -\frac{A_{0}\left(
Z,Z^{\prime }\right) }{v\left\langle T\right\rangle }+\left( \frac{%
w_{1}w_{1}^{\prime }}{\lambda _{-}}+\frac{w_{2}w_{2}^{\prime }}{\lambda _{+}}%
\right) \frac{A_{1}\left( Z,Z^{\prime }\right) \hat{A}}{\left\langle A_{1}%
\hat{A}\right\rangle }\left( \frac{1-\frac{\left\langle \frac{s}{uv}%
\right\rangle A\left\Vert \Delta \Gamma \right\Vert ^{2}}{\left\langle
T\right\rangle }}{\left\langle \frac{w_{1}^{2}}{\lambda _{-}}+\frac{w_{2}^{2}%
}{\lambda _{+}}\right\rangle \left\Vert \Delta \Gamma \right\Vert ^{2}}%
\right) \right) \left\Vert \Delta \Gamma \right\Vert ^{2}\left\langle \Delta
T\right\rangle  \notag
\end{eqnarray}%
With our approximations and using (\ref{fht}) this becomes:%
\begin{eqnarray*}
\Delta T\left( Z,Z^{\prime }\right) &=&\frac{A_{1}\left( Z,Z^{\prime
}\right) A\left( Z,Z^{\prime }\right) \left\langle v^{2}\right\rangle }{%
\left\langle A_{1}\left( Z,Z^{\prime }\right) \right\rangle \left\langle
A_{0}\left( Z,Z^{\prime }\right) \right\rangle v^{2}}\left\langle \Delta
T\right\rangle \\
\Delta \hat{T}\left( Z,Z^{\prime }\right) &=&\frac{A_{0}\left( Z,Z^{\prime
}\right) }{\left\langle A_{0}\left( Z,Z^{\prime }\right) \right\rangle }%
\left\langle \Delta \hat{T}\right\rangle
\end{eqnarray*}

\subsection{Minimization of (\protect\ref{fvm})}

The minimization of (\ref{fvm}) with constraint (\ref{cnT}) is thus:%
\begin{equation}
U^{\prime }\left( \left\vert \Gamma \left( \theta ,Z,Z^{\prime }\right)
\right\vert ^{2}\right) +\Delta T_{0}-\frac{u+v}{2}+\lambda \frac{\delta
h\left( \Delta \Gamma \left( Z,Z^{\prime }\right) ,\left( Z,Z^{\prime
}\right) \right) }{\delta \Delta \Gamma \left( Z,Z^{\prime }\right) }=0
\label{frN}
\end{equation}%
and the constraint:%
\begin{equation}
h\left( \Delta \Gamma \left( Z,Z^{\prime }\right) ,\left( Z,Z^{\prime
}\right) \right) =\frac{u+v}{2}+\left( \frac{w_{1}^{2}}{\lambda _{+}}+\frac{%
w_{2}^{2}}{\lambda _{-}}\right) V^{2}\left( Z,Z^{\prime }\right) -\alpha
_{0}+U^{\prime }\left( \left\vert \Delta \Gamma \left( Z,Z^{\prime }\right)
\right\vert ^{2}\right) =0  \label{Scn}
\end{equation}%
with:%
\begin{equation*}
V\left( Z,Z^{\prime }\right) =A_{1}\left( Z,Z^{\prime }\right) \left\langle
\Delta \hat{T}\right\rangle \left\Vert \Delta \Gamma \right\Vert ^{2}\left(
1+A_{2}\left( Z,Z^{\prime }\right) \frac{\left\langle \Delta T\right\rangle 
}{\left\langle T\right\rangle }\left\Vert \Delta \Gamma \right\Vert
^{2}\right)
\end{equation*}%
so that:%
\begin{equation*}
\frac{\delta h\left( \Delta \Gamma \left( Z,Z^{\prime }\right) ,\left(
Z,Z^{\prime }\right) \right) }{\delta \Delta \Gamma \left( Z,Z^{\prime
}\right) }=U^{\prime \prime }\left( \left\vert \Delta \Gamma \left(
Z,Z^{\prime }\right) \right\vert ^{2}\right)
\end{equation*}%
Equation (\ref{frN}) allows to compute the Lagrange multiplier $\lambda $:%
\begin{equation*}
\lambda ^{-1}=-\frac{U^{\prime \prime }\left( \left\vert \Delta \Gamma
\left( Z,Z^{\prime }\right) \right\vert ^{2}\right) }{U^{\prime }\left(
\left\vert \Gamma \left( \theta ,Z,Z^{\prime }\right) \right\vert
^{2}\right) +\Delta T_{0}-\frac{u+v}{2}}
\end{equation*}%
while (\ref{cnT}) yields the solution for $\left\vert \Delta \Gamma \left(
Z,Z^{\prime }\right) \right\vert ^{2}$ as a function of $\alpha _{0}$:%
\begin{equation*}
\left\vert \Delta \Gamma \left( Z,Z^{\prime },\alpha _{0}\right) \right\vert
^{2}=\left( U^{\prime }\right) ^{-1}\left( -\left( \frac{u+v}{2}+\left( 
\frac{w_{1}^{2}}{\lambda _{+}}+\frac{w_{2}^{2}}{\lambda _{-}}\right)
V^{2}\left( Z,Z^{\prime }\right) -\alpha _{0}\right) \right)
\end{equation*}%
This solution, written $\left\vert \Delta \Gamma \left( Z,Z^{\prime },\alpha
_{0}\right) \right\vert ^{2}$, is then used to compute $\alpha _{0}$:%
\begin{equation*}
\int \left\vert \Delta \Gamma \left( Z,Z^{\prime },\alpha _{0}\right)
\right\vert ^{2}d\left( Z,Z^{\prime }\right) =\left\Vert \Delta \Gamma
\right\Vert ^{2}
\end{equation*}%
Once the solution $\left\vert \Delta \Gamma \left( Z,Z^{\prime },\alpha
_{0}\right) \right\vert ^{2}$ is found, the action for the solution becomes:%
\begin{equation}
S=U\left( \left\vert \Gamma \left( \theta ,Z,Z^{\prime }\right) \right\vert
^{2}\right) +\left( \Delta T_{0}-\frac{u+v}{2}\right) \left\vert \Delta
\Gamma \left( Z,Z^{\prime }\right) \right\vert ^{2}
\end{equation}%
and given (\ref{frN}), this writes:%
\begin{equation}
S=U\left( \left\vert \Gamma \left( \theta ,Z,Z^{\prime }\right) \right\vert
^{2}\right) -U^{\prime }\left( \left\vert \Gamma \left( \theta ,Z,Z^{\prime
}\right) \right\vert ^{2}\right) \left\vert \Delta \Gamma \left( Z,Z^{\prime
}\right) \right\vert ^{2}-\lambda \frac{\delta h\left( \Delta \Gamma \left(
Z,Z^{\prime }\right) ,\left( Z,Z^{\prime }\right) \right) }{\delta \Delta
\Gamma \left( Z,Z^{\prime }\right) }\left\vert \Delta \Gamma \left(
Z,Z^{\prime }\right) \right\vert ^{2}
\end{equation}%
i.e.%
\begin{eqnarray}
S &=&U\left( \left\vert \Gamma \left( \theta ,Z,Z^{\prime }\right)
\right\vert ^{2}\right) -U^{\prime }\left( \left\vert \Gamma \left( \theta
,Z,Z^{\prime }\right) \right\vert ^{2}\right) \left\vert \Delta \Gamma
\left( Z,Z^{\prime }\right) \right\vert ^{2}  \label{shc} \\
&&+\frac{\left\vert \Delta \Gamma \left( Z,Z^{\prime }\right) \right\vert
^{2}}{U^{\prime \prime }\left( \left\vert \Delta \Gamma \left( Z,Z^{\prime
}\right) \right\vert ^{2}\right) }\left( U^{\prime }\left( \left\vert \Gamma
\left( \theta ,Z,Z^{\prime }\right) \right\vert ^{2}\right) +\Delta T_{0}-%
\frac{u+v}{2}\right)  \notag
\end{eqnarray}%
Expression (\ref{shc}) yields the condition for shifted state. At points $%
\left( Z,Z^{\prime }\right) $ such that (\ref{shc}) is negative, a shifted
state exists.

Given (\ref{VRP}), we have:%
\begin{eqnarray*}
\left\langle \Delta T_{0}\right\rangle &=&\left\langle \frac{s}{uv}%
\right\rangle A\frac{\left\langle \Delta T\right\rangle }{\left\langle
T\right\rangle }\left\Vert \Delta \Gamma \right\Vert ^{2} \\
&\simeq &\frac{\left\langle \frac{s}{u}\right\rangle \left\langle
T\right\rangle \left\langle \left\vert \bar{\Psi}_{0}\left( Z,Z^{\prime
}\right) \right\vert ^{2}\omega _{0}\left( Z\right) \right\rangle }{D\left(
\theta \right) \left\langle \hat{T}\right\rangle \left\vert \Psi _{0}\left(
Z^{\prime }\right) \right\vert ^{2}k\underline{A_{1}}A\left\Vert \Delta
\Gamma \right\Vert ^{4}}<<\frac{u+v}{2}
\end{eqnarray*}%
As a consequence, the most relevant parameter for emergence of some shifted
connectivity due to interactions is :%
\begin{equation*}
\frac{u+v}{2}=\frac{\left\vert \Psi _{0}\left( Z\right) \right\vert ^{2}}{%
\tau \omega _{0}\left( Z\right) }+\rho \left\vert \bar{\Psi}_{0}\left(
Z,Z^{\prime }\right) \right\vert ^{2}
\end{equation*}%
This is a decreasing function of background activity at point $Z$. Lower
activity favors a switching in connectivity.

\section{Application 1, last: Extension to n interacting fields}

The extension to a system of $n$ interacting field (see (\cite{GLr})) is
straightforward. It amounts to replace:%
\begin{eqnarray*}
\omega _{0}\left( Z\right) &\rightarrow &\omega _{0i}\left( Z\right) \\
\omega _{0}\left( Z^{\prime }\right) &\rightarrow &\omega _{0j}\left(
Z^{\prime }\right) \\
\delta \omega \left( \theta ,Z,\left\vert \Psi \right\vert ^{2}\right)
&\rightarrow &\delta \omega _{i}\left( \theta ,Z,\left\vert \Psi \right\vert
^{2}\right) \\
\delta \omega \left( \theta -\frac{\left\vert Z-Z^{\prime }\right\vert }{c}%
,Z^{\prime },\left\vert \Psi \right\vert ^{2}\right) &\rightarrow &\delta
\omega _{j}\left( \theta -\frac{\left\vert Z-Z^{\prime }\right\vert }{c}%
,Z^{\prime },\left\vert \Psi \right\vert ^{2}\right)
\end{eqnarray*}%
and:%
\begin{eqnarray*}
\left\vert \Psi _{0}\left( Z\right) \right\vert ^{2} &\rightarrow
&\left\vert \Psi _{0i}\left( Z\right) \right\vert ^{2} \\
\left\vert \Psi _{0}\left( Z^{\prime }\right) \right\vert ^{2} &\rightarrow
&\left\vert \Psi _{0j}\left( Z^{\prime }\right) \right\vert ^{2}
\end{eqnarray*}%
along with:%
\begin{eqnarray*}
T &\rightarrow &T_{ij},\left\langle T\right\rangle \rightarrow \left\langle
T_{ij}\right\rangle \\
\hat{T} &\rightarrow &\hat{T}_{ij},\left\langle \hat{T}\right\rangle
\rightarrow \left\langle \hat{T}_{ij}\right\rangle
\end{eqnarray*}%
in the expressions for the activities and action functionals.

\subsection{Effective action}

The effective action for $\Gamma \left( T,\hat{T},\theta ,Z,Z^{\prime
}\right) $ is obtained directly by modifying (\ref{fcp}) and (\ref{pnl}):%
\begin{eqnarray}
&&S\left( \Delta \Gamma \left( T,\hat{T},\theta ,Z,Z^{\prime }\right) \right)
\label{fcg} \\
&=&\Delta \Gamma ^{\dag }\left( T,\hat{T},\theta ,Z,Z^{\prime }\right)
\left( \nabla _{T}\left( \nabla _{T}+\frac{\left( T-\left\langle
T\right\rangle \right) }{\tau \omega _{0i}\left( Z\right) }\left\vert \Psi
\left( \theta ,Z\right) \right\vert ^{2}\right) \right) \Delta \Gamma \left(
T,\hat{T},\theta ,Z,Z^{\prime }\right)  \notag \\
&&+\Delta \Gamma ^{\dag }\left( T,\hat{T},\theta ,Z,Z^{\prime }\right)
\nabla _{\hat{T}}\left( \nabla _{\hat{T}}+\rho \left\vert \bar{\Psi}%
_{0ij}\left( Z,Z^{\prime }\right) \right\vert ^{2}\left( \hat{T}%
-\left\langle \hat{T}\right\rangle \right) \right) \Delta \Gamma \left( T,%
\hat{T},\theta ,Z,Z^{\prime }\right)  \notag \\
&&+V\left( \Delta \Gamma ,\Delta \Gamma ^{\dag }\right)  \notag
\end{eqnarray}

\begin{eqnarray}
&&V\left( \Delta \Gamma ,\Delta \Gamma ^{\dag }\right) =-\Delta \Gamma
^{\dag }\left( T,\hat{T},\theta ,Z,Z^{\prime }\right)  \label{Tlrr} \\
&&\times \nabla _{\hat{T}}\left( \frac{\rho D\left( \theta \right)
\left\langle \hat{T}\right\rangle \left\vert \Psi _{0j}\left( Z^{\prime
}\right) \right\vert ^{2}}{\omega _{0i}^{2}\left( Z\right) }\left( \omega
_{0i}\left( Z\right) \delta \omega _{j}\left( \theta -\frac{\left\vert
Z-Z^{\prime }\right\vert }{c},Z^{\prime },\left\vert \Psi \right\vert
^{2}\right) -\omega _{0j}\left( Z^{\prime }\right) \delta \omega _{i}\left(
\theta ,Z,\left\vert \Psi \right\vert ^{2}\right) \right) \right)  \notag \\
&&\times \Delta \Gamma \left( T,\hat{T},\theta ,Z,Z^{\prime }\right)  \notag
\end{eqnarray}%
whr:%
\begin{equation*}
\left\vert \bar{\Psi}_{0ij}\left( Z,Z^{\prime }\right) \right\vert ^{2}=%
\frac{C\left( \theta \right) \left\vert \Psi _{0i}\left( Z\right)
\right\vert ^{2}\omega _{0i}\left( Z\right) +D\left( \theta \right) \hat{T}%
\left\vert \Psi _{0j}\left( Z^{\prime }\right) \right\vert ^{2}\omega
_{0j}\left( Z^{\prime }\right) }{\omega _{0i}\left( Z\right) }
\end{equation*}

\subsection{Saddle point equation}

The saddle point equation for the background state $\Delta \Gamma \left(
T_{ij},\hat{T}_{ij},\theta ,Z,Z^{\prime }\right) $:%
\begin{eqnarray}
0 &=&\left( -\sigma _{\hat{T}_{ij}}^{2}\nabla _{\hat{T}_{ij}^{\prime }}^{2}+%
\frac{\lambda _{+}^{2}}{4\sigma _{\hat{T}_{ij}}^{2}}\left( \Delta \hat{T}%
_{ij}^{\prime }-\Delta \hat{T}_{0ij}^{\prime }-\frac{w_{2}}{\lambda _{+}}%
V\right) ^{2}\right) \Delta \Gamma \left( T_{ij},\hat{T}_{ij},\theta
,Z,Z^{\prime }\right)  \label{sdll} \\
&&+\left( -\sigma _{T_{ij}}^{2}\nabla _{T_{ij}^{\prime }}^{2}+\frac{\lambda
_{-}^{2}}{\sigma _{T_{ij}}^{2}}\left( \Delta T_{ij}^{\prime }-\Delta
T_{0ij}^{\prime }-\frac{w_{1}}{\lambda _{-}}V\right) ^{2}\right) \Delta
\Gamma \left( T_{ij},\hat{T}_{ij},\theta ,Z,Z^{\prime }\right)  \notag \\
&&-\left( u+v+\left( \frac{w_{1}^{2}}{\lambda _{+}}V^{2}+\frac{w_{2}^{2}}{%
\lambda _{-}}V^{2}\right) -\alpha \right) \Delta \Gamma \left( T_{ij},\hat{T}%
_{ij},\theta ,Z,Z^{\prime }\right)  \notag
\end{eqnarray}%
where:%
\begin{equation*}
\lambda _{\pm }=\sqrt{\frac{1}{2}\left( u^{2}+v^{2}\right) +s^{2}\pm \frac{%
\left( u+v\right) }{2}\sqrt{\left( u-v\right) ^{2}+4s^{2}}}
\end{equation*}%
\begin{eqnarray*}
u &=&\frac{\left\vert \Psi _{0i}\left( Z\right) \right\vert ^{2}}{\tau
\omega _{0i}\left( Z\right) } \\
v &=&\rho \left\vert \bar{\Psi}_{0ij}\left( Z,Z^{\prime }\right) \right\vert
^{2} \\
s &=&-\frac{\lambda \left\vert \Psi _{0i}\left( Z\right) \right\vert ^{2}}{%
\omega _{0i}\left( Z\right) }\frac{\sigma _{\hat{T}_{ij}}}{\sigma _{T_{ij}}}
\end{eqnarray*}%
\begin{equation}
\left( \Delta T_{0ij},\Delta \hat{T}_{0ij}\right) \simeq \left( -\frac{%
\lambda \tau V_{0}\omega _{0i}\left( Z\right) }{\sigma _{T}\left\vert \bar{%
\Psi}_{0ij}\left( Z,Z^{\prime }\right) \right\vert ^{2}},\frac{\Delta T_{0ij}%
}{\lambda \tau }\frac{\sigma _{\hat{T}_{ij}}}{\sigma _{T_{ij}}}\right)
\label{Sfhh}
\end{equation}%
and $\left( X^{\prime },\hat{X}^{\prime }\right) $ are the coordinates of
any vector in the diagonal basis of the potential:%
\begin{equation*}
\left( X^{\prime },\hat{X}^{\prime }\right) ^{t}=P^{-1}\left( X,\hat{X}%
\right)
\end{equation*}%
\begin{equation*}
P=\left( 
\begin{array}{cc}
w_{1} & w_{2} \\ 
w_{1}^{\prime } & w_{2}^{\prime }%
\end{array}%
\right)
\end{equation*}%
\begin{equation*}
V\left( Z,Z^{\prime }\right) =V_{1}\left( Z,Z^{\prime }\right) \left(
1+V_{2}\left( Z,Z^{\prime }\right) \right)
\end{equation*}%
where:%
\begin{eqnarray*}
V_{1}\left( Z,Z^{\prime }\right) &=&A_{1}\left( Z,Z^{\prime }\right)
\left\langle \Delta \hat{T}_{ij}\right\rangle \left\Vert \Delta \Gamma
\right\Vert ^{2} \\
V_{2}\left( Z,Z^{\prime }\right) &=&A_{2}\left( Z,Z^{\prime }\right) \frac{%
\left\langle \Delta T_{ij}\right\rangle }{\left\langle T_{ij}\right\rangle }%
\left\Vert \Delta \Gamma \right\Vert ^{2}
\end{eqnarray*}%
We also define:%
\begin{equation}
V_{0}\left( Z,Z^{\prime }\right) =A_{0}\left( Z,Z^{\prime }\right) \frac{%
\left\langle \Delta T\right\rangle }{\left\langle T\right\rangle }\left\Vert
\Delta \Gamma \right\Vert ^{2}
\end{equation}%
with the various functions defined by:%
\begin{equation}
A_{0}\left( Z,Z^{\prime }\right) =F\left( Z,Z^{\prime }\right) \left\langle 
\check{T}\left( 1-\left( 1+\left\langle \left\vert \Psi _{0}\right\vert
^{2}\right\rangle \left( 1+\frac{\Delta T}{\left\langle T\right\rangle }%
\right) \left\Vert \Delta \Gamma \right\Vert ^{2}\right) \check{T}\right)
^{-1}O\right\rangle ^{\left( T,\hat{T},\theta ,Z,Z^{\prime }\right) }
\end{equation}%
\begin{equation}
A_{1}\left( Z,Z^{\prime }\right) \simeq \left\langle \left[ F\left(
Z_{2},Z_{2}^{\prime }\right) \left[ \check{T}\left( 1-\left\langle
\left\vert \Psi _{0}\right\vert ^{2}\right\rangle \frac{\left\langle \Delta
T\right\rangle }{T}\left\Vert \Delta \Gamma \right\Vert ^{2}\right) ^{-1}O%
\right] \right] _{\left( T,\hat{T},\theta ,Z,Z^{\prime }\right)
}\right\rangle
\end{equation}%
and:%
\begin{equation}
A_{2}\left( Z,Z^{\prime }\right) =\left\langle \left[ \check{T}\left(
1-\left( 1+\left\langle \left\vert \Psi _{0}\right\vert ^{2}\right\rangle 
\frac{\left\langle \Delta T\right\rangle }{T}\left\Vert \Delta \Gamma
\right\Vert ^{2}\right) \check{T}\right) ^{-1}\right] ^{\left( T,\hat{T}%
,\theta ,Z,Z^{\prime }\right) }\right\rangle  \label{Vxpp}
\end{equation}%
with the intermediate function defined by:%
\begin{equation}
F\left( Z,Z^{\prime }\right) =\frac{\rho D\left( \theta \right) \left\langle 
\hat{T}\right\rangle \left\vert \Psi _{0}\left( Z^{\prime }\right)
\right\vert ^{2}}{\omega _{0}\left( Z\right) }  \label{Vzp1}
\end{equation}

\subsubsection{Solutions of (\protect\ref{sdll})}

The solutions (\ref{Stn}) become:%
\begin{eqnarray}
&&\Delta \Gamma _{\delta }\left( T_{ij},\hat{T}_{ij},\theta ,Z,Z^{\prime
}\right)  \label{Sdnn} \\
&=&\exp \left( -\frac{1}{2}\left( \mathbf{\Delta T}_{ij}\mathbf{-\Delta \bar{%
T}}_{ij}\right) ^{t}\hat{U}\left( \mathbf{\Delta T}_{ij}\mathbf{-\Delta \bar{%
T}}_{ij}\right) \right)  \notag \\
&&\times H_{p}\left( \left( \mathbf{\Delta T}_{ij}^{\prime }\mathbf{-\Delta 
\bar{T}}_{ij}^{\prime }\right) _{1}\frac{\sigma _{T}\lambda _{+}}{2\sqrt{2}}%
\left( \mathbf{\Delta T}_{ij}^{\prime }\mathbf{-\Delta \bar{T}}_{ij}^{\prime
}\right) _{1}\right) H_{p-\delta }\left( \left( \mathbf{\Delta T}^{\prime }%
\mathbf{-\Delta \bar{T}}^{\prime }\right) _{2}\frac{\sigma _{\hat{T}}\lambda
_{-}}{2\sqrt{2}}\left( \mathbf{\Delta T}^{\prime }\mathbf{-\Delta \bar{T}}%
^{\prime }\right) _{2}\right)  \notag
\end{eqnarray}%
and (\ref{Stt}):%
\begin{eqnarray}
&&\Delta \Gamma _{\delta }^{\dagger }\left( T,\hat{T},\theta ,Z,Z^{\prime
}\right)  \label{Sddd} \\
&=&H_{p}\left( \left( \mathbf{\Delta T}_{ij}^{\prime }\mathbf{-\Delta \bar{T}%
}_{ij}^{\prime }\right) _{2}\frac{\sigma _{T}\lambda _{+}}{2\sqrt{2}}\left( 
\mathbf{\Delta T}_{ij}^{\prime }\mathbf{-\Delta \bar{T}}_{ij}^{\prime
}\right) _{2}\right) H_{p-\delta }\left( \left( \mathbf{\Delta T}%
_{ij}^{\prime }\mathbf{-\Delta \bar{T}}_{ij}^{\prime }\right) _{2}\frac{%
\sigma _{\hat{T}}\lambda _{-}}{2\sqrt{2}}\left( \mathbf{\Delta T}%
_{ij}^{\prime }\mathbf{-\Delta \bar{T}}_{ij}^{\prime }\right) _{2}\right) 
\notag
\end{eqnarray}%
where $H_{p}$ and $H_{p-\delta }$ are Hermite polynomials and the variables
are:%
\begin{eqnarray}
\mathbf{\Delta T-\Delta \bar{T}} &=&\left( 
\begin{array}{c}
\Delta T_{ij}-\left\langle \Delta T_{ij}\right\rangle \\ 
\Delta \hat{T}_{ij}-\left\langle \Delta \hat{T}_{ij}\right\rangle%
\end{array}%
\right)  \label{shnn} \\
\mathbf{\Delta T}_{ij}^{\prime }\mathbf{-\Delta \bar{T}}_{ij}^{\prime }
&=&P^{-1}\left( \mathbf{\Delta T}_{ij}\mathbf{-\Delta \bar{T}}_{ij}\right) 
\notag
\end{eqnarray}%
and the matrix $\hat{U}$ given by:%
\begin{equation*}
\hat{U}=\left( 
\begin{array}{cc}
\frac{1}{\sigma _{T}} & 0 \\ 
0 & \frac{1}{\sigma _{\hat{T}}}%
\end{array}%
\right) U\left( 
\begin{array}{cc}
\frac{1}{\sigma _{T}} & 0 \\ 
0 & \frac{1}{\sigma _{\hat{T}}}%
\end{array}%
\right) =\left( 
\begin{array}{cc}
\frac{s^{2}+u^{2}}{\sigma _{T}^{2}} & -\frac{s\left( u+v\right) }{\sigma
_{T}\sigma _{\hat{T}}} \\ 
-\frac{s\left( u+v\right) }{\sigma _{T}\sigma _{\hat{T}}} & \frac{s^{2}+v^{2}%
}{\sigma _{\hat{T}}^{2}}%
\end{array}%
\right)
\end{equation*}%
The potential background field of the sytem are thus defined by:%
\begin{equation*}
\prod\limits_{W}\Delta \Gamma _{\delta \left( Z,Z^{\prime }\right) }\left(
T,\hat{T},\theta ,Z,Z^{\prime }\right)
\end{equation*}%
and:%
\begin{equation*}
\prod\limits_{W}\Delta \Gamma _{\delta \left( Z,Z^{\prime }\right)
}^{\dagger }\left( T,\hat{T},\theta ,Z,Z^{\prime }\right)
\end{equation*}

\subsubsection{Estimation of $\left\langle \Delta T\right\rangle $, $%
\left\langle \Delta \hat{T}\right\rangle $}

At each point of $W$, the shift in connectivity $\mathbf{\Delta \bar{T}}$ is
defined by (\ref{shnn}):

\begin{eqnarray*}
\Delta T_{ij}\left( Z,Z^{\prime }\right) &=&\frac{A_{1}\left( Z,Z^{\prime
}\right) A_{0}\left( Z,Z^{\prime }\right) \left\langle v^{2}\right\rangle }{%
\left\langle A_{1}\left( Z,Z^{\prime }\right) \right\rangle \left\langle
A_{0}\left( Z,Z^{\prime }\right) \right\rangle v^{2}}\left\langle \Delta
T_{ij}\right\rangle \\
\Delta \hat{T}_{ij}\left( Z,Z^{\prime }\right) &=&\frac{A_{0}\left(
Z,Z^{\prime }\right) }{\left\langle A_{0}\left( Z,Z^{\prime }\right)
\right\rangle }\left\langle \Delta \hat{T}_{ij}\right\rangle
\end{eqnarray*}%
\begin{equation}
\left\langle \Delta T_{ij}\right\rangle \simeq \frac{\omega _{0}\left(
Z\right) \left\langle T_{ij}\right\rangle }{\rho D\left( \theta \right)
\left\langle \hat{T}_{ij}\right\rangle \left\vert \Psi _{0}\left( Z^{\prime
}\right) \right\vert ^{2}k\underline{A_{1}}\left\Vert \Delta \Gamma
\right\Vert ^{6}}\left\langle \rho \frac{\left\vert \bar{\Psi}_{0ij}\left(
Z,Z^{\prime }\right) \right\vert ^{2}}{A}\right\rangle ^{2}\left\langle
T_{ij}\right\rangle
\end{equation}%
\begin{equation}
\left\langle \Delta \hat{T}_{ij}\right\rangle =-\frac{1}{v}A\frac{\left\Vert
\Delta \Gamma \right\Vert ^{2}}{\left\langle T_{ij}\right\rangle }%
\left\langle \Delta T_{ij}\right\rangle
\end{equation}%
where averages of structural parameters are:%
\begin{equation}
\left\langle A_{0}\left( Z,Z^{\prime }\right) \right\rangle =\left\langle 
\frac{\rho D\left( \theta \right) \left\langle \hat{T}\right\rangle
\left\vert \Psi _{0}\left( Z^{\prime }\right) \right\vert ^{2}}{\omega
_{0}\left( Z\right) }\right\rangle \left\langle \check{T}\left( 1-\left(
1+\left\langle \left\vert \Psi _{0}\right\vert ^{2}\right\rangle \left( 1+%
\frac{\Delta T}{\left\langle T\right\rangle }\right) \left\Vert \Delta
\Gamma \right\Vert ^{2}\right) \check{T}\right) ^{-1}O\right\rangle
\end{equation}%
\begin{equation}
\left\langle A_{1}\left( Z,Z^{\prime }\right) \right\rangle \simeq
\left\langle \left[ F\left( Z_{2},Z_{2}^{\prime }\right) \left[ \check{T}%
\left( 1-\left\langle \left\vert \Psi _{0}\right\vert ^{2}\right\rangle 
\frac{\left\langle \Delta T\right\rangle }{T}\left\Vert \Delta \Gamma
\right\Vert ^{2}\right) ^{-1}O\right] \right] \right\rangle
\end{equation}%
along with the constant $\underline{A_{1}}$: 
\begin{equation}
\underline{A_{1}}\simeq \left\langle \left[ \check{T}\left( 1-\left\langle
\left\vert \Psi _{0}\right\vert ^{2}\right\rangle \frac{\left\langle \Delta
T\right\rangle }{T}\left\Vert \Delta \Gamma \right\Vert ^{2}\right) ^{-1}O%
\right] \right\rangle
\end{equation}

\subsubsection{Stability and condition for shifting state}

The possibility for a shifted state, i.e. a state for which $\left\vert
\Delta \Gamma \left( Z,Z^{\prime }\right) \right\vert ^{2}>0$ depends on the
value of the action for this state. If the corresponding action is negative,
the state $\Delta \Gamma \left( T,\hat{T},Z,Z^{\prime }\right) $ is the
minimum of the system. Otherwise, the state $\Delta \Gamma \left( T,\hat{T}%
,Z,Z^{\prime }\right) =0$ is the background state at point $\left(
Z,Z^{\prime }\right) $.

The condition of existence for the shift is:%
\begin{eqnarray}
S &=&U\left( \left\vert \Gamma \left( \theta ,Z,Z^{\prime }\right)
\right\vert ^{2}\right) -U^{\prime }\left( \left\vert \Gamma \left( \theta
,Z,Z^{\prime }\right) \right\vert ^{2}\right) \left\vert \Delta \Gamma
\left( Z,Z^{\prime }\right) \right\vert ^{2}  \label{shcc} \\
&&+\frac{\left\vert \Delta \Gamma \left( Z,Z^{\prime }\right) \right\vert
^{2}}{U^{\prime \prime }\left( \left\vert \Delta \Gamma \left( Z,Z^{\prime
}\right) \right\vert ^{2}\right) }\left( U^{\prime }\left( \left\vert \Gamma
\left( \theta ,Z,Z^{\prime }\right) \right\vert ^{2}\right) +\Delta T_{0}-%
\frac{u+v}{2}\right)  \notag
\end{eqnarray}%
Expression (\ref{shcc}) yields the condition for shifted state. At points $%
\left( Z,Z^{\prime }\right) $ such that (\ref{shcc}) is negative, a shifted
state exists.

In our order of approximation $\Delta T_{0}<<\frac{u+v}{2}$ and: 
\begin{equation*}
u+v=\frac{\left\vert \Psi _{0i}\left( Z\right) \right\vert ^{2}}{\tau \omega
_{0i}\left( Z\right) }+\rho \left\vert \bar{\Psi}_{0ij}\left( Z,Z^{\prime
}\right) \right\vert ^{2}
\end{equation*}%
Background activity $\omega _{0i}\left( Z\right) $ is lower with inhibitory
interactions rather than without.

\section{Application 2: Dynamics between $T\left( Z,Z^{\prime }\right) $ and 
$T\left( Z^{\prime },Z\right) $}

We study the interactions between $T\left( Z,Z^{\prime }\right) $ and $%
T\left( Z^{\prime },Z\right) $, i.e. the connectivity in both direction, by
computing the transition function:%
\begin{equation*}
G\left( \Delta T_{i}\left( Z,Z^{\prime }\right) ,\Delta T_{1}\left(
Z,Z^{\prime }\right) ,\Delta T_{i}^{\prime }\left( Z^{\prime },Z\right)
,\Delta T_{f}\left( Z^{\prime },Z\right) \right)
\end{equation*}%
Neglecting the interaction, this is at the zeroth order given by a product
of two transition function. We use the large $t$ approximation, and we have:%
\begin{eqnarray*}
&&G\left( \Delta T_{i}\left( Z,Z^{\prime }\right) ,\Delta T_{1}\left(
Z,Z^{\prime }\right) ,\Delta T_{i}^{\prime }\left( Z^{\prime },Z\right)
,\Delta T_{f}\left( Z^{\prime },Z\right) \right) \\
&\simeq &G_{0}\left( \Delta T_{i}\left( Z,Z^{\prime }\right) ,\Delta
T_{f}\left( Z,Z^{\prime }\right) ,\Delta T_{i}\left( Z^{\prime },Z\right)
,\Delta T_{f}\left( Z^{\prime },Z\right) \right) G_{0}\left( \Delta
T_{i}\left( Z^{\prime },Z\right) ,\Delta T_{f}\left( Z^{\prime },Z\right)
\right)
\end{eqnarray*}%
This formula can be corrected by using the formula (\ref{tdr}) for the
interaction term:

\begin{eqnarray}
&&-\Delta \Gamma ^{\dag }\left( T,\hat{T},\theta ,Z,Z^{\prime }\right) \times
\label{ftrr} \\
&&\times \nabla _{\hat{T}}\left( \frac{\rho \left( D\left( \theta \right)
\left\langle \hat{T}\right\rangle \left\vert \Psi _{0}\left( Z^{\prime
}\right) \right\vert ^{2}\left( \omega _{0}\left( Z\right) \delta \omega
\left( \theta -\frac{\left\vert Z-Z^{\prime }\right\vert }{c},Z^{\prime
},\left\vert \Psi \right\vert ^{2}\right) -\omega _{0}\left( Z^{\prime
}\right) \delta \omega \left( \theta ,Z,\left\vert \Psi \right\vert
^{2}\right) \right) \right) }{\omega _{0}^{2}\left( Z\right) }\right) \Delta
\Gamma \left( T,\hat{T},\theta ,Z,Z^{\prime }\right)  \notag
\end{eqnarray}%
We then replace $\delta \omega \left( \theta -\frac{\left\vert Z-Z^{\prime
}\right\vert }{c},Z^{\prime },\left\vert \Psi \right\vert ^{2}\right) $ and $%
\delta \omega \left( \theta ,Z,\left\vert \Psi \right\vert ^{2}\right) $\
using (\ref{tdr}) at lowest order:%
\begin{eqnarray*}
\delta \omega \left( \theta -\frac{\left\vert Z-Z^{\prime }\right\vert }{c}%
,Z^{\prime },\left\vert \Psi \right\vert ^{2}\right) &\simeq &\frac{\omega
_{0}\left( Z\right) \Delta T\left( Z^{\prime },Z\right) \left\vert \Delta
\Gamma \left( \theta -2\frac{\left\vert Z-Z^{\prime }\right\vert }{c}%
,Z\right) \right\vert ^{2}}{\left\langle T\right\rangle } \\
\delta \omega \left( \theta ,Z,\left\vert \Psi \right\vert ^{2}\right)
&\simeq &\frac{\omega _{0}\left( Z^{\prime }\right) \Delta T\left(
Z,Z^{\prime }\right) \left\vert \Delta \Gamma \left( \theta -\frac{%
\left\vert Z-Z^{\prime }\right\vert }{c},Z^{\prime }\right) \right\vert ^{2}%
}{\left\langle T\right\rangle }
\end{eqnarray*}%
and (\ref{ftrr}) writes:

\begin{eqnarray*}
&&-\Delta \Gamma ^{\dag }\left( T,\hat{T},\theta ,Z,Z^{\prime }\right) \\
&&\nabla _{\hat{T}}\frac{\rho D\left( \theta \right) \left\langle \hat{T}%
\right\rangle \left\vert \Psi _{0}\left( Z^{\prime }\right) \right\vert ^{2}%
}{\omega _{0}^{2}\left( Z\right) }\times \\
&&\times \left( \frac{\omega _{0}\left( Z\right) \Delta T\left( Z^{\prime
},Z\right) \left\vert \Delta \Gamma \left( \theta -2\frac{\left\vert
Z-Z^{\prime }\right\vert }{c},Z\right) \right\vert ^{2}}{\left\langle
T\right\rangle }-\frac{\omega _{0}\left( Z^{\prime }\right) \Delta T\left(
Z,Z^{\prime }\right) \left\vert \Delta \Gamma \left( \theta -\frac{%
\left\vert Z-Z^{\prime }\right\vert }{c},Z^{\prime }\right) \right\vert ^{2}%
}{\left\langle T\right\rangle }\right) \\
&&\times \Delta \Gamma \left( T,\hat{T},\theta ,Z,Z^{\prime }\right) \\
&=&-\Delta \Gamma ^{\dag }\left( T,\hat{T},\theta ,Z,Z^{\prime }\right) \\
&&\nabla _{\hat{T}}\left( a\left( Z^{\prime },Z\right) \Delta T\left(
Z^{\prime },Z\right) \left\vert \Delta \Gamma \left( \theta -2\frac{%
\left\vert Z-Z^{\prime }\right\vert }{c},Z\right) \right\vert ^{2}-b\left(
Z,Z^{\prime }\right) \Delta T\left( Z,Z^{\prime }\right) \left\vert \Delta
\Gamma \left( \theta -\frac{\left\vert Z-Z^{\prime }\right\vert }{c}%
,Z^{\prime }\right) \right\vert ^{2}\right) \\
&&\times \Delta \Gamma \left( T,\hat{T},\theta ,Z,Z^{\prime }\right)
\end{eqnarray*}%
with:%
\begin{eqnarray}
a\left( Z^{\prime },Z\right) &=&\frac{\rho D\left( \theta \right)
\left\langle \hat{T}\right\rangle \left\vert \Psi _{0}\left( Z^{\prime
}\right) \right\vert ^{2}}{\omega _{0}\left( Z\right) \left\langle
T\right\rangle }  \label{bcf} \\
b\left( Z,Z^{\prime }\right) &=&\frac{\rho D\left( \theta \right)
\left\langle \hat{T}\right\rangle \left\vert \Psi _{0}\left( Z^{\prime
}\right) \right\vert ^{2}\omega _{0}\left( Z^{\prime }\right) }{\omega
_{0}^{2}\left( Z\right) \left\langle T\right\rangle }  \notag
\end{eqnarray}%
The graphs that compute mutual interactions between $T\left( Z,Z^{\prime
}\right) $ and $T\left( Z^{\prime },Z\right) $ at the lowest order are given
by the squared interaction term averaged between an initial and a final $2$-
state that writes in an expanded form:

\begin{eqnarray}
&&\left\langle \Delta T_{i}\left( Z,Z^{\prime }\right) ,\Delta T_{i}\left(
Z^{\prime },Z\right) \right\vert \left\{ \Delta \Gamma ^{\dag }\left( T,\hat{%
T},\theta ,Z,Z^{\prime }\right) \right.  \label{crn} \\
&&\nabla _{\hat{T}}\left( a\left( Z^{\prime },Z\right) \Delta T\left(
Z^{\prime },Z\right) \left\vert \Delta \Gamma \left( \theta -2\frac{%
\left\vert Z-Z^{\prime }\right\vert }{c},Z\right) \right\vert ^{2}-b\left(
Z,Z^{\prime }\right) \Delta T\left( Z,Z^{\prime }\right) \left\vert \Delta
\Gamma \left( \theta -\frac{\left\vert Z-Z^{\prime }\right\vert }{c}%
,Z^{\prime }\right) \right\vert ^{2}\right)  \notag \\
&&\left. \Delta \Gamma \left( T,\hat{T},\theta ,Z,Z^{\prime }\right)
\right\} \left\{ \Delta \Gamma ^{\dag }\left( T,\hat{T},\theta ,Z^{\prime
},Z\right) \right.  \notag \\
&&\times \nabla _{\hat{T}}\left( a\left( Z,Z^{\prime }\right) \Delta T\left(
Z,Z^{\prime }\right) \left\vert \Delta \Gamma \left( \theta -2\frac{%
\left\vert Z-Z^{\prime }\right\vert }{c},Z^{\prime }\right) \right\vert
^{2}-b\left( Z^{\prime },Z\right) \Delta T\left( Z,Z^{\prime }\right)
\left\vert \Delta \Gamma \left( \theta -\frac{\left\vert Z-Z^{\prime
}\right\vert }{c},Z\right) \right\vert ^{2}\right)  \notag \\
&&\left. \Delta \Gamma \left( T,\hat{T},\theta ,Z^{\prime },Z\right) \right\}
\notag \\
&&\left\vert \Delta T_{f}\left( Z,Z^{\prime }\right) ,\Delta T_{f}\left(
Z^{\prime },Z\right) \right\rangle  \notag
\end{eqnarray}%
developping the square leads to three contributions that are computed in
appendix 3.

Writing $\Delta \mathbf{T}_{i}$ for $\Delta T_{i}\left( Z,Z^{\prime }\right) 
$, $\Delta \mathbf{T}_{i}^{\prime }$ for $\Delta T_{i}\left( Z^{\prime
},Z\right) $ and similarly for $\Delta \mathbf{T}_{1}$, $\Delta \mathbf{T}%
_{1}^{\prime }$ we compute the effect of some fluctuations in the
connectivity, by assume that $\Delta \mathbf{T}_{i}=0$, so that we study the
impact of a deviation $\Delta \mathbf{T}_{1}\neq 0$ on both final states $%
\Delta \mathbf{T}_{f}$ and $\Delta \mathbf{T}_{f}^{\prime }$. Correction (%
\ref{crn}) adds to the free transition function and leads to:

\begin{eqnarray}
&&G\left( \Delta T_{i}\left( Z,Z^{\prime }\right) ,\Delta T_{1}\left(
Z,Z^{\prime }\right) ,\Delta T_{i}^{\prime }\left( Z^{\prime },Z\right)
,\Delta T_{f}\left( Z^{\prime },Z\right) \right)  \label{grpp} \\
&=&G_{0}\left( \Delta T_{i}\left( Z,Z^{\prime }\right) ,\Delta T_{1}\left(
Z,Z^{\prime }\right) ,\Delta T_{i}^{\prime }\left( Z^{\prime },Z\right)
,\Delta T_{f}\left( Z^{\prime },Z\right) \right)  \notag \\
&&+\exp \left( -\frac{1}{2}\left( \Delta \mathbf{T}_{1}-\left( 
\begin{array}{cc}
e^{-tu} & s\frac{e^{-tu}-e^{-tv}}{u-v} \\ 
0 & e^{-tv}%
\end{array}%
\right) \left( \Delta \mathbf{T}_{i}\right) \right) ^{t}\sigma ^{-1}\left(
t\right) \left( \Delta \mathbf{T}_{1}-\left( 
\begin{array}{cc}
e^{-tu} & s\frac{e^{-tu}-e^{-tv}}{u-v} \\ 
0 & e^{-tv}%
\end{array}%
\right) \left( \Delta \mathbf{T}_{i}\right) \right) \right)  \notag \\
&&\times \exp \left( -\frac{1}{2}\left( \left( \Delta \mathbf{T}_{f}\right)
-\left( 
\begin{array}{cc}
e^{-tu} & s\frac{e^{-tu}-e^{-tv}}{u-v} \\ 
0 & e^{-tv}%
\end{array}%
\right) \Delta \mathbf{T}_{1}\right) ^{t}\sigma ^{-1}\left( t\right) \left(
\left( \Delta \mathbf{T}_{f}\right) -\left( 
\begin{array}{cc}
e^{-tu} & s\frac{e^{-tu}-e^{-tv}}{u-v} \\ 
0 & e^{-tv}%
\end{array}%
\right) \Delta \mathbf{T}_{1}\right) \right)  \notag \\
&&\times \left( a\left( Z^{\prime },Z\right) \Delta T_{1}-b\left(
Z,Z^{\prime }\right) \Delta T_{1}^{\prime }\right) \left( a\left(
Z,Z^{\prime }\right) \Delta T_{1}^{\prime }-b\left( Z^{\prime },Z\right)
\Delta T_{1}\right)  \notag \\
&&\times \exp \left( -\frac{1}{2}\left( \Delta \mathbf{T}_{1}^{\prime
}\right) ^{t}\sigma ^{-1}\left( t\right) \left( \Delta \mathbf{T}%
_{1}^{\prime }\right) \right)  \notag \\
&&\times \exp \left( -\frac{1}{2}\left( \left( \Delta \mathbf{T}_{f}^{\prime
}\right) -\left( 
\begin{array}{cc}
e^{-tu} & s\frac{e^{-tu}-e^{-tv}}{u-v} \\ 
0 & e^{-tv}%
\end{array}%
\right) \Delta \mathbf{T}_{1}^{\prime }\right) ^{t}\sigma ^{-1}\left(
t\right) \left( \left( \Delta \mathbf{T}_{f}^{\prime }\right) -\left( 
\begin{array}{cc}
e^{-tu} & s\frac{e^{-tu}-e^{-tv}}{u-v} \\ 
0 & e^{-tv}%
\end{array}%
\right) \Delta \mathbf{T}_{1}^{\prime }\right) \right)  \notag
\end{eqnarray}%
with:%
\begin{eqnarray*}
&&G_{0}\left( \Delta T_{i}\left( Z,Z^{\prime }\right) ,\Delta T_{1}\left(
Z,Z^{\prime }\right) ,\Delta T_{i}^{\prime }\left( Z^{\prime },Z\right)
,\Delta T_{f}\left( Z^{\prime },Z\right) \right) \\
&=&\exp \left( -\frac{1}{2}\left( \Delta \mathbf{T}_{1}^{\prime }\right)
^{t}\sigma ^{-1}\left( t\right) \left( \Delta \mathbf{T}_{1}^{\prime
}\right) \right) \\
&&\times \exp \left( -\frac{1}{2}\left( \left( \Delta \mathbf{T}_{f}^{\prime
}\right) -\left( 
\begin{array}{cc}
e^{-tu} & s\frac{e^{-tu}-e^{-tv}}{u-v} \\ 
0 & e^{-tv}%
\end{array}%
\right) \Delta \mathbf{T}_{1}^{\prime }\right) ^{t}\sigma ^{-1}\left(
t\right) \left( \left( \Delta \mathbf{T}_{f}^{\prime }\right) -\left( 
\begin{array}{cc}
e^{-tu} & s\frac{e^{-tu}-e^{-tv}}{u-v} \\ 
0 & e^{-tv}%
\end{array}%
\right) \Delta \mathbf{T}_{1}^{\prime }\right) \right)
\end{eqnarray*}%
Appendix 3, shows that the maximum of the correction (\ref{grpp})\ is
obtained for:%
\begin{eqnarray*}
\left( \Delta \mathbf{T}_{f}\right) &\simeq &\left( 
\begin{array}{cc}
e^{-tu} & s\frac{e^{-tu}-e^{-tv}}{u-v} \\ 
0 & e^{-tv}%
\end{array}%
\right) \Delta \mathbf{T}_{1} \\
\left( \Delta \mathbf{T}_{f}^{\prime }\right) &\simeq &\left( 
\begin{array}{cc}
e^{-tu} & s\frac{e^{-tu}-e^{-tv}}{u-v} \\ 
0 & e^{-tv}%
\end{array}%
\right) \overline{\left( \Delta T_{1}^{\prime }\right) }
\end{eqnarray*}%
with $\overline{\left( \Delta T_{1}^{\prime }\right) }$ a given value:

\begin{equation*}
\overline{\left( \Delta T_{1}^{\prime }\right) }\in \left[ \inf \left( \frac{%
\omega _{0}\left( Z^{\prime }\right) }{\omega _{0}\left( Z\right) },\frac{%
\omega _{0}\left( Z\right) }{\omega _{0}\left( Z^{\prime }\right) }\right)
,\sup \left( \frac{\omega _{0}\left( Z^{\prime }\right) }{\omega _{0}\left(
Z\right) },\frac{\omega _{0}\left( Z\right) }{\omega _{0}\left( Z^{\prime
}\right) }\right) \right]
\end{equation*}%
As a consequence, the most likely configuration for the system is:%
\begin{eqnarray*}
\left( \Delta \mathbf{T}_{f}\right) &\simeq &\left( 
\begin{array}{cc}
e^{-tu} & s\frac{e^{-tu}-e^{-tv}}{u-v} \\ 
0 & e^{-tv}%
\end{array}%
\right) \Delta \mathbf{T}_{1} \\
\left( \Delta \mathbf{T}_{f}^{\prime }\right) &\simeq &\left( 
\begin{array}{cc}
e^{-tu} & s\frac{e^{-tu}-e^{-tv}}{u-v} \\ 
0 & e^{-tv}%
\end{array}%
\right) \overline{\left( \Delta T_{1}^{\prime }\right) }
\end{eqnarray*}%
which means that the connectivity fluctuations leads then the system from a
state:%
\begin{equation*}
\left( \Delta \mathbf{T}_{i}\right) ,\left( \Delta \mathbf{T}_{i}^{\prime
}\right) =0
\end{equation*}%
to a state:%
\begin{equation*}
\left( \Delta \mathbf{T}_{f}\right) ,\left( \Delta \mathbf{T}_{f}^{\prime
}\right) \simeq \left( 
\begin{array}{cc}
e^{-tu} & s\frac{e^{-tu}-e^{-tv}}{u-v} \\ 
0 & e^{-tv}%
\end{array}%
\right) \overline{\left( \Delta T_{1}^{\prime }\right) }
\end{equation*}%
The fluctuation has propagated to $\left( \Delta \mathbf{T}_{f}^{\prime
}\right) $ with the tendency to symetrize the connection from $Z$ to $%
Z^{\prime }$ and $Z^{\prime }$ to $Z$.

\section{Conclusion}

The use of effective theory to analyze fluctuations in the connectivity
field above a background state has enabled us to comprehend the emergence of
specific collective states that interact with one another. The concept of a
state above the background state corresponds to additional activity in
comparison to an average, persistent baseline. This arises from a
description in which individual neurons may participate in various connected
states, meaning that cells may exhibit different activation patterns.
Furthermore, within this perspective, this implies that we should consider
families of collective states, taking into account the possibility of
multiple activations or deactivations of such states. Consequently, an
effective field theory for emerging and interacting states should be
considered on its own. This is the objective of the fourth paper in this
series, which aims to develop such a formalism.

\section*{Appendix 1. Effective action for $\Gamma \left( T,\hat{T}%
,C,D\right) $: interactions}

\subsection*{1.1 Effective action for connectivities}

\subsubsection*{1.1.1 Full conributions}

As explained in the text, the effective action for $\Gamma \left( T,\hat{T}%
,C,D\right) $ is written by replacing $C$ and $D$ with their averages and
disregarding the threshold term $\eta H\left( \delta -T\right) $. The result
of these simplifications is the action (\ref{tCD}). 
\begin{eqnarray}
S\left( \Gamma ,\Gamma ^{\dag }\right) &=&\Gamma ^{\dag }\left( T,\hat{T}%
,\theta ,Z,Z^{\prime }\right) \left( \nabla _{T}\left( \nabla _{T}-\left( -%
\frac{1}{\tau \omega }T+\frac{\lambda }{\omega }\hat{T}\right) \left\vert
\Psi \left( \theta ,Z\right) \right\vert ^{2}\right) \right) \Gamma \left( T,%
\hat{T},\theta ,Z,Z^{\prime }\right)  \label{FC} \\
&&+\Gamma ^{\dag }\left( T,\hat{T},\theta ,Z,Z^{\prime }\right) \left(
\nabla _{\hat{T}}\left( \nabla _{\hat{T}}-\frac{\rho }{\omega \left(
J,\theta ,Z,\left\vert \Psi \right\vert ^{2}\right) }\left( \left( h\left(
Z,Z^{\prime }\right) -\hat{T}\right) C\left( \theta \right) \left\vert \Psi
\left( \theta ,Z\right) \right\vert ^{2}h_{C}\left( \omega \left( \theta
,Z,\left\vert \Psi \right\vert ^{2}\right) \right) \right. \right. \right. 
\notag \\
&&\times \left. \left. \left. -D\left( \theta \right) \hat{T}\left\vert \Psi
\left( \theta -\frac{\left\vert Z-Z^{\prime }\right\vert }{c},Z^{\prime
}\right) \right\vert ^{2}h_{D}\left( \omega \left( \theta -\frac{\left\vert
Z-Z^{\prime }\right\vert }{c},Z^{\prime },\left\vert \Psi \right\vert
^{2}\right) \right) \right) \right) \right) \Gamma \left( T,\hat{T},\theta
,Z,Z^{\prime }\right)  \notag
\end{eqnarray}

We also use (\ref{sdn}) to replace $\omega \left( \theta ,Z,\left\vert \Psi
\right\vert ^{2}\right) $ by $\omega _{0}\left( Z\right) +\delta \omega
\left( \theta ,Z,\left\vert \Psi \right\vert ^{2}\right) $ where $\delta
\omega \left( \theta ,Z,\left\vert \Psi \right\vert ^{2}\right) $ is given
by (\ref{RM}) and (\ref{rnll}): 
\begin{eqnarray*}
&&\delta \omega \left( \theta ,Z,\left\vert \Psi \right\vert ^{2}\right) \\
&=&\int^{\theta _{i}}\check{T}\left( 1-\left( 1+\left\vert \Psi \left(
Z,\theta \right) \right\vert ^{2}-\frac{\frac{\check{T}}{\left( 1-\left(
1+\left\vert \Psi \right\vert ^{2}\right) \check{T}\right) }\left[
\left\vert \Psi \left( Z,\theta \right) \right\vert ^{2}\frac{\omega
_{0}\left( \theta ,Z\right) }{\Lambda ^{2}}\right] }{\omega _{0}\left(
Z\right) +\frac{\check{T}}{\left( 1-\left( 1+\left\vert \Psi \right\vert
^{2}\right) \check{T}\right) }\left[ \left\vert \Psi \left( Z,\theta \right)
\right\vert ^{2}\frac{\omega _{0}\left( \theta ,Z\right) }{\Lambda ^{2}}%
\right] }\right) \check{T}\right) ^{-1}\left( Z,\theta ,Z_{i},\theta
_{i}\right) \\
&&\times \left[ \left\vert \Psi \left( Z_{i},\theta _{i}\right) \right\vert
^{2}\frac{\omega _{0}\left( \theta _{i},Z_{i}\right) }{\Lambda ^{2}}\right]
\\
&\equiv &\sum_{i}\int K\left( Z,\theta ,Z_{i},\theta _{i}\right) \left\{
\left\vert \Psi \left( Z,\theta _{i}\right) \right\vert ^{2}\frac{\omega
_{0}\left( \theta _{i},Z\right) }{\Lambda ^{2}}\right\} d\theta _{i}
\end{eqnarray*}

As explained in the text, we decompose the fields $\Gamma $ and $\Gamma
^{\dag }$ as sums:%
\begin{eqnarray*}
\Gamma \left( T,\hat{T},\theta ,Z,Z^{\prime }\right) &=&\Gamma _{0}\left( T,%
\hat{T},\theta ,Z,Z^{\prime }\right) +\Delta \Gamma \left( T,\hat{T},\theta
,Z,Z^{\prime }\right) \\
\Gamma ^{\dag }\left( T,\hat{T},\theta ,Z,Z^{\prime }\right) &=&\Gamma
_{0}^{\dag }\left( T,\hat{T},\theta ,Z,Z^{\prime }\right) +\Delta \Gamma
^{\dag }\left( T,\hat{T},\theta ,Z,Z^{\prime }\right)
\end{eqnarray*}%
In the sequel, the expression $\delta \omega \left( \theta ,Z,\left\vert
\Psi \right\vert ^{2}\right) $ will stand for the part of $\delta \omega
\left( \theta ,Z,\left\vert \Psi \right\vert ^{2}\right) $ depending on $%
\Delta \Gamma \left( T,\hat{T},\theta ,Z,Z^{\prime }\right) $ and $\Delta
\Gamma ^{\dag }\left( T,\hat{T},\theta ,Z,Z^{\prime }\right) $, while the
constant part is written $\delta \omega _{0}$. In other words:%
\begin{equation*}
\delta \omega \left( \theta ,Z,\left\vert \Psi \right\vert ^{2}\right)
-\delta \omega \left( \theta ,Z,\left\vert \Psi \right\vert ^{2},\Delta
\Gamma =\Delta \Gamma ^{\dag }=0\right) \rightarrow \delta \omega \left(
\theta ,Z,\left\vert \Psi \right\vert ^{2}\right)
\end{equation*}%
and: 
\begin{equation*}
\delta \omega _{0}=\delta \omega \left( \theta ,Z,\left\vert \Psi
\right\vert ^{2},\Delta \Gamma =\Delta \Gamma ^{\dag }=0\right)
\end{equation*}%
Moreover, the constant $\delta \omega _{0}$ will be now included in the
background activity $\omega _{0}$.

The second order expansion in $\Delta \Gamma \left( T,\hat{T},\theta
,Z,Z^{\prime }\right) $ and $\Delta \Gamma ^{\dag }\left( T,\hat{T},\theta
,Z,Z^{\prime }\right) $ of action (\ref{FC}) around the background state:%
\begin{equation*}
\Gamma _{0}\left( T,\hat{T},\theta ,Z,Z^{\prime }\right) ,\Gamma _{0}^{\dag
}\left( T,\hat{T},\theta ,Z,Z^{\prime }\right)
\end{equation*}%
writes:%
\begin{equation*}
S\left( \Gamma ,\Gamma ^{\dag }\right) \simeq S\left( \Gamma _{0},\Gamma
_{0}^{\dag }\right) +\int \Delta \Gamma ^{\dag }\left( T,\hat{T},\theta
,Z,Z^{\prime }\right) \frac{\delta ^{2}}{\delta \Gamma _{0}\left( T,\hat{T}%
,\theta ,Z,Z^{\prime }\right) \delta \Gamma _{0}^{\dag }\left( T,\hat{T}%
,\theta ,Z,Z^{\prime }\right) }\Delta \Gamma \left( T,\hat{T},\theta
,Z,Z^{\prime }\right)
\end{equation*}%
is decomposed in two parts. The first one, noted $C_{1}$computes the "free"
transition functions in the background state 
\begin{eqnarray}
&&C_{1}=\Delta \Gamma ^{\dag }\left( T,\hat{T},\theta ,Z,Z^{\prime }\right)
\left( \nabla _{T}\left( \nabla _{T}-\left( -\frac{1}{\tau \omega _{0}\left(
Z\right) }T+\frac{\lambda }{\omega _{0}}\hat{T}\right) \left\vert \Psi
\left( \theta ,Z\right) \right\vert ^{2}\right) \right) \Delta \Gamma \left(
T,\hat{T},\theta ,Z,Z^{\prime }\right) \\
&&+\Delta \Gamma ^{\dag }\left( T,\hat{T},\theta ,Z,Z^{\prime }\right)
\left( \nabla _{\hat{T}}\left( \nabla _{\hat{T}}-\frac{\rho }{\omega
_{0}\left( Z\right) }\right. \right.  \notag \\
&&\times \left. \left. \left( \left( h\left( Z,Z^{\prime }\right) -\hat{T}%
\right) C\left( \theta \right) \left\vert \Psi _{0}\left( Z\right)
\right\vert ^{2}\omega _{0}\left( Z\right) -D\left( \theta \right) \hat{T}%
\left\vert \Psi _{0}\left( Z^{\prime }\right) \right\vert ^{2}\omega
_{0}\left( Z^{\prime }\right) \right) \right) \right) \Delta \Gamma \left( T,%
\hat{T},\theta ,Z,Z^{\prime }\right)  \notag
\end{eqnarray}%
The second term includes two perturbative contributions:%
\begin{eqnarray}
&&C_{2,1}=\Gamma _{0}^{\dag }\left( T,\hat{T},\theta ,Z,Z^{\prime }\right) \\
&&\times \left( \nabla _{T}\left( \left( -\frac{\delta \omega \left( \theta
,Z,\left\vert \Psi \right\vert ^{2}\right) }{\tau \omega _{0}^{2}\left(
Z\right) }T+\frac{\lambda \delta \omega \left( \theta ,Z,\left\vert \Psi
\right\vert ^{2}\right) }{\omega _{0}^{2}}\hat{T}\right) \left\vert \Psi
_{0}\left( Z\right) \right\vert ^{2}\right) \right) \Gamma _{0}\left( T,\hat{%
T},\theta ,Z,Z^{\prime }\right)  \notag \\
&&-\Gamma _{0}^{\dag }\left( T,\hat{T},\theta ,Z,Z^{\prime }\right) \times
\nabla _{\hat{T}}\left( \frac{\rho }{\omega _{0}\left( Z\right) }\left(
\left( h\left( Z,Z^{\prime }\right) -\hat{T}\right) C\left( \theta \right)
\left\vert \Psi _{0}\left( Z\right) \right\vert ^{2}\delta \omega \left(
\theta ,Z,\left\vert \Psi \right\vert ^{2}\right) \right. \right.  \notag \\
&&\left. \left. -D\left( \theta \right) \hat{T}\left\vert \Psi _{0}\left(
Z^{\prime }\right) \right\vert ^{2}\delta \omega \left( \theta -\frac{%
\left\vert Z-Z^{\prime }\right\vert }{c},Z^{\prime },\left\vert \Psi
\right\vert ^{2}\right) \right) \right) \times \Gamma _{0}\left( T,\hat{T}%
,\theta ,Z,Z^{\prime }\right)  \notag
\end{eqnarray}%
and:%
\begin{eqnarray}
C_{2,2} &=&\Delta \Gamma ^{\dag }\left( T,\hat{T},\theta ,Z,Z^{\prime
}\right) \\
&&\times \left( \nabla _{T}\left( \left( -\frac{\delta \omega \left( \theta
,Z,\left\vert \Psi \right\vert ^{2}\right) }{\tau \omega _{0}^{2}\left(
Z\right) }T+\frac{\lambda \delta \omega \left( \theta ,Z,\left\vert \Psi
\right\vert ^{2}\right) }{\omega _{0}^{2}}\hat{T}\right) \left\vert \Psi
_{0}\left( Z\right) \right\vert ^{2}\right) \right) \Delta \Gamma \left( T,%
\hat{T},\theta ,Z,Z^{\prime }\right)  \notag \\
&&-\Delta \Gamma ^{\dag }\left( T,\hat{T},\theta ,Z,Z^{\prime }\right)
\times \nabla _{\hat{T}}\left( \frac{\rho }{\omega _{0}\left( Z\right) }%
\left( \left( h\left( Z,Z^{\prime }\right) -\hat{T}\right) C\left( \theta
\right) \left\vert \Psi _{0}\left( Z\right) \right\vert ^{2}\delta \omega
\left( \theta ,Z,\left\vert \Psi \right\vert ^{2}\right) \right. \right. 
\notag \\
&&\times \left. \left. -D\left( \theta \right) \hat{T}\left\vert \Psi
_{0}\left( Z^{\prime }\right) \right\vert ^{2}\delta \omega \left( \theta -%
\frac{\left\vert Z-Z^{\prime }\right\vert }{c},Z^{\prime },\left\vert \Psi
\right\vert ^{2}\right) \right) \right) \Delta \Gamma \left( T,\hat{T}%
,\theta ,Z,Z^{\prime }\right)  \notag
\end{eqnarray}

We use that in the background state:%
\begin{equation*}
\Gamma _{0}\left( T,\hat{T},\theta ,Z,Z^{\prime }\right) ,\Gamma _{0}^{\dag
}\left( T,\hat{T},\theta ,Z,Z^{\prime }\right)
\end{equation*}%
the following relations stand:%
\begin{equation*}
-\frac{\delta \omega \left( \theta ,Z,\left\vert \Psi \right\vert
^{2}\right) }{\tau \omega _{0}\left( Z\right) }\left\langle T\right\rangle +%
\frac{\lambda \delta \omega \left( \theta ,Z,\left\vert \Psi \right\vert
^{2}\right) }{\omega _{0}}\left\langle \hat{T}\right\rangle =0
\end{equation*}%
and:%
\begin{equation*}
\left( h\left( Z,Z^{\prime }\right) -\hat{T}\right) C\left( \theta \right)
\left\vert \Psi _{0}\left( Z\right) \right\vert ^{2}\omega _{0}\left(
Z\right) -D\left( \theta \right) \hat{T}\left\vert \Psi _{0}\left( Z^{\prime
}\right) \right\vert ^{2}\omega _{0}\left( Z^{\prime }\right) =0
\end{equation*}%
As a consequence we have the following identities:%
\begin{equation*}
-\frac{\delta \omega \left( \theta ,Z,\left\vert \Psi \right\vert
^{2}\right) }{\tau \omega _{0}^{2}\left( Z\right) }T+\frac{\lambda \delta
\omega \left( \theta ,Z,\left\vert \Psi \right\vert ^{2}\right) }{\omega
_{0}^{2}}\hat{T}=-\frac{\delta \omega \left( \theta ,Z,\left\vert \Psi
\right\vert ^{2}\right) }{\tau \omega _{0}^{2}\left( Z\right) }\left(
T-\left\langle T\right\rangle \right) +\frac{\lambda \delta \omega \left(
\theta ,Z,\left\vert \Psi \right\vert ^{2}\right) }{\omega _{0}^{2}}\left( 
\hat{T}-\left\langle \hat{T}\right\rangle \right)
\end{equation*}%
and:%
\begin{eqnarray*}
&&\left( h\left( Z,Z^{\prime }\right) -\hat{T}\right) C\left( \theta \right)
\left\vert \Psi _{0}\left( Z\right) \right\vert ^{2}\delta \omega \left(
\theta ,Z,\left\vert \Psi \right\vert ^{2}\right) -D\left( \theta \right) 
\hat{T}\left\vert \Psi _{0}\left( Z^{\prime }\right) \right\vert ^{2}\delta
\omega \left( \theta -\frac{\left\vert Z-Z^{\prime }\right\vert }{c}%
,Z^{\prime },\left\vert \Psi \right\vert ^{2}\right) \\
&=&D\left( \theta \right) \hat{T}\left\vert \Psi _{0}\left( Z^{\prime
}\right) \right\vert ^{2}\left( \frac{\omega _{0}\left( Z^{\prime }\right) }{%
\omega _{0}\left( Z\right) }\delta \omega \left( \theta ,Z,\left\vert \Psi
\right\vert ^{2}\right) -\delta \omega \left( \theta -\frac{\left\vert
Z-Z^{\prime }\right\vert }{c},Z^{\prime },\left\vert \Psi \right\vert
^{2}\right) \right)
\end{eqnarray*}%
These relations enable to rewrite the various contributions to the action.
The first contribution $C_{1}$ becomes:%
\begin{eqnarray}
&&C_{1}=\Delta \Gamma ^{\dag }\left( T,\hat{T},\theta ,Z,Z^{\prime }\right)
\\
&&\times \left( \nabla _{T}\left( \nabla _{T}+\frac{\left( T-\left\langle
T\right\rangle \right) -\lambda \left( \hat{T}-\left\langle \hat{T}%
\right\rangle \right) }{\tau \omega _{0}\left( Z\right) }\left\vert \Psi
\left( \theta ,Z\right) \right\vert ^{2}\right) \right) \Delta \Gamma \left(
T,\hat{T},\theta ,Z,Z^{\prime }\right)  \notag \\
&&+\Delta \Gamma ^{\dag }\left( T,\hat{T},\theta ,Z,Z^{\prime }\right)
\nabla _{\hat{T}}\left( \nabla _{\hat{T}}+\rho \left\vert \bar{\Psi}%
_{0}\left( Z,Z^{\prime }\right) \right\vert ^{2}\left( \hat{T}-\left\langle 
\hat{T}\right\rangle \right) \right) \Delta \Gamma \left( T,\hat{T},\theta
,Z,Z^{\prime }\right)  \notag
\end{eqnarray}%
where:%
\begin{equation*}
\left\vert \bar{\Psi}_{0}\left( Z,Z^{\prime }\right) \right\vert ^{2}=\frac{%
\left( C\left( \theta \right) \left\vert \Psi _{0}\left( Z\right)
\right\vert ^{2}\omega _{0}\left( Z\right) +D\left( \theta \right) \hat{T}%
\left\vert \Psi _{0}\left( Z^{\prime }\right) \right\vert ^{2}\omega
_{0}\left( Z^{\prime }\right) \right) }{\omega _{0}\left( Z\right) }
\end{equation*}%
The contribution $C_{2,1}$ is: 
\begin{eqnarray}
C_{2,1} &=&\Gamma _{0}^{\dag }\left( T,\hat{T},\theta ,Z,Z^{\prime }\right)
\label{frr} \\
&&\times \left( \nabla _{\hat{T}}\left( \frac{\rho }{\omega _{0}^{2}\left(
Z\right) }\left( D\left( \theta \right) \left\langle \hat{T}\right\rangle
\left\vert \Psi _{0}\left( Z^{\prime }\right) \right\vert ^{2}\left( \omega
_{0}\left( Z\right) \delta \omega \left( \theta -\frac{\left\vert
Z-Z^{\prime }\right\vert }{c},Z^{\prime },\left\vert \Psi \right\vert
^{2}\right) -\omega _{0}\left( Z^{\prime }\right) \delta \omega \left(
\theta ,Z,\left\vert \Psi \right\vert ^{2}\right) \right) \right) \right)
\right)  \notag \\
&&\times \Gamma _{0}\left( T,\hat{T},\theta ,Z,Z^{\prime }\right)  \notag
\end{eqnarray}%
While. $C_{2,2}$ writes:%
\begin{eqnarray}
&&\Delta \Gamma ^{\dag }\left( T,\hat{T},\theta ,Z,Z^{\prime }\right)
\label{sdr} \\
&&\times \left( \nabla _{T}\left( \left( -\frac{\delta \omega \left( \theta
,Z,\left\vert \Psi \right\vert ^{2}\right) }{\tau \omega _{0}^{2}\left(
Z\right) }\left( T-\left\langle T\right\rangle \right) +\frac{\lambda \delta
\omega \left( \theta ,Z,\left\vert \Psi \right\vert ^{2}\right) }{\omega
_{0}^{2}}\left( \hat{T}-\left\langle \hat{T}\right\rangle \right) \right)
\left\vert \Psi _{0}\left( Z\right) \right\vert ^{2}\right) \right) \Delta
\Gamma \left( T,\hat{T},\theta ,Z,Z^{\prime }\right)  \notag \\
&&+\Delta \Gamma ^{\dag }\left( T,\hat{T},\theta ,Z,Z^{\prime }\right) 
\notag \\
&&\times \left( \nabla _{\hat{T}}\left( \frac{\rho }{\omega _{0}^{2}\left(
Z\right) }\left( D\left( \theta \right) \left\langle \hat{T}\right\rangle
\left\vert \Psi _{0}\left( Z^{\prime }\right) \right\vert ^{2}\left( \omega
_{0}\left( Z\right) \delta \omega \left( \theta -\frac{\left\vert
Z-Z^{\prime }\right\vert }{c},Z^{\prime },\left\vert \Psi \right\vert
^{2}\right) -\omega _{0}\left( Z^{\prime }\right) \delta \omega \left(
\theta ,Z,\left\vert \Psi \right\vert ^{2}\right) \right) \right) \right)
\right.  \notag \\
&&\left. +\nabla _{\hat{T}}\left( \frac{\rho \left( C\left( \theta \right)
\delta \omega \left( \theta ,Z,\left\vert \Psi \right\vert ^{2}\right)
+D\left( \theta \right) \delta \omega \left( \theta -\frac{\left\vert
Z-Z^{\prime }\right\vert }{c},Z^{\prime },\left\vert \Psi \right\vert
^{2}\right) \right) }{\omega _{0}\left( Z\right) }\left\vert \Psi _{0}\left(
Z\right) \right\vert ^{2}\left( \hat{T}-\left\langle \hat{T}\right\rangle
\right) \left\vert \Psi _{0}\left( Z^{\prime }\right) \right\vert
^{2}\right) \right)  \notag \\
&&\times \Delta \Gamma \left( T,\hat{T},\theta ,Z,Z^{\prime }\right)  \notag
\end{eqnarray}

\subsubsection*{1. 1.2 Several approximations for interaction terms}

The contribution $C_{2,1}$ describes the modification of the background by
the fluctuations. In first approximation it can be neglected.

Moreover, while studying the internal dynamics of connectivities, the
activities oscillations $\delta \omega \left( \theta ,Z,\left\vert \Psi
\right\vert ^{2}\right) $ is proportional to $\left( T-\left\langle
T\right\rangle \right) $. As a consequence, the first and last terms in (\ref%
{sdr}) can also be neglected for\ small oscillations. As a consequence, the
interaction terms to be considered reduce to:%
\begin{eqnarray}
&&C_{2,2}=\Delta \Gamma ^{\dag }\left( T,\hat{T},\theta ,Z,Z^{\prime }\right)
\label{tdr} \\
&&\times \nabla _{\hat{T}}\left( \frac{\rho D\left( \theta \right)
\left\langle \hat{T}\right\rangle \left\vert \Psi _{0}\left( Z^{\prime
}\right) \right\vert ^{2}}{\omega _{0}^{2}\left( Z\right) }\left( \omega
_{0}\left( Z\right) \delta \omega \left( \theta -\frac{\left\vert
Z-Z^{\prime }\right\vert }{c},Z^{\prime },\left\vert \Psi \right\vert
^{2}\right) -\omega _{0}\left( Z^{\prime }\right) \delta \omega \left(
\theta ,Z,\left\vert \Psi \right\vert ^{2}\right) \right) \right) \Delta
\Gamma \left( T,\hat{T},\theta ,Z,Z^{\prime }\right)  \notag
\end{eqnarray}

In first approximation we can assume that in the fluctuation state $\Delta
\Gamma \left( T,\hat{T},\theta ,Z,Z^{\prime }\right) $, $\left(
T-\left\langle T\right\rangle \right) =\left( \hat{T}-\left\langle \hat{T}%
\right\rangle \right) $. We approximate:%
\begin{eqnarray*}
&&\omega _{0}\left( Z\right) \delta \omega \left( \theta -\frac{\left\vert
Z-Z^{\prime }\right\vert }{c},Z^{\prime },\left\vert \Psi \right\vert
^{2}\right) -\omega _{0}\left( Z^{\prime }\right) \delta \omega \left(
\theta ,Z,\left\vert \Psi \right\vert ^{2}\right) \\
&\simeq &\omega _{0}\left( Z\right) \left( \left( Z^{\prime }-Z\right)
\nabla _{Z}-\frac{\left\vert Z-Z^{\prime }\right\vert }{c}\nabla _{\theta
}-\left( Z^{\prime }-Z\right) \frac{\nabla _{Z}\omega _{0}\left( Z\right) }{%
\omega _{0}\left( Z\right) }\right. \\
&&+\left. \frac{1}{2}\left( \left( Z^{\prime }-Z\right) _{i}\left( Z^{\prime
}-Z\right) _{j}\nabla _{Z_{i}}\nabla _{Z_{j}}+\frac{\left\vert Z-Z^{\prime
}\right\vert ^{2}}{c^{2}}\nabla _{\theta }^{2}-\left( Z^{\prime }-Z\right)
_{i}\left( Z^{\prime }-Z\right) _{j}\frac{\nabla _{Z_{i}}\nabla
_{Z_{j}}\omega _{0}\left( Z\right) }{\omega _{0}\left( Z\right) }\right)
\right) \delta \omega \left( \theta ,Z,\left\vert \Psi \right\vert
^{2}\right)
\end{eqnarray*}%
Inserted in integrals, the first order odd term $\left( Z^{\prime }-Z\right)
\nabla _{Z}\omega _{0}\left( Z\right) $ cancel in first approximations.
Similarly the\ terms:%
\begin{equation*}
\left( Z^{\prime }-Z\right) _{i}\left( Z^{\prime }-Z\right) _{j}\frac{\nabla
_{Z_{i}}\nabla _{Z_{j}}\omega _{0}\left( Z\right) }{\omega _{0}\left(
Z\right) }
\end{equation*}%
and:%
\begin{equation*}
\left( Z^{\prime }-Z\right) _{i}\left( Z^{\prime }-Z\right) _{j}\nabla
_{Z_{i}}\nabla _{Z_{j}}
\end{equation*}%
also cancel for $i\neq j$, and previous formula reduces to:%
\begin{eqnarray}
&&\omega _{0}\left( Z\right) \delta \omega \left( \theta -\frac{\left\vert
Z-Z^{\prime }\right\vert }{c},Z^{\prime },\left\vert \Psi \right\vert
^{2}\right) -\omega _{0}\left( Z^{\prime }\right) \delta \omega \left(
\theta ,Z,\left\vert \Psi \right\vert ^{2}\right)  \label{cmp} \\
&\simeq &\omega _{0}\left( Z\right) \left( \left( Z^{\prime }-Z\right)
\nabla _{Z}-\frac{\left\vert Z-Z^{\prime }\right\vert }{c}\nabla _{\theta
}\right.  \notag \\
&&+\left. \frac{1}{2}\left( \left( Z^{\prime }-Z\right) _{i}\left( Z^{\prime
}-Z\right) _{j}\nabla _{Z_{i}}\nabla _{Z_{j}}+\frac{\left\vert Z-Z^{\prime
}\right\vert ^{2}}{c^{2}}\nabla _{\theta }^{2}-\frac{\nabla _{Z}^{2}\omega
_{0}\left( Z\right) }{\omega _{0}\left( Z\right) }\right) \right) \delta
\omega \left( \theta ,Z,\left\vert \Psi \right\vert ^{2}\right)  \notag
\end{eqnarray}%
and we have the interaction term:%
\begin{eqnarray}
&&C_{2,2}=\Delta \Gamma ^{\dag }\left( T,\hat{T},\theta ,Z,Z^{\prime }\right)
\label{cmq} \\
&&\times \nabla _{\hat{T}}\left( \frac{\rho }{\omega _{0}\left( Z\right) }%
\left( D\left( \theta \right) \left\langle \hat{T}\right\rangle \left\vert
\Psi _{0}\left( Z^{\prime }\right) \right\vert ^{2}\left( \left( -\frac{%
\left\vert Z-Z^{\prime }\right\vert }{c}\nabla _{\theta }+\frac{\left(
Z^{\prime }-Z\right) ^{2}}{2}\left( \nabla _{Z}^{2}+\frac{\nabla _{\theta
}^{2}}{c^{2}}-\frac{\nabla _{Z}^{2}\omega _{0}\left( Z\right) }{\omega
_{0}\left( Z\right) }\right) \right) \delta \omega \left( \theta
,Z,\left\vert \Psi \right\vert ^{2}\right) \right) \right) \right)  \notag \\
&&\Delta \Gamma \left( T,\hat{T},\theta ,Z,Z^{\prime }\right)  \notag
\end{eqnarray}%
Note that the term (\ref{frr}) can also be written similarly, if we want to
keep its contribution:%
\begin{eqnarray}
C_{2,1} &=&\Gamma _{0}^{\dag }\left( T,\hat{T},\theta ,Z,Z^{\prime }\right)
\label{frw} \\
&&\times \nabla _{\hat{T}}\left( \frac{\rho D\left( \theta \right)
\left\langle \hat{T}\right\rangle \left\vert \Psi _{0}\left( Z^{\prime
}\right) \right\vert ^{2}}{\omega _{0}\left( Z\right) }\left( -\frac{%
\left\vert Z-Z^{\prime }\right\vert }{c}\nabla _{\theta }+\frac{\left(
Z^{\prime }-Z\right) ^{2}}{2}\left( \nabla _{Z}^{2}+\frac{\nabla _{\theta
}^{2}}{c^{2}}-\frac{\nabla _{Z}^{2}\omega _{0}\left( Z\right) }{\omega
_{0}\left( Z\right) }\right) \right) \delta \omega \left( \theta
,Z,\left\vert \Psi \right\vert ^{2}\right) \right)  \notag \\
&&\times \Gamma _{0}\left( T,\hat{T},\theta ,Z,Z^{\prime }\right)  \notag
\end{eqnarray}

\subsection*{1.2 Derivation of $\protect\delta \protect\omega \left( \protect%
\theta ,Z,\left\vert \Psi \right\vert ^{2}\right) $ as a function of
connectivities fluctuations}

\subsubsection*{1.2.1 Compact formula for the first order}

In a fluctuating state:%
\begin{equation*}
\Gamma _{0}\left( T,\hat{T},\theta ,Z,Z^{\prime }\right) +\Delta \Gamma
\left( T,\hat{T},\theta ,Z,Z^{\prime }\right)
\end{equation*}%
the activities $\delta \omega \left( \theta ,Z,\left\vert \Psi \right\vert
^{2}\right) $ are modified by fluctuations $\Delta \Gamma \left( T,\hat{T}%
,\theta ,Z,Z^{\prime }\right) $. Actually, the averages connectivities in
the background:%
\begin{equation*}
T\left( Z,Z_{1},\theta \right) =\left\langle T\right\rangle \left\vert
\Gamma _{0}\left( T,\hat{T},\theta ,Z,Z^{\prime }\right) \right\vert
^{2}=\int T\left\vert \Gamma \left( T,\hat{T},\theta ,Z,Z^{\prime }\right)
\right\vert ^{2}dTd\hat{T}
\end{equation*}%
become:%
\begin{eqnarray*}
&&\left\langle T\right\rangle \left\vert \Gamma _{0}\left( T,\hat{T},\theta
,Z,Z^{\prime }\right) \right\vert ^{2}+\left( T-\left\langle T\right\rangle
\right) \left\vert \Delta \Gamma \left( T,\hat{T},\theta ,Z,Z^{\prime
}\right) \right\vert ^{2} \\
&=&T\left( Z,Z_{1},\theta \right) \left( 1+\frac{\left( T-\left\langle
T\right\rangle \right) \left\vert \Delta \Gamma \left( T,\hat{T},\theta
,Z,Z^{\prime }\right) \right\vert ^{2}}{T\left( Z,Z_{1},\theta \right) }%
\right)
\end{eqnarray*}%
As a consequence, the activity equation defined by (\ref{dng}):%
\begin{equation}
\omega ^{-1}\left( \theta ,Z\right) =G\left( J\left( \theta \right) +\frac{%
\kappa }{N}\int T\left( Z,Z_{1},\theta \right) \frac{\omega \left( \theta -%
\frac{\left\vert Z-Z_{1}\right\vert }{c},Z_{1}\right) }{\omega \left( \theta
,Z\right) }\left\vert \Psi \left( \theta -\frac{\left\vert
Z-Z_{1}\right\vert }{c},Z_{1}\right) \right\vert ^{2}dZ_{1}\right)
\end{equation}%
is modified by replacing the background value: 
\begin{equation*}
T\left( Z,Z_{1},\theta \right) \left\vert \Psi \left( \theta -\frac{%
\left\vert Z-Z_{1}\right\vert }{c},Z_{1}\right) \right\vert ^{2}
\end{equation*}
with:%
\begin{equation*}
\left( 1+\frac{\left( T-\left\langle T\left( Z\right) \right\rangle \right)
\left\vert \Delta \Gamma \left( T,\hat{T},\theta ,Z,Z_{1}\right) \right\vert
^{2}}{T\left( Z,Z_{1},\theta \right) }\right) \left\vert \Psi \left( \theta -%
\frac{\left\vert Z-Z_{1}\right\vert }{c},Z_{1}\right) \right\vert ^{2}
\end{equation*}%
Since we are interested in the self interactions of $\Delta \Gamma \left( T,%
\hat{T},\theta ,Z,Z_{1}\right) $, we will approximate $\left\vert \Psi
\left( \theta -\frac{\left\vert Z-Z_{1}\right\vert }{c},Z_{1}\right)
\right\vert ^{2}$ by its static value, so that in (\ref{dng}), we replace: 
\begin{equation*}
T\left( Z,Z_{1},\theta \right) \left\vert \Psi \left( \theta -\frac{%
\left\vert Z-Z_{1}\right\vert }{c},Z_{1}\right) \right\vert ^{2}\rightarrow
\left( 1+\frac{\left( T-\left\langle T\left( Z\right) \right\rangle \right)
\left\vert \Delta \Gamma \left( T,\hat{T},\theta ,Z,Z_{1}\right) \right\vert
^{2}}{T\left( Z,Z_{1},\theta \right) }\right) \left\vert \Psi _{0}\left(
Z_{1}\right) \right\vert ^{2}
\end{equation*}%
Thus, at the first order, including the corrections to the activities due to
the fluctuations $\Delta \Gamma \left( T,\hat{T},\theta ,Z,Z_{1}\right) $
leads to:%
\begin{eqnarray}
&&\delta \omega _{f}\left( \theta ,Z,\left\vert \Psi \right\vert ^{2}\right)
\notag \\
&=&\int \check{T}\left( 1-\left( 1+\left\vert \Psi _{\Gamma }\right\vert
^{2}-\frac{\frac{\check{T}}{\left( 1-\left( 1+\left\vert \Psi _{\Gamma
}\right\vert ^{2}\right) \check{T}\right) }\left[ \left\vert \Psi _{\Gamma
}\left( Z,\theta \right) \right\vert ^{2}\frac{\omega _{0}\left( \theta
,Z\right) }{\Lambda ^{2}}\right] }{\omega _{0}\left( Z\right) +\frac{\check{T%
}}{\left( 1-\left( 1+\left\vert \Psi _{\Gamma }\right\vert ^{2}\right) 
\check{T}\right) }\left[ \left\vert \Psi _{\Gamma }\left( Z,\theta \right)
\right\vert ^{2}\frac{\omega _{0}\left( \theta ,Z\right) }{\Lambda ^{2}}%
\right] }\right) \check{T}\right) ^{-1}\left( Z,\theta ,Z_{i},\theta
_{i}\right)  \notag \\
&&\times \left[ \left\vert \Psi _{\Gamma }\left( Z_{i},\theta _{i}\right)
\right\vert ^{2}\frac{\omega _{0}\left( \theta _{i},Z_{i}\right) }{\Lambda
^{2}}\right] d\left( Z_{i},\theta _{i}\right)  \label{dmf}
\end{eqnarray}

where:%
\begin{equation*}
=\left\vert \Psi _{\Gamma }\left( Z,\theta \right) \right\vert ^{2}=\left( 1+%
\frac{\Delta T\left\vert \Delta \Gamma \right\vert ^{2}}{T}\right)
\left\vert \Psi \right\vert ^{2}
\end{equation*}%
\begin{equation*}
\frac{\Delta T\left\vert \Delta \Gamma \right\vert ^{2}}{T}=\frac{\left(
T-\left\langle T\left( Z\right) \right\rangle \right) \left\vert \Delta
\Gamma \left( T,\hat{T},\theta ,Z,Z_{1}\right) \right\vert ^{2}}{T\left(
Z,Z_{1},\theta \right) }
\end{equation*}%
and the formula has to inserted in formula for the interaction terms of the
effective action (\ref{cmq}) or (\ref{frw}).

\subsubsection*{1.2.2 Series expansion of interaction terms}

\paragraph*{1.2.2.1 Expression for $\protect\delta \protect\omega \left( 
\protect\theta ,Z,\left\vert \Psi \right\vert ^{2}\right) $}

More generally, we can expand $\delta \omega _{f}\left( \theta ,Z,\left\vert
\Psi \right\vert ^{2}\right) $ in series. The operator:%
\begin{equation*}
\frac{\check{T}}{\left( 1-\left( 1+\left\vert \Psi _{\Gamma }\right\vert
^{2}\right) \check{T}\right) }
\end{equation*}%
arising in (\ref{dmf}) has the following series expansion: 
\begin{equation*}
\sum \frac{\check{T}}{\left( 1-\left( 1+\right) \check{T}\right) }\left\vert
\Psi _{\Gamma }\left( \theta _{1},Z_{1}\right) \right\vert ^{2}\frac{\check{T%
}}{\left( 1-\left( 1+\right) \check{T}\right) }...\left\vert \Psi _{\Gamma
}\left( \theta _{n},Z_{n}\right) \right\vert ^{2}\frac{\check{T}}{\left(
1-\left( 1+\right) \check{T}\right) }
\end{equation*}%
with the kernel $\frac{\check{T}}{\left( 1-\left( 1+\right) \check{T}\right) 
}$ estimated previously (see (\cite{GLs})):%
\begin{equation}
\frac{\check{T}}{\left( 1-\left( 1+\right) \check{T}\right) }\left(
Z,Z_{1},l_{1}\right) \simeq \frac{\exp \left( -cl_{1}-\alpha \left( \left(
cl_{1}\right) ^{2}-\left\vert Z-Z_{1}\right\vert ^{2}\right) \right) }{D}%
H\left( cl_{1}-\left\vert Z-Z_{1}\right\vert \right)  \label{KL}
\end{equation}%
where the constant $D$ depends on the average values of $\check{T}$ in the
background field. Then the interaction term becomes:%
\begin{eqnarray}
&&\delta \omega \left( \theta ,Z,\left\vert \Psi \right\vert ^{2}\right)
\label{DG} \\
&=&\frac{\check{T}}{\left( 1-\left( 1+\right) \check{T}\right) }\left\vert
\Psi _{\Gamma }\left( \theta _{1},Z_{1}\right) \right\vert ^{2}\frac{\check{T%
}}{\left( 1-\left( 1+\right) \check{T}\right) }...\left\vert \Psi _{\Gamma
}\left( \theta _{1},Z_{1}\right) \right\vert ^{2}\frac{\check{T}}{\left(
1-\left( 1+\right) \check{T}\right) }\left\vert \Psi _{\Gamma }\left(
Z_{i},\theta _{i}\right) \right\vert ^{2}\frac{\omega _{0}\left( \theta
_{i},Z_{i}\right) }{\Lambda ^{2}}  \notag \\
&=&\frac{\exp \left( -c\left( \theta -\theta _{1}\right) -\alpha \left(
\left( c\left( \theta -\theta _{1}\right) \right) ^{2}-\left\vert
Z-Z_{1}\right\vert ^{2}\right) \right) }{D}  \notag \\
&&\times \left\vert \Psi _{\Gamma }\left( \theta _{1},Z_{1}\right)
\right\vert ^{2}\frac{\exp \left( -c\left( \theta _{1}-\theta _{2}\right)
-\alpha \left( \left( c\left( \theta _{1}-\theta _{2}\right) \right)
^{2}-\left\vert Z_{1}-Z_{2}\right\vert ^{2}\right) \right) }{D}  \notag \\
&&...\left\vert \Psi _{\Gamma }\left( \theta _{n},Z_{n}\right) \right\vert
^{2}\frac{\exp \left( -c\left( \theta _{n}-\theta _{i}\right) -\alpha \left(
\left( c\left( \theta _{n}-\theta _{i}\right) \right) ^{2}-\left\vert
Z_{n}-Z_{i}\right\vert ^{2}\right) \right) }{D}\left\vert \Psi _{\Gamma
}\left( Z_{i},\theta _{i}\right) \right\vert ^{2}\frac{\omega _{0}\left(
\theta _{i},Z_{i}\right) }{\Lambda ^{2}}  \notag \\
&=&G\left( \theta -\theta _{1},Z-Z_{1}\right) \left[ \prod \left\vert \Psi
_{\Gamma }\left( \theta _{j},Z_{j}\right) \right\vert ^{2}G\left( \theta
_{j}-\theta _{j+1},Z_{j}-Z_{j+1}\right) \right] G\left( \theta _{n}-\theta
_{i},Z_{n}-Z_{i}\right) \left\vert \Psi _{\Gamma }\left( Z_{i},\theta
_{i}\right) \right\vert ^{2}\frac{\omega _{0}\left( \theta _{i},Z_{i}\right) 
}{\Lambda ^{2}}  \notag
\end{eqnarray}

Keeping only the fluctuations corrections $\frac{\Delta T\left\vert \Delta
\Gamma \right\vert ^{2}}{T}\left\vert \Psi \left( Z_{i},\theta _{i}\right)
\right\vert ^{2}$ inthe series yields the interaction terms:%
\begin{eqnarray}
&&G\left( \theta -\theta _{1},Z-Z_{1}\right) \left[ \prod \frac{\Delta
T\left\vert \Delta \Gamma \left( \theta _{j},Z_{j},Z_{j+1}\right)
\right\vert ^{2}}{T}\left\vert \Psi \left( \theta _{j},Z_{j}\right)
\right\vert ^{2}G\left( \theta _{j}-\theta _{j+1},Z_{j}-Z_{j+1}\right) %
\right]  \label{trc} \\
&&\times G\left( \theta _{n-1}-\theta _{n},Z_{n-1}-Z_{n}\right) \left\vert
\Psi _{0}\left( Z_{n},\theta _{n}\right) \right\vert ^{2}\frac{\omega
_{0}\left( \theta _{n},Z_{n}\right) }{\Lambda ^{2}}  \notag
\end{eqnarray}

where:%
\begin{equation}
G\left( \theta _{j}-\theta _{j+1},Z_{j}-Z_{j+1}\right) =\frac{\exp \left(
-c\left( \theta _{j}-\theta _{j+1}\right) -\alpha \left( \left( c\left(
\theta _{j}-\theta _{j+1}\right) \right) ^{2}-\left\vert
Z_{j}-Z_{j+1}\right\vert ^{2}\right) \right) }{D}  \label{gtm}
\end{equation}%
The insertion of corrections of the same type amounts to branch such series
at some $\left( \theta _{k},Z_{k}\right) $ and leads to sum of terms of the
form:%
\begin{eqnarray}
&&\sum \prod\limits_{p}\underset{\left( Z_{j}^{\left( p\right) },\theta
_{j}^{\left( p\right) }\right) }{\ast }\left[ G\left( \theta -\theta
_{1}^{\left( p\right) },Z-Z_{1}^{\left( p\right) }\right) \right.
\label{tcn} \\
&&\left[ \prod \frac{\Delta T\left\vert \Delta \Gamma \left( \theta
_{j}^{\left( p\right) },Z_{j}^{\left( p\right) },Z_{j+1}^{\left( p\right)
}\right) \right\vert ^{2}}{T}\left\vert \Psi \left( \theta _{j}^{\left(
p\right) },Z_{j}^{\left( p\right) }\right) \right\vert ^{2}G\left( \theta
_{j}^{\left( p\right) }-\theta _{j+1}^{\left( p\right) },Z_{j}^{\left(
p\right) }-Z_{j+1}^{\left( p\right) }\right) \right]  \notag \\
&&\left. \times G\left( \theta _{n}^{\left( p\right) }-\theta _{i}^{\left(
p\right) },Z_{n}^{\left( p\right) }-Z_{i}^{\left( p\right) }\right)
\left\vert \Psi _{\Gamma }\left( Z_{i}^{\left( p\right) },\theta
_{i}^{\left( p\right) }\right) \right\vert ^{2}\frac{\omega _{0}\left(
\theta _{i}^{\left( p\right) },Z_{i}^{\left( p\right) }\right) }{\Lambda ^{2}%
}\right]  \notag \\
&=&\sum \prod\limits_{p}\underset{\left( Z_{j}^{\left( p\right) },\theta
_{j}^{\left( p\right) }\right) }{\ast }V\left( \left( Z_{j}^{\left( p\right)
},\theta _{j}^{\left( p\right) }\right) \right)  \notag
\end{eqnarray}%
where the sign $\underset{\left( Z_{j}^{\left( p\right) },\theta
_{j}^{\left( p\right) }\right) }{\ast }$ denotes the branching of lines at
any points. The sum takes into acount all the possibility of branching lines.

\paragraph*{1.2.2.Inserting $\protect\delta \protect\omega \left( \protect%
\theta ,Z,\left\vert \Psi \right\vert ^{2}\right) $\ in the effective action}

Once the series for $\delta \omega \left( \theta ,Z,\left\vert \Psi
\right\vert ^{2}\right) $ obtained, it can be inserted in the interaction
terms (\ref{frw}) and (\ref{cmq}):%
\begin{eqnarray}
&&\Gamma _{0}^{\dag }\left( T,\hat{T},\theta ,Z,Z^{\prime }\right) \\
&&\times \left( \nabla _{\hat{T}}\left( \frac{\rho }{\omega _{0}\left(
Z\right) }\left( D\left( \theta \right) \left\langle \hat{T}\right\rangle
\left\vert \Psi _{0}\left( Z^{\prime }\right) \right\vert ^{2}\left( \left( -%
\frac{\left\vert Z-Z^{\prime }\right\vert }{c}\nabla _{\theta }+\frac{\left(
Z^{\prime }-Z\right) ^{2}}{2}\left( \nabla _{Z}^{2}+\frac{\nabla _{\theta
}^{2}}{c^{2}}-\frac{\nabla _{Z}^{2}\omega _{0}\left( Z\right) }{\omega
_{0}\left( Z\right) }\right) \right) \right) \right. \right. \right.  \notag
\\
&&\times \left. \left. \left. \left( \sum \prod\limits_{p}\underset{\left(
Z_{j}^{\left( p\right) },\theta _{j}^{\left( p\right) }\right) }{\ast }%
V\left( \left( Z_{j}^{\left( p\right) },\theta _{j}^{\left( p\right)
}\right) \right) \right) \right) \right) \right) \times \Gamma _{0}\left( T,%
\hat{T},\theta ,Z,Z^{\prime }\right)  \notag
\end{eqnarray}%
\begin{eqnarray}
&&\Delta \Gamma ^{\dag }\left( T,\hat{T},\theta ,Z,Z^{\prime }\right)
\label{ntm} \\
&&\times \left( \nabla _{\hat{T}}\left( \frac{\rho }{\omega _{0}\left(
Z\right) }\left( D\left( \theta \right) \left\langle \hat{T}\right\rangle
\left\vert \Psi _{0}\left( Z^{\prime }\right) \right\vert ^{2}\left( \left( -%
\frac{\left\vert Z-Z^{\prime }\right\vert }{c}\nabla _{\theta }+\frac{\left(
Z^{\prime }-Z\right) ^{2}}{2}\left( \nabla _{Z}^{2}+\frac{\nabla _{\theta
}^{2}}{c^{2}}-\frac{\nabla _{Z}^{2}\omega _{0}\left( Z\right) }{\omega
_{0}\left( Z\right) }\right) \right) \right) \right. \right. \right.  \notag
\\
&&\times \left. \left. \left. \left( \sum \prod\limits_{p}\underset{\left(
Z_{j}^{\left( p\right) },\theta _{j}^{\left( p\right) }\right) }{\ast }%
V\left( \left( Z_{j}^{\left( p\right) },\theta _{j}^{\left( p\right)
}\right) \right) \right) \right) \right) \right) \times \Delta \Gamma \left(
T,\hat{T},\theta ,Z,Z^{\prime }\right)  \notag
\end{eqnarray}

\subsubsection*{1.2.3 Graphs expansion}

\paragraph*{1.2.3.1 Vertices expansion and amplitudes}

In formula (\ref{ntm}), we replace the term:%
\begin{equation*}
\left( \sum \prod\limits_{p}\underset{\left( Z_{j}^{\left( p\right)
},\theta _{j}^{\left( p\right) }\right) }{\ast }V\left( \left( Z_{j}^{\left(
p\right) },\theta _{j}^{\left( p\right) }\right) \right) \right)
\end{equation*}%
by the sum of vertices:%
\begin{equation*}
\sum \prod\limits_{p}\underset{\left( Z_{j}^{\left( p\right) },\theta
_{j}^{\left( p\right) }\right) }{\ast }V\left( \left( Z_{j}^{\left( p\right)
},\theta _{j}^{\left( p\right) }\right) \right)
\end{equation*}%
defined in (\ref{tcn}).

These terms allow to compute the transitions from one state $\left( \Delta
T_{j}^{\left( i\right) }\left( Z_{j}^{\left( i\right) },Z_{j}^{\prime \left(
i\right) }\right) \right) _{j\leqslant n}$ of $n$ connections to an other $%
\left( \Delta T_{j}^{\left( f\right) }\left( Z_{j}^{\left( f\right)
},Z_{j}^{\prime \left( f\right) }\right) \right) _{j\leqslant n}$.

The amplitudes given by products of $k$ vertices:%
\begin{eqnarray}
&&\left\langle \left( \Delta T_{j}^{\left( i\right) }\left( Z_{j}^{\left(
i\right) },Z_{j}^{\prime \left( i\right) }\right) \right) _{j\leqslant
n}\right\vert \left\{ \int \Delta \Gamma ^{\dag }\left( T,\hat{T},\theta
,Z,Z^{\prime }\right) \right.  \label{vrs} \\
&&\times \nabla _{\hat{T}}\left( \frac{\rho }{\omega _{0}\left( Z\right) }%
D\left( \theta \right) \left\langle \hat{T}\right\rangle \left\vert \Psi
_{0}\left( Z^{\prime }\right) \right\vert ^{2}\left( \left( \alpha
\left\vert Z-Z_{1}\right\vert ^{2}+\frac{\left\vert Z-Z^{\prime }\right\vert 
}{c}\right) -\left( Z^{\prime }-Z\right) \nabla _{Z}\omega _{0}\left(
Z\right) \right) \right.  \notag \\
&&\left. \left( \sum \prod\limits_{p}\underset{\left( Z_{j}^{\left(
p\right) },\theta _{j}^{\left( p\right) }\right) }{\ast }V\left( \left(
Z_{j}^{\left( p\right) },\theta _{j}^{\left( p\right) }\right) \right)
\right) \right) \left. \Delta \Gamma \left( T,\hat{T},\theta ,Z,Z^{\prime
}\right) \right\} ^{k}\left\vert \left( \Delta T_{j}^{\left( f\right)
}\left( Z_{j}^{\left( f\right) },Z_{j}^{\prime \left( f\right) }\right)
\right) _{j\leqslant n}\right\rangle  \notag
\end{eqnarray}

\paragraph*{1.2.3.2 Amplitudes computation}

The calculus of:%
\begin{equation*}
\left\langle \left( \Delta T_{j}^{\left( i\right) }\left( Z_{j}^{\left(
i\right) },Z_{j}^{\prime \left( i\right) }\right) \right) _{j\leqslant
n}\right\vert \prod V\left( \left( Z_{j}^{\left( p\right) },\theta
_{j}^{\left( p\right) }\right) \right) \left\vert \left( \Delta
T_{j}^{\left( i\right) }\left( Z_{j}^{\left( i\right) },Z_{j}^{\prime \left(
i\right) }\right) \right) _{j\leqslant n}\right\rangle
\end{equation*}%
for a given $\prod V\left( \left( Z_{j}^{\left( p\right) },\theta
_{j}^{\left( p\right) }\right) \right) $ is obtained by wick theorem. $%
\prod V\left( \left( Z_{j}^{\left( p\right) },\theta _{j}^{\left( p\right)
}\right) \right) $ is represented as before by branched lines, with inserted
points along the lines, corresponding to the:%
\begin{equation*}
\Delta T\left\vert \Delta \Gamma \left( \Delta T,\theta _{j}^{\left(
p\right) },Z_{j}^{\left( p\right) },Z_{j+1}^{\left( p\right) }\right)
\right\vert ^{2}
\end{equation*}

The $n$ external lines $\left\langle \left( \Delta T_{j}^{\left( i\right)
}\left( Z_{j}^{\left( i\right) },Z_{j}^{\prime \left( i\right) }\right)
\right) _{j\leqslant n}\right\vert $, are contracted with $n$ conjugate
fields $\Delta \Gamma ^{\dag }\left( T,\hat{T},\theta ,Z,Z^{\prime }\right) $%
, and the $\left\vert \left( \Delta T_{j}^{\left( f\right) }\left(
Z_{j}^{\left( f\right) },Z_{j}^{\prime \left( f\right) }\right) \right)
_{j\leqslant n}\right\rangle $ are contracted with $n$ of the $\Delta \Gamma
\left( T,\hat{T},\theta ,Z,Z^{\prime }\right) $.

The remaining fields in (\ref{vrs}) are then contracted to produce internal
lines joining the external vertices. The contracted vertices produce all
possible graphs with others internal vertices. Once a graph is drawn, some
vertices may remain disconnected from the graph. These vertices can be
removed, since their contributions are cancelled while normalizing the Green
function by dividing by the partition function. We impose that the external
vertices in (\ref{vrs}) are part of the set of contracted vertices.

The graph produced may include some internal loops: due to the branching
points, some edges may start from the same initial point, and end at the
same final point. However we will see that the contributions of such graphs
can be neglected.

The quantity associated to a graph is obtained by associating to each
internal vertex a factor $\frac{\Delta T\left\vert \Delta \Gamma \left(
\Delta T,\theta _{j}^{\left( p\right) },Z_{j}^{\left( p\right)
},Z_{j+1}^{\left( p\right) }\right) \right\vert ^{2}}{T}$ and a propagator
to each internal line. A propagator is associated to each external line,
with a given initial value $\Delta T_{j}^{\left( i\right) }\left(
Z_{j}^{\left( i\right) },Z_{j}^{\prime \left( i\right) }\right) $ or final
value $\Delta T_{j}^{\left( f\right) }\left( Z_{j}^{\left( f\right)
},Z_{j}^{\prime \left( f\right) }\right) $. The vertices are connected by
the lines in the developpement of:%
\begin{equation*}
\prod\limits_{p}\underset{\left( Z_{j}^{\left( p\right) },\theta
_{j}^{\left( p\right) }\right) }{\ast }V\left( \left( Z_{j}^{\left( p\right)
},\theta _{j}^{\left( p\right) }\right) \right)
\end{equation*}%
A factor $G\left( \theta -\theta _{1}^{\left( p\right) },Z-Z_{1}^{\left(
p\right) }\right) $ is associated to each of these line.

\paragraph*{1.2.3.3 Simplifications}

Some simplifications arise.

First, as for the activities graphs, the loop produced by contracting the
fields along different lines is negligible. Actually, considering two lines
and contracting two fields:%
\begin{eqnarray*}
&&\left( \Delta T\left\vert \Delta \Gamma \left( \Delta T,\theta
_{j}^{\left( p\right) },\left( Z_{j}^{\left( p\right) },Z_{j+1}^{\left(
p\right) }\right) \right) \right\vert ^{2}\right) \left( \Delta T^{\prime
}\left\vert \Delta \Gamma \left( \Delta T^{\prime },\left( \theta
_{j}^{\left( p\right) }\right) ^{\prime },\left( Z_{j}^{\left( p\right)
},Z_{j+1}^{\left( p\right) }\right) ^{\prime }\right) \right\vert ^{2}\right)
\\
&\rightarrow &\Delta T\Delta T^{\prime }\overbrace{\Delta \Gamma \left(
\Delta T,\theta _{j}^{\left( p\right) },\left( Z_{j}^{\left( p\right)
},Z_{j+1}^{\left( p\right) }\right) \right) \Delta \Gamma ^{\dagger }\left(
\Delta T^{\prime },\left( \theta _{j}^{\left( p\right) }\right) ^{\prime
},\left( Z_{j}^{\left( p\right) },Z_{j+1}^{\left( p\right) }\right) ^{\prime
}\right) } \\
&&\overbrace{\Delta \Gamma ^{\dagger }\left( \Delta T,\theta _{j}^{\left(
p\right) },\left( Z_{j}^{\left( p\right) },Z_{j+1}^{\left( p\right) }\right)
\right) \Delta \Gamma \left( \Delta T^{\prime },\left( \theta _{j}^{\left(
p\right) }\right) ^{\prime },\left( Z_{j}^{\left( p\right) },Z_{j+1}^{\left(
p\right) }\right) ^{\prime }\right) }
\end{eqnarray*}%
implies that: $\left( Z_{j}^{\left( p\right) },Z_{j+1}^{\left( p\right)
}\right) =\left( Z_{j}^{\left( p\right) },Z_{j+1}^{\left( p\right) }\right)
^{\prime }$ and $\theta _{j}^{\left( p\right) }=\left( \theta _{j}^{\left(
p\right) }\right) ^{\prime }$. This implies that the loop sums over the set
of doublet of lines with the same length. The measure of this set is nul,
and the loops can be neglected.

As a consequence, the sum of graphs is computed for tree graphs.

Second, if the graph is computed with some perturbation arising at some "far
in the past"-points, and if we assume that the fluctuations $\Delta \Gamma
\left( \Delta T,\theta ,\left( Z,Z^{\prime }\right) \right) $ cancel before
this perturbations, the graphs to consider are trees made of branched lines
joining some initial modifications.

For these graphs:

1) the internal factors $\Delta T\left\vert \Delta \Gamma \left( \Delta
T,\theta _{j}^{\left( p\right) },\left( Z_{j}^{\left( p\right)
},Z_{j+1}^{\left( p\right) }\right) \right) \right\vert ^{2}$ are contracted
and yield a factor $\left\langle \Delta T\right\rangle _{\Delta \Gamma
\left( \theta _{j}^{\left( p\right) },\left( Z_{j}^{\left( p\right)
},Z_{j+1}^{\left( p\right) }\right) \right) }$ where this symbol denotes the
average of $\Delta T$ in the state defined by the field $\Delta \Gamma $ at
the point $\left( \theta _{j}^{\left( p\right) },\left( Z_{j}^{\left(
p\right) },Z_{j+1}^{\left( p\right) }\right) \right) $.

2) The terminal points of the graphs are contracted with the perturbations
and between themselves through propagators.

\bigskip

These conditions allows also to rewrite the interaction terms in the
following form. Each line in (\ref{trc}):%
\begin{eqnarray}
&&G\left( \theta -\theta _{1},Z-Z_{1}\right) \left[ \prod \frac{\Delta
T\left\vert \Delta \Gamma \left( \theta _{j},Z_{j},Z_{j+1}\right)
\right\vert ^{2}}{T}\left\vert \Psi \left( \theta _{j},Z_{j}\right)
\right\vert ^{2}G\left( \theta _{j}-\theta _{j+1},Z_{j}-Z_{j+1}\right) %
\right] \\
&&\times G\left( \theta _{n-1}-\theta _{n},Z_{n-1}-Z_{n}\right) \left\vert
\Psi _{0}\left( Z_{n},\theta _{n}\right) \right\vert ^{2}\frac{\omega
_{0}\left( \theta _{n},Z_{n}\right) }{\Lambda ^{2}}  \notag
\end{eqnarray}

and these terms can be summed to produce a factor:%
\begin{equation*}
\frac{\check{T}}{\left( 1-\left( 1+\left\langle \left\vert \Psi _{\Gamma
}\right\vert ^{2}\right\rangle \right) \check{T}\right) }\frac{\Delta
T\left\vert \Delta \Gamma \left( \theta _{1},Z_{1},Z_{1}\right) \right\vert
^{2}}{T}
\end{equation*}%
with:%
\begin{equation*}
\left\langle \left\vert \Psi _{\Gamma }\right\vert ^{2}\right\rangle =\left(
1+\frac{\left\langle \Delta T\right\rangle _{\Delta \Gamma }}{T}\right)
\left\vert \Psi \right\vert ^{2}
\end{equation*}%
As a consequence, the analysis of the fluctuations in activities applies, so
that we can replace the sum of graph by:%
\begin{equation*}
\delta \omega ^{-1}\left( \theta ,Z,\left\vert \Psi \right\vert ^{2}\right) =%
\frac{\int \check{T}\Lambda ^{\dag }\left( Z,\theta \right) \exp \left(
-S\left( \Lambda \right) +\int \Lambda \left( X,\theta \right) \omega
_{0}^{-1}\left( J,\theta ,Z\right) \frac{\Delta T\left\vert \Delta \Gamma
\left( \theta _{1},Z_{1},Z_{1}\right) \right\vert ^{2}}{T}d\left( X,\theta
\right) \right) \mathcal{D}\Lambda }{\int \exp \left( -S\left( \Lambda
\right) \right) \mathcal{D}\Lambda }
\end{equation*}%
and the action for the auxiliary field $S\left( \Lambda \right) $ is
obtained by replacing $\left\vert \Psi \left( \theta ,Z\right) \right\vert
^{2}$ with $\left\langle \left\vert \Psi _{\Gamma }\right\vert
^{2}\right\rangle $:%
\begin{eqnarray*}
S\left( \Lambda \right) &=&\int \Lambda \left( Z,\theta \right) \left(
1-\left\langle \left\vert \Psi _{\Gamma }\right\vert ^{2}\right\rangle 
\check{T}\right) \Lambda ^{\dag }\left( Z,\theta \right) d\left( Z,\theta
\right) -\int \Lambda \left( Z,\theta \right) \check{T}\left( \theta -\frac{%
\left\vert Z-Z^{\left( 1\right) }\right\vert }{c},Z,Z^{\left( 1\right)
},\omega _{0}^{-1}+\check{T}\Lambda ^{\dag }\right) \\
&&\times \Lambda ^{\dag }\left( Z^{\left( 1\right) },\theta -\frac{%
\left\vert Z-Z^{\left( 1\right) }\right\vert }{c}\right) dZdZ^{\left(
1\right) }d\theta
\end{eqnarray*}%
As the derivation of (\ref{sdn}) we can approximate the saddle point
equation by%
\begin{equation*}
\check{T}\left( Z,Z^{\prime },\omega +\hat{T}\Lambda ^{\dag }\right) -\check{%
T}\simeq -\frac{\frac{\check{T}}{\left( 1-\left( 1+\left\langle \left\vert
\Psi _{\Gamma }\right\vert ^{2}\right\rangle \right) \check{T}\right) }\left[
\frac{\Delta T\left\vert \Delta \Gamma \left( \theta _{1},Z_{1},Z_{1}\right)
\right\vert ^{2}}{T}\right] }{\omega _{0}\left( Z\right) +\frac{\check{T}}{%
\left( 1-\left( 1+\left\langle \left\vert \Psi _{\Gamma }\right\vert
^{2}\right\rangle \right) \check{T}\right) }\left[ \frac{\Delta T\left\vert
\Delta \Gamma \left( \theta _{1},Z_{1},Z_{1}\right) \right\vert ^{2}}{T}%
\right] }
\end{equation*}%
This allows to solve the saddle point equation at the zeroth order:%
\begin{equation}
\int^{\theta _{i}}\check{T}\left( 1-\left( 1+\left\langle \left\vert \Psi
_{\Gamma }\right\vert ^{2}\right\rangle \right) \check{T}\right) ^{-1}\left(
Z,\theta ,Z_{1},\theta _{1}\right) \left[ \frac{\Delta T\left\vert \Delta
\Gamma \left( \theta _{1},Z_{1},Z_{1}\right) \right\vert ^{2}}{T}d\theta _{1}%
\right]  \label{lrd}
\end{equation}%
then at the first order:%
\begin{eqnarray}
&&\int^{\theta _{i}}\check{T}\left( 1-\left( 1+\left\langle \left\vert \Psi
_{\Gamma }\right\vert ^{2}\right\rangle -\frac{\frac{\check{T}}{\left(
1-\left( 1+\left\langle \left\vert \Psi _{\Gamma }\right\vert
^{2}\right\rangle \right) \check{T}\right) }\left[ \frac{\Delta T\left\vert
\Delta \Gamma \left( \theta _{1},Z_{1},Z_{1}\right) \right\vert ^{2}}{T}%
\right] }{\omega _{0}\left( Z\right) +\frac{\check{T}}{\left( 1-\left(
1+\left\langle \left\vert \Psi _{\Gamma }\right\vert ^{2}\right\rangle
\right) \check{T}\right) }\left[ \frac{\Delta T\left\vert \Delta \Gamma
\left( \theta _{1},Z_{1},Z_{1}\right) \right\vert ^{2}}{T}\right] }\right) 
\check{T}\right) ^{-1}\left( Z,\theta ,Z_{1},\theta _{1}\right)  \label{Pnn}
\\
&&\times \left[ \frac{\Delta T\left\vert \Delta \Gamma \left( \theta
_{1},Z_{1},Z_{1}\right) \right\vert ^{2}}{T}d\theta _{1}\right]  \notag \\
&\equiv &\sum_{i}\int K\left( Z,\theta ,Z_{1},\theta _{1}\right) \left\{ 
\frac{\Delta T\left\vert \Delta \Gamma \left( \theta _{1},Z_{1},Z_{1}\right)
\right\vert ^{2}}{T}\right\} d\theta _{1}  \notag
\end{eqnarray}%
The next orders,of approximation for $K\left( Z,\theta ,Z_{i},\theta
_{i}\right) $ have been detailed in (\cite{GLs}).

\paragraph*{1.2.3.4 Approximation of interaction terms}

In the effective action (\ref{ntm}), the term:%
\begin{eqnarray*}
&&\omega _{0}\left( Z\right) \delta \omega \left( \theta -\frac{\left\vert
Z-Z^{\prime }\right\vert }{c},Z^{\prime },\left\vert \Psi \right\vert
^{2}\right) -\omega _{0}\left( Z^{\prime }\right) \delta \omega \left(
\theta ,Z,\left\vert \Psi \right\vert ^{2}\right) \\
&\simeq &\left( -\frac{\left\vert Z-Z^{\prime }\right\vert }{c}\nabla
_{\theta }+\frac{\left( Z^{\prime }-Z\right) ^{2}}{2}\left( \nabla _{Z}^{2}+%
\frac{\nabla _{\theta }^{2}}{c^{2}}-\frac{\nabla _{Z}^{2}\omega _{0}\left(
Z\right) }{\omega _{0}\left( Z\right) }\right) \right) \delta \omega \left(
\theta ,Z,\left\vert \Psi \right\vert ^{2}\right)
\end{eqnarray*}%
can be approximated which implies that we can go further in the computation
of \ (\ref{cmq}). Given (\ref{DG}), (\ref{gtm}) and (\ref{tcn}), we have in
first approximation:%
\begin{equation*}
\delta \omega \left( \theta ,Z,\left\vert \Psi \right\vert ^{2}\right)
\simeq \int^{Z,\theta }\frac{\exp \left( -c\left( \theta -\theta _{1}\right)
-\alpha \left( \left( c\left( \theta -\theta _{1}\right) \right)
^{2}-\left\vert Z-Z_{1}\right\vert ^{2}\right) \right) }{D}\frac{\omega
_{0}\left( \theta _{1},Z_{1}\right) \left\vert \Psi _{\Gamma }\left(
Z_{1},\theta _{1}\right) \right\vert ^{2}}{\Lambda ^{2}}dZ_{1}d\theta _{1}
\end{equation*}%
Since we are interested in the fluctuations $\Delta T$, we can replace $%
\frac{\omega _{0}\left( \theta _{1},Z_{1}\right) \left\vert \Psi _{\Gamma
}\left( Z_{1},\theta _{1}\right) \right\vert ^{2}}{\Lambda ^{2}}$ by:%
\begin{equation*}
\frac{\omega _{0}\left( \theta _{1},Z_{1}\right) \Delta T\left\vert \Delta
\Gamma \right\vert ^{2}}{T\Lambda ^{2}}
\end{equation*}%
Thus we can write:%
\begin{eqnarray}
&&\left( \left( Z-Z^{\prime }\right) \nabla _{Z}-\frac{\left\vert
Z-Z^{\prime }\right\vert }{c}\nabla _{\theta }\right) \delta \omega \left(
\theta ,Z,\left\vert \Psi \right\vert ^{2}\right)  \label{tfn} \\
&=&\left( \left( Z-Z^{\prime }\right) \nabla _{Z}-\frac{\left\vert
Z-Z^{\prime }\right\vert }{c}\nabla _{\theta }\right)  \notag \\
&&\int^{Z,\theta }\frac{\exp \left( -c\left( \theta -\theta _{1}\right)
-\alpha \left( \left( c\left( \theta -\theta _{1}\right) \right)
^{2}-\left\vert Z-Z_{1}\right\vert ^{2}\right) \right) }{D}\frac{\omega
_{0}\left( \theta _{1},Z_{1}\right) \Delta T\left\vert \Delta \Gamma
\right\vert ^{2}}{T\Lambda ^{2}}dZ_{1}d\theta _{1}  \notag
\end{eqnarray}%
Expression (\ref{tfn}) includes implicitely an heaviside function $H\left(
\left( c\left( \theta -\theta _{1}\right) \right) ^{2}-\left\vert
Z-Z^{\prime }\right\vert ^{2}\right) $. As a consequence, this becomes:%
\begin{eqnarray*}
&&\left( \left( Z-Z^{\prime }\right) \nabla _{Z}-\frac{\left\vert
Z-Z^{\prime }\right\vert }{c}\nabla _{\theta }\right) \delta \omega \left(
\theta ,Z,\left\vert \Psi \right\vert ^{2}\right) \\
&=&\int^{Z,\theta }\left( \left( Z-Z^{\prime }\right) \nabla _{Z}-\frac{%
\left\vert Z-Z^{\prime }\right\vert }{c}\nabla _{\theta }\right) \frac{\exp
\left( -c\left( \theta -\theta _{1}\right) -\alpha \left( \left( c\left(
\theta -\theta _{1}\right) \right) ^{2}-\left\vert Z-Z_{1}\right\vert
^{2}\right) \right) }{D}\frac{\omega _{0}\left( \theta _{1},Z_{1}\right)
\Delta T\left\vert \Delta \Gamma \right\vert ^{2}}{T\Lambda ^{2}}%
dZ_{1}d\theta _{1} \\
&=&-\int^{Z,\theta }\left( \left( Z-Z^{\prime }\right) \nabla _{Z_{1}}-\frac{%
\left\vert Z-Z^{\prime }\right\vert }{c}\nabla _{\theta _{1}}\right) \\
&&\times \frac{\exp \left( -c\left( \theta -\theta _{1}\right) -\alpha
\left( \left( c\left( \theta -\theta _{1}\right) \right) ^{2}-\left\vert
Z-Z_{1}\right\vert ^{2}\right) \right) }{D}\frac{\omega _{0}\left( \theta
_{1},Z_{1}\right) \Delta T\left\vert \Delta \Gamma \right\vert ^{2}}{%
T\Lambda ^{2}}dZ_{1}d\theta _{1}
\end{eqnarray*}%
and this is equal to:%
\begin{eqnarray*}
&&\left( \left( Z-Z^{\prime }\right) \nabla _{Z}-\frac{\left\vert
Z-Z^{\prime }\right\vert }{c}\nabla _{\theta }\right) \delta \omega \left(
\theta ,Z,\left\vert \Psi \right\vert ^{2}\right) \\
&=&\int^{Z,\theta }\frac{\exp \left( -c\left( \theta -\theta _{1}\right)
-\alpha \left( \left( c\left( \theta -\theta _{1}\right) \right)
^{2}-\left\vert Z-Z_{1}\right\vert ^{2}\right) \right) }{D}\left( \left(
Z-Z^{\prime }\right) \nabla _{Z_{1}}-\frac{\left\vert Z-Z^{\prime
}\right\vert }{c}\nabla _{\theta _{1}}\right) \frac{\omega _{0}\left( \theta
_{1},Z_{1}\right) \Delta T\left\vert \Delta \Gamma \right\vert ^{2}}{%
T\Lambda ^{2}}dZ_{1}d\theta _{1}
\end{eqnarray*}%
Similarly, we find:%
\begin{eqnarray*}
&&\frac{1}{2}\left( \left( Z^{\prime }-Z\right) _{i}\left( Z^{\prime
}-Z\right) _{j}\nabla _{Z_{i}}\nabla _{Z_{j}}+\frac{\left\vert Z-Z^{\prime
}\right\vert ^{2}}{c^{2}}\nabla _{\theta }^{2}\right) \delta \omega \left(
\theta ,Z,\left\vert \Psi \right\vert ^{2}\right) \\
&=&\frac{1}{2}\int^{Z,\theta }\frac{\exp \left( -c\left( \theta -\theta
_{1}\right) -\alpha \left( \left( c\left( \theta -\theta _{1}\right) \right)
^{2}-\left\vert Z-Z_{1}\right\vert ^{2}\right) \right) }{D}\left( \left(
\left( Z^{\prime }-Z\right) ^{2}\right) \nabla _{Z_{1}}^{2}+\frac{\left\vert
Z-Z^{\prime }\right\vert ^{2}}{c^{2}}\nabla _{\theta _{1}}^{2}\right) \\
&&\times \frac{\omega _{0}\left( \theta _{1},Z_{1}\right) \Delta T\left\vert
\Delta \Gamma \right\vert ^{2}}{T\Lambda ^{2}}dZ_{1}d\theta _{1}
\end{eqnarray*}

Thus, the action of:%
\begin{equation*}
\left( -\frac{\left\vert Z-Z^{\prime }\right\vert }{c}\nabla _{\theta }+%
\frac{\left( Z^{\prime }-Z\right) ^{2}}{2}\left( \nabla _{Z}^{2}+\frac{%
\nabla _{\theta }^{2}}{c^{2}}-\frac{\nabla _{Z}^{2}\omega _{0}\left(
Z\right) }{\omega _{0}\left( Z\right) }\right) \right)
\end{equation*}%
consists in inserting the operator:%
\begin{equation*}
\left( \left( Z-Z^{\prime }\right) \nabla _{Z_{1}}-\frac{\left\vert
Z-Z^{\prime }\right\vert }{c}\nabla _{\theta _{1}}+\frac{\left( Z^{\prime
}-Z\right) ^{2}}{2}\nabla _{Z_{1}}^{2}+\frac{\left\vert Z-Z^{\prime
}\right\vert ^{2}}{2c^{2}}\nabla _{\theta _{1}}^{2}-\frac{\left( Z^{\prime
}-Z\right) ^{2}\nabla _{Z}^{2}\omega _{0}\left( Z\right) }{2}\right)
\end{equation*}%
on the source term. Using then the general form for $\delta \omega \left(
\theta ,Z,\left\vert \Psi \right\vert ^{2}\right) $:%
\begin{equation*}
\delta \omega \left( \theta ,Z,\left\vert \Psi \right\vert ^{2}\right) =\int
K\left( Z,\theta ,Z_{1},\theta _{1}\right) \left\{ \frac{\omega _{0}\left(
\theta _{1},Z_{1}\right) \Delta T\left\vert \Delta \Gamma \left( \theta
_{1},Z_{1},Z_{1}^{\prime }\right) \right\vert ^{2}}{T\Lambda ^{2}}\right\}
\end{equation*}%
where the notation corresponds to inserting a factor $\left\{ \frac{\Delta
T\left\vert \Delta \Gamma \left( \theta _{1},Z_{1},Z_{1}\right) \right\vert
^{2}}{T}\right\} $ at each end of the tree graph expansion of $K\left(
Z,\theta ,Z_{1},\theta _{1}\right) $ (see (\ref{Pnn}) and (\ref{tfn})), we
can replace:%
\begin{eqnarray*}
&&\left( -\frac{\left\vert Z-Z^{\prime }\right\vert }{c}\nabla _{\theta }+%
\frac{\left( Z^{\prime }-Z\right) ^{2}}{2}\left( \nabla _{Z}^{2}+\frac{%
\nabla _{\theta }^{2}}{c^{2}}-\frac{\nabla _{Z}^{2}\omega _{0}\left(
Z\right) }{\omega _{0}\left( Z\right) }\right) \right) \delta \omega \left(
\theta ,Z,\left\vert \Psi \right\vert ^{2}\right) \\
&=&\int \bar{K}\left( Z,Z^{\prime },\theta ,Z_{1},\theta _{1}\right) \left\{ 
\frac{\omega _{0}\left( \theta _{1},Z_{1}\right) \Delta T\left\vert \Delta
\Gamma \left( \theta _{1},Z_{1},Z_{1}\right) \right\vert ^{2}}{T\Lambda ^{2}}%
\right\}
\end{eqnarray*}%
with:%
\begin{equation*}
\bar{K}\left( Z,Z^{\prime },\theta ,Z_{1},\theta _{1}\right) =K\left(
Z,\theta ,Z_{1},\theta _{1}\right) O\left( Z,Z^{\prime },Z_{1}\right)
\end{equation*}%
and where:%
\begin{equation*}
O\left( Z,Z^{\prime },Z_{1}\right) =-\frac{\left\vert Z-Z^{\prime
}\right\vert }{c}\nabla _{\theta _{1}}+\frac{\left( Z^{\prime }-Z\right) ^{2}%
}{2}\nabla _{Z_{1}}^{2}+\frac{\left\vert Z-Z^{\prime }\right\vert ^{2}}{%
2c^{2}}\nabla _{\theta _{1}}^{2}-\frac{\left( Z^{\prime }-Z\right)
^{2}\nabla _{Z}^{2}\omega _{0}\left( Z\right) }{2}
\end{equation*}%
This generalization is the consequence of the expansion (\ref{KL}) arising
in (\ref{Pnn}) and (\ref{tfn}). The convolution of kernels:%
\begin{equation*}
\frac{\exp \left( -c\left( \theta -\theta _{1}\right) -\alpha \left( \left(
c\left( \theta -\theta _{1}\right) \right) ^{2}-\left\vert
Z-Z_{1}\right\vert ^{2}\right) \right) }{D}
\end{equation*}%
allow to recursively move the operator $O\left( Z,Z^{\prime },Z_{1}\right) $
to the right of the kernel $K\left( Z,\theta ,Z_{1},\theta _{1}\right) $.

While integrating over $Z^{\prime }$ we can assume that the first order term
cancels, so that the interaction terms are:%
\begin{eqnarray}
&&\Gamma _{0}^{\dag }\left( T,\hat{T},\theta ,Z,Z^{\prime }\right)
\label{prn} \\
&&\times \left( \nabla _{\hat{T}}\left( \frac{\rho }{\omega _{0}\left(
Z\right) }\left( D\left( \theta \right) \left\langle \hat{T}\right\rangle
\left\vert \Psi _{0}\left( Z^{\prime }\right) \right\vert ^{2}\int \bar{K}%
\left( Z,Z^{\prime },\theta ,Z_{1},\theta _{1}\right) \left\{ \frac{\omega
_{0}\left( \theta _{1},Z_{1}\right) \Delta T\left\vert \Delta \Gamma \left(
\theta _{1},Z_{1},Z_{1}^{\prime }\right) \right\vert ^{2}}{T\Lambda ^{2}}%
\right\} \right) \right) \right)  \notag \\
&&\times \Gamma _{0}\left( T,\hat{T},\theta ,Z,Z^{\prime }\right)  \notag \\
&&+\Delta \Gamma ^{\dag }\left( T,\hat{T},\theta ,Z,Z^{\prime }\right) 
\notag \\
&&\times \left( \nabla _{\hat{T}}\left( \frac{\rho }{\omega _{0}\left(
Z\right) }\left( D\left( \theta \right) \left\langle \hat{T}\right\rangle
\left\vert \Psi _{0}\left( Z^{\prime }\right) \right\vert ^{2}\int \bar{K}%
\left( Z,Z^{\prime },\theta ,Z_{1},\theta _{1}\right) \left\{ \frac{\omega
_{0}\left( \theta _{1},Z_{1}\right) \Delta T\left\vert \Delta \Gamma \left(
\theta _{1},Z_{1},Z_{1}^{\prime }\right) \right\vert ^{2}}{T\Lambda ^{2}}%
\right\} \right) \right) \right)  \notag \\
&&\times \Delta \Gamma \left( T,\hat{T},\theta ,Z,Z^{\prime }\right)  \notag
\end{eqnarray}%
or, if we neglect the background displacement:%
\begin{eqnarray}
&&\Delta \Gamma ^{\dag }\left( T,\hat{T},\theta ,Z,Z^{\prime }\right)
\label{prs} \\
&&\times \left( \nabla _{\hat{T}}\left( \frac{\rho }{\omega _{0}\left(
Z\right) }\left( D\left( \theta \right) \left\langle \hat{T}\right\rangle
\left\vert \Psi _{0}\left( Z^{\prime }\right) \right\vert ^{2}\int \bar{K}%
\left( Z,Z^{\prime },\theta ,Z_{1},\theta _{1}\right) \left\{ \frac{\omega
_{0}\left( \theta _{1},Z_{1}\right) \Delta T\left\vert \Delta \Gamma \left(
\theta _{1},Z_{1},Z_{1}^{\prime }\right) \right\vert ^{2}}{T\Lambda ^{2}}%
\right\} \right) \right) \right)  \notag \\
&&\times \Delta \Gamma \left( T,\hat{T},\theta ,Z,Z^{\prime }\right)  \notag
\end{eqnarray}%
At the lowest order approximation (\ref{lrd}), this writes in a compact
operatorial form:%
\begin{eqnarray}
&&\Delta \Gamma ^{\dag }\left( T,\hat{T},\theta ,Z,Z^{\prime }\right)
\label{prl} \\
&&\times \nabla _{\hat{T}}\left( \frac{\rho }{\omega _{0}\left( Z\right) }%
\left( D\left( \theta \right) \left\langle \hat{T}\right\rangle \left\vert
\Psi _{0}\left( Z^{\prime }\right) \right\vert ^{2}\hat{T}\left( 1-\left(
1+\left\langle \left\vert \Psi _{\Gamma }\right\vert ^{2}\right\rangle
\right) \hat{T}\right) ^{-1}\left[ O\frac{\Delta T\left\vert \Delta \Gamma
\left( \theta _{1},Z_{1},Z_{1}^{\prime }\right) \right\vert ^{2}}{T\Lambda
^{2}}\right] \right) \right)  \notag \\
&&\times \Delta \Gamma \left( T,\hat{T},\theta ,Z,Z^{\prime }\right)  \notag
\end{eqnarray}

\section*{Appendix 2. Application: Background states for connectivity field
in interaction}

\subsection*{2.1 Solving for the Background field}

We solve equation (\ref{Sdt}) by the same method as in appendix 2. Starting
by writing (\ref{Sdt}):%
\begin{equation}
\left( \mathbf{\nabla }^{2}+\left( \mathbf{\nabla }\right) ^{t}\left( \gamma 
\mathbf{\Delta T}+V_{0}\mathbf{a}_{0}\right) +V\left( \mathbf{a}\right) ^{t}%
\mathbf{\Delta T}+\alpha \right) \Gamma \left( T,\hat{T},\theta ,Z,Z^{\prime
}\right) =0
\end{equation}%
with:%
\begin{eqnarray*}
\left( \mathbf{\Delta T}\right) ^{t} &=&\left( \Delta T,\Delta \hat{T}\right)
\\
\left( \mathbf{a}_{0}\right) ^{t} &=&\left( 0,1\right) \\
\left( \mathbf{a}\right) ^{t} &=&\left( 1,0\right)
\end{eqnarray*}%
\begin{equation*}
\gamma =\left( 
\begin{array}{cc}
u & s \\ 
0 & v%
\end{array}%
\right)
\end{equation*}%
and:%
\begin{eqnarray*}
u &=&\frac{\left\vert \Psi _{0}\left( Z\right) \right\vert ^{2}}{\tau \omega
_{0}\left( Z\right) } \\
v &=&\rho C\frac{\left\vert \Psi _{0}\left( Z\right) \right\vert
^{2}h_{C}\left( \omega _{0}\left( Z\right) \right) }{\omega _{0}\left(
Z\right) }+\rho D\frac{\left\vert \Psi _{0}\left( Z^{\prime }\right)
\right\vert ^{2}h_{D}\left( \omega _{0}\left( Z^{\prime }\right) \right) }{%
\omega _{0}\left( Z\right) } \\
s &=&-\frac{\lambda \left\vert \Psi _{0}\left( Z\right) \right\vert ^{2}}{%
\omega _{0}\left( Z\right) }
\end{eqnarray*}%
\begin{equation*}
V_{0}=\left( \frac{\rho D\left( \theta \right) \left\langle \hat{T}%
\right\rangle \left\vert \Psi _{0}\left( Z^{\prime }\right) \right\vert ^{2}%
}{\omega _{0}\left( Z\right) }\hat{T}\left( 1-\left( 1+\left\langle
\left\vert \Psi _{\Gamma }\right\vert ^{2}\right\rangle \right) \hat{T}%
\right) ^{-1}\left[ O\frac{\Delta T\left\vert \Delta \Gamma \left( \theta
_{1},Z_{1},Z_{1}^{\prime }\right) \right\vert ^{2}}{T}\right] \right)
\end{equation*}%
shifting variable:%
\begin{eqnarray}
\mathbf{\Delta T}+\gamma ^{-1}V_{0}\mathbf{a}_{0} &\rightarrow &\mathbf{%
\Delta T} \\
-V\left( \mathbf{a}\right) ^{t}\gamma ^{-1}V_{0}\mathbf{a}_{0}+\alpha
&\rightarrow &\alpha  \notag
\end{eqnarray}%
equation (\ref{mdt}) writes:%
\begin{equation}
\left( \mathbf{\nabla }^{2}+\left( \mathbf{\nabla }\right) ^{t}\gamma 
\mathbf{\Delta T}+V\left( \mathbf{a}\right) ^{t}\mathbf{\Delta T}+\alpha
\right) \Gamma \left( T,\hat{T},\theta ,Z,Z^{\prime }\right) =0
\end{equation}%
This is solved by considering the Fourier transform of this equation:%
\begin{equation}
\left( -\mathbf{k}^{2}-\left( \mathbf{k}\right) ^{t}\gamma \mathbf{\nabla }_{%
\mathbf{k}}-iV\left( \mathbf{a}\right) ^{t}\mathbf{\nabla }_{\mathbf{k}%
}\right) \Gamma \left( \mathbf{k},\theta ,Z,Z^{\prime }\right) =0
\end{equation}%
and writing the solution:%
\begin{equation*}
\Gamma \left( \mathbf{k},\theta ,Z,Z^{\prime }\right) =\exp \left( -\frac{1}{%
2}\mathbf{k}^{t}N\mathbf{k}\right) \hat{\Gamma}\left( \mathbf{k},\theta
,Z,Z^{\prime }\right)
\end{equation*}%
where the matrix $N$ satisifies:%
\begin{equation*}
-\mathbf{k}^{2}+\left( \mathbf{k}\right) ^{t}\gamma N\mathbf{k}=0
\end{equation*}%
with solution:%
\begin{equation*}
N=\left( 
\begin{array}{cc}
\frac{1}{u}\left( 1+\frac{s^{2}}{v\left( u+v\right) }\right) & -\frac{s}{%
v\left( u+v\right) } \\ 
-\frac{s}{v\left( u+v\right) } & \frac{1}{v}%
\end{array}%
\right)
\end{equation*}%
This factorization yields the equation for $\hat{\Gamma}\left( \mathbf{k}%
,\theta ,Z,Z^{\prime }\right) $:%
\begin{equation}
\left( \left( -\left( \mathbf{k}\right) ^{t}\gamma -iV\left( \mathbf{a}%
\right) ^{t}\right) \mathbf{\nabla }_{\mathbf{k}}+\alpha \right) \hat{\Gamma}%
\left( \mathbf{k},\theta ,Z,Z^{\prime }\right) =0
\end{equation}%
that is:%
\begin{equation}
\left( \left( -\left( \mathbf{k+}iV\left( \gamma ^{t}\right) ^{-1}\mathbf{a}%
\right) ^{t}\gamma \right) \mathbf{\nabla }_{\mathbf{k}}+\alpha \right) \hat{%
\Gamma}\left( \mathbf{k},\theta ,Z,Z^{\prime }\right) =0
\end{equation}%
The solution is similar to appendix 2, shifting:%
\begin{equation*}
\mathbf{k}\rightarrow \mathbf{k}+iV\left( \gamma ^{t}\right) ^{-1}\mathbf{a}=%
\mathbf{k}+\left( 
\begin{array}{c}
\frac{iV}{u} \\ 
-\frac{isV}{uv}%
\end{array}%
\right) =\mathbf{k}^{\prime }
\end{equation*}%
and defining:%
\begin{eqnarray*}
P &=&\left( 
\begin{array}{cc}
1 & 1 \\ 
0 & \frac{v-u}{s}%
\end{array}%
\right) \\
\mathbf{\hat{k}}^{\prime } &=&P^{t}\mathbf{k}^{\prime }
\end{eqnarray*}%
We find:%
\begin{equation*}
\hat{\Gamma}\left( \mathbf{k},\theta ,Z,Z^{\prime }\right) =\hat{k}%
_{1}^{\prime \frac{\alpha \delta }{u}}\hat{k}_{2}^{\prime \frac{\left(
1-\delta \right) \alpha }{v}}
\end{equation*}%
where $\hat{k}_{1}$ and $\hat{k}_{2}$ are\ the component of $\mathbf{\hat{k}}
$:%
\begin{eqnarray*}
\hat{k}_{1}^{\prime } &=&k_{1}^{\prime }=k_{1}+\frac{iV}{u} \\
\hat{k}_{1}^{\prime } &=&k_{1}^{\prime }+\frac{v-u}{s}k_{2}^{\prime }=k_{1}+%
\frac{v-u}{s}k_{2}+i\frac{V}{v}
\end{eqnarray*}%
Due to the presence of the gaussian factor $\exp \left( -\frac{1}{2}\mathbf{k%
}^{t}N\mathbf{k}\right) $ we aim at expanding around $k_{1}\rightarrow 0$,
so that we write:%
\begin{equation*}
\left( k_{1}+\frac{iV}{u}\right) ^{\frac{\alpha \delta }{u}}\left( k_{1}+%
\frac{v-u}{s}k_{2}+i\frac{V}{v}\right) ^{\frac{\left( 1-\delta \right)
\alpha }{v}}=i^{\frac{\alpha \delta }{u}+\frac{\left( 1-\delta \right)
\alpha }{v}}\left( \frac{V}{u}-ik_{1}\right) ^{\frac{\alpha \delta }{u}%
}\left( \frac{V}{v}-i\left( k_{1}+\frac{v-u}{s}k_{2}\right) \right) ^{\frac{%
\left( 1-\delta \right) \alpha }{v}}
\end{equation*}%
and ultimately the solution of (\ref{Fsp}) is:%
\begin{eqnarray*}
\Gamma _{\delta }\left( \mathbf{k},\theta ,Z,Z^{\prime }\right) &=&i^{\frac{%
\alpha \delta }{u}+\frac{\left( 1-\delta \right) \alpha }{v}}\exp \left( -%
\frac{1}{2}\mathbf{k}^{t}N\mathbf{k}\right) \left( \frac{V}{u}-ik_{1}\right)
^{\frac{\alpha \delta }{u}}\left( \frac{V}{v}-i\left( k_{1}+\frac{v-u}{s}%
k_{2}\right) \right) ^{\frac{\left( 1-\delta \right) \alpha }{v}} \\
&=&\exp \left( -\frac{1}{2}\mathbf{k}^{t}N\mathbf{k}\right) \left(
k_{1}^{2}+\left( \frac{V}{u}\right) ^{2}\right) ^{\frac{\alpha \delta }{2u}%
}\left( \left( k_{1}+\frac{v-u}{s}k_{2}\right) ^{2}+\left( \frac{V}{v}%
\right) ^{2}\right) ^{\frac{\left( 1-\delta \right) \alpha }{2v}} \\
&&\times \exp \left( -i\left( \frac{\alpha \delta }{u}\arctan \left( \frac{%
k_{1}u}{V}\right) -\frac{\left( 1-\delta \right) \alpha }{v}\arctan \left( 
\frac{\left( k_{1}+\frac{v-u}{s}k_{2}\right) v}{V}\right) \right) \right)
\end{eqnarray*}%
In the limit of relatively large interactions $V>1$, the last exponential
becomes:%
\begin{eqnarray*}
&&\exp \left( -i\left( \frac{\alpha \delta }{u}\arctan \left( \frac{k_{1}u}{V%
}\right) -\frac{\left( 1-\delta \right) \alpha }{v}\arctan \left( \frac{%
\left( k_{1}+\frac{v-u}{s}k_{2}\right) v}{V}\right) \right) \right) \\
&=&\exp \left( -i\left( \frac{\alpha \delta }{u}\left( \frac{k_{1}u}{V}%
\right) -\frac{\left( 1-\delta \right) \alpha }{v}\left( \frac{\left( k_{1}+%
\frac{v-u}{s}k_{2}\right) v}{V}\right) \right) \right)
\end{eqnarray*}%
and up to a constant, the solution to (\ref{Sdt}) is:

\begin{eqnarray*}
&&\Gamma _{\delta }\left( T,\hat{T},\theta ,Z,Z^{\prime }\right) \\
&=&i^{\frac{\alpha \delta }{u}+\frac{\left( 1-\delta \right) \alpha }{v}%
}\int \exp \left( -\frac{1}{2}\mathbf{k}^{t}N\mathbf{k}-i\mathbf{k}\left( 
\mathbf{\Delta T-}\overline{\mathbf{\Delta T}}\right) \right) \left(
k_{1}^{2}+\left( \frac{V}{u}\right) ^{2}\right) ^{\frac{\alpha \delta }{2u}%
}\left( \left( k_{1}+\frac{v-u}{s}k_{2}\right) ^{2}+\left( \frac{V}{v}%
\right) ^{2}\right) ^{\frac{\left( 1-\delta \right) \alpha }{2v}}\frac{d%
\mathbf{k}}{2\pi }
\end{eqnarray*}%
where:%
\begin{eqnarray*}
\mathbf{\Delta T} &\mathbf{=}&\left( 
\begin{array}{c}
T-\left\langle T\right\rangle \\ 
\hat{T}-\left\langle \hat{T}\right\rangle%
\end{array}%
\right) \\
\overline{\mathbf{\Delta T}} &=&\left( 
\begin{array}{c}
-\frac{\alpha }{V} \\ 
-\frac{\left( 1-\delta \right) \left( v-u\right) \alpha }{Vs}%
\end{array}%
\right)
\end{eqnarray*}

The estimation of the integral uses the diagonalization of $N=PDP^{-1}$.
This is done in appendix 2, we find:%
\begin{equation*}
D=\left( 
\begin{array}{cc}
\lambda _{+} & 0 \\ 
0 & \lambda _{-}%
\end{array}%
\right)
\end{equation*}%
and the eigenvalues:%
\begin{equation*}
\lambda _{\pm }=\frac{\frac{1}{u}\left( 1+\frac{s^{2}}{v\left( u+v\right) }%
\right) +\frac{1}{v}}{2}\pm \sqrt{\left( \frac{\frac{1}{u}\left( 1+\frac{%
s^{2}}{v\left( u+v\right) }\right) -\frac{1}{v}}{2}\right) ^{2}+\left( \frac{%
s}{v\left( u+v\right) }\right) ^{2}}
\end{equation*}%
The matrix $P$ is orthogonal:%
\begin{equation*}
P=\left( 
\begin{array}{cc}
\cos x & -\sin x \\ 
\sin x & \cos x%
\end{array}%
\right)
\end{equation*}%
and $x$ satifies:%
\begin{equation*}
\tan 2x=\frac{-\frac{2s}{v\left( u+v\right) }}{\frac{\frac{1}{u}\left( 1+%
\frac{s^{2}}{v\left( u+v\right) }\right) -\frac{1}{v}}{2}}=-\frac{4su}{%
v^{2}-u^{2}+s^{2}}
\end{equation*}%
so that:%
\begin{equation*}
x=-\frac{1}{2}\arctan \frac{4su}{v^{2}-u^{2}+s^{2}}
\end{equation*}%
It thus implies that $\Gamma _{\delta }\left( T,\hat{T},\theta ,Z,Z^{\prime
}\right) $ is given by:

\begin{eqnarray*}
\Gamma _{\delta }\left( T,\hat{T},\theta ,Z,Z^{\prime }\right) &=&\int \exp
\left( -\frac{1}{2}\mathbf{k}^{t}D\mathbf{k}-i\mathbf{k}\left( \mathbf{%
\Delta T}^{\prime }-\overline{\mathbf{\Delta T}}^{\prime }\right) \right)
\times \\
&&\times \left( \left( k_{1}\cos x-k_{2}\sin x\right) ^{2}+\left( \frac{V}{u}%
\right) ^{2}\right) ^{\frac{\alpha \delta }{2u}} \\
&&\times \left( \left( k_{1}\left( \cos x+\frac{v-u}{s}\sin x\right) +\left( 
\frac{v-u}{s}\cos x-\sin x\right) k_{2}\right) ^{2}+\left( \frac{V}{u}%
\right) ^{2}\right) ^{\frac{\left( 1-\delta \right) \alpha }{2v}}\frac{d%
\mathbf{k}}{2\pi }
\end{eqnarray*}%
with:%
\begin{equation*}
\mathbf{\Delta T}^{\prime }-\overline{\mathbf{\Delta T}}^{\prime
}=P^{t}\left( \mathbf{\Delta T-}\overline{\mathbf{\Delta T}}\right)
\end{equation*}%
In the approximation given in the text, we have $s<<1$ so that $x<<1$ and
the computations of appendix 2 apply. We rewrite $\Gamma _{\delta }\left( T,%
\hat{T},\theta ,Z,Z^{\prime }\right) $ in this approximation:%
\begin{eqnarray*}
&&\Gamma _{\delta }\left( T,\hat{T},\theta ,Z,Z^{\prime }\right) \\
&\simeq &\int \exp \left( -\frac{1}{2}\mathbf{k}^{t}D\mathbf{k}-i\mathbf{%
k\Delta T}^{\prime }\right) \left( \left( k_{1}-xk_{2}\right) ^{2}+\left( 
\frac{V}{u}\right) ^{2}\right) ^{\frac{\alpha \delta }{2u}} \\
&&\times \left( \left( k_{1}\left( 1+x\frac{v-u}{s}\right) +\left( \frac{v-u%
}{s}-x\right) k_{2}\right) ^{2}+\left( \frac{V}{u}\right) ^{2}\right) ^{%
\frac{\left( 1-\delta \right) \alpha }{2v}}\frac{d\mathbf{k}}{2\pi } \\
&\simeq &\int \exp \left( -\frac{1}{2}\mathbf{k}^{t}D\mathbf{k}-i\mathbf{%
k\Delta T}^{\prime }\right) \left( k_{1}^{2}+\left( \frac{V}{u}\right)
^{2}\right) ^{\frac{\alpha \delta }{2u}} \\
&&\times \left( \left( \frac{v-u}{s}k_{2}\right) ^{2}+\left( \frac{V}{u}%
\right) ^{2}\right) ^{\frac{\left( 1-\delta \right) \alpha }{2v}}\left( 1-x%
\frac{\alpha \delta }{u}\frac{k_{2}k_{1}}{k_{1}^{2}+\left( \frac{V}{u}%
\right) ^{2}}\right) \left( 1+\frac{\left( 1-\delta \right) \alpha }{vs}%
\frac{\left( v-u\right) k_{1}k_{2}}{\left( \frac{v-u}{s}k_{2}\right)
^{2}+\left( \frac{V}{u}\right) ^{2}}\right) \frac{d\mathbf{k}}{2\pi } \\
&=&\left( \frac{v-u}{s}\right) ^{\frac{\left( 1-\delta \right) \alpha }{v}%
}\int \exp \left( -\frac{1}{2}\mathbf{k}^{t}D\mathbf{k}-i\mathbf{k\Delta T}%
^{\prime }\right) \left( k_{1}^{2}+\left( \frac{V}{u}\right) ^{2}\right) ^{%
\frac{\alpha \delta }{2u}}\left( k_{2}^{2}+\left( \frac{sV}{\left(
v-u\right) u}\right) ^{2}\right) ^{\frac{\left( 1-\delta \right) \alpha }{2v}%
} \\
&&\times \left( 1-x\frac{\alpha \delta }{u}\frac{k_{2}k_{1}}{%
k_{1}^{2}+\left( \frac{V}{u}\right) ^{2}}\right) \left( 1+\frac{\left(
1-\delta \right) \alpha }{v\left( v-u\right) }\frac{sk_{1}k_{2}}{%
k_{2}^{2}+\left( \frac{sV}{\left( v-u\right) u}\right) ^{2}}\right) \frac{d%
\mathbf{k}}{2\pi } \\
&\simeq &\left( \frac{v-u}{s}\right) ^{\frac{\left( 1-\delta \right) \alpha 
}{v}}\int \exp \left( -\frac{1}{2}\mathbf{k}^{t}D\mathbf{k}-i\mathbf{k\Delta
T}^{\prime }\right) \left( k_{1}^{2}+\left( \frac{V}{u}\right) ^{2}\right) ^{%
\frac{\alpha \delta }{2u}}\times \\
&&\times \left( k_{2}^{2}+\left( \frac{sV}{\left( v-u\right) u}\right)
^{2}\right) ^{\frac{\left( 1-\delta \right) \alpha }{2v}}\left( 1+\left( 
\frac{\left( 1-\delta \right) \alpha }{v\left( v-u\right) }\frac{s}{%
k_{2}^{2}+\left( \frac{sV}{\left( v-u\right) u}\right) ^{2}}-x\frac{\alpha
\delta }{u}\frac{1}{k_{1}^{2}+\left( \frac{V}{u}\right) ^{2}}\right)
k_{1}k_{2}\right) \frac{d\mathbf{k}}{2\pi }
\end{eqnarray*}%
In the approximation given in the text, we have $s<<1$ and the computations
of appendix 2 apply. We find:

\begin{eqnarray*}
&&\Gamma _{\delta }\left( T,\hat{T},\theta ,Z,Z^{\prime }\right) \\
&\simeq &\left( \frac{v-u}{s}\right) ^{\frac{\left( 1-\delta \right) \alpha 
}{v}}2^{\frac{\alpha }{u}+1}\prod\limits_{i=1}^{2}\exp \left( -\left(
\left( \frac{D^{-\frac{1}{2}}P^{t}\left( \mathbf{\Delta T}-\overline{\mathbf{%
\Delta T}}^{\prime }\right) }{2}\right) _{i}\right) ^{2}\right) \\
&&\times \left\{ \prod\limits_{i=1}^{2}\hat{D}_{p_{i}}^{m_{i}}\left( \left( 
\frac{D^{-\frac{1}{2}}P^{t}\left( \mathbf{\Delta T}-\overline{\mathbf{\Delta
T}}^{\prime }\right) }{4}\right) _{i}\right) \right. \\
&&\left. +\nabla _{\left( \Delta T^{\prime }\right) _{1}}\nabla _{\left(
\Delta T^{\prime }\right) _{2}}\left\{ x\alpha \frac{\delta
\prod\limits_{i=1}^{2}\hat{D}_{p_{i}^{\left( 1\right) }}^{m_{i}}\left(
\left( \frac{D^{-\frac{1}{2}}P^{t}\left( \mathbf{\Delta T}-\overline{\mathbf{%
\Delta T}}^{\prime }\right) }{4}\right) _{i}\right) }{u}-\frac{s\alpha
\left( 1-\delta \right) \prod\limits_{i=1}^{2}\hat{D}_{p_{i}^{\left(
1\right) }}^{m_{i}}\left( \left( \frac{D^{-\frac{1}{2}}P^{t}\left( \mathbf{%
\Delta T}-\overline{\mathbf{\Delta T}}^{\prime }\right) }{4}\right)
_{i}\right) }{v\left( u-v\right) }\right\} \right\}
\end{eqnarray*}%
where:%
\begin{eqnarray*}
p_{1} &=&\frac{\alpha \delta }{u},p_{2}=\frac{\left( 1-\delta \right) \alpha 
}{v} \\
p_{1}^{\left( 1\right) } &=&\frac{\alpha \delta }{u}-1,p_{2}^{\left(
1\right) }=\frac{\left( 1-\delta \right) \alpha }{v}+1 \\
p_{1}^{\left( 1\right) } &=&\frac{\alpha \delta }{u}+1,p_{2}^{\left(
1\right) }=\frac{\left( 1-\delta \right) \alpha }{v}-1
\end{eqnarray*}%
and:%
\begin{eqnarray*}
m_{1} &=&\frac{V}{u} \\
m_{2} &=&\frac{sV}{\left( v-u\right) u}
\end{eqnarray*}%
and the function $\hat{D}_{p}^{m}$ are "massive" parabolic cylinder function
defined by the integral representation, up to some irrelevant constant:%
\begin{equation*}
\hat{D}_{p}^{m}\left( x\right) =\frac{2^{p+1}}{\sqrt{\pi }}\exp \left( -%
\frac{\pi }{2}pi\right) \exp \left( \frac{x^{2}}{4}\right) \int \left(
x^{2}+m^{2}\right) ^{\frac{p}{2}}\exp \left( -2k^{2}+2ixk\right) dk
\end{equation*}

\subsection*{2.2 Equation for shift in connectivity functions}

Given (\ref{svl}), the shift $\overline{\mathbf{\Delta T}}$ is solution of:%
\begin{eqnarray}
\Delta T\left( Z,Z^{\prime }\right) &=&-\frac{\alpha }{V\left( Z,Z^{\prime
}\right) } \\
\Delta \hat{T}\left( Z,Z^{\prime }\right) &=&-\frac{\left( 1-\delta \right)
\left( v-u\right) \alpha }{V\left( Z,Z^{\prime }\right) s}  \notag
\end{eqnarray}%
however, taking into account (\ref{svb}), we replace:%
\begin{eqnarray}
\mathbf{\Delta T} &\rightarrow &\mathbf{\Delta T}+\gamma ^{-1}V_{0}\mathbf{a}%
_{0} \\
\alpha &\rightarrow &\alpha -V\left( \mathbf{a}\right) ^{t}\gamma ^{-1}V_{0}%
\mathbf{a}_{0}  \notag
\end{eqnarray}%
The first equation amounts to replace $\Delta T\left( Z,Z^{\prime }\right) $
by $\Delta T\left( Z,Z^{\prime }\right) -\gamma ^{-1}V_{0}\mathbf{a}_{0}$.
Given that:%
\begin{equation*}
\gamma ^{-1}V_{0}\mathbf{a}_{0}=\left( 
\begin{array}{c}
-\frac{s}{uv}V_{0} \\ 
\frac{1}{v}V_{0}%
\end{array}%
\right)
\end{equation*}%
and:%
\begin{equation*}
-V\left( \mathbf{a}\right) ^{t}\gamma ^{-1}V_{0}\mathbf{a}_{0}=V\frac{s}{uv}%
V_{0}
\end{equation*}%
the shift equation writes:%
\begin{eqnarray}
\Delta T\left( Z,Z^{\prime }\right) &=&-\frac{\alpha }{V\left( Z,Z^{\prime
}\right) }  \label{hfnn} \\
\Delta \hat{T}\left( Z,Z^{\prime }\right) &=&-\left( \frac{1}{v}+\frac{%
\left( 1-\delta \right) \left( v-u\right) }{uv}\right) V_{0}-\frac{\left(
1-\delta \right) \left( v-u\right) \alpha }{V\left( Z,Z^{\prime }\right) s} 
\notag
\end{eqnarray}

Given our assumptions, the terms $V_{0}$ and $V$ are relatively large.
Moreover, $V_{0}$ measures the modification due to sources terms, and $V$
the backreaction of the system on the sources.

\subsubsection*{2.2.1 Average shift}

To solve (\ref{hfnn}) we first compute the averages:%
\begin{eqnarray*}
\left\langle \Delta T\right\rangle &=&\left\langle \Delta T\left(
Z_{i},Z_{i}^{\prime }\right) \right\rangle _{\left( Z_{i},Z_{i}^{\prime
}\right) } \\
\left\langle \Delta \hat{T}\right\rangle &=&\left\langle \Delta \hat{T}%
\left( Z_{i},Z_{i}^{\prime }\right) \right\rangle _{\left(
Z_{i},Z_{i}^{\prime }\right) }
\end{eqnarray*}%
by averaging all quantities in (\ref{hfnn}). We start with:

\begin{equation}
V_{0}\left( Z,Z^{\prime }\right) =\left( \frac{\rho D\left( \theta \right)
\left\langle \hat{T}\right\rangle \left\vert \Psi _{0}\left( Z^{\prime
}\right) \right\vert ^{2}}{\omega _{0}\left( Z\right) }\hat{T}\left(
1-\left( 1+\left\langle \left\vert \Psi _{\Gamma }\right\vert
^{2}\right\rangle \right) \hat{T}\right) ^{-1}\left[ O\frac{\Delta
T\left\vert \Delta \Gamma \left( T_{1},\hat{T}_{1},\theta
_{1},Z_{1},Z_{1}^{\prime }\right) \right\vert ^{2}}{T}\right] \right)
\label{frv}
\end{equation}%
In (\ref{frv}), we replace the integrated functions by their average in
state $\Delta \Gamma $. To do so we write:%
\begin{equation}
\int \frac{\Delta T\left\Vert \Delta \Gamma \left( T_{1},\hat{T}_{1},\theta
_{1},Z_{1},Z_{1}^{\prime }\right) \right\Vert ^{2}}{T}d\left( T_{1},\hat{T}%
_{1},\theta _{1},Z_{1},Z_{1}^{\prime }\right) \simeq \frac{\left\langle
\Delta T\right\rangle }{\left\langle T\right\rangle }\left\Vert \Delta
\Gamma \right\Vert ^{2}  \label{Grv}
\end{equation}%
so that:%
\begin{eqnarray}
V_{0}\left( Z,Z^{\prime }\right) &=&\frac{\rho D\left( \theta \right)
\left\langle \hat{T}\right\rangle \left\vert \Psi _{0}\left( Z^{\prime
}\right) \right\vert ^{2}}{\omega _{0}\left( Z\right) }\left\langle \hat{T}%
\left( 1-\left( 1+\left\langle \left\vert \Psi _{\Gamma }\right\vert
^{2}\right\rangle \right) \hat{T}\right) ^{-1}\left[ O\frac{\Delta
T\left\vert \Delta \Gamma \left( \theta _{1},Z_{1},Z_{1}^{\prime }\right)
\right\vert ^{2}}{T}\right] \right\rangle ^{\left( T,\hat{T},\theta
,Z,Z^{\prime }\right) }  \label{Vzll} \\
&\simeq &F\left( Z,Z^{\prime }\right) A_{0}\left( Z,Z^{\prime }\right) \frac{%
\left\langle \Delta T\right\rangle }{\left\langle T\right\rangle }\left\Vert
\Delta \Gamma \right\Vert ^{2}  \notag
\end{eqnarray}%
with:%
\begin{eqnarray}
F\left( Z,Z^{\prime }\right) &=&\frac{\rho D\left( \theta \right)
\left\langle \hat{T}\right\rangle \left\vert \Psi _{0}\left( Z^{\prime
}\right) \right\vert ^{2}}{\omega _{0}\left( Z\right) }  \label{Vzpp} \\
A_{0}\left( Z,Z^{\prime }\right) &=&\left\langle \hat{T}\left( 1-\left(
1+\left\langle \left\vert \Psi _{0}\right\vert ^{2}\right\rangle \left( 1+%
\frac{\Delta T}{\left\langle T\right\rangle }\right) \left\Vert \Delta
\Gamma \right\Vert ^{2}\right) \hat{T}\right) ^{-1}O\right\rangle ^{\left( T,%
\hat{T},\theta ,Z,Z^{\prime }\right) }  \notag
\end{eqnarray}%
and where the notation:%
\begin{equation*}
\left\langle \left[ O\right] _{\left( X\right) }\right\rangle \text{, }%
\left\langle \left[ O\right] ^{\left( X\right) }\right\rangle
\end{equation*}%
for an operator with kernel $O\left( X,Y\right) $ denotes $\int O\left(
X,Y\right) dY$ and $\int O\left( Y,X\right) dY$ respectively.

The corrections depending on $V_{1}$ and $V_{2}$ arising in (\ref{hfn}) are
computed similarly.

First, given the formal solutions, as well as the form of the operators $%
V_{1}$ and $V_{2}$, the kernel intervening in the integrals do not dpend on $%
T_{2}$ and $\hat{T}_{2}$. The quantities:%
\begin{equation*}
\int \Delta \Gamma ^{\dag }\left( T_{2},\hat{T}_{2},\theta
_{2},Z_{2},Z_{2}^{\prime }\right) \nabla _{\hat{T}_{2}}\Delta \Gamma \left(
T_{2},\hat{T}_{2},\theta _{2},Z_{2},Z_{2}^{\prime }\right) d\left( T_{2},%
\hat{T}_{2}\right)
\end{equation*}%
can be computed, and given the formal solutions, they are proportional to
the shift $\left\langle \Delta \hat{T}\left( Z_{2},Z_{2}^{\prime }\right)
\right\rangle $. Given that solutions for the background field are
approximatively gaussian functions, the proportionality is of negative sign.
Performing the integration over the variables $\left( T_{i},\hat{T}%
_{i}\right) $ and using (\ref{Grv}) we thus replace:%
\begin{eqnarray}
&&V_{1}\left( \theta ,Z,Z^{\prime },\Delta \Gamma \right)  \label{Vnf} \\
&=&-k\int \Delta \hat{T}_{0}\left( Z_{2},Z_{2}^{\prime }\right) \left( \frac{%
\rho D\left( \theta \right) \left\langle \hat{T}_{2}\right\rangle \left\vert
\Psi _{0}\left( Z_{2}^{\prime }\right) \right\vert ^{2}}{\omega _{0}\left(
Z_{2}\right) }\left[ \hat{T}\left( 1-\left( 1+\left\langle \left\vert \Psi
_{\Gamma }\right\vert ^{2}\right\rangle \right) \hat{T}\right) ^{-1}O\right]
_{\left( T,\hat{T},\theta ,Z,Z^{\prime }\right) }^{\left( T_{2},\hat{T}%
_{2},\theta _{2},Z_{2},Z_{2}^{\prime }\right) }\right)  \notag \\
&&\times \left\Vert \Delta \Gamma \left( \theta _{2},Z_{2},Z_{2}^{\prime
}\right) \right\Vert ^{2}d\left( \theta _{2},Z_{2},Z_{2}^{\prime }\right) 
\notag
\end{eqnarray}%
with $k>0$ and:%
\begin{equation*}
\left\Vert \Delta \Gamma \left( \theta _{2},Z_{2},Z_{2}^{\prime }\right)
\right\Vert ^{2}=\int \left\vert \Delta \Gamma \left( T_{2},\hat{T}%
_{2},\theta _{2},Z_{2},Z_{2}^{\prime }\right) \right\vert ^{2}d\left( T_{2},%
\hat{T}_{2}\right)
\end{equation*}

\begin{equation}
V_{2}\left( \theta ,Z,Z^{\prime },\Delta \Gamma \right) =\int \left[ \hat{T}%
\left( 1-\left( 1+\left\langle \left\vert \Psi _{\Gamma }\right\vert
^{2}\right\rangle \right) \hat{T}\right) ^{-1}\right] _{\left( T_{1},\hat{T}%
_{1},\theta _{1},Z_{1},Z_{1}^{\prime }\right) }^{\left( T,\hat{T},\theta
,Z,Z^{\prime }\right) }\left[ \frac{\Delta T_{0}\left( Z_{1},Z_{1}^{\prime
}\right) }{T}\right] \left\Vert \Delta \Gamma \left( \theta
_{2},Z_{1},Z_{1}^{\prime }\right) \right\Vert ^{2}d\left( \theta
_{1},Z_{1},Z_{1}^{\prime }\right)  \label{Vtf}
\end{equation}

We replace the quantities multiplied by the squared field by their average
over $\left( Z_{i},Z_{i}^{\prime }\right) $, so that using:%
\begin{equation*}
\int \frac{\Delta T\left\Vert \Delta \Gamma \left( Z_{1},Z_{1}^{\prime
}\right) \right\Vert ^{2}}{T}d\left( Z_{1},Z_{1}^{\prime }\right) \simeq 
\frac{\left\langle \Delta T\right\rangle }{\left\langle T\right\rangle }%
\left\Vert \Delta \Gamma \right\Vert ^{2}
\end{equation*}%
we have:%
\begin{eqnarray}
&&V_{1}\left( Z,Z^{\prime },\Delta \Gamma \right)  \label{vfn} \\
&\simeq &-k\left\langle \Delta \hat{T}\right\rangle \left\langle \left[
F\left( Z_{2},Z_{2}^{\prime }\right) \left[ \hat{T}\left( 1-\left\langle
\left\vert \Psi _{0}\right\vert ^{2}\right\rangle \frac{\left\langle \Delta
T\right\rangle }{T}\left\Vert \Delta \Gamma \right\Vert ^{2}\right) ^{-1}O%
\right] \right] _{\left( T,\hat{T},\theta ,Z,Z^{\prime }\right)
}\right\rangle \left\Vert \Delta \Gamma \right\Vert ^{2}  \notag \\
&=&A_{1}\left( Z,Z^{\prime }\right) \left\langle \Delta \hat{T}\right\rangle
\left\Vert \Delta \Gamma \right\Vert ^{2}  \notag
\end{eqnarray}%
\begin{eqnarray}
V_{2}\left( Z,Z^{\prime },\Delta \Gamma \right) &=&\left\langle \left[ \hat{T%
}\left( 1-\left( 1+\left\langle \left\vert \Psi _{0}\right\vert
^{2}\right\rangle \frac{\left\langle \Delta T\right\rangle }{T}\left\Vert
\Delta \Gamma \right\Vert ^{2}\right) \hat{T}\right) ^{-1}\right] ^{\left( T,%
\hat{T},\theta ,Z,Z^{\prime }\right) }\right\rangle \frac{\left\langle
\Delta T\right\rangle }{\left\langle T\right\rangle }\left\Vert \Delta
\Gamma \right\Vert ^{2}  \label{vft} \\
&=&A_{2}\left( Z,Z^{\prime }\right) \frac{\left\langle \Delta T\right\rangle 
}{\left\langle T\right\rangle }\left\Vert \Delta \Gamma \right\Vert ^{2} 
\notag
\end{eqnarray}

Then, taking the average of equation (\ref{hfn}) yields the defining
equation for $\left\langle \Delta T\right\rangle $, $\left\langle \Delta 
\hat{T}\right\rangle $:%
\begin{eqnarray}
\left\langle \Delta T\right\rangle &\simeq &-\frac{\alpha }{%
A_{1}\left\langle \Delta \hat{T}\right\rangle \left\Vert \Delta \Gamma
\right\Vert ^{2}\left( 1+A_{2}\frac{\left\langle \Delta T\right\rangle }{%
\left\langle T\right\rangle }\left\Vert \Delta \Gamma \right\Vert
^{2}\right) }  \label{vrt} \\
\left\langle \Delta \hat{T}\right\rangle &\simeq &-\left( \frac{1}{%
\left\langle v\right\rangle }+\frac{\left( 1-\delta \right) \left(
v-u\right) }{\left\langle u\right\rangle \left\langle v\right\rangle }%
\right) A\frac{\left\langle \Delta T\right\rangle }{\left\langle
T\right\rangle }\left\Vert \Delta \Gamma \right\Vert ^{2}-\alpha \frac{%
\left\langle \frac{\left( 1-\delta \right) \left( v-u\right) }{s}%
\right\rangle }{A_{1}\left\langle \Delta \hat{T}\right\rangle \left\Vert
\Delta \Gamma \right\Vert ^{2}\left( 1+A_{2}\frac{\left\langle \Delta
T\right\rangle }{\left\langle T\right\rangle }\left\Vert \Delta \Gamma
\right\Vert ^{2}\right) }  \notag
\end{eqnarray}%
with:%
\begin{eqnarray*}
A &=&\left\langle F\left( Z,Z^{\prime }\right) A_{0}\left( Z,Z\right)
\right\rangle \\
A_{1} &=&\left\langle A_{1}\left( Z,Z^{\prime }\right) \right\rangle \\
A_{2} &=&\left\langle A_{2}\left( Z,Z^{\prime }\right) \right\rangle
\end{eqnarray*}

Let define the parameters:%
\begin{eqnarray*}
d &=&-\frac{\alpha }{A_{1}\left\Vert \Delta \Gamma \right\Vert ^{2}} \\
f &=&A_{2}\frac{\left\Vert \Delta \Gamma \right\Vert ^{2}}{\left\langle
T\right\rangle } \\
g &=&-\left( \frac{1}{v}+\frac{\left( 1-\delta \right) \left( v-u\right) }{uv%
}\right) A\frac{\left\Vert \Delta \Gamma \right\Vert ^{2}}{\left\langle
T\right\rangle } \\
h &=&-\left\langle \frac{\left( 1-\delta \right) \left( v-u\right) }{s}%
\right\rangle \frac{\alpha }{A_{1}\left\Vert \Delta \Gamma \right\Vert ^{2}}%
=\left\langle \frac{\left( 1-\delta \right) \left( v-u\right) }{s}%
\right\rangle d
\end{eqnarray*}

to rewrite the system:%
\begin{eqnarray}
\left\langle \Delta T\right\rangle &=&\frac{d}{\left\langle \Delta \hat{T}%
\right\rangle \left( 1+f\left\langle \Delta T\right\rangle \right) } \\
\left\langle \Delta \hat{T}\right\rangle &=&g\left\langle \Delta
T\right\rangle +\frac{h}{\left\langle \Delta \hat{T}\right\rangle \left(
1+f\left\langle \Delta T\right\rangle \right) }  \notag
\end{eqnarray}%
which can be solved for $\left\langle \Delta \hat{T}\right\rangle $ as a
function of $\left\langle \Delta T\right\rangle $: 
\begin{equation}
\left\langle \Delta \hat{T}\right\rangle =\left\langle \Delta T\right\rangle 
\frac{h+dg}{d}  \label{Rsl}
\end{equation}%
and $\left\langle \Delta T\right\rangle $ satifies: 
\begin{equation*}
f\left\langle \Delta T\right\rangle ^{3}+\left\langle \Delta T\right\rangle
^{2}-\frac{d^{2}}{\left( h+dg\right) }=0
\end{equation*}%
i.e.:%
\begin{equation}
\left\langle \Delta \tilde{T}\right\rangle ^{3}+\left\langle \Delta \tilde{T}%
\right\rangle ^{2}-\frac{d^{2}f^{2}}{\left( h+dg\right) }=0  \label{qncc}
\end{equation}%
with:%
\begin{equation*}
\Delta \tilde{T}=f\Delta T
\end{equation*}%
From equation (\ref{qncc}) we find the conditions for the solutions. A
particular case arises when $\left\Vert \Delta \Gamma \right\Vert ^{2}>>1$.
In such case:%
\begin{equation*}
-\frac{d^{2}f^{2}}{\left( h+dg\right) }<0
\end{equation*}%
and equation (\ref{qncc}) has a single negative root. Too many fluctuations
in connectivities leads ultimately to a lower shift in this variable. The
solutions are studied in the text.

\subsubsection*{2.2.2 Solving equation (\protect\ref{hfn}) for $\Delta
T\left( Z,Z^{\prime }\right) $ and $\Delta \hat{T}\left( Z,Z^{\prime
}\right) $}

We use (\ref{Vzll}) and (\ref{Vzpp}), (\ref{vfn}), (\ref{vft}), to write(\ref%
{hfn}):%
\begin{eqnarray}
\Delta T\left( Z,Z^{\prime }\right) &=&-\frac{\alpha }{\left( 1+\frac{%
A_{2}\left( Z,Z^{\prime }\right) \left\langle \Delta T\right\rangle }{%
\left\langle T\right\rangle }\left\Vert \Delta \Gamma \right\Vert
^{2}\right) A_{1}\left( Z,Z^{\prime }\right) \left\langle \Delta \hat{T}%
\right\rangle \left\Vert \Delta \Gamma \right\Vert ^{2}}  \label{pdt} \\
\Delta \hat{T}\left( Z,Z^{\prime }\right) &=&-\left( \frac{1}{v}+\frac{%
\left( 1-\delta \right) \left( v-u\right) }{uv}\right) \frac{F\left(
Z,Z^{\prime }\right) A_{0}\left( Z,Z^{\prime }\right) \left\langle \Delta
T\right\rangle }{\left\langle T\right\rangle }\left\Vert \Delta \Gamma
\right\Vert ^{2}  \notag \\
&&-\frac{\left( 1-\delta \right) \left( v-u\right) \alpha }{s\left( 1+\frac{%
A_{2}\left( Z,Z^{\prime }\right) \left\langle \Delta T\right\rangle }{%
\left\langle T\right\rangle }\left\Vert \Delta \Gamma \right\Vert
^{2}\right) A_{1}\left( Z,Z^{\prime }\right) \left\langle \Delta \hat{T}%
\right\rangle \left\Vert \Delta \Gamma \right\Vert ^{2}}  \notag
\end{eqnarray}

\subsubsection*{2.2.3 Condition for shifted state and constraint.}

For states (\ref{Slgg}) and (\ref{Sld}) the action is:%
\begin{equation*}
S\left( \Delta \Gamma \left( T,\hat{T},\theta ,Z,Z^{\prime }\right) \right)
+U_{\Delta \Gamma }\left( \left\Vert \Delta \Gamma \left( Z,Z^{\prime
}\right) \right\Vert ^{2}\right)
\end{equation*}%
where the first term is given by (\ref{fcp}) and the potential by (\ref{Ptm}%
). Given (\ref{Sdt}), this reduces to:%
\begin{eqnarray}
&&S\left( \Delta \Gamma \left( T,\hat{T},\theta ,Z,Z^{\prime }\right)
\right) +U_{\Delta \Gamma }\left( \left\Vert \Delta \Gamma \left(
Z,Z^{\prime }\right) \right\Vert ^{2}\right)  \label{cnf} \\
&=&\int \Delta \Gamma ^{\dagger }\left( T,\hat{T},\theta ,Z,Z^{\prime
}\right) \left( \left( V_{1}\left( \theta ,Z,Z^{\prime },\Delta \Gamma
\right) \left( 1+V_{2}\left( \theta ,Z,Z^{\prime },\Delta \Gamma \right)
\right) \right) \Delta T\right) \Delta \Gamma \left( T,\hat{T},\theta
,Z,Z^{\prime }\right)  \notag \\
&&+U_{\Delta \Gamma }\left( \left\Vert \Delta \Gamma \left( Z,Z^{\prime
}\right) \right\Vert ^{2}\right) +\int \left( \alpha _{0}-\frac{\delta
U_{\Delta \Gamma }\left( \left\Vert \Delta \Gamma \left( Z,Z^{\prime
}\right) \right\Vert ^{2}\right) }{\delta \left\Vert \Delta \Gamma \left(
Z,Z^{\prime }\right) \right\Vert ^{2}}\right) \left\Vert \Delta \Gamma
\left( Z,Z^{\prime }\right) \right\Vert ^{2}  \notag
\end{eqnarray}%
By the same computation as previous paragraph:%
\begin{eqnarray*}
&&\int \Delta \Gamma ^{\dagger }\left( T,\hat{T},\theta ,Z,Z^{\prime
}\right) \left( \left( V_{1}\left( \theta ,Z,Z^{\prime },\Delta \Gamma
\right) \left( 1+V_{2}\left( \theta ,Z,Z^{\prime },\Delta \Gamma \right)
\right) \right) \Delta T\right) \Delta \Gamma \left( T,\hat{T},\theta
,Z,Z^{\prime }\right) \\
&=&\int \Delta \Gamma ^{\dagger }\left( T,\hat{T},\theta ,Z,Z^{\prime
}\right) V\left( \theta ,Z,Z^{\prime },\Delta \Gamma \right) \Delta T\Delta
\Gamma \left( T,\hat{T},\theta ,Z,Z^{\prime }\right) \\
&\simeq &A_{1}\left( Z,Z^{\prime }\right) \left\langle \Delta \hat{T}%
\right\rangle \left\Vert \Delta \Gamma \right\Vert ^{2}\left( 1+A_{2}\left(
Z,Z^{\prime }\right) \frac{\left\langle \Delta T\right\rangle }{\left\langle
T\right\rangle }\left\Vert \Delta \Gamma \right\Vert ^{2}\right)
\left\langle \Delta T\right\rangle
\end{eqnarray*}%
Using (\ref{pdt}):%
\begin{equation}
\Delta T\left( Z,Z^{\prime }\right) =-\frac{\alpha }{\left( 1+\frac{%
A_{2}\left( Z,Z^{\prime }\right) \left\langle \Delta T\right\rangle }{%
\left\langle T\right\rangle }\left\Vert \Delta \Gamma \right\Vert
^{2}\right) A_{1}\left( Z,Z^{\prime }\right) \left\langle \Delta \hat{T}%
\right\rangle \left\Vert \Delta \Gamma \right\Vert ^{2}}  \label{TDp}
\end{equation}%
and given that:%
\begin{equation*}
\alpha =-\frac{\delta U_{\Delta \Gamma }\left( \left\Vert \Delta \Gamma
\left( Z,Z^{\prime }\right) \right\Vert ^{2}\right) }{\delta \left\Vert
\Delta \Gamma \left( Z,Z^{\prime }\right) \right\Vert ^{2}}+\alpha _{0}
\end{equation*}%
formula (\ref{cnf}) becomes:%
\begin{equation*}
S\left( \Delta \Gamma \left( T,\hat{T},\theta ,Z,Z^{\prime }\right) \right)
=\int U_{\Delta \Gamma }\left( \left\Vert \Delta \Gamma \left( Z,Z^{\prime
}\right) \right\Vert ^{2}\right)
\end{equation*}%
and the minimization of:%
\begin{equation*}
\int U_{\Delta \Gamma }\left( \left\Vert \Delta \Gamma \left( Z,Z^{\prime
}\right) \right\Vert ^{2}\right)
\end{equation*}%
yields $\left\Vert \Delta \Gamma \left( Z,Z^{\prime }\right) \right\Vert
^{2} $.

However, a constraint has to be included. Actually, since $\Delta T$ and $%
\Delta \hat{T}$ can be both positive or negative, we impose: 
\begin{equation*}
p_{1}=\frac{\alpha \delta }{u},p_{2}=\frac{\left( 1-\delta \right) \alpha }{v%
}
\end{equation*}%
to belong to $\frac{1}{2}+%
%TCIMACRO{\U{2115} }%
%BeginExpansion
\mathbb{N}
%EndExpansion
$. This allow to obtain integrable solutions $\Delta \Gamma \left( T,\hat{T}%
,Z,Z^{\prime }\right) $ over $%
%TCIMACRO{\U{211d} }%
%BeginExpansion
\mathbb{R}
%EndExpansion
^{2}$, we have the condition:%
\begin{equation*}
\frac{\left( \alpha +\left( V\frac{s}{uv}V_{0}\right) \right) \delta }{u},%
\frac{\left( 1-\delta \right) \left( \alpha +\left( V\frac{s}{uv}%
V_{0}\right) \right) }{v}\in \frac{1}{2}+%
%TCIMACRO{\U{2115}}%
%BeginExpansion
\mathbb{N}%
%EndExpansion
\end{equation*}%
\begin{eqnarray*}
V\frac{s}{uv}V_{0} &=&\frac{uk}{\delta }-\alpha \\
V\frac{s}{uv}V_{0} &=&\frac{vl}{1-\delta }-\alpha
\end{eqnarray*}%
\begin{equation*}
\frac{uk}{\delta }=\frac{vl}{1-\delta }
\end{equation*}%
\begin{eqnarray*}
\delta &=&\frac{ku}{ku+lv} \\
1-\delta &=&\frac{lv}{ku+lv}
\end{eqnarray*}%
\begin{equation*}
V\frac{s}{uv}V_{0}=ku+lv-\alpha
\end{equation*}%
$\allowbreak $Rewrite the constraint as:%
\begin{equation*}
V\frac{s}{uv}V_{0}-\left( ku+lv-\alpha \right)
\end{equation*}%
Then use that;%
\begin{eqnarray*}
V\left( Z,Z^{\prime }\right) &=&-\frac{\alpha }{\Delta T\left( Z,Z^{\prime
}\right) } \\
V_{0}\left( Z,Z^{\prime }\right) &=&\frac{-\Delta \hat{T}\left( Z,Z^{\prime
}\right) +\frac{\left( 1-\delta \right) \left( v-u\right) \Delta T\left(
Z,Z^{\prime }\right) }{s}}{\frac{1}{v}+\frac{\left( 1-\delta \right) \left(
v-u\right) }{uv}}
\end{eqnarray*}%
and (\ref{Rsl}):%
\begin{equation*}
\left\langle \Delta \hat{T}\right\rangle =\left\langle \Delta T\right\rangle
\left( \frac{\left( 1-\delta \right) \left( v-u\right) }{s}-\left( \frac{1}{v%
}+\frac{\left( 1-\delta \right) \left( v-u\right) }{uv}\right) A\frac{%
\left\Vert \Delta \Gamma \right\Vert ^{2}}{\left\langle T\right\rangle }%
\right)
\end{equation*}%
yields:%
\begin{eqnarray*}
VsV_{0} &=&\frac{\alpha }{\Delta T\left( Z,Z^{\prime }\right) }\frac{s\Delta 
\hat{T}\left( Z,Z^{\prime }\right) -\left( 1-\delta \right) \left(
v-u\right) \Delta T\left( Z,Z^{\prime }\right) }{\frac{1}{v}+\frac{\left(
1-\delta \right) \left( v-u\right) }{uv}} \\
&=&\alpha \frac{s\frac{\Delta \hat{T}\left( Z,Z^{\prime }\right) }{\Delta
T\left( Z,Z^{\prime }\right) }-\left( 1-\delta \right) \left( v-u\right) }{%
\frac{1}{v}+\frac{\left( 1-\delta \right) \left( v-u\right) }{uv}}=-\frac{%
\alpha sA\left\Vert \Delta \Gamma \right\Vert ^{2}}{\left\langle
T\right\rangle }
\end{eqnarray*}%
constraint to add to the potentil:%
\begin{equation*}
-\lambda \left( \frac{\alpha sA\left\Vert \Delta \Gamma \right\Vert ^{2}}{%
\left\langle T\right\rangle }+\left( ku+lv-\alpha \right) \right)
\end{equation*}%
mnmz:%
\begin{equation*}
\int U_{\Delta \Gamma }\left( \left\Vert \Delta \Gamma \left( Z,Z^{\prime
}\right) \right\Vert ^{2}\right) -\lambda \left( \frac{\alpha sA\left\Vert
\Delta \Gamma \right\Vert ^{2}}{\left\langle T\right\rangle }+\left(
ku+lv-\alpha \right) \right)
\end{equation*}%
and replace $\alpha $ by:%
\begin{equation*}
\alpha _{0}-\frac{\delta U_{\Delta \Gamma }\left( \left\Vert \Delta \Gamma
\left( Z,Z^{\prime }\right) \right\Vert ^{2}\right) }{\delta \left\Vert
\Delta \Gamma \left( Z,Z^{\prime }\right) \right\Vert ^{2}}
\end{equation*}%
We thus find:%
\begin{equation*}
\frac{\delta U_{\Delta \Gamma }\left( \left\Vert \Delta \Gamma \left(
Z,Z^{\prime }\right) \right\Vert ^{2}\right) }{\delta \left\Vert \Delta
\Gamma \left( Z,Z^{\prime }\right) \right\Vert ^{2}}-\lambda \frac{\delta
^{2}U_{\Delta \Gamma }\left( \left\Vert \Delta \Gamma \left( Z,Z^{\prime
}\right) \right\Vert ^{2}\right) }{\delta ^{2}\left\Vert \Delta \Gamma
\left( Z,Z^{\prime }\right) \right\Vert ^{2}}\left( 1-\frac{sA\left\Vert
\Delta \Gamma \right\Vert ^{2}}{\left\langle T\right\rangle }\right) =0
\end{equation*}%
and the constraint:%
\begin{equation*}
\frac{\left( \alpha _{0}-\frac{\delta U_{\Delta \Gamma }\left( \left\Vert
\Delta \Gamma \left( Z,Z^{\prime }\right) \right\Vert ^{2}\right) }{\delta
\left\Vert \Delta \Gamma \left( Z,Z^{\prime }\right) \right\Vert ^{2}}%
\right) sA\left\Vert \Delta \Gamma \right\Vert ^{2}}{\left\langle
T\right\rangle }+ku+lv-\left( \alpha _{0}-\frac{\delta U_{\Delta \Gamma
}\left( \left\Vert \Delta \Gamma \left( Z,Z^{\prime }\right) \right\Vert
^{2}\right) }{\delta \left\Vert \Delta \Gamma \left( Z,Z^{\prime }\right)
\right\Vert ^{2}}\right) =0
\end{equation*}%
that leads to:%
\begin{equation*}
\lambda =\frac{\frac{\delta U_{\Delta \Gamma }\left( \left\Vert \Delta
\Gamma \left( Z,Z^{\prime }\right) \right\Vert ^{2}\right) }{\delta
\left\Vert \Delta \Gamma \left( Z,Z^{\prime }\right) \right\Vert ^{2}}%
-\alpha _{0}}{\frac{\delta ^{2}U_{\Delta \Gamma }\left( \left\Vert \Delta
\Gamma \left( Z,Z^{\prime }\right) \right\Vert ^{2}\right) }{\delta
^{2}\left\Vert \Delta \Gamma \left( Z,Z^{\prime }\right) \right\Vert ^{2}}%
\left( 1-\frac{sA\left\Vert \Delta \Gamma \right\Vert ^{2}}{\left\langle
T\right\rangle }\right) }
\end{equation*}%
and:%
\begin{equation}
\frac{\delta U_{\Delta \Gamma }\left( \left\Vert \Delta \Gamma \left(
Z,Z^{\prime }\right) \right\Vert ^{2}\right) }{\delta \left\Vert \Delta
\Gamma \left( Z,Z^{\prime }\right) \right\Vert ^{2}}=-\frac{ku+lv}{1-\frac{%
sA\left\Vert \Delta \Gamma \right\Vert ^{2}}{\left\langle T\right\rangle }}%
+\alpha _{0}  \label{nsll}
\end{equation}%
\begin{equation*}
\left\Vert \Delta \Gamma \left( Z,Z^{\prime }\right) \right\Vert ^{2}=\left(
U_{\Delta \Gamma }^{\prime }\right) ^{-1}\left( -\frac{ku+lv}{1-\frac{%
sA\left\Vert \Delta \Gamma \right\Vert ^{2}}{\left\langle T\right\rangle }}%
+\alpha _{0}\right)
\end{equation*}%
and $\alpha _{0}$ obtained by average:%
\begin{equation*}
U_{\Delta \Gamma }^{\prime }\left( \frac{\left\Vert \Delta \Gamma
\right\Vert ^{2}}{Vol}\right) \simeq \left\langle -\frac{ku+lv}{1-\frac{%
sA\left\Vert \Delta \Gamma \right\Vert ^{2}}{\left\langle T\right\rangle }}%
+\alpha _{0}\right\rangle
\end{equation*}%
\begin{equation*}
\alpha _{0}\simeq \left\langle \frac{ku+lv}{1-\frac{sA\left\Vert \Delta
\Gamma \right\Vert ^{2}}{\left\langle T\right\rangle }}\right\rangle
+U_{\Delta \Gamma }^{\prime }\left( \frac{\left\Vert \Delta \Gamma
\right\Vert ^{2}}{Vol}\right)
\end{equation*}%
However, denote $\left\Vert \Delta \Gamma \left( Z,Z^{\prime }\right)
\right\Vert _{\min }^{2}$ the minimum of $U_{\Delta \Gamma }\left(
\left\Vert \Delta \Gamma \left( Z,Z^{\prime }\right) \right\Vert ^{2}\right) 
$. We expand (\ref{nsll}) around $\left\Vert \Delta \Gamma \left(
Z,Z^{\prime }\right) \right\Vert _{\min }^{2}$:%
\begin{equation*}
\frac{\delta U_{\Delta \Gamma }\left( \frac{\left\Vert \Delta \Gamma \left(
Z,Z^{\prime }\right) \right\Vert _{\min }^{2}}{V}+\left( \left\Vert \Delta
\Gamma \left( Z,Z^{\prime }\right) \right\Vert ^{2}-\left\Vert \Delta \Gamma
\left( Z,Z^{\prime }\right) \right\Vert _{\min }^{2}\right) \right) }{\delta
\left\Vert \Delta \Gamma \left( Z,Z^{\prime }\right) \right\Vert ^{2}}=-%
\frac{ku+lv}{1-\frac{sA\left\Vert \Delta \Gamma \right\Vert ^{2}}{%
\left\langle T\right\rangle }}+\alpha _{0}
\end{equation*}%
which leads to:%
\begin{equation}
-\frac{ku+lv}{1-\frac{sA\left\Vert \Delta \Gamma \right\Vert ^{2}}{%
\left\langle T\right\rangle }}+\alpha _{0}=U_{\Delta \Gamma }^{\prime \prime
}\left( \left\Vert \Delta \Gamma \left( Z,Z^{\prime }\right) \right\Vert
_{\min }^{2}\right) \left( \left\Vert \Delta \Gamma \left( Z,Z^{\prime
}\right) \right\Vert ^{2}-\left\Vert \Delta \Gamma \left( Z,Z^{\prime
}\right) \right\Vert _{\min }^{2}\right)  \label{qnd}
\end{equation}%
Moreover, since:%
\begin{equation*}
\frac{\left\Vert \Delta \Gamma \right\Vert ^{2}}{V}=\left\langle \left\Vert
\Delta \Gamma \left( Z,Z^{\prime }\right) \right\Vert ^{2}\right\rangle
\end{equation*}%
we find:%
\begin{equation}
\alpha _{0}\simeq \left\langle \frac{ku+lv}{1-\frac{sA\left\Vert \Delta
\Gamma \right\Vert ^{2}}{\left\langle T\right\rangle }}\right\rangle
+\left\langle U_{\Delta \Gamma }^{\prime \prime }\left( \left\Vert \Delta
\Gamma \left( Z,Z^{\prime }\right) \right\Vert _{\min }^{2}\right)
\right\rangle \left( \frac{\left\Vert \Delta \Gamma \right\Vert ^{2}}{V}%
-\left\langle \left\Vert \Delta \Gamma \left( Z,Z^{\prime }\right)
\right\Vert _{\min }^{2}\right\rangle \right)  \label{lpr}
\end{equation}%
\bigskip and the equation (\ref{qnd}) for $\left\Vert \Delta \Gamma \left(
Z,Z^{\prime }\right) \right\Vert ^{2}$ writes:%
\begin{equation}
-\Delta \left( \frac{ku+lv}{1-\frac{sA\left\Vert \Delta \Gamma \right\Vert
^{2}}{\left\langle T\right\rangle }}\right) \simeq U_{\Delta \Gamma
}^{\prime \prime }\left( \left\Vert \Delta \Gamma \left( Z,Z^{\prime
}\right) \right\Vert _{\min }^{2}\right) \left( \left\Vert \Delta \Gamma
\left( Z,Z^{\prime }\right) \right\Vert ^{2}-\frac{\left\Vert \Delta \Gamma
\right\Vert ^{2}}{V}-\Delta \left\Vert \Delta \Gamma \left( Z,Z^{\prime
}\right) \right\Vert _{\min }^{2}\right)
\end{equation}%
with:%
\begin{eqnarray*}
\Delta \left( \frac{ku+lv}{1-\frac{sA\left\Vert \Delta \Gamma \right\Vert
^{2}}{\left\langle T\right\rangle }}\right) &=&\frac{ku+lv}{1-\frac{%
sA\left\Vert \Delta \Gamma \right\Vert ^{2}}{\left\langle T\right\rangle }}%
-\left\langle \frac{ku+lv}{1-\frac{sA\left\Vert \Delta \Gamma \right\Vert
^{2}}{\left\langle T\right\rangle }}\right\rangle \\
\Delta \left\Vert \Delta \Gamma \left( Z,Z^{\prime }\right) \right\Vert
_{\min }^{2} &=&\left\Vert \Delta \Gamma \left( Z,Z^{\prime }\right)
\right\Vert _{\min }^{2}-\left\langle \left\Vert \Delta \Gamma \left(
Z,Z^{\prime }\right) \right\Vert _{\min }^{2}\right\rangle
\end{eqnarray*}%
We find:%
\begin{equation}
\left\Vert \Delta \Gamma \left( Z,Z^{\prime }\right) \right\Vert ^{2}\simeq
\Delta \left\Vert \Delta \Gamma \left( Z,Z^{\prime }\right) \right\Vert
_{\min }^{2}+\frac{\left\Vert \Delta \Gamma \right\Vert ^{2}}{V}-\frac{%
\Delta \left( \frac{ku+lv}{1-\frac{sA\left\Vert \Delta \Gamma \right\Vert
^{2}}{\left\langle T\right\rangle }}\right) }{U_{\Delta \Gamma }^{\prime
\prime }\left( \left\Vert \Delta \Gamma \left( Z,Z^{\prime }\right)
\right\Vert _{\min }^{2}\right) }
\end{equation}%
By consistency, we may assume that:%
\begin{equation}
\frac{\left\Vert \Delta \Gamma \right\Vert ^{2}}{V}=\left\langle \left\Vert
\Delta \Gamma \left( Z,Z^{\prime }\right) \right\Vert _{\min
}^{2}\right\rangle  \label{CN}
\end{equation}%
so that:%
\begin{equation*}
-\Delta \left( \frac{ku+lv}{1-\frac{sA\left\Vert \Delta \Gamma \right\Vert
^{2}}{\left\langle T\right\rangle }}\right) =U_{\Delta \Gamma }^{\prime
\prime }\left( \left\Vert \Delta \Gamma \left( Z,Z^{\prime }\right)
\right\Vert _{\min }^{2}\right) \left( \left\Vert \Delta \Gamma \left(
Z,Z^{\prime }\right) \right\Vert ^{2}-\left\Vert \Delta \Gamma \left(
Z,Z^{\prime }\right) \right\Vert _{\min }^{2}\right)
\end{equation*}%
This implies that, at lowest order:%
\begin{equation*}
\left\Vert \Delta \Gamma \left( Z,Z^{\prime }\right) \right\Vert
^{2}=\left\Vert \Delta \Gamma \left( Z,Z^{\prime }\right) \right\Vert _{\min
}^{2}-\frac{1}{U_{\Delta \Gamma }^{\prime \prime }\left( \left\Vert \Delta
\Gamma \left( Z,Z^{\prime }\right) \right\Vert _{\min }^{2}\right) }\Delta
\left( \frac{ku+lv}{1-\frac{sA\left\Vert \Delta \Gamma \right\Vert ^{2}}{%
\left\langle T\right\rangle }}\right)
\end{equation*}%
and the action for the field is:%
\begin{eqnarray}
&&\int U_{\Delta \Gamma }\left( \left\Vert \Delta \Gamma \left( Z,Z^{\prime
}\right) \right\Vert ^{2}\right)  \label{ptr} \\
&=&\int U_{\Delta \Gamma }\left( \left\Vert \Delta \Gamma \left( Z,Z^{\prime
}\right) \right\Vert _{\min }^{2}+\frac{\left\Vert \Delta \Gamma \right\Vert
^{2}}{V}-\left\langle \left\Vert \Delta \Gamma \left( Z,Z^{\prime }\right)
\right\Vert _{\min }^{2}\right\rangle -\frac{1}{U_{\Delta \Gamma }^{\prime
\prime }\left( \left\Vert \Delta \Gamma \left( Z,Z^{\prime }\right)
\right\Vert _{\min }^{2}\right) }\Delta \left( \frac{ku+lv}{1-\frac{%
sA\left\Vert \Delta \Gamma \right\Vert ^{2}}{\left\langle T\right\rangle }}%
\right) \right)  \notag \\
&\simeq &\int U_{\Delta \Gamma }\left( \left\Vert \Delta \Gamma \left(
Z,Z^{\prime }\right) \right\Vert _{\min }^{2}\right)  \notag \\
&&+\frac{1}{2}\int U_{\Delta \Gamma }^{\prime \prime }\left( \left\Vert
\Delta \Gamma \left( Z,Z^{\prime }\right) \right\Vert _{\min }^{2}\right)
\left( \frac{\left\Vert \Delta \Gamma \right\Vert ^{2}}{V}-\left\langle
\left\Vert \Delta \Gamma \left( Z,Z^{\prime }\right) \right\Vert _{\min
}^{2}\right\rangle -\frac{\Delta \left( \frac{ku+lv}{1-\frac{sA\left\Vert
\Delta \Gamma \right\Vert ^{2}}{\left\langle T\right\rangle }}\right) }{%
U_{\Delta \Gamma }^{\prime \prime }\left( \left\Vert \Delta \Gamma \left(
Z,Z^{\prime }\right) \right\Vert _{\min }^{2}\right) }\right) ^{2}  \notag
\end{eqnarray}

Assuming a $U$ shape form for the potential so that $U_{\Delta \Gamma
}^{\prime \prime }\left( \left\Vert \Delta \Gamma \left( Z,Z^{\prime
}\right) \right\Vert _{\min }^{2}\right) >0$, implies that state with $%
\left\Vert \Delta \Gamma \left( Z,Z^{\prime }\right) \right\Vert ^{2}>0$
exists if:%
\begin{equation*}
U_{\Delta \Gamma }\left( \left\Vert \Delta \Gamma \left( Z,Z^{\prime
}\right) \right\Vert _{\min }^{2}\right) +\frac{1}{2}U_{\Delta \Gamma
}^{\prime \prime }\left( \left\Vert \Delta \Gamma \left( Z,Z^{\prime
}\right) \right\Vert _{\min }^{2}\right) \left( \frac{\left\Vert \Delta
\Gamma \right\Vert ^{2}}{V}-\left\langle \left\Vert \Delta \Gamma \left(
Z,Z^{\prime }\right) \right\Vert _{\min }^{2}\right\rangle -\frac{\Delta
\left( \frac{ku+lv}{1-\frac{sA\left\Vert \Delta \Gamma \right\Vert ^{2}}{%
\left\langle T\right\rangle }}\right) }{U_{\Delta \Gamma }^{\prime \prime
}\left( \left\Vert \Delta \Gamma \left( Z,Z^{\prime }\right) \right\Vert
_{\min }^{2}\right) }\right) ^{2}<0
\end{equation*}%
i.e.:%
\begin{equation*}
\left\vert \frac{\left\Vert \Delta \Gamma \right\Vert ^{2}}{V}-\left\langle
\left\Vert \Delta \Gamma \left( Z,Z^{\prime }\right) \right\Vert _{\min
}^{2}\right\rangle -\frac{\Delta \left( \frac{ku+lv}{1-\frac{sA\left\Vert
\Delta \Gamma \right\Vert ^{2}}{\left\langle T\right\rangle }}\right) }{%
U_{\Delta \Gamma }^{\prime \prime }\left( \left\Vert \Delta \Gamma \left(
Z,Z^{\prime }\right) \right\Vert _{\min }^{2}\right) }\right\vert <\sqrt{-%
\frac{2U_{\Delta \Gamma }\left( \left\Vert \Delta \Gamma \left( Z,Z^{\prime
}\right) \right\Vert _{\min }^{2}\right) }{U_{\Delta \Gamma }^{\prime \prime
}\left( \left\Vert \Delta \Gamma \left( Z,Z^{\prime }\right) \right\Vert
_{\min }^{2}\right) }}
\end{equation*}%
Expression (\ref{ptr}) is minimal for: $\left( k,l\right) $ minimizing:%
\begin{equation*}
\left\vert \frac{\left\Vert \Delta \Gamma \right\Vert ^{2}}{V}-\left\langle
\left\Vert \Delta \Gamma \left( Z,Z^{\prime }\right) \right\Vert _{\min
}^{2}\right\rangle -\frac{\Delta \left( \frac{ku+lv}{1-\frac{sA\left\Vert
\Delta \Gamma \right\Vert ^{2}}{\left\langle T\right\rangle }}\right) }{%
U_{\Delta \Gamma }^{\prime \prime }\left( \left\Vert \Delta \Gamma \left(
Z,Z^{\prime }\right) \right\Vert _{\min }^{2}\right) }\right\vert
\end{equation*}%
Under the consitency assumption (\ref{CN}), this reduces to minimize:%
\begin{equation*}
\left\vert \frac{\Delta \left( \frac{ku+lv}{1-\frac{sA\left\Vert \Delta
\Gamma \right\Vert ^{2}}{\left\langle T\right\rangle }}\right) }{U_{\Delta
\Gamma }^{\prime \prime }\left( \left\Vert \Delta \Gamma \left( Z,Z^{\prime
}\right) \right\Vert _{\min }^{2}\right) }\right\vert \simeq \frac{\Delta
\left( ku+lv\right) }{\left\langle \left( 1-\frac{sA\left\Vert \Delta \Gamma
\right\Vert ^{2}}{\left\langle T\right\rangle }\right) U_{\Delta \Gamma
}^{\prime \prime }\left( \left\Vert \Delta \Gamma \left( Z,Z^{\prime
}\right) \right\Vert _{\min }^{2}\right) \right\rangle }
\end{equation*}%
Using that:%
\begin{equation*}
\Delta \left( ku+lv\right) =\left( ku+lv\right) -\left\langle \left(
ku+lv\right) \right\rangle
\end{equation*}%
The minimal configuration for $k=l=\frac{1}{2}$ at every point $\left(
Z,Z^{\prime }\right) $, for which:%
\begin{equation*}
\Delta \left( ku+lv\right) =\frac{1}{2}\left( \left( u+v\right)
-\left\langle u+v\right\rangle \right)
\end{equation*}%
As a consequence, the points such that:%
\begin{equation*}
\left\vert \left( u+l\right) -\left\langle u+v\right\rangle \right\vert <%
\sqrt{-8U_{\Delta \Gamma }\left( \left\Vert \Delta \Gamma \left( Z,Z^{\prime
}\right) \right\Vert _{\min }^{2}\right) U_{\Delta \Gamma }^{\prime \prime
}\left( \left\Vert \Delta \Gamma \left( Z,Z^{\prime }\right) \right\Vert
_{\min }^{2}\right) }
\end{equation*}%
have a shifted states, while others present $\Delta \Gamma \left(
Z,Z^{\prime }\right) =0$.

If:%
\begin{equation*}
\frac{\left\Vert \Delta \Gamma \right\Vert ^{2}}{V}-\left\langle \left\Vert
\Delta \Gamma \left( Z,Z^{\prime }\right) \right\Vert _{\min
}^{2}\right\rangle \neq 0
\end{equation*}%
two cases are possible.

\ If $\frac{\left\Vert \Delta \Gamma \right\Vert ^{2}}{V}-\left\langle
\left\Vert \Delta \Gamma \left( Z,Z^{\prime }\right) \right\Vert _{\min
}^{2}\right\rangle >0$, \ the minimum may be reached for $k$ and $l\neq 
\frac{1}{2}$ at some points mainly if $u>\left\langle u\right\rangle $ and $%
v>\left\langle v\right\rangle $. Points such that $u<\left\langle
u\right\rangle $ and $v<\left\langle v\right\rangle $ are rather driven
towards no shift $\Delta \Gamma \left( Z,Z^{\prime }\right) =0$.

But if $\frac{\left\Vert \Delta \Gamma \right\Vert ^{2}}{V}-\left\langle
\left\Vert \Delta \Gamma \left( Z,Z^{\prime }\right) \right\Vert _{\min
}^{2}\right\rangle <0$, the minimum may be reached for $k$ and $l\neq \frac{1%
}{2}$ at some points mainly if $u<\left\langle u\right\rangle $ and $%
v<\left\langle v\right\rangle $. \ For other points, most often, no shift
occurs.

\subsubsection*{2.2.4 Values of and averages shifts}

Using (\ref{lpr}), we have:%
\begin{eqnarray*}
\alpha &=&\alpha _{0}-U_{\Delta \Gamma }^{\prime }\left( \Delta \Gamma
\left( Z,Z^{\prime }\right) \right) \simeq \left\langle \frac{ku+lv}{1-\frac{%
sA\left\Vert \Delta \Gamma \right\Vert ^{2}}{\left\langle T\right\rangle }}%
\right\rangle +\left\langle U_{\Delta \Gamma }^{\prime \prime }\left(
\left\Vert \Delta \Gamma \left( Z,Z^{\prime }\right) \right\Vert _{\min
}^{2}\right) \right\rangle \left( \frac{\left\Vert \Delta \Gamma \right\Vert
^{2}}{V}-\left\langle \left\Vert \Delta \Gamma \left( Z,Z^{\prime }\right)
\right\Vert _{\min }^{2}\right\rangle \right) \\
&&-U_{\Delta \Gamma }^{\prime }\left( \left\Vert \Delta \Gamma \left(
Z,Z^{\prime }\right) \right\Vert _{\min }^{2}+\frac{\left\Vert \Delta \Gamma
\right\Vert ^{2}}{V}-\left\langle \left\Vert \Delta \Gamma \left(
Z,Z^{\prime }\right) \right\Vert _{\min }^{2}\right\rangle -\frac{1}{%
U_{\Delta \Gamma }^{\prime \prime }\left( \left\Vert \Delta \Gamma \left(
Z,Z^{\prime }\right) \right\Vert _{\min }^{2}\right) }\Delta \left( \frac{%
ku+lv}{1-\frac{sA\left\Vert \Delta \Gamma \right\Vert ^{2}}{\left\langle
T\right\rangle }}\right) \right) \\
&=&\frac{ku+lv}{1-\frac{sA\left\Vert \Delta \Gamma \right\Vert ^{2}}{%
\left\langle T\right\rangle }}
\end{eqnarray*}

and as a consequence the shifts are:%
\begin{eqnarray}
\left\langle \Delta T\right\rangle &\simeq &-\frac{\alpha \left(
ku+lv\right) }{\left( 1-\frac{sA\left\Vert \Delta \Gamma \right\Vert ^{2}}{%
\left\langle T\right\rangle }\right) A_{1}\left\langle \Delta \hat{T}%
\right\rangle \left\Vert \Delta \Gamma \right\Vert ^{2}\left( 1+A_{2}\frac{%
\left\langle \Delta T\right\rangle }{\left\langle T\right\rangle }\left\Vert
\Delta \Gamma \right\Vert ^{2}\right) } \\
\left\langle \Delta \hat{T}\right\rangle &\simeq &-\left( \frac{1}{%
\left\langle v\right\rangle }+\frac{\left( 1-\delta \right) \left(
v-u\right) }{\left\langle u\right\rangle \left\langle v\right\rangle }%
\right) A\frac{\left\langle \Delta T\right\rangle }{\left\langle
T\right\rangle }\left\Vert \Delta \Gamma \right\Vert ^{2}-\frac{ku+lv}{1-%
\frac{sA\left\Vert \Delta \Gamma \right\Vert ^{2}}{\left\langle
T\right\rangle }}\frac{\left\langle \frac{\left( 1-\delta \right) \left(
v-u\right) }{s}\right\rangle }{A_{1}\left\langle \Delta \hat{T}\right\rangle
\left\Vert \Delta \Gamma \right\Vert ^{2}\left( 1+A_{2}\frac{\left\langle
\Delta T\right\rangle }{\left\langle T\right\rangle }\left\Vert \Delta
\Gamma \right\Vert ^{2}\right) }  \notag
\end{eqnarray}%
\bigskip

\subsection*{2.3 First approximation approach}

\subsubsection*{2.3.1 Change of variable}

We perform the following change of variables plus shift that yields the
approximate states\footnote{%
These changes of variables are similar to those defined in (\cite{GLs}).
More about the associated approximations and their validity can be found in
this work.}:%
\begin{eqnarray}
\Delta \Gamma \left( T,\hat{T},\theta ,Z,Z^{\prime }\right) &\rightarrow
&\exp \left( -\frac{\rho \left( \omega \left( \theta ,Z,\left\vert \Psi
\right\vert ^{2}\right) \left\vert \bar{\Psi}_{0}\left( Z,Z^{\prime }\right)
\right\vert ^{2}\left( \hat{T}-\left\langle \hat{T}\right\rangle \right)
^{2}\right) +2V_{0}\left( \hat{T}-\left\langle \hat{T}\right\rangle \right) 
}{4\sigma _{\hat{T}}^{2}\omega \left( \theta ,Z,\left\vert \Psi \right\vert
^{2}\right) }\right)  \label{Cgv} \\
&&\times \exp \left( -\frac{\left( \left( T-\left\langle T\right\rangle
\right) ^{2}-2\lambda \tau \left( \hat{T}-\left\langle \hat{T}\right\rangle
\right) \left( T-\left\langle T\right\rangle \right) \right) }{4\sigma
_{T}^{2}\tau \omega }\right) \Delta \Gamma \left( T,\hat{T},\theta
,Z,Z^{\prime }\right)  \notag
\end{eqnarray}%
\begin{eqnarray}
\Delta \Gamma ^{\dag }\left( T,\hat{T},\theta ,Z,Z^{\prime }\right)
&\rightarrow &\exp \left( \frac{\rho \left( \omega \left( \theta
,Z,\left\vert \Psi \right\vert ^{2}\right) \left\vert \bar{\Psi}_{0}\left(
Z,Z^{\prime }\right) \right\vert ^{2}\left( \hat{T}-\left\langle \hat{T}%
\right\rangle \right) ^{2}\right) +2V_{0}\left( \hat{T}-\left\langle \hat{T}%
\right\rangle \right) }{4\sigma _{\hat{T}}^{2}\omega \left( \theta
,Z,\left\vert \Psi \right\vert ^{2}\right) }\right)  \label{Cgr} \\
&&\times \exp \left( \frac{\left( \frac{\left( T-\left\langle T\right\rangle
\right) ^{2}}{\tau }-2\lambda \left( \hat{T}-\left\langle \hat{T}%
\right\rangle \right) \left( T-\left\langle T\right\rangle \right) \right) }{%
4\sigma _{T}^{2}\tau \omega }\right) \Delta \Gamma ^{\dag }\left( T,\hat{T}%
,\theta ,Z,Z^{\prime }\right)  \notag
\end{eqnarray}%
with:%
\begin{equation*}
V_{0}=\left( \frac{\rho D\left( \theta \right) \left\langle \hat{T}%
\right\rangle \left\vert \Psi _{0}\left( Z^{\prime }\right) \right\vert ^{2}%
}{\omega _{0}\left( Z\right) }\hat{T}\left( 1-\left( 1+\left\langle
\left\vert \Psi _{\Gamma }\right\vert ^{2}\right\rangle \right) \hat{T}%
\right) ^{-1}\left[ O\frac{\Delta T\left\vert \Delta \Gamma \left( \theta
_{1},Z_{1},Z_{1}^{\prime }\right) \right\vert ^{2}}{T}\right] \right)
\end{equation*}%
and (\ref{Sdt}) writes:

\begin{eqnarray}
&&0=\left( -\sigma _{\hat{T}}^{2}\nabla _{\hat{T}}^{2}+\frac{1}{4\sigma _{%
\hat{T}}^{2}}\left( \rho \left\vert \bar{\Psi}_{0}\left( Z,Z^{\prime
}\right) \right\vert ^{2}\Delta \hat{T}+\frac{\rho \left( V_{0}-\frac{\sigma
_{\hat{T}}^{2}}{\sigma _{T}^{2}}\lambda \Delta T\left\vert \Psi \left(
Z\right) \right\vert ^{2}\right) }{\omega _{0}\left( Z\right) }\right)
^{2}\right) \Delta \Gamma \left( T,\hat{T},\theta ,Z,Z^{\prime }\right) 
\notag \\
&&+\left( -\sigma _{T}^{2}\nabla _{T}^{2}+\frac{1}{4\sigma _{T}^{2}}\left( 
\frac{\Delta T-\lambda \tau \Delta \hat{T}}{\tau \omega _{0}\left( Z\right) }%
\left\vert \Psi \left( Z\right) \right\vert ^{2}\right) ^{2}\right) \Delta
\Gamma \left( T,\hat{T},\theta ,Z,Z^{\prime }\right)  \notag \\
&&-\left( \frac{\rho \left\vert \bar{\Psi}_{0}\left( Z,Z^{\prime }\right)
\right\vert ^{2}}{2}+\frac{\left\vert \Psi \left( Z\right) \right\vert ^{2}}{%
2\tau \omega _{0}\left( Z\right) }+V\left( \theta ,Z,Z^{\prime },\Delta
\Gamma \right) \Delta T\right) \Delta \Gamma \left( T,\hat{T},\theta
,Z,Z^{\prime }\right)  \label{Sdp}
\end{eqnarray}%
Doing so, the change of variable misses a term:%
\begin{equation*}
\nabla _{\hat{T}}\frac{\sigma _{\hat{T}}^{2}\lambda \Delta T}{2\sigma
_{T}^{2}\tau \omega }\Delta \Gamma \left( T,\hat{T},\theta ,Z,Z^{\prime
}\right) =\nabla _{\hat{T}^{\prime }}\frac{\sigma _{\hat{T}}\lambda \Delta
T^{\prime }}{2\sigma _{T}\tau \omega }\Delta \Gamma \left( T,\hat{T},\theta
,Z,Z^{\prime }\right)
\end{equation*}%
where: 
\begin{equation*}
\Delta T^{\prime }=\frac{\Delta T}{\sigma _{T}}
\end{equation*}%
and: 
\begin{equation*}
\Delta \hat{T}^{\prime }=\frac{\Delta \hat{T}}{\sigma _{\hat{T}}}
\end{equation*}%
\ are normalized variables. Given our assumption that $\frac{\sigma _{\hat{T}%
}^{2}}{\sigma _{T}^{2}}<<1$, the error can be neglected.

\subsubsection{2.3.2 Diagonalization}

The potential:%
\begin{equation*}
\frac{1}{\sigma _{\hat{T}}^{2}}\left( \rho \left\vert \bar{\Psi}_{0}\left(
Z,Z^{\prime }\right) \right\vert ^{2}\Delta \hat{T}+\frac{\rho \left( V_{0}-%
\frac{\sigma _{\hat{T}}^{2}}{\sigma _{T}^{2}}\lambda \Delta T\left\vert \Psi
\left( Z\right) \right\vert ^{2}\right) }{\omega _{0}\left( Z\right) }%
\right) ^{2}+\frac{1}{\sigma _{T}^{2}}\left( \frac{\Delta T-\lambda \tau
\Delta \hat{T}}{\tau \omega _{0}\left( Z\right) }\left\vert \Psi \left(
Z\right) \right\vert ^{2}\right) ^{2}
\end{equation*}%
writes in term of the normalized variables (with $\left( \Delta T^{\prime
},\Delta \hat{T}^{\prime }\right) \rightarrow \left( \Delta T,\Delta \hat{T}%
\right) $):%
\begin{equation*}
\left( 
\begin{array}{cc}
\Delta T-\Delta T_{0} & \Delta \hat{T}-\Delta \hat{T}_{0}%
\end{array}%
\right) U\left( 
\begin{array}{c}
\Delta T-\Delta T_{0} \\ 
\Delta \hat{T}-\Delta \hat{T}_{0}%
\end{array}%
\right)
\end{equation*}%
where:%
\begin{eqnarray*}
u &=&\frac{\left\vert \Psi _{0}\left( Z\right) \right\vert ^{2}}{\tau \omega
_{0}\left( Z\right) } \\
v &=&\rho \left\vert \bar{\Psi}_{0}\left( Z,Z^{\prime }\right) \right\vert
^{2} \\
s &=&-\frac{\lambda \left\vert \Psi _{0}\left( Z\right) \right\vert ^{2}}{%
\omega _{0}\left( Z\right) }\frac{\sigma _{\hat{T}}}{\sigma _{T}}
\end{eqnarray*}%
\begin{eqnarray*}
\Delta T_{0} &\simeq &-\frac{\lambda \tau V_{0}}{\sigma _{T}\omega
_{0}\left( Z\right) \left\vert \bar{\Psi}_{0}\left( Z,Z^{\prime }\right)
\right\vert ^{2}} \\
\Delta \hat{T}_{0} &\simeq &\frac{\Delta T_{0}}{\lambda \tau }\frac{\sigma
_{T}}{\sigma _{\hat{T}}}
\end{eqnarray*}%
and:%
\begin{equation*}
U=\left( 
\begin{array}{cc}
u^{2}+s^{2} & -\left( u+v\right) s \\ 
-\left( u+v\right) s & v^{2}+s^{2}%
\end{array}%
\right)
\end{equation*}%
under the assumption that $\frac{\sigma _{\hat{T}}^{2}}{\sigma _{T}^{2}}<<1$.

Performing the diagonalization of $U=PDP^{-1}$ with:%
\begin{equation*}
P=\left( 
\begin{array}{cc}
\cos x & \sin x \\ 
-\sin x & \cos x%
\end{array}%
\right)
\end{equation*}%
and setting:%
\begin{equation*}
D=\left( 
\begin{array}{cc}
\cos x & -\sin x \\ 
\sin x & \cos x%
\end{array}%
\right) \left( 
\begin{array}{cc}
u^{2}+s^{2} & -\left( u+v\right) s \\ 
-\left( u+v\right) s & v^{2}+s^{2}%
\end{array}%
\right) \left( 
\begin{array}{cc}
\cos x & \sin x \\ 
-\sin x & \cos x%
\end{array}%
\right)
\end{equation*}%
we obtain:

\begin{equation*}
D=\left( 
\begin{array}{cc}
\lambda _{+} & 0 \\ 
0 & \lambda _{-}%
\end{array}%
\right) ,P=\left( 
\begin{array}{cc}
w_{1} & w_{2} \\ 
w_{1}^{\prime } & w_{2}^{\prime }%
\end{array}%
\right)
\end{equation*}%
$\allowbreak \allowbreak $with:%
\begin{equation*}
\lambda _{\pm }^{2}=\frac{1}{2}\left( u^{2}+v^{2}\right) +s^{2}\pm \frac{%
\left( u+v\right) }{2}\sqrt{\left( u-v\right) ^{2}+4s^{2}}
\end{equation*}%
and:%
\begin{eqnarray*}
w_{1} &=&\sqrt{\frac{1}{2}\left( 1+\sqrt{\frac{\left( \frac{\left(
v-u\right) }{2s}\right) ^{2}}{1+\left( \frac{\left( v-u\right) }{2s}\right)
^{2}}}\right) }\text{, }w_{2}=\sqrt{\frac{1}{2}\left( 1-\sqrt{\frac{\left( 
\frac{\left( v-u\right) }{2s}\right) ^{2}}{1+\left( \frac{\left( v-u\right) 
}{2s}\right) ^{2}}}\right) } \\
w_{1}^{\prime } &=&-\sqrt{\frac{1}{2}\left( 1-\sqrt{\frac{\left( \frac{%
\left( v-u\right) }{2s}\right) ^{2}}{1+\left( \frac{\left( v-u\right) }{2s}%
\right) ^{2}}}\right) }\text{, }w_{2}^{\prime }=\sqrt{\frac{1}{2}\left( 1+%
\sqrt{\frac{\left( \frac{\left( v-u\right) }{2s}\right) ^{2}}{1+\left( \frac{%
\left( v-u\right) }{2s}\right) ^{2}}}\right) }
\end{eqnarray*}%
These relations lead to replace in (\ref{Sdp}):%
\begin{equation*}
\Delta T=w_{1}\Delta T^{\prime }+w_{2}\Delta \hat{T}^{\prime }=\left( \frac{%
w_{1}^{2}}{\lambda _{-}}+\frac{w_{2}^{2}}{\lambda _{+}}\right) V
\end{equation*}%
\begin{equation*}
\Delta \hat{T}=w_{1}^{\prime }\Delta T^{\prime }+w_{2}^{\prime }\Delta \hat{T%
}^{\prime }=\left( \frac{w_{1}^{\prime }w_{1}}{\lambda _{-}}+\frac{%
w_{2}^{\prime }w_{2}}{\lambda _{+}}\right) V
\end{equation*}%
\begin{equation*}
\Delta \hat{T}=\frac{\left( \frac{w_{1}^{\prime }w_{1}}{\lambda _{-}}+\frac{%
w_{2}^{\prime }w_{2}}{\lambda _{+}}\right) }{\left( \frac{w_{1}^{2}}{\lambda
_{-}}+\frac{w_{2}^{2}}{\lambda _{+}}\right) }\Delta T
\end{equation*}%
where $\Delta T^{\prime }$ are the coordinates in the diagonal basis that
satisfies the relation: $\left( 
\begin{array}{c}
\Delta T^{\prime } \\ 
\Delta \hat{T}^{\prime }%
\end{array}%
\right) =P^{-1}\left( 
\begin{array}{c}
\Delta T \\ 
\Delta \hat{T}%
\end{array}%
\right) $.

As a consequence, the background state equations becomes: 
\begin{eqnarray}
0 &=&\left( -\sigma _{\hat{T}}^{2}\nabla _{\hat{T}^{\prime }}^{2}+\frac{%
\lambda _{+}}{4\sigma _{\hat{T}}^{2}}\left( \Delta \hat{T}^{\prime }-\Delta 
\hat{T}_{0}^{\prime }\right) ^{2}\right) \Delta \Gamma \left( T,\hat{T}%
,\theta ,Z,Z^{\prime }\right)  \label{sdl} \\
&&+\left( -\sigma _{T}^{2}\nabla _{T^{\prime }}^{2}+\frac{\lambda _{-}}{%
\sigma _{T}^{2}}\left( \Delta T^{\prime }-\Delta T_{0}^{\prime }\right)
^{2}\right) \Delta \Gamma \left( T,\hat{T},\theta ,Z,Z^{\prime }\right) 
\notag \\
&&-\left( u+v+V\left( w_{1}\Delta T^{\prime }+w_{2}\Delta \hat{T}^{\prime
}\right) \right) \Delta \Gamma \left( T,\hat{T},\theta ,Z,Z^{\prime }\right)
\notag \\
&=&\left( -\sigma _{\hat{T}}^{2}\nabla _{\hat{T}^{\prime }}^{2}+\frac{%
\lambda _{+}}{4\sigma _{\hat{T}}^{2}}\left( \Delta \hat{T}^{\prime }-\Delta 
\hat{T}_{0}^{\prime }-\frac{w_{2}}{\lambda _{+}}V\right) ^{2}\right) \Delta
\Gamma \left( T,\hat{T},\theta ,Z,Z^{\prime }\right)  \notag \\
&&+\left( -\sigma _{T}^{2}\nabla _{T^{\prime }}^{2}+\frac{\lambda _{-}}{%
\sigma _{T}^{2}}\left( \Delta T^{\prime }-\Delta T_{0}^{\prime }-\frac{w_{1}%
}{\lambda _{-}}V\right) ^{2}\right) \Delta \Gamma \left( T,\hat{T},\theta
,Z,Z^{\prime }\right)  \notag \\
&&-\left( u+v+\left( \frac{w_{1}^{2}}{\lambda _{+}}V^{2}+\frac{w_{2}^{2}}{%
\lambda _{-}}V^{2}\right) \right) \Delta \Gamma \left( T,\hat{T},\theta
,Z,Z^{\prime }\right)  \notag
\end{eqnarray}%
where we defined:%
\begin{equation*}
\left( 
\begin{array}{c}
\Delta T_{0}^{\prime } \\ 
\Delta \hat{T}_{0}^{\prime }%
\end{array}%
\right) =P^{-1}\left( 
\begin{array}{c}
\Delta T_{0} \\ 
\Delta \hat{T}_{0}%
\end{array}%
\right)
\end{equation*}

\section*{Appendix 3 Example of application: dynamics between $T\left(
Z,Z^{\prime }\right) $ and $T\left( Z^{\prime },Z\right) $}

We study the interactions between $T\left( Z,Z^{\prime }\right) $ and $%
T\left( Z^{\prime },Z\right) $, i.e. the connectivity in both direction, by
computing the transition function:%
\begin{equation*}
G\left( \Delta T_{i}\left( Z,Z^{\prime }\right) ,\Delta T_{1}\left(
Z,Z^{\prime }\right) ,\Delta T_{i}^{\prime }\left( Z^{\prime },Z\right)
,\Delta T_{f}\left( Z^{\prime },Z\right) \right)
\end{equation*}%
At the zeroth order in perturbation, that is, neglecting the interaction,
the transition function is given by a product of two "free" transition
functions:%
\begin{eqnarray*}
&&G\left( \Delta T_{i}\left( Z,Z^{\prime }\right) ,\Delta T_{1}\left(
Z,Z^{\prime }\right) ,\Delta T_{i}^{\prime }\left( Z^{\prime },Z\right)
,\Delta T_{f}\left( Z^{\prime },Z\right) \right) \\
&\simeq &G_{0}\left( \Delta T_{i}\left( Z,Z^{\prime }\right) ,\Delta
T_{f}\left( Z,Z^{\prime }\right) ,\Delta T_{i}\left( Z^{\prime },Z\right)
,\Delta T_{f}\left( Z^{\prime },Z\right) \right) G_{0}\left( \Delta
T_{i}\left( Z^{\prime },Z\right) ,\Delta T_{f}\left( Z^{\prime },Z\right)
\right)
\end{eqnarray*}%
These transitions where computed in (\cite{GLs}). Defining $\mathbf{T-}%
\left\langle \mathbf{T}\right\rangle $ to be the vector with components:%
\begin{equation*}
\left( T-\left\langle T\right\rangle ,\hat{T}-\left\langle \hat{T}%
\right\rangle \right)
\end{equation*}%
the transition between $\mathbf{T-}\left\langle \mathbf{T}\right\rangle $
and $\mathbf{T}^{\prime }\mathbf{-}\left\langle \mathbf{T}\right\rangle $
during a time $t$, written $G_{0}\left( \mathbf{T-}\left\langle \mathbf{T}%
\right\rangle ,\mathbf{T}^{\prime }\mathbf{-}\left\langle \mathbf{T}%
\right\rangle ,t\right) $, is given by:%
\begin{eqnarray}
&&G_{0}\left( \mathbf{T-}\left\langle \mathbf{T}\right\rangle ,\mathbf{T}%
^{\prime }\mathbf{-}\left\langle \mathbf{T}\right\rangle ,t\right)
\label{trss} \\
&=&\left( 2\pi \right) ^{-1}\left( Det\left( \sigma \left( t\right) \right)
\right) ^{-\frac{1}{2}}  \notag \\
&&\times \exp \left( -\left( \left( \mathbf{T-}\left\langle \mathbf{T}%
\right\rangle \right) -M\left( t\right) \left( \mathbf{T}^{\prime }\mathbf{-}%
\left\langle \mathbf{T}\right\rangle \right) \right) ^{t}\frac{\sigma
^{-1}\left( t\right) }{2}\left( \left( \mathbf{T-}\left\langle \mathbf{T}%
\right\rangle \right) -M\left( t\right) \left( \mathbf{T}^{\prime }\mathbf{-}%
\left\langle \mathbf{T}\right\rangle \right) \right) \right)  \notag
\end{eqnarray}%
where the matrices $M\left( t\right) $ and $\sigma \left( t\right) $ are
defined by:%
\begin{eqnarray*}
M\left( t\right) &=&\left( 
\begin{array}{cc}
e^{-tu} & s\frac{e^{-tu}-e^{-tv}}{u-v} \\ 
0 & e^{-tv}%
\end{array}%
\right) \\
\sigma \left( t\right) &=&\left( 
\begin{array}{cc}
\frac{1-e^{-2tu}}{u}+s^{2}\frac{\frac{\left( u-v\right) ^{2}}{uv\left(
u+v\right) }-\left( \frac{e^{-2tu}}{u}-4\frac{e^{-t\left( u+v\right) }}{u+v}+%
\frac{e^{-2tv}}{v}\right) }{\left( u-v\right) ^{2}} & s\frac{\frac{v-u}{%
v\left( u+v\right) }-\left( 2\frac{e^{-t\left( u+v\right) }}{u+v}-\frac{%
e^{-2tv}}{v}\right) }{u-v} \\ 
s\frac{\frac{v-u}{v\left( u+v\right) }-\left( 2\frac{e^{-t\left( u+v\right) }%
}{u+v}-\frac{e^{-2tv}}{v}\right) }{u-v} & \frac{1-e^{-2tv}}{v}%
\end{array}%
\right)
\end{eqnarray*}%
In the large $t$ approximation, the transition can be approximated by: 
\begin{eqnarray}
G_{0}\left( \mathbf{T-}\left\langle \mathbf{T}\right\rangle ,\mathbf{T}%
^{\prime }\mathbf{-}\left\langle \mathbf{T}\right\rangle \right) &=&\left(
2\pi \right) ^{-1}\left( Det\left( \sigma \left( \infty \right) \right)
\right) ^{-\frac{1}{2}}  \label{tsrr} \\
&&\times \exp \left( -\frac{1}{2}\left( \left( \mathbf{T-}\left\langle 
\mathbf{T}\right\rangle \right) \right) ^{t}\sigma ^{-1}\left( \infty
\right) \left( \left( \mathbf{T-}\left\langle \mathbf{T}\right\rangle
\right) \right) \right)  \notag
\end{eqnarray}%
with:%
\begin{equation*}
\sigma \left( \infty \right) =\left( 
\begin{array}{cc}
\frac{1}{u}+\frac{s^{2}}{uv\left( u+v\right) } & -\frac{s}{v\left(
u+v\right) } \\ 
-\frac{s}{v\left( u+v\right) } & \frac{1-e^{-2tv}}{v}%
\end{array}%
\right)
\end{equation*}%
As explained in the text the graphs that compute mutual interactions between 
$T\left( Z,Z^{\prime }\right) $ and $T\left( Z^{\prime },Z\right) $ at the
lowest order are given by the squared interaction term averaged between an
initial and a final $2$- state that writes in an expanded form:

\begin{eqnarray}
&&\left\langle \Delta T_{i}\left( Z,Z^{\prime }\right) ,\Delta T_{i}\left(
Z^{\prime },Z\right) \right\vert \left\{ \Delta \Gamma ^{\dag }\left( T,\hat{%
T},\theta ,Z,Z^{\prime }\right) \right.  \label{CTR} \\
&&\nabla _{\hat{T}}\left( a\left( Z^{\prime },Z\right) \Delta T\left(
Z^{\prime },Z\right) \left\vert \Delta \Gamma \left( \theta -2\frac{%
\left\vert Z-Z^{\prime }\right\vert }{c},Z\right) \right\vert ^{2}-b\left(
Z,Z^{\prime }\right) \Delta T\left( Z,Z^{\prime }\right) \left\vert \Delta
\Gamma \left( \theta -\frac{\left\vert Z-Z^{\prime }\right\vert }{c}%
,Z^{\prime }\right) \right\vert ^{2}\right)  \notag \\
&&\left. \Delta \Gamma \left( T,\hat{T},\theta ,Z,Z^{\prime }\right) \right\}
\notag \\
&&\left\{ \Delta \Gamma ^{\dag }\left( T,\hat{T},\theta ,Z^{\prime
},Z\right) \right. \nabla _{\hat{T}}\left( a\left( Z,Z^{\prime }\right)
\Delta T\left( Z,Z^{\prime }\right) \left\vert \Delta \Gamma \left( \theta -2%
\frac{\left\vert Z-Z^{\prime }\right\vert }{c},Z^{\prime }\right)
\right\vert ^{2}\right.  \notag \\
&&\left. -b\left( Z^{\prime },Z\right) \Delta T\left( Z,Z^{\prime }\right)
\left\vert \Delta \Gamma \left( \theta -\frac{\left\vert Z-Z^{\prime
}\right\vert }{c},Z\right) \right\vert ^{2}\right) \left. \Delta \Gamma
\left( T,\hat{T},\theta ,Z^{\prime },Z\right) \right\} \left\vert \Delta
T_{f}\left( Z,Z^{\prime }\right) ,\Delta T_{f}\left( Z^{\prime },Z\right)
\right\rangle  \notag
\end{eqnarray}%
Developping the square leads to three contributions to (\ref{CTR}). Each of
them is derived independently:%
\begin{eqnarray*}
&&\left\langle \Delta T_{i}\left( Z,Z^{\prime }\right) ,\Delta T_{i}\left(
Z^{\prime },Z\right) \right\vert \\
&&\left\{ \Delta \Gamma ^{\dag }\left( T,\hat{T},\theta ,Z,Z^{\prime
}\right) \right. \nabla _{\hat{T}}\left( a\left( Z^{\prime },Z\right) \Delta
T\left( Z^{\prime },Z\right) \left\vert \Delta \Gamma \left( \theta -2\frac{%
\left\vert Z-Z^{\prime }\right\vert }{c},Z\right) \right\vert ^{2}\right) \\
&&\left. \Delta \Gamma \left( T,\hat{T},\theta ,Z,Z^{\prime }\right) \right\}
\\
&&\times \left\{ \Delta \Gamma ^{\dag }\left( T,\hat{T},\theta ,Z^{\prime
},Z\right) \right. \nabla _{\hat{T}}\left( a\left( Z,Z^{\prime }\right)
\Delta T\left( Z,Z^{\prime }\right) \left\vert \Delta \Gamma \left( \theta -2%
\frac{\left\vert Z-Z^{\prime }\right\vert }{c},Z^{\prime }\right)
\right\vert ^{2}\right) \\
&&\left. \Delta \Gamma \left( T,\hat{T},\theta ,Z^{\prime },Z\right) \right\}
\\
&&\left\vert \Delta T_{f}\left( Z,Z^{\prime }\right) ,\Delta T_{f}\left(
Z^{\prime },Z\right) \right\rangle \\
&=&G\left( \Delta T_{i}\left( Z,Z^{\prime }\right) ,\Delta T_{1}\left(
Z,Z^{\prime }\right) \right) a\left( Z^{\prime },Z\right) \Delta T_{1}\left(
Z^{\prime },Z\right) \nabla _{\hat{T}}G\left( \Delta T_{1}\left( Z,Z^{\prime
}\right) ,\Delta T_{f}\left( Z,Z^{\prime }\right) \right) \\
&&\times G\left( \Delta T_{i}\left( Z^{\prime },Z\right) ,\Delta T_{1}\left(
Z^{\prime },Z\right) \right) a\left( Z,Z^{\prime }\right) \Delta T_{1}\left(
Z,Z^{\prime }\right) \nabla _{\hat{T}}G\left( \Delta T_{1}\left( Z^{\prime
},Z\right) ,\Delta T_{f}\left( Z^{\prime },Z\right) \right)
\end{eqnarray*}%
\begin{eqnarray*}
&&\left\langle \Delta T_{i}\left( Z,Z^{\prime }\right) ,\Delta T_{i}\left(
Z^{\prime },Z\right) \right\vert \left\{ \Delta \Gamma ^{\dag }\left( T,\hat{%
T},\theta ,Z,Z^{\prime }\right) \right. \\
&&\nabla _{\hat{T}}\left( b\left( Z,Z^{\prime }\right) \Delta T\left(
Z,Z^{\prime }\right) \left\vert \Delta \Gamma \left( \theta -\frac{%
\left\vert Z-Z^{\prime }\right\vert }{c},Z^{\prime }\right) \right\vert
^{2}\right) \\
&&\left. \Delta \Gamma \left( T,\hat{T},\theta ,Z,Z^{\prime }\right)
\right\} ^{2}\left\vert \Delta T_{f}\left( Z,Z^{\prime }\right) ,\Delta
T_{f}\left( Z^{\prime },Z\right) \right\rangle \\
&=&G\left( \Delta T_{i}\left( Z,Z^{\prime }\right) ,\Delta T_{1}\left(
Z,Z^{\prime }\right) \right) b\left( Z^{\prime },Z\right) \Delta T_{1}\left(
Z^{\prime },Z\right) \nabla _{\hat{T}}G\left( \Delta T_{1}\left( Z,Z^{\prime
}\right) ,\Delta T_{f}\left( Z,Z^{\prime }\right) \right) \\
&&\times G\left( \Delta T_{i}\left( Z^{\prime },Z\right) ,\Delta T_{1}\left(
Z^{\prime },Z\right) \right) b\left( Z,Z^{\prime }\right) \Delta T_{1}\left(
Z,Z^{\prime }\right) \nabla _{\hat{T}}G\left( \Delta T_{1}\left( Z^{\prime
},Z\right) ,\Delta T_{f}\left( Z^{\prime },Z\right) \right)
\end{eqnarray*}%
and:%
\begin{eqnarray*}
&&-2\left\langle \Delta T_{i}\left( Z,Z^{\prime }\right) ,\Delta T_{i}\left(
Z^{\prime },Z\right) \right\vert \\
&&\left\{ \Delta \Gamma ^{\dag }\left( T,\hat{T},\theta ,Z,Z^{\prime
}\right) \nabla _{\hat{T}}\left( a\left( Z^{\prime },Z\right) \Delta T\left(
Z^{\prime },Z\right) \left\vert \Delta \Gamma \left( \theta -2\frac{%
\left\vert Z-Z^{\prime }\right\vert }{c},Z\right) \right\vert ^{2}\right)
\Delta \Gamma \left( T,\hat{T},\theta ,Z,Z^{\prime }\right) \right\} \\
&&\times \left\{ \Delta \Gamma ^{\dag }\left( T,\hat{T},\theta ,Z,Z^{\prime
}\right) \nabla _{\hat{T}}\left( b\left( Z,Z^{\prime }\right) \Delta T\left(
Z,Z^{\prime }\right) \left\vert \Delta \Gamma \left( \theta -\frac{%
\left\vert Z-Z^{\prime }\right\vert }{c},Z^{\prime }\right) \right\vert
^{2}\right) \Delta \Gamma \left( T,\hat{T},\theta ,Z,Z^{\prime }\right)
\right\} \\
&&\left. \Delta \Gamma \left( T,\hat{T},\theta ,Z,Z^{\prime }\right)
\right\} ^{2}\left\vert \Delta T_{f}\left( Z,Z^{\prime }\right) ,\Delta
T_{f}\left( Z^{\prime },Z\right) \right\rangle \\
&=&G\left( \Delta T_{i}\left( Z,Z^{\prime }\right) ,\Delta T_{1}\left(
Z,Z^{\prime }\right) \right) b\left( Z^{\prime },Z\right) \Delta T_{1}\left(
Z^{\prime },Z\right) \nabla _{\hat{T}}G\left( \Delta T_{1}\left( Z,Z^{\prime
}\right) ,\Delta T_{f}\left( Z,Z^{\prime }\right) \right) \\
&&\times G\left( \Delta T_{i}\left( Z^{\prime },Z\right) ,\Delta T_{1}\left(
Z^{\prime },Z\right) \right) b\left( Z,Z^{\prime }\right) \Delta T_{1}\left(
Z,Z^{\prime }\right) \nabla _{\hat{T}}G\left( \Delta T_{1}\left( Z^{\prime
},Z\right) ,\Delta T_{f}\left( Z^{\prime },Z\right) \right) +\left(
Z\leftrightarrow Z^{\prime }\right)
\end{eqnarray*}%
Gathering all contribution leads to the contribution of (\ref{CTR}) to the
transition function:%
\begin{eqnarray*}
&=&G\left( \Delta T_{i}\left( Z,Z^{\prime }\right) ,\Delta T_{1}\left(
Z,Z^{\prime }\right) \right) \left( a\left( Z^{\prime },Z\right) \Delta
T_{1}\left( Z^{\prime },Z\right) -b\left( Z,Z^{\prime }\right) \Delta
T_{1}\left( Z,Z^{\prime }\right) \right) \nabla _{\hat{T}}G\left( \Delta
T_{1}\left( Z,Z^{\prime }\right) ,\Delta T_{f}\left( Z,Z^{\prime }\right)
\right) \\
&&\times G\left( \Delta T_{i}^{\prime }\left( Z^{\prime },Z\right) ,\Delta
T_{1}\left( Z^{\prime },Z\right) \right) \left( a\left( Z,Z^{\prime }\right)
\Delta T_{1}\left( Z,Z^{\prime }\right) -b\left( Z^{\prime },Z\right) \Delta
T_{1}\left( Z^{\prime },Z\right) \right) \nabla _{\hat{T}}G\left( \Delta
T_{1}\left( Z^{\prime },Z\right) ,\Delta T_{f}\left( Z^{\prime },Z\right)
\right)
\end{eqnarray*}%
Using the formula (\ref{trss}) for the transition functions leads to the
correction to the free amplitude:%
\begin{eqnarray*}
&&\Delta G\left( \Delta T_{i}\left( Z,Z^{\prime }\right) ,\Delta T_{1}\left(
Z,Z^{\prime }\right) ,\Delta T_{i}^{\prime }\left( Z^{\prime },Z\right)
,\Delta T_{f}\left( Z^{\prime },Z\right) \right) \\
&=&\exp \left( -\frac{1}{2}\left( \Delta \mathbf{T}_{1}-M\left( t\right)
\left( \Delta \mathbf{T}_{i}\right) \right) ^{t}\sigma ^{-1}\left( t\right)
\left( \Delta \mathbf{T}_{1}-M\left( t\right) \left( \Delta \mathbf{T}%
_{i}\right) \right) \right) \\
&&\times \exp \left( -\frac{1}{2}\left( \left( \Delta \mathbf{T}_{f}\right)
-M\left( t\right) \Delta \mathbf{T}_{1}\right) ^{t}\sigma ^{-1}\left(
t\right) \left( \left( \Delta \mathbf{T}_{f}\right) -M\left( t\right) \Delta 
\mathbf{T}_{1}\right) \right) \\
&&\times \left( a\left( Z^{\prime },Z\right) \Delta T_{1}-b\left(
Z,Z^{\prime }\right) \Delta T_{1}^{\prime }\right) \left( a\left(
Z,Z^{\prime }\right) \Delta T_{1}^{\prime }-b\left( Z^{\prime },Z\right)
\Delta T_{1}\right) \\
&&\times \exp \left( -\frac{1}{2}\left( \Delta \mathbf{T}_{1}^{\prime
}-M\left( t\right) \left( \Delta \mathbf{T}_{i}^{\prime }\right) \right)
^{t}\sigma ^{-1}\left( t\right) \left( \Delta \mathbf{T}_{1}^{\prime
}-M\left( t\right) \left( \Delta \mathbf{T}_{i}^{\prime }\right) \right)
\right) \\
&&\times \exp \left( -\frac{1}{2}\left( \left( \Delta \mathbf{T}_{f}^{\prime
}\right) -M\left( t\right) \Delta \mathbf{T}_{1}^{\prime }\right) ^{t}\sigma
^{-1}\left( t\right) \left( \left( \Delta \mathbf{T}_{f}^{\prime }\right)
-M\left( t\right) \Delta \mathbf{T}_{1}^{\prime }\right) \right)
\end{eqnarray*}%
where $\Delta \mathbf{T}_{i}$ stands for $\Delta T_{i}\left( Z,Z^{\prime
}\right) $, $\Delta \mathbf{T}_{i}^{\prime }$ for $\Delta T_{i}\left(
Z^{\prime },Z\right) $ and similarly for $\Delta \mathbf{T}_{1}$, $\Delta 
\mathbf{T}_{1}^{\prime }$.

To compute the effect of some fluctuations in the connectivity, we assume
that $\Delta \mathbf{T}_{i}=0$, so that we study the impact of a deviation $%
\Delta \mathbf{T}_{1}\neq 0$ on both final states $\Delta \mathbf{T}_{f}$
and $\Delta \mathbf{T}_{f}^{\prime }$. This correction modifies the free
transition function by the contribution:

\begin{eqnarray}
&&G\left( \Delta T_{i}\left( Z,Z^{\prime }\right) ,\Delta T_{1}\left(
Z,Z^{\prime }\right) ,\Delta T_{i}^{\prime }\left( Z^{\prime },Z\right)
,\Delta T_{f}\left( Z^{\prime },Z\right) \right)  \label{grp} \\
&=&G_{0}\left( \Delta T_{i}\left( Z,Z^{\prime }\right) ,\Delta T_{1}\left(
Z,Z^{\prime }\right) ,\Delta T_{i}^{\prime }\left( Z^{\prime },Z\right)
,\Delta T_{f}\left( Z^{\prime },Z\right) \right)  \notag \\
&&+\exp \left( -\frac{1}{2}\left( \Delta \mathbf{T}_{1}-M\left( t\right)
\left( \Delta \mathbf{T}_{i}\right) \right) ^{t}\sigma ^{-1}\left( t\right)
\left( \Delta \mathbf{T}_{1}-M\left( t\right) \left( \Delta \mathbf{T}%
_{i}\right) \right) \right)  \notag \\
&&\times \exp \left( -\frac{1}{2}\left( \left( \Delta \mathbf{T}_{f}\right)
-M\left( t\right) \Delta \mathbf{T}_{1}\right) ^{t}\sigma ^{-1}\left(
t\right) \left( \left( \Delta \mathbf{T}_{f}\right) -M\left( t\right) \Delta 
\mathbf{T}_{1}\right) \right)  \notag \\
&&\times \left( a\left( Z^{\prime },Z\right) \Delta T_{1}-b\left(
Z,Z^{\prime }\right) \Delta T_{1}^{\prime }\right) \left( a\left(
Z,Z^{\prime }\right) \Delta T_{1}^{\prime }-b\left( Z^{\prime },Z\right)
\Delta T_{1}\right)  \notag \\
&&\times \exp \left( -\frac{1}{2}\left( \Delta \mathbf{T}_{1}^{\prime
}\right) ^{t}\sigma ^{-1}\left( t\right) \left( \Delta \mathbf{T}%
_{1}^{\prime }\right) \right)  \notag \\
&&\times \exp \left( -\frac{1}{2}\left( \left( \Delta \mathbf{T}_{f}^{\prime
}\right) -M\left( t\right) \Delta \mathbf{T}_{1}^{\prime }\right) ^{t}\sigma
^{-1}\left( t\right) \left( \left( \Delta \mathbf{T}_{f}^{\prime }\right)
-M\left( t\right) \Delta \mathbf{T}_{1}^{\prime }\right) \right)  \notag
\end{eqnarray}%
with:%
\begin{eqnarray*}
&&G_{0}\left( \Delta T_{i}\left( Z,Z^{\prime }\right) ,\Delta T_{1}\left(
Z,Z^{\prime }\right) ,\Delta T_{i}^{\prime }\left( Z^{\prime },Z\right)
,\Delta T_{f}\left( Z^{\prime },Z\right) \right) \\
&=&\exp \left( -\frac{1}{2}\left( \Delta \mathbf{T}_{1}^{\prime }\right)
^{t}\sigma ^{-1}\left( t\right) \left( \Delta \mathbf{T}_{1}^{\prime
}\right) \right) \\
&&\times \exp \left( -\frac{1}{2}\left( \left( \Delta \mathbf{T}_{f}^{\prime
}\right) -M\left( t\right) \Delta \mathbf{T}_{1}^{\prime }\right) ^{t}\sigma
^{-1}\left( t\right) \left( \left( \Delta \mathbf{T}_{f}^{\prime }\right)
-M\left( t\right) \Delta \mathbf{T}_{1}^{\prime }\right) \right)
\end{eqnarray*}%
The maximum of the correction (\ref{grp})\ is obtained for:%
\begin{eqnarray*}
\left( \Delta \mathbf{T}_{f}\right) &\simeq &M\left( t\right) \Delta \mathbf{%
T}_{1}=\left( 
\begin{array}{cc}
e^{-tu} & s\frac{e^{-tu}-e^{-tv}}{u-v} \\ 
0 & e^{-tv}%
\end{array}%
\right) \Delta \mathbf{T}_{1} \\
\left( \Delta \mathbf{T}_{f}^{\prime }\right) &\simeq &\Delta \mathbf{T}%
_{1}\Delta \mathbf{T}_{1}^{\prime }=\left( 
\begin{array}{cc}
e^{-tu} & s\frac{e^{-tu}-e^{-tv}}{u-v} \\ 
0 & e^{-tv}%
\end{array}%
\right) \Delta \mathbf{T}_{1}^{\prime }
\end{eqnarray*}%
and the transition function becomes for these values:%
\begin{eqnarray}
&&G\left( \Delta T_{i}\left( Z,Z^{\prime }\right) ,\Delta T_{1}\left(
Z,Z^{\prime }\right) ,\Delta T_{i}^{\prime }\left( Z^{\prime },Z\right)
,\Delta T_{f}\left( Z^{\prime },Z\right) \right)  \label{GN} \\
&=&\left( 1+\left( a\left( Z^{\prime },Z\right) \Delta T_{1}-b\left(
Z,Z^{\prime }\right) \Delta T_{1}^{\prime }\right) \left( a\left(
Z,Z^{\prime }\right) \Delta T_{1}^{\prime }-b\left( Z^{\prime },Z\right)
\Delta T_{1}\right) \right) \exp \left( -\frac{1}{2}\left( \Delta \mathbf{T}%
_{1}^{\prime }\right) ^{t}\sigma ^{-1}\left( t\right) \left( \Delta \mathbf{T%
}_{1}^{\prime }\right) \right)  \notag
\end{eqnarray}%
The correction:%
\begin{equation*}
\left( a\left( Z^{\prime },Z\right) \Delta T_{1}-b\left( Z,Z^{\prime
}\right) \Delta T_{1}^{\prime }\right) \left( a\left( Z,Z^{\prime }\right)
\Delta T_{1}^{\prime }-b\left( Z^{\prime },Z\right) \Delta T_{1}\right)
\end{equation*}%
in (\ref{GN}) is positive and maximal for a value:%
\begin{equation*}
\overline{\left( \Delta T_{1}^{\prime }\right) }\in \left[ \inf \left( \frac{%
b\left( Z,Z^{\prime }\right) }{a\left( Z^{\prime },Z\right) },\frac{b\left(
Z^{\prime },Z\right) }{a\left( Z,Z^{\prime }\right) }\right) ,\sup \left( 
\frac{b\left( Z,Z^{\prime }\right) }{a\left( Z^{\prime },Z\right) },\frac{%
b\left( Z^{\prime },Z\right) }{a\left( Z,Z^{\prime }\right) }\right) \right]
\end{equation*}%
that is, given (\ref{bcf}) for $\overline{\left( \Delta T_{1}^{\prime
}\right) }$\ satisfying:

\begin{equation*}
\overline{\left( \Delta T_{1}^{\prime }\right) }\in \left[ \inf \left( \frac{%
\omega _{0}\left( Z^{\prime }\right) }{\omega _{0}\left( Z\right) },\frac{%
\omega _{0}\left( Z\right) }{\omega _{0}\left( Z^{\prime }\right) }\right)
,\sup \left( \frac{\omega _{0}\left( Z^{\prime }\right) }{\omega _{0}\left(
Z\right) },\frac{\omega _{0}\left( Z\right) }{\omega _{0}\left( Z^{\prime
}\right) }\right) \right]
\end{equation*}%
As a consequence, the most likely configuration for the system is given by
the following final values:%
\begin{eqnarray*}
\left( \Delta \mathbf{T}_{f}\right) &\simeq &\left( 
\begin{array}{cc}
e^{-tu} & s\frac{e^{-tu}-e^{-tv}}{u-v} \\ 
0 & e^{-tv}%
\end{array}%
\right) \Delta \mathbf{T}_{1} \\
\left( \Delta \mathbf{T}_{f}^{\prime }\right) &\simeq &\left( 
\begin{array}{cc}
e^{-tu} & s\frac{e^{-tu}-e^{-tv}}{u-v} \\ 
0 & e^{-tv}%
\end{array}%
\right) \overline{\left( \Delta T_{1}^{\prime }\right) }
\end{eqnarray*}%
as claimed in the text.

\end{document}